\documentclass[
superscriptaddress,
twocolumn,
nofootinbib,
 amsmath,amssymb,
 aps,
prd,
showkeys
]{revtex4-2}

\usepackage{graphicx}

\usepackage[labelfont = bf]{caption}
\usepackage{subcaption}
\usepackage[export]{adjustbox}  

\usepackage[normalem]{ulem}
\usepackage{dcolumn}   
\usepackage{bm}        
\usepackage{mathrsfs}  

\usepackage{outlines}
\usepackage[english]{babel}

\usepackage[dvipsnames]{xcolor}
\usepackage{hyperref}
\hypersetup{
    colorlinks = true,
    linkcolor  = Blue,
    urlcolor   = Blue,
    citecolor  = Blue
}

\newcommand{\aap}{Astronomy \& Astrophysics}
\newcommand{\apjl}{The Astrophysical Journal Letters}
\newcommand{\apjs}{The Astrophysical Journal Supplement Series}
\newcommand{\araa}{Annual Review of Astronomy and Astrophysics}
\newcommand{\mnras}{Monthly Notices of the Royal Astronomical Society}
\newcommand{\pasj}{Publications of the Astronomical Society of Japan}

\newcommand{\orcid}[1]{\href{https://orcid.org/#1}{\includegraphics[height=\fontcharht\font`\B]{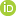}}}

\begin{document}


\title{Relativistic gas accretion onto supermassive black hole binaries\\from inspiral through merger}

\thanks{This manuscript has been authored in part by UT-Battelle, LLC, under contract DE-AC05-00OR22725 with the US Department of Energy (DOE). The US government retains and the publisher, by accepting the article for publication, acknowledges that the US government retains a nonexclusive, paid-up, irrevocable, worldwide license to publish or reproduce the published form of this manuscript, or allow others to do so, for US government purposes. DOE will provide public access to these results of federally sponsored research in accordance with the DOE Public Access Plan (http://energy.gov/downloads/doe-public-access-plan).}


\author{Lorenzo Ennoggi\:\orcid{0000-0002-2771-5765}}
\affiliation{Center for Computational Relativity and Gravitation \& School of Physics and Astronomy, Rochester Institute of Technology, Rochester, New York 14623, USA}

\author{Manuela Campanelli\:\orcid{0000-0002-8659-6591}}
\affiliation{Center for Computational Relativity and Gravitation \& School of Mathematics and Statistics, Rochester Institute of Technology, Rochester, New York 14623, USA}

\author{Yosef Zlochower\:\orcid{0000-0002-7541-6612}}
\affiliation{Center for Computational Relativity and Gravitation \& School of Mathematics and Statistics, Rochester Institute of Technology, Rochester, New York 14623, USA}

\author{Scott C. Noble\:\orcid{0000-0003-3547-8306}}
\affiliation{Gravitational Astrophysics Lab, NASA Goddard Space Flight Center, Greenbelt, Maryland 20771, USA}

\author{Julian Krolik\:\orcid{0000-0002-2995-7717}} \affiliation{Physics and Astronomy Department, Johns Hopkins University, Baltimore, Maryland 21218, USA}

\author{Federico Cattorini\:\orcid{0000-0002-3907-9583}}
\affiliation{INFN Sezione di Milano-Bicocca \& Dipartimento di Fisica Giuseppe Occhialini, Universit\`a di Milano-Bicocca, Piazza della Scienza 3, I-20126 Milano, Italy}

\author{Jay V. Kalinani\:\orcid{0000-0002-2945-1142}}
\affiliation{Center for Computational Relativity and Gravitation \& School of Mathematics and Statistics, Rochester Institute of Technology, Rochester, New York 14623, USA}

\author{Vassilios Mewes\:\orcid{0000-0001-5869-8542}}
\affiliation{National Center for Computational Sciences, Oak Ridge National Laboratory, P.O. Box 2008, Oak Ridge, Tennessee 37831-6164, USA}

\author{Michail Chabanov\:\orcid{0000-0001-9008-0267}}
\affiliation{Center for Computational Relativity and Gravitation \& School of Mathematics and Statistics, Rochester Institute of Technology, Rochester, New York 14623, USA}

\author{Liwei Ji\:\orcid{0000-0001-9008-0267}}
\affiliation{Center for Computational Relativity and Gravitation \& School of Mathematics and Statistics, Rochester Institute of Technology, Rochester, New York 14623, USA}

\author{Maria Chiara de Simone\:\orcid{0009-0008-8088-1392}}
\affiliation{Center for Computational Relativity and Gravitation \& School of Physics and Astronomy, Rochester Institute of Technology, Rochester, New York 14623, USA}

\begin{abstract}

Accreting supermassive black hole binaries are powerful multimessenger sources emitting both gravitational and electromagnetic (EM) radiation. Understanding the accretion dynamics of these systems and predicting their distinctive EM signals is crucial to informing and guiding upcoming efforts aimed at detecting gravitational waves produced by these binaries. To this end, accurate numerical modeling is required to describe both the spacetime and the magnetized gas around the black holes. In this paper, we present two key advances in this field of research.

First, we have developed a novel 3D general relativistic magnetohydrodynamics (GRMHD) framework that combines multiple numerical codes to simulate the inspiral and merger of supermassive black hole binaries starting from realistic initial data and running all the way through merger. Throughout the evolution, we adopt a simple but functional prescription to account for gas cooling through photon emission.

Next, we have applied our new computational method to follow the time evolution of a circular, equal-mass, nonspinning black hole binary for ${\sim\!200}$ orbits, starting from a separation of ${20\,r_g}$ and reaching the postmerger evolutionary stage of the system. We have shown how mass continues to flow toward the binary even after the binary ``decouples" from its surrounding disk, but the accretion rate onto the black holes diminishes. We have identified how the minidisks orbiting each black hole are slowly drained and eventually dissolve as the binary compresses. We confirm previous findings that the system's luminosity decreases by a factor of a few during inspiral; however, we observe an abrupt increase by ${\sim\!50\%}$ in this quantity at the time of merger, likely accompanied by an equally abrupt change in spectrum. Finally, we have demonstrated that during the inspiral, fluid ram pressure regulates the fraction of the magnetic flux transported to the binary that attaches to the black holes' horizons.

\end{abstract}


\maketitle

\section{Introduction}
\label{sec: Introduction}

The ${\lambda\text{CDM}}$ cosmological model predicts that present-day galaxies have formed through hierarchical mergers of smaller galaxies~\cite{Schramm1992, Peebles1993}. Since galaxies often host supermassive black holes (SMBH) at their centers~\cite{Schneider2006, KormendyHo2013}, these mergers typically result in the central objects gradually moving closer together. Various mechanisms, such as dynamical friction, asymmetries in the stellar distribution, and gas accretion, can drive these black holes to form a bound binary system. Once bound, interactions with both stars and gas can potentially reduce the orbital separation to the point where the emission of gravitational radiation becomes the dominant channel of energy and angular momentum loss~\cite{Bogdanovic2022}. The end-result is black hole coalescence, producing a burst of detectable gravitational radiation, and, if in a gas-rich environment, potentially detectable electromagnetic (EM) radiation as well. Therefore, SMBH mergers play a crucial role in astrophysics and cosmology, in terms of the development of the SMBH population, their effects on the growth of their host galaxies, and their observability as transients~\cite{Volonteri2003}.

At the time of merger, the gravitational wave (GW) luminosity can potentially outshine the entire observable photon universe~\cite{Campanelli2010}. Consequently, supermassive binary black hole (SMBBH) systems will be key targets for future GW detection campaigns involving pulsar timing arrays (PTA)~\cite{Antoniadis2023} and the space-based interferometer LISA~\cite{Colpi2024, AmaroSeoane2023}. Any photons radiated greatly augment the scientific value of GW detections. If the spectrum or lightcurve of the EM radiation from a binary can be distinguished from that of a single accreting SMBH, observations in the EM sector could more accurately localize GW emission events in the sky~\cite{DalCanton2019}. EM observations could also identify SMBBHs \textit{before} their GW radiation is detectable, potentially enabling studies defining their population or identifying sources that should merge in the near future.

\smallskip
The current understanding of the dynamics of matter and EM fields surrounding SMBH binaries has largely been shaped by many numerical studies conducted over the past two decades. In particular, a large number of hydrodynamical (HD) simulations of SMBBH systems have been conducted that reduced the 3D problem to 2D by vertical integration; these studies also assumed Newtonian gravity with internal stresses described by the phenomenological ``${\alpha}$'' model~\cite{Duffell2024}. However, the physical outcome of 2D HD simulations with unphysical internal stresses can be substantially different from those employing 3D magnetohydrodynamics (MHD), the physical mechanism actually producing the stress (see, e.g.,~\cite{Gutierrez2024, NobleKrolik2025}).

Nonetheless, there are some findings so robust that all numerical studies produce them---whether employing HD or MHD, conducted in 2D or 3D, or using general relativistic or Newtonian dynamics. If the orbiting gas rotates in the same sense as the binary and the binary's mass ratio is not too far from 1, the gravitational torque exerted by the binary transfers angular momentum to the surrounding accretion disk, clearing a gap whose radius approximately equals twice the orbital separation ~\cite{Pringle1991, ArtymowiczLubow1994, MacFadyenMilosavljevic2008, Shi2012, Noble2012}. Under the same conditions, in the circumbinary disk (CBD) just outside the gap, a \textit{``lump''} can develop that modulates mass transfer to the binary at a frequency comparable to the binary's orbital frequency~\cite{Shi2012, Noble2012}. Mass accretion onto the binary flows through a pair of streams at a time-averaged rate matching the inflow rate through the CBD~\cite{Farris2014, ShiKrolik2015, Bowen2017, Bowen2019}. The streams feed individual \textit{``minidisks''} around each black hole~\cite{ArtymowiczLubow1996}, which exchange mass between each other in a process known as \textit{``sloshing''}~\cite{Bowen2017, Bowen2019, Avara2024, Westernacher-Schneider2024}. Finally, the bulk of the CBD, the minidisks, and possibly the accretion streams produce a roughly thermal EM spectrum~\cite{Roedig2014, Farris2015, dAscoli2018, Bowen2019, Gutierrez2022,Nagele2025}, and inverse Compton scattering of photons generates hard X-rays in the minidisks' coron\ae, similar to what is described in~\cite{Kinch2020}.

\smallskip
To treat the late inspiral and merger in the presence of gas, one must certainly use general relativistic dynamics; in addition, because the ``${\alpha}$"-model for internal stresses rests on the assumption of time-steady conditions distant from disk edges, genuine MHD is also essential. However, such calculations face a practical difficulty: genuine numerical relativity, essential to describing binaries with semimajor axes ${a\lesssim 10\text{--}15\,r_g}$, is exceedingly computationally expensive at larger separations because the timescale for orbital evolution is ${\propto a^4}$; on the other hand, to trace both the way the inner edge of the circumbinary disk creeps inward as the binary shrinks and how the minidisks evolve from genuine miniature accretion disks to stray gas parcels subjected to a dynamical spacetime requires starting from larger separations, where previous work has given a general idea of the gas's initial state.

As a result of this difficulty, previous work employing 3D general relativistic MHD has chosen one of two approaches. One of these~\cite{Farris2012, Gold2014a, Gold2014b, Paschalidis2021, Bright2023, Cattorini2021, Cattorini2022, Fedrigo2023, Cattorini2024, Manikantan2024} begins with the gas and/or the binary already close. For example, in \cite{Farris2012, Gold2014a, Gold2014b} the initial semimajor axis ${a_0=10\,M}$ and the inner edge of surrounding gas is at ${r_\text{in} = 18\,M}$, while in \cite{Cattorini2021, Cattorini2022, Fedrigo2023, Cattorini2024} ${a_0 = 12\,M}$ and the binary is immersed in gas whose initial density in the equatorial plane (and sometimes the entire volume) is uniform. Although ${a_0 = 20\,M}$ in~\cite{Paschalidis2021, Bright2023}, ${r_\text{in} = 18\,M}$.

Simulations with full numerical relativity are then feasible, but there is little in the way of independent guidance about the initial state of the gas. Some of these efforts~\cite{Farris2012, Gold2014a, Gold2014b} have tried to mitigate this problem by relaxing the dynamical state of gas close to the binary while fixing the orbital separation, but because the orbital evolution when the separation is this small can be faster than the gas's radial distribution evolution, the result remains sensitive to the initial guess. The time to merger is ${\sim\!1000\,M}$ (see Fig.~\ref{fig: binary separation}), which is only ${\sim\!2}$ orbits at the inner edge of the CBD, and therefore a very small fraction of the timescale over which internal stresses can rearrange the CBD's radial mass profile, generically orders of magnitude larger than the orbital period. Nonetheless, work using this general approach has explored what happens to gas going through the late inspiral and merger for a variety of cooling prescriptions (including no cooling, as in~\cite{Farris2012, Gold2014a, Cattorini2021, Cattorini2022, Cattorini2024}), binary mass ratios~\cite{Gold2014a, Gold2014b} and spin configurations~\cite{Paschalidis2021, Bright2023, Cattorini2021, Cattorini2022, Cattorini2024, Fedrigo2023}. This approach has also been employed to study the earlier stages of inspiral without cooling~\cite{Paschalidis2021, Bright2023}, but assuming an initial gas density distribution identical to that used for simulations beginning at half the separation. In addition, a first exploration of jet-launching was conducted in~\cite{Gold2014b}.

The other approach~\cite{Noble2012, Bowen2017, Bowen2019, Noble2021, LopezArmengol2021, Combi2022, Avara2024} begins at a larger binary separation (frequently ${20\,M}$), places the initial ${r_\text{in}}$ at ${3a_0}$ (i.e., ${r_\text{in} = 60\,M}$ if ${a_0 = 20\,M}$), and uses one of several different sorts of approximate, but high-order, binary spacetimes~\cite{Mundim2014, Ireland2016, Nakano2016, Combi2021, CombiRessler2024}. Using this method it is possible to equilibrate the circumbinary disk while holding the binary orbit fixed and then follow the evolution of both the CBD and the minidisks using well-understood dynamics. All the work to date using this approach has employed a cooling rate fast enough to radiate nearly all the heat produced by dissipation in the MHD flow. Unfortunately, without solving the full Einstein equations, it cannot treat the final stage of inspiral or the merger.

\smallskip
This work aims to combine the best aspects of these two approaches to support studies of black hole long-term inspiral and merger with surrounding gas that seamlessly cover the entire span of regimes, from inspiral to merger to postmerger relaxation. Here we provide a first example: a circular, equal-mass, nonspinning SMBBH surrounded by a CBD. In order to overcome the limitations described above, we use a novel technique involving three main steps. First, we run a CBD simulation with the GRMHD code \textsc{SphericalNR}~\cite{Mewes2018, Mewes2020, Ji2023} adopting curvilinear coordinates and an approximate spacetime metric until the gas becomes turbulent and steadily accretes onto the binary. The second step, and the core of our method, is the \textit{``hand-off''}: once the CBD has reached the desired regime, the fluid and EM fields are interpolated onto a Cartesian grid with box-in-box mesh refinement and an appropriate numerical binary black hole spacetime is set up.\footnote{An analogous method was used in~\cite{Farris2012, Gold2014a, Gold2014b}.}
As a third and final step, we evolve the system through the inspiral and merger stages using the full-NR Cartesian GRMHD code \textsc{IllinoisGRMHD}~\cite{Etienne2015, Werneck2023}.

\smallskip
This paper is organized as follows. In Sec.~\ref{sec: Methods}, we briefly outline the evolution equations solved by \textsc{SphericalNR} and \textsc{IllinoisGRMHD} (\ref{subsec: Evolution equations}), describe the initial data, evolution methods, and radiation cooling prescriptions for both the CBD (\ref{subsec: Circumbinary disk evolution}) and NR (\ref{subsec: Inspiral and merger evolution}) portions of our simulation, and detail all the steps involved in performing the hand-off (\ref{subsec: Hand-off to the inspiral stage}). In Sec.~\ref{sec: Results}, we present the results from the CBD run (\ref{subsec: Circumbinary disk equilibration}), provide solid evidence of the quality of our hand-off method (\ref{subsec: Hand-off consistency}), and give a thorough account of the matter and EM field dynamics during the inspiral, merger, and postmerger phases of the binary evolution (\ref{subsec: Inspiral and merger accretion dynamics} and~\ref{subsec: Electromagnetic phenomena}). In Sec.~\ref{sec: Discussion}, we discuss and comment the results presented in Sec.~\ref{sec: Results}, with particular attention to the motion of the gas during the late inspiral to merger (\ref{subsec: Dynamics unique to late inspiral and merger}), the predicted lightcurve (\ref{subsec: Bolometric lightcurve}) and EM spectrum (\ref{subsec: Spectrum}), and the character of the relativistic jet launched from the merger remnant and the individual spiraling black holes if they had significant spin (\ref{subsec: Jet-launching}). Finally, in Sec.~\ref{sec: Conclusions and outlook}, we draw our conclusions and discuss some ways our work can be improved and extended.

We adopt Einstein's summation convention throughout the text and, unless otherwise stated, we label the components of 4(3)-dimensional objects (e.g., tensors) with Greek (Latin) indices. We present all results in geometric units, whereby ${G = c = 1}$\,.

\section{Methods}
\label{sec: Methods}

\subsection{Evolution equations}
\label{subsec: Evolution equations}

This work focuses on highly conductive magnetized fluids on a curved spacetime under the assumption that fluids do not self-gravitate, thereby leaving the spacetime dynamics unaltered. This assumption is a very good one for gas distributions surrounding SMBH binaries because in any reasonable astrophysical scenario the mass of the gas is tiny compared to the mass of the black holes. The evolution of such a system is described by Einstein's equations in vacuum,
\begin{equation}
    \label{eq: Einstein's equations in vacuum}
    R_{\mu\nu} = 0\;,
\end{equation}
coupled to the local conservation of baryon number density, the evolution of matter energy-momentum, and the EM induction equation,
\begin{align}
    \label{eq: conservation of the baryon number density}
    \nabla_\mu\left(\rho u^\mu\right) &= 0\;,\\
    \label{eq: evolution of the energy-momentum}
    \nabla_\mu T^{\mu\nu} &= -\mathcal{L}_\text{cool}u^\mu\;,\\
    \label{eq: homogeneous Maxwell equations}
    \nabla_\nu {^{*}\!F^{\mu\nu}} &= 0\;.
\end{align}
Maxwell's equations with sources are excluded, as the ideal MHD limit ${F^{\mu\nu}u_\nu = 0}$ (i.e., an observer comoving with the fluid measures no electric field) is assumed. In Eqs.~\eqref{eq: Einstein's equations in vacuum}--\eqref{eq: homogeneous Maxwell equations}, ${R_{\mu\nu}}$ is the Ricci tensor, ${T_{\mu\nu}}$ is the stress-energy tensor, ${\rho}$ and ${u^\mu}$ are the fluid's mass density and 4-velocity, and ${^{*}\!F_{\mu\nu}}$ is the Hodge dual of the Faraday 2-form
\begin{equation}
    F_{\mu\nu}\equiv\nabla_\mu\mathcal{A}_\nu - \nabla_\nu\mathcal{A}_\mu = \partial_\mu\mathcal{A}_\nu - \partial_\nu\mathcal{A}_\mu\;,
\end{equation}
where ${\mathcal{A}_\mu}$ is the EM 4-potential. We account for energy and momentum lost by the fluid in the form of photons via the sink term ${-\mathcal{L}_\text{cool}u^\mu}$ in the RHS of~\eqref{eq: evolution of the energy-momentum}, where ${\mathcal{L}_\text{cool}}$ is the fluid-frame cooling rate; more on this in Secs.~\ref{subsubsec: Radiative cooling (CBD)} and~\ref{subsubsec: Radiative cooling (IGM)}. In the ideal MHD limit, the MHD stress-energy tensor can be written as
\begin{equation}
    T_{\mu\nu} = \left(\rho h + 2p_\text{mag}\right)u_\mu u_\nu + \left(p + p_\text{mag}\right)g_{\mu\nu} - b_\mu b_\nu\;,
\end{equation}
where ${g_{\mu\nu}}$ is the spacetime metric, ${h\equiv 1 + \epsilon + p/\rho}$ is the specific enthalpy of the fluid, ${\epsilon}$ its specific internal energy, and ${p}$ its pressure; ${p_\text{mag}\equiv b^2/2}$ is the magnetic pressure, where ${b^2\equiv b^\mu b_\mu}$ and ${b^\mu\equiv -^{*}\!F^{\mu\nu}u_\nu}$ is the magnetic field measured by a \textit{Lagrangian} observer, i.e., an observer comoving with the fluid.

\smallskip
Numerical solutions of the coupled Einstein and GRMHD equations are typically found by a 3\texttt{+}1 splitting of spacetime (see, e.g.,~\cite{Darmois1927, FouresBruhat1952, Arnowitt2008, Gourgoulhon2012}). The spacetime is foliated with a set of nonintersecting spacelike hypersurfaces ${\{\Sigma_t\}_{t\in\mathbb{R}}}$\,, coordinates ${x^i\equiv\left(x^1, x^2, x^3\right)}$ are chosen on each slice ${\Sigma_t}$ and, assuming the spatial coordinates ${x^i}$ vary smoothly between neighboring hypersurfaces, ${x^\mu\equiv\left(t, x^i\right)}$ is a well-behaved coordinate system on the spacetime. The coordinate time derivative operator ${\partial_t}$ can then be decomposed as
\begin{equation}
    \left(\partial_t\right)^\mu\equiv\alpha n^\mu + \beta^\mu\;,
\end{equation}
where ${n^\mu}$ is the timelike unit 4-vector orthogonal to the foliation, ${\alpha}$ is the lapse function, and ${\beta^i}$ (note that it is spacelike) is the shift vector. Because it is timelike and has unit norm, the 4-vector ${n^\mu}$ can be regarded as the 4-velocity of the observer aligned with the foliation, also called the \textit{Eulerian} observer. In terms of lapse and shift, ${n^\mu}$ is expressed as
\begin{align}
    n^\mu &= (\frac{1}{\alpha}, -\frac{\beta^i}{\alpha})\;,\\
    n_\mu &= (-\alpha, 0^i)\;.
\end{align}
Projecting ${g_{\mu\nu}}$ onto the foliation yields the spatial metric (or 3-metric),
\begin{equation}
    \gamma_{\mu\nu}\equiv g_{\mu\nu} + n_\mu n_\nu\;;
\end{equation}
with this, the spacetime metric can be expressed as 
\begin{align}
    g_{\mu\nu} &=
        \begin{pmatrix}
            -\alpha^2 + \beta_k\beta^k & \beta_j\\
            \beta_i & \gamma_{ij}
        \end{pmatrix}\;,\\
    g^{\mu\nu} &=
        \begin{pmatrix}
            -\frac{1}{\alpha^2} & \frac{\beta^j}{\alpha^2}\\
            \frac{\beta^i}{\alpha^2} & \gamma^{ij} - \frac{\beta^i \beta^j}{\alpha^2}
        \end{pmatrix}\;,
\end{align}
and the relation between the metric determinants ${g\equiv\det\left(g_{\mu\nu}\right)}$ and ${\gamma\equiv\det\left(\gamma_{ij}\right)}$ is ${\sqrt{-g} = \alpha\sqrt{\gamma}}$\,.

In the BSSN formalism~\cite{BaumgarteShapiro1998, ShibataNakamura1995}, the spatial metric is conformally rescaled according to
\begin{align}
    \bar{\gamma}_{ij}&\equiv e^{-4\phi}\gamma_{ij}\;,\\
    e^{4\phi}&\equiv\left(\gamma/\bar{\gamma}\right)^{1/3}\;,
\end{align}
where ${e^\phi}$ is the conformal factor and ${\bar{\gamma}}$ is the determinant of the conformally related spatial metric. In the full-NR portion of the simulation presented in this work, we track the evolution of the binary black hole spacetime using the BSSN approach coupled to the ``moving-puncture'' gauge (i.e., slicing) choice~\cite{Campanelli2006, Baker2006}.

\smallskip
Within the 3\texttt{+}1 formulation of general relativity, the matter-field equations~\eqref{eq: conservation of the baryon number density} and~\eqref{eq: evolution of the energy-momentum} can be written in conservative form according to the ``Valencia'' formulation~\cite{Banyuls1997}, i.e., as a system of the type
\begin{equation}
    \label{eq: GRMHD equations}
    \partial_t\mathbf{C} + \partial_i\mathbf{F}^i = \mathbf{S}\;,
\end{equation}
where ${\mathbf{C}}$, ${\mathbf{F}^i}$, and ${\mathbf{S}}$ are the vectors of \textit{conserved variables} and the associated \textit{flux} and \textit{source} terms, respectively. In \textsc{IllinoisGRMHD}, the vectors ${\mathbf{C}}$, ${\mathbf{F}^i}$, and ${\mathbf{S}}$ are defined as~\cite{Etienne2015}
\begin{align}
    \label{eq: IGM conservatives}
    \mathbf{C}&\equiv\sqrt{\gamma}
    \begin{bmatrix}
        \rho W\\
        \alpha^2 T^{00} - \rho W\\
        \alpha {T^0}_j
    \end{bmatrix}\;,\\
    \mathbf{F}^i&\equiv\sqrt{\gamma}
    \begin{bmatrix}
        \rho W\tilde{v}^i\\
        \alpha^2 T^{0i} - \rho W\tilde{v}^i\\
        \alpha {T^i}_j
    \end{bmatrix}\;,\\
    \label{eq: IGM sources}
    \mathbf{S}&\equiv\alpha\sqrt{\gamma}
    \begin{bmatrix}
        0\\
        s - \mathcal{L}_\text{cool}W\\
        \frac{1}{2}\,T^{\mu\nu}\partial_i g_{\mu\nu} - \mathcal{L}_\text{cool} W v_j
    \end{bmatrix}\;,
\end{align}
where
\begin{equation}
    \begin{split}
        s &\equiv\left(T^{00}\beta^k\beta^l+ 2T^{0k}\beta^l + T^{kl}\right)K_{kl} +\\
          &- \left(T^{00}\beta^k + T^{0K}\right)\partial_i\alpha\;.
    \end{split}
\end{equation}
Here,
\begin{equation}
    \tilde{v}^i\equiv dx^i/dt = u^i/u^0 = \alpha v^i - \beta^i
\end{equation}
is the fluid's coordinate velocity, where ${v^i}$ is the fluid's velocity in the Eulerian frame;
\begin{equation}
    W\equiv-n_\mu u^\mu = \alpha u^0 = \left(1 - v_i v^i\right)^{-1/2}
\end{equation}
is the Lorentz factor relating the Eulerian and fluid's frames; and ${K_{ij}\equiv -\left(1/2\right)\mathscr{L}_n\gamma_{ij}}$ is the extrinsic curvature, where ${\mathscr{L}_n\gamma_{ij}}$ is the Lie derivative of the 3-metric along ${n^\mu}$. Because \textsc{IllinoisGRMHD} originally evolves ${\nabla_\mu T^{\mu\nu} = 0}$ instead of~\eqref{eq: evolution of the energy-momentum}, the source contributions containing ${\mathcal{L}_\text{cool}}$ are not originally accounted for in the code; further details in Sec.~\ref{subsubsec: Radiative cooling (IGM)}.

\smallskip
In the ideal MHD limit, the electric field ${E^\mu\equiv F^{\mu\nu}n_\nu}$ in the Eulerian frame is related to the magnetic field ${B^\mu\equiv -^{*}\!F^{\mu\nu}u_\nu}$ in the same frame by
\begin{equation}
    \label{eq: electric field in Eulerian frame in ideal MHD}
    E^i = -\epsilon^{ijk}v_j B_k
\end{equation}
(note that both ${E^i}$ and ${B^i}$ are spacelike), where ${\epsilon^{ijk}\equiv -\left[ijk\right]/\sqrt{\gamma}}$ or ${\epsilon_{ijk}\equiv\sqrt{\gamma}\left[ijk\right]}$ is the Levi-Civita pseudotensor on each spacetime slice and ${\left[ijk\right]}$ equals ${1}$ (${-1}$) if ${\{i,j,k\}}$ is an even (odd) permutation of ${\{1,2,3\}}$ and is ${0}$ otherwise. Thus, we only need to solve for ${B^i}$ via~\eqref{eq: homogeneous Maxwell equations}, which can be recast as the system
\begin{align}
    \label{eq: B evolution in ideal MHD}
    \partial_t\left(\sqrt{\gamma}\,B^i\right) + \partial_j\left(\sqrt{\gamma}\left(\tilde{v}^i B^j - \tilde{v}^j B^i\right)\right) &= 0\;,\\
    \label{eq: solenoidal constraint}
    \partial_i\left(\sqrt{\gamma}\,B^i\right) &= 0\;.
\end{align}
However, in order to ensure the solenoidal constraint~\eqref{eq: solenoidal constraint} is satisfied, both \textsc{SphericalNR} and \textsc{IllinoisGRMHD} decompose the EM 4-potential as
\begin{equation}
    \mathcal{A}_\mu\equiv\varphi\,n_\mu + A_\mu\;,
\end{equation}
with ${\varphi\equiv -n^\nu\mathcal{A}_\nu}$ and ${A_\mu\equiv\gamma_{\mu\nu}\mathcal{A}^\nu}$, and express the magnetic field in the Eulerian frame as
\begin{equation}
    B^i = \epsilon^{ijk}\partial_j A_k\;;
\end{equation}
then, Eq.~\eqref{eq: B evolution in ideal MHD} can be recast using~\eqref{eq: electric field in Eulerian frame in ideal MHD} as an evolution equation for the vector potential ${A_i}$\,,
\begin{equation}
    \label{eq: evolution of the vector potential}
    \partial_t A_i + \epsilon_{ijk}\,\tilde{v}^j B^k = \partial_i\left(\beta^j A_j - \alpha\varphi\right)\;.
\end{equation}
Using the invariance of the Faraday 2-form under a transformation ${\mathcal{A}_\mu\mapsto\mathcal{A}_\mu + \partial_\mu\psi}$ for any (reasonably behaved) scalar function ${\psi}$ on the spacetime, we may fix an EM gauge by choosing ${\mathcal{A}_\mu}$\,. In this work, we pick an EM potential belonging to the \textit{generalized Lorenz gauge}, where
\begin{equation}
    \label{eq: Lorenz condition}
    \nabla_\mu\mathcal{A}^\mu = \zeta n_\mu\mathcal{A}^\mu
\end{equation}
and ${\zeta}$ is a damping term (possibly equal to zero) in the advection of ${\mathcal{A}^\mu}$~\cite{Farris2012}. The Lorenz condition~\eqref{eq: Lorenz condition} can be shown to give rise to an evolution equation for the scalar potential ${\varphi}$\,,
\begin{equation}
    \partial_t\left(\sqrt{\gamma}\,\varphi\right) + \partial_i\left(\sqrt{\gamma}\left(\alpha A^i - \beta^i\varphi\right)\right) = -\zeta\alpha\sqrt{\gamma}\,\varphi\;.
\end{equation}

\smallskip
In \textsc{SphericalNR}, the GRMHD equations are also solved in the Valencia formulation, but using a reference metric formalism~\cite{Gourgoulhon2012, Bonazzola2004, Shibata2004, BrownCovariant2009, Montero2012} and evolving all tensorial fields in an orthonormal basis with respect to the spherical background reference metric~\cite{Baumgarte2013, Montero2014, SanchisGual2014, Baumgarte2015, Mewes2018, Mewes2020, Ji2023}. See~\cite{Mewes2020, Ji2023} for the full details of \textsc{SphericalNR}'s GRMHD implementation.

\subsection{Circumbinary disk evolution}
\label{subsec: Circumbinary disk evolution}

The EM appearance of the system depends significantly on the amount and distribution of mass, so it is important to start the merger calculation with reasonable initial conditions. Since the binary has likely evolved over a period longer than we can afford to simulate, we assume that the merger takes place in an accretion flow that has reached a relatively steady state as described in~\cite{Noble2012}. To achieve this quasi-steady state in the CBD region in a computationally affordable manner, we first evolve the flow with the binary excised from the domain for ${165}$ binary orbits using \textsc{SphericalNR}. We have found that this many orbits is sufficient to achieve a quasi-steady state from our initial conditions~\cite{Farris2012, Noble2012, LopezArmengol2021, Noble2021}.

\subsubsection{Initial data}

The initial setup for the CBD simulation features a weakly magnetized Keplerian gas torus in quasi-hydrostatic equilibrium, following the model described in Section 3.2 and Appendix A of~\cite{Noble2012}. However, unlike in~\cite{Noble2012}, we do not average the metric components of the spacetime at ${t=0}$ over the azimuthal direction; instead, in order to set up the initial gas configuration, we approximate the binary spacetime by a single black hole spacetime in Boyer-Lindquist coordinates with dimensionless spin ${\chi=1.25}$~\cite{Zilhao2015}. The single black hole spacetime is used only for this initial set up; the subsequent evolution of the system is tracked using a binary black hole spacetime (see Sec.~\ref{subsubsec: Evolution methods (CBD)}).

The torus surrounds a circular, equal-mass, nonspinning SMBBH system of total mass ${M}$ and is supported against the gravitational pull exerted by the latter (actually, by a point source having the same mass as the binary) via fluid pressure and bulk rotational motion. The binary separation ${a}$ is artificially kept fixed in time at the value ${20\,r_g\equiv 20\,GM/c^2\equiv 20\,M}$ in coordinate distance; this is a well-justified approximation, since at this separation the inspiral would be much slower than the orbital thermal timescales, and comparable to the binary orbital evolution timescale~\cite{Noble2012, Noble2021, Avara2024}. In fact, the ratio of the GW-driven orbital evolution timescale to the nominal inflow timescale at the CBD inner edge is
\begin{equation}
    \begin{split}
        &{\sim\!1\left(\alpha/0.1\right)\left[\left(H/r\right)\!/\,0.1\right]^2\left[\left(d\log\!\left(\Sigma\right)\!/d\log\!\left(r\right)\right)\!/\,6\right]\times}\\
        &{\times\left(r_\text{in}/2a\right)^{-\frac{3}{2}}\left(a/20r_g\right)^\frac{5}{2}\;.}
    \end{split}
\end{equation}
For simplicity, the gas is assumed to not self-gravitate, i.e., the stress-energy tensor is set to zero when solving Einstein's equations, so that both the back-reaction of the torus on the binary and the accumulation of mass and angular momentum on the black holes resulting from gas accretion are neglected.

The torus initially extends from a minimum radius ${r_\text{in} = 3a}$ to a maximum radius ${r_\text{max}\simeq 12a}$. This choice of inner radius for the torus, coupled with the initial binary separation of ${20\,M}$, ensures that the gap region around the binary forms dynamically as the system evolves in time. The maximum gas pressure is located at a radius ${r_p = 5a}$, where the aspect ratio of the torus is ${0.1}$\,.

At each point in the computational domain, we close the system of GRMHD equations~\eqref{eq: GRMHD equations} by relating the gas pressure ${P}$ to the mass density ${\rho}$ and specific internal energy ${\epsilon}$ via the ideal-fluid equation of state (EOS)
\begin{equation}
    \label{eq: ideal-fluid EOS}
    P = \left(\Gamma - 1\right)\rho\epsilon
\end{equation}
with adiabatic index ${\Gamma = 5/3}$\,, consistent with a plasma that is not relativistically hot (i.e., its specific thermal energy is smaller than the electron's mass). The specific entropy of the gas,
\begin{equation}
    s = \frac{P}{\rho^{\Gamma}}\;,
\end{equation}
is constant everywhere in the torus and has the initial value ${s_0 = 0.01}$\,.

\smallskip
A purely poloidal magnetic field is added on top of the equilibrium configuration as a set of dipolar loops that follow density contours in the disk's interior. Specifically, the initial magnetic field distribution is obtained from an EM vector potential ${\mathbf{A}}$ with azimuthal component
\begin{equation}
    A_\phi = A_{\phi,0}\max\left[\left(\rho - \frac{1}{4}\rho_\text{max}\right), 0\right]\;,
\end{equation}
where ${\rho_\text{max}}$ is the initial maximum mass density (see ${\text{Eq.}\left(20\right)}$ in~\cite{Noble2012}); the constant ${A_{\phi,0}}$ is chosen in such a way that the ratio of the fluid's total internal energy to total magnetic energy equals ${100}$\,.

A tenuous, unmagnetized, radially dependent atmosphere is added outside the torus in order to keep the fluid's mass density and pressure positive, which is required by the numerical evolution scheme in use. The two latter quantities scale with the radial distance ${r}$ from the binary's center of mass as ${r^{-\frac{3}{2}}}$ and ${\left(\Gamma-1\right)r^{-\frac{3}{2}\Gamma}}\equiv\left(2/3\right)r^{-\frac{5}{2}}$, respectively; the maximum floor values (reached at the inner edge of the computational domain) for the mass density, energy density ${u\equiv\rho\epsilon}$, and pressure are ${\rho_\text{floor} = 2\cdot 10^{-10}M^{-2}}$, ${u_\text{floor} = 2\cdot 10^{-12}M^{-2}}$, and ${P_\text{floor} = \left(\Gamma - 1\right)u_\text{floor} = \left(2/3\right)u_\text{floor}}$\,.

\subsubsection{Evolution methods}
\label{subsubsec: Evolution methods (CBD)}

The initial gas configuration is evolved using the \textsc{SphericalNR} code~\cite{Mewes2018, Mewes2020, Ji2023} on a spherical-like mesh, which conforms to the approximate symmetries of the CBD scenario. This code is built as a module---or \textit{`thorn'}---for the \textsc{Cactus}/\textsc{Einstein Toolkit} framework~\cite{GoodaleCactus2003, Loffler2012, Zilhao2013}, which provides the basic parallelization and utility tools (e.g., parallel I/O and checkpointing), and uses \textsc{Carpet}~\cite{Schnetter2004, Schnetter2006, CarpetDocs, CarpetBitBucket} as the driver code delivering fixed (or ``box-in-box'') mesh refinement capabilities, although \textsc{SphericalNR} grids are uniform and unchanging.

\textsc{SphericalNR} solves the GRMHD equations~\eqref{eq: GRMHD equations} via a finite-volume scheme whereby the MHD \textit{primitive variables}---i.e., the fluid's mass density, pressure, and 3-velocity and the EM scalar and vector potentials---are reconstructed at cell faces using the corresponding values at cell centers via the custom-built ninth-order WENO-Z9 reconstruction scheme described in~\cite{Ji2023}, which uses building blocks from~\cite{Balsara2016, Castro2011, Tchekhovskoy2007}. The main reason for using a higher order reconstruction is reduced numerical dissipation in the evolution (see, e.g.~\cite{Rembiasz2016, Rembiasz2017, Mahlmann2021}). The reconstructed quantities are then used to compute the fluxes of the conserved fields across neighboring cells via the Harten-Lax-Van Leer (HLLE) method~\cite{HLLE1983}. After taking one timestep, the primitive variables are recovered from the evolved ones via the robust prescription extensively described in Sec.~II of~\cite{Ji2023} and are thus made available for the next evolution timestep. Maxwell's equations are tackled in the ideal MHD approximation using a vector potential-based approach coupled to the generalized Lorenz prescription~\eqref{eq: Lorenz condition}; we choose the damping factor in Eq.~\eqref{eq: Lorenz condition} as ${\zeta = 1.5/\Delta t}$\,, where ${\Delta t}$ is the evolution timestep (see below)~\cite{Farris2012}. Kreiss-Oliger dissipation~\cite{KreissOligerBook, DissipationDocs} is applied to the time evolution equations of the EM potentials in order to maintain numerical stability. All equations are solved via the method of lines using the \textsc{Method of Lines (MoL)} thorn~\cite{MoLBitBucket, MoLDocs} adopting a strong-stability-preserving, five-stage, fourth-order accurate Runge-Kutta time integrator~\cite{SpiteriRuuth2002}. See~\cite{Ji2023} for details about recent updates to \textsc{SphericalNR}.

\smallskip
In the setup presented here, \textsc{SphericalNR} adopts ``exponential fisheye'' coordinates, i.e., spherical coordinates with logarithmic spacing along the radial direction; explicitly, the code uses coordinates ${x^1\equiv\log\!\left(r\right)}$, ${\theta}$, ${\phi}$ which are related to the Cartesian coordinates ${x}$, ${y}$, ${z}$ by the transformations
\begin{align}
    r &= \exp\!\left(x^1\right) = \sqrt{x^2 + y^2 + z^2}\;,\\
    x &= r\sin\!\left(\theta\right)\cos\!\left(\phi\right)\;,\\
    y &= r\sin\!\left(\theta\right)\sin\!\left(\phi\right)\;,\\
    z &= r\cos\!\left(\theta\right)\;.
\end{align}
This way, the cell size along ${r}$ increases linearly with the radial distance of the cell from the origin, thus alleviating the computational burden of the simulation by progressively reducing the spatial resolution away from the binary. The numerical mesh is composed by 724, 140, and 280 points along ${r}$, ${\theta}$, and ${\phi}$, respectively. The computational domain extends up to a maximum radius of ${20000\,M}$, which is far larger than the initial radius of the torus to allow the latter to expand over time and completely embed the Cartesian domain used for the inspiral and merger part of the simulation (see Sec.~\ref{subsec: Hand-off to the inspiral stage}).

\smallskip
In order to handle the singularities intrinsic to spherical-like coordinates, \textsc{SphericalNR} evolves all tensorial quantities in a frame that is orthonormal with respect to a chosen reference 3-metric ${\hat{\gamma}_{ij}}$ and expresses the conformal 3-metric as ${\bar{\gamma}_{ij}\equiv\hat{\gamma}_{ij} + \epsilon_{ij}}$\,, where the correction terms ${\epsilon_{ij}}$ need not be ``small'' compared to ${\hat{\gamma}_{ij}}$ (see~\cite{Mewes2018, Mewes2020} and references therein). In the present work, the reference 3-metric is set to the flat 3-metric in ``exponential fisheye'' coordinates:
\begin{equation}
    \hat{\gamma}_{ij}\equiv\text{diag}\left(r^2, r^2, r^2\sin^2\!\left(\theta\right)\right)\;.
\end{equation}
As a result of using the reference-metric approach, both the BSSN evolution equations and the MHD equations {\eqref{eq: IGM conservatives}--\eqref{eq: IGM sources}},~\eqref{eq: B evolution in ideal MHD},~\eqref{eq: solenoidal constraint}, and~\eqref{eq: evolution of the vector potential} are different from their original form; however, all of these equations reduce to their original counterparts when employing Cartesian coordinates~\cite{Mewes2018, Mewes2020}.

For the present CBD evolution, we do not evolve the BSSN system of equations; instead, the binary is evolved via a post-Newtonian (PN), time-dependent approximation to a nonspinning, equal-mass binary black hole spacetime using the in-house developed \textsc{PNSpacetime} thorn. The approximation is based on the asymptotically matched perturbative spacetime described in~\cite{Mundim2014,Zilhao2015,Zlochower2016, Ireland2016,Nakano2016,Ireland2019}\footnote{Future work may use a more general approximate binary black hole spacetime provided by \textsc{PNSpacetime}, one created by superimposing two Kerr-Schild black holes that inspiral according to 3.5PN theory~\cite{Combi2021, LopezArmengol2021, CombiRessler2024}. This approach is accurate for ${a\gtrsim 10\,M}$ and can accommodate arbitrary mass ratios and eccentricities as well as spins up to ${0.99}$ in dimensionless units.}.

\smallskip
Since the \textsc{SphericalNR} simulation is aimed at achieving a quasi-steady state in the CBD for the subsequent NR evolution, we remove a spherical region of coordinate radius ${r = 15\,M}$ around the binary from the computational domain. This improves computational efficiency by avoiding prohibitively small timesteps due to the Courant-Friedrich-Lewy (CFL) condition (in 3D spherical coordinates, the timestep ${\Delta t\sim r\sin\left(\Delta\theta/2\right)\Delta\phi}$\,) and is anyway required by the \textsc{PNSpacetime} metric being inaccurate close to the binary.

While the central cutout in the domain helps to alleviate the CFL restriction to some extent, the ${\sin\left(\Delta\theta/2\right)\Delta\phi}$ part still results in prohibitively small timesteps with high angular resolutions. To overcome this CFL restriction, \textsc{SphericalNR} adopts a fully MPI/OpenMP parallel Fast Fourier Transform (FFT) filtering algorithm capable of damping CFL-unstable modes in the evolved fields along the angular directions while maintaining numerical accuracy~\cite{Ji2023}. However, since the central part of our numerical grid is excised from the computational domain, the minimum cell size along ${\theta}$ is large enough that we only need to apply the filter along ${\phi}$. In addition, the FFT algorithm automatically switches from an exponential to a Gaussian filter over steep gradients in the filtered fields. By applying this filtering technique to our CBD simulation, we can increase the timestep ${\Delta t}$ by a factor of a few tens compared to an analogous no-filter scenario while keeping the evolution stable: thus, we set ${\Delta t = 0.15\,M}$.

\subsubsection{Radiative cooling}
\label{subsubsec: Radiative cooling (CBD)}

During the CBD evolution, gas cooling through the emission of EM radiation is accounted for as done in~\cite{Noble2012} and other subsequent works (e.g.,~\cite{Farris2012, Bowen2017, Bowen2019, LopezArmengol2021, Combi2022, Avara2024}). The cooling timescale ${t_\text{cool}}$ of a fluid element at a distance ${r}$ from the center of mass of the binary, whose gravitational mass is ${M}$, is approximated by the Keplerian period of the circular orbit of radius ${r}$ around a point mass ${M}$,
\begin{equation}
    \label{eq: cooling timescale}
    t_\text{cool} = 2\pi\sqrt{\frac{r^3}{M}}\;.
\end{equation}
We define the \textit{cooling rate} as
\begin{equation}
    \label{eq: cooling rate}
    \mathcal{L}_\text{cool} = \frac{\rho\epsilon}{t_\text{cool}}\left(\frac{s - s_0}{s_0} + \left\lvert\frac{s - s_0}{s_0}\right\rvert\right)^q\;,
\end{equation}
where ${q}$ is set to ${1/2}$ and ${s_0 = 0.01}$ is both the initial specific entropy of the torus and the target specific entropy for cooling, meaning that ${\mathcal{L}_\text{cool} = 0}$ when ${s = s_0}$\,; in other words, our cooling prescription keeps the specific entropy fixed at the value ${s_0}$\,. The cooling rate~\eqref{eq: cooling rate} is then used to build the sink term ${-\mathcal{L}_\text{cool}u^\mu}$ on the RHS of~\eqref{eq: evolution of the energy-momentum} (see also~\eqref{eq: IGM sources}). However, we only cool fluid elements that are bound according to Bernoulli's criterion~\cite{Noble2012, KastaunGaleazzi2015, Foucart2021},
\begin{equation}
    \label{eq: Bernoulli criterion}
    \left(\rho h + b^2\right)u_0 < -\rho\;;
\end{equation}
this prescription avoids cooling material that has been injected in order to maintain the density and pressure floors~\cite{Noble2012}.

\subsection{Hand-off to the inspiral stage}
\label{subsec: Hand-off to the inspiral stage}

As anticipated at the end of Sec.~\ref{sec: Introduction}, a snapshot of the CBD evolution taken at the beginning of the quasi-steady accretion stage is used as the initial condition for an inspiral and merger simulation in full NR performed with the Cartesian GRMHD code \textsc{IllinoisGRMHD}. This section describes the software tools we developed in order to transfer---or\textit{``hand-off''}---all the relevant information about the CBD evolution to the inspiral stage and thus be able to successfully continue the simulation. The hand-off process involves two steps:
\begin{enumerate}
    \item A new spacetime must be set up that is suitable for the full numerical-relativistic solution of Einstein's equation through the inspiral and merger phases.
    \item All the MHD primitive fields, i.e., the fluid's density, pressure, and velocity and the EM scalar and vector potential, must be interpolated from the spherical-like grid into a Cartesian grid with a box-in-box mesh refinement structure adapted to the geometry of the problem.
\end{enumerate}

The following three sections detail each of the above two steps and the criteria used to set up the destination grid.

\subsubsection{Spacetime setup}

The PN approximation to the circular, equal-mass, nonspinning binary black hole spacetime adopted for the CBD evolution breaks down close to the black holes and is therefore unsuitable for the subsequent numerical relativity evolution; instead, equal-mass, low-eccentricity initial data are generated via the Bowen-York / puncture~\cite{Bowen1980, Brandt1997, Cook2000} approach using a private version of the publicly available \textsc{TwoPunctures} solver~\cite{Ansorg2004} that can handle punctures at arbitrary locations on the numerical grid.

\smallskip
The PN trajectories from the CBD simulation cannot be used to feed \textsc{TwoPunctures} appropriate initial conditions, as the PN spacetime evolution routine used to equilibrate the CBD only returns the locations and orbital frequency of the black holes, but not their momenta. Therefore, the black hole locations, momenta, and spins are provided to \textsc{TwoPunctures} via a separate code that (again) computes a PN approximation of a circular, equal-mass, nonspinning binary black hole spacetime given either the orbital separation or the frequency resulting from the previous PN evolution; this code is currently not public. We choose to keep the same orbital frequency (a gauge-independent quantity) in the two PN codes rather than fix the orbital separation at the value ${20\,M}$, and we keep the angle the binary forms with the ${x}$ axis of the computational domain unchanged. Because the two PN codes adopt slightly different spacetime gauges, the binary separation calculated from the second code is lower than ${20\,M}$ by ${\sim\!0.25\%}$\,; since this work is focused on studying the dynamics of matter and EM fields around the binary, rather than evolving the binary orbit with very high accuracy, we deem this difference acceptable for our purposes.

\subsubsection{Interpolation of the MHD primitives}
\label{subsubsec: Interpolation of the MHD primitives}

In order to interpolate the MHD primitives from the spherical-like mesh onto the Cartesian grid with box-in-box mesh refinement, the capabilities of the publicly available \textsc{ReadInterpolate}~\cite{ReadInterpolateDocs, ReadInterpolateBitBucket} \textsc{Cactus} thorn were extended as follows. \textsc{ReadInterpolate} was originally designed to read \textsc{Cactus} output files in HDF5 format~\cite{HDF5} representing data on a Cartesian grid with box-in-box mesh refinement and interpolate those data onto a new grid of the same type. After all, though, the spherical-like mesh used by \textsc{SphericalNR} is logically Cartesian (due to the absence of grid-points on the polar axis), meaning that the (logarithmic) radial coordinate ${x^1\equiv\log\!\left(r\right)}$ and the two angular coordinates ${\theta}$ and ${\phi}$ span a box-like domain in the computer's memory. Therefore, expressing the Cartesian coordinates of a given point in the destination grid in terms of the curvilinear coordinates ${x^1}$, ${\theta}$, ${\phi}$ of the source grid effectively results in interpolations between two logically Cartesian meshes, which \textsc{ReadInterpolate} can handle.

\smallskip
The excised cavity of coordinate radius ${r = 15\,M}$ where the two black holes lie during the CBD evolution is part of the computational domain during the inspiral and merger stages and must therefore be filled with MHD data that do not exist in the previous simulation. The prescription we adopt here to do so aims at filling the cavity with a tenuous atmosphere consistent with the \textsc{SphericalNR} evolution, while at the same time avoiding the formation of strong shock waves at the boundary of the excised region as a result of steep gradients in the mass density, pressure, velocity, and magnetic field. To that end, the field values ${f\!\left(x, y, z\right)}$ for the fluid's density, pressure, and velocity are copied from the boundary of the cavity along radial paths terminating at the grid's origin and are then multiplied by a third-order polynomial ${P\!\left(r\right)}$ that ensures a ``smooth'' (technically, ${C^2}$) transition between some constant, field-dependent floor value ${f_1}$ reached at a distance ${r_1 = 13\,M}$ from the domain's origin and the value ${f_2\left(x, y, z\right)}$ of the field at the radial boundary of the \textsc{SphericalNR} grid, ${r_2 = 15\,M}$:
\begin{equation}
\label{eq: smoothing polynomial}
    \begin{aligned}
        rr &\equiv\frac{2r - r_1 - r_2}{r_2 - r_1}\;,\\
        P\!\left(r\right) &\equiv
        \begin{cases}
            0 & r < r_1 \\
            \frac{1}{2} + \frac{3}{4}rr\left(1 - \frac{rr^2}{3}\right) & r_1\leq r\leq r_2
        \end{cases}\;,\\[1. mm]
        f\!\left(x, y, z\right) &\mapsto P\!\left(r\right)\left[f\!\left(x, y, z\right) - f_1\right] + f_1 =\\
        &= P\!\left(r\right)\left[f_2\left(x, y, z\right) - f_1\right] + f_1\;.
    \end{aligned}
\end{equation}
The equality ${f\!\left(x, y, z\right) = f_2\left(x, y, z\right)}$ in the last equation stems from the radial copy described above. Consistent with the CBD evolution, the floor values for the mass density, energy density ${u\equiv\rho\epsilon}$, and pressure are ${\rho_\text{floor} = 2\cdot 10^{-10}M^{-2}}$, ${u_\text{floor} = 2\cdot 10^{-12}M^{-2}}$, and ${P_\text{floor} = \left(\Gamma - 1\right)u_\text{floor} = \left(2/3\right)u_\text{floor}}$ (see~\eqref{eq: ideal-fluid EOS}), and the velocity floor is zero. The density floor is about ${10^{-8}\times}$ smaller than the maximum density in the CBD at the time of hand-off.

\smallskip
Handling the components of the EM vector potential ${A_i}$ (${i = x, y, z}$) inside the cavity requires special care. In principle, ${A_i}$ (and thus the magnetic field ${B^i = \varepsilon^{ijk}\partial_j A_k}$) should be constant in the excised region, as we would like the magnetic field to ``naturally'' flow towards the black holes from outside the cavity. On the other hand, an abrupt jump in the values of ${A_i}$ at the boundary of the cavity would produce a thin layer of extremely high magnetic field there, which in turn would generate a very strong shock wave propagating both towards the black holes and the CBD, unphysically perturbing the former two and destroying the latter. Additionally, the ``copy-and-smooth'' strategy adopted for the hydrodynamic fields would eliminate the radial gradient from each vector potential component inside the cavity, thus introducing Cartesian gradients and, again, generating unphysical contributions to the magnetic field. Therefore, instead of copying the values of ${A_i}$ from the boundary of the cavity radially inwards, the vector potential is set to zero inside the cavity and the prescription~\eqref{eq: smoothing polynomial} is applied using ${r_1 = 15\,M}$, ${r_2 = 35\,M}$, and ${f_1 = 0}$, so that
\begin{equation}
    A_i\!\left(x, y, z\right)\mapsto P\!\left(r\right)A_i\!\left(x, y, z\right)
\end{equation}
in the spherical shell bounded by ${r_1}$ and ${r_2}$\,. This approach aims at preserving the topology of the magnetic field just outside the cavity as much as possible, while ensuring a sufficiently smooth transition from zero to nonzero field values when moving radially from inside to outside the cavity. We choose ${r_A = 35\,M}$ as a tradeoff value between having a wide transition region, which mitigates the artificial magnetic field introduced by setting ${A_i = 0}$ inside the cavity, and avoiding unphysical modifications to the magnetic field in the bulk of the CBD.

\smallskip
The MHD variables are interpolated using a third-order Lagrange stencil, which is accurate where the fields are smooth, but occasionally produces unphysical results in the proximity of steep gradients. Unphysical values for the interpolated fields include mass density or fluid pressure lower than the respective floor values and negative or luminal/superluminal fluid velocity. We modified \textsc{ReadInterpolate} so that, in numerical cells where any of these conditions occur, the mass density and fluid pressure are set to the corresponding floor values, while the fluid velocity is set to zero; subsequently, the specific internal energy ${\epsilon}$ is initialized via the ideal-fluid EOS~\eqref{eq: ideal-fluid EOS} with adiabatic index ${\Gamma = 5/3}$\,. The floor prescription enforced by \textsc{ReadInterpolate} is the same in use during the CBD evolution, namely, the mass density and fluid pressure respectively decrease with increasing radial distance ${r}$ from the origin of the computational domain as ${r^{-\frac{3}{2}}}$ and ${\left(\Gamma-1\right)r^{-\frac{3}{2}\Gamma}\equiv\left(2/3\right)r^{-\frac{5}{2}}}$.

\subsubsection{Destination grid setup}
\label{subsubsec: Destination grid setup}

The extent and mesh refinement structure of the Cartesian grid used in the inspiral and merger portions of the simulation are the outcome of a few considerations, which are discussed below.

\smallskip
First, even if this work does not aim to produce accurate gravitational waveforms, the apparent horizon of each black hole must still be resolved with a sufficient number of grid cells so that numerical errors leading to imperfect conservation of black hole mass and spin and inaccuracies in the orbital evolution do not significantly affect the dynamics of matter and EM fields in the CBD, minidisks, and jet regions. We choose to cover each black hole with a cubical grid of half-edge length ${0.7\,M}$ and resolution ${M/64}$\,, so that each apparent horizon lies well inside the finest refinement level and is covered by ${\sim\!59\,\text{--}\,60}$ (${\sim\!55\,\text{--}\,57}$) points along its largest (smallest) dimension; the larger numbers refer to the early inspiral, while the smaller ones refer to ${\sim\!1000\,M}$ in time before merger. Note that, even though the black holes are not spinning, tidal interactions can distort the horizons, which are therefore not perfectly spherical in this context~\cite{Prasad2022, Prasad2024}. Our resolution choice allows us to conserve the black holes' irreducible mass and spin (zero) to approximately one part in ${10^5}$ during the inspiral and one part in ${10^4}$ after the merger.

We also conducted a preliminary numerical relativity simulation in which the black holes were enclosed within a refinement level with a half-edge length ${0.5\,M}$, which is too close to the surface of the apparent horizon to yield meaningful physical results. Nonetheless, although the time to merger in the preliminary simulation is ${\sim\!2000\,M}$ shorter than in the production run, the merger dynamics do not change significantly; for instance, the total mass of the gas near the merging binary is essentially unchanged.

\smallskip
Second, consistency of the hand-off process requires that the grid resolution at the inner edge of the computational domain in the CBD simulation---located at a radial distance ${15\,M}$ from the origin---is matched in the Cartesian grid. In the CBD simulation, the grid resolution at that location is approximately ${0.15\,M}$ along ${r}$ and ${0.33\,M}$ along both ${\theta}$ and ${\phi}$ (in the equatorial plane). On the other hand, on the Cartesian domain, the resolution differs by a factor of ${2}$ across refinement levels. Given that, and the fact that the finest resolution on the Cartesian grid is ${M/64}$\,, the resolution in a cubical box centered at the origin with half-edge length ${20\,M}$ is set to ${M/8 = 0.125\,M}$\,. This means that two additional, nested refinement levels must be set up that are fully contained by the latter and that, in turn, fully contain the one covering the two black holes. We then set up two cubical refinement levels with half-edge length ${3M}$ and ${6M}$\,, respectively, tracking the black holes; this also ensures that the expected radial extent of the minidisks and a large part of the accretion streams are fully covered by grids with a minimum resolution of ${M/16}$\,.

\smallskip
Third, the CBD is by far more extended along the orbital plane of the binary than it is along the polar axis, so the refinement levels are compressed to adapt to the shape of the CBD in order to save computational time and memory. At the same time, there should be enough room for relativistic jets to propagate along the polar direction without hitting the domain boundary. Additionally, the outer boundary must be located sufficiently far away from the black holes to be causally disconnected from them for the majority of the simulation runtime in order to prevent potential reflected waves (in particular, gravitational and EM gauge waves) from affecting the evolution of the matter and EM fields. However, the outer boundary must also be limited in size in order to be fully contained in the source spherical domain, which extends to a maximum radius of ${20000\,M}$.

\smallskip
Taking all of the above factors into account, the Cartesian domain of choice is a cubical box with half-edge length ${8192\,M}$ and resolution ${64\,M}$\,, inside of which eight refinement levels are placed with half-edge lengths ${4096\,M}$, ${2048\,M}$, ${1024\,M}$, ${512\,M}$, ${256\,M}$, ${128\,M}$, ${64\,M}$, ${32\,M}$ on the equatorial plane of the binary and ${960\,M}$, ${512\,M}$, ${304\,M}$, ${144\,M}$, ${120\,M}$, ${72\,M}$, ${48\,M}$, ${24\,M}$ along the polar direction.

\begin{figure*}[htp]
    \centering
    \begin{subfigure}[c]{0.49\textwidth}
        \includegraphics[width = \textwidth]{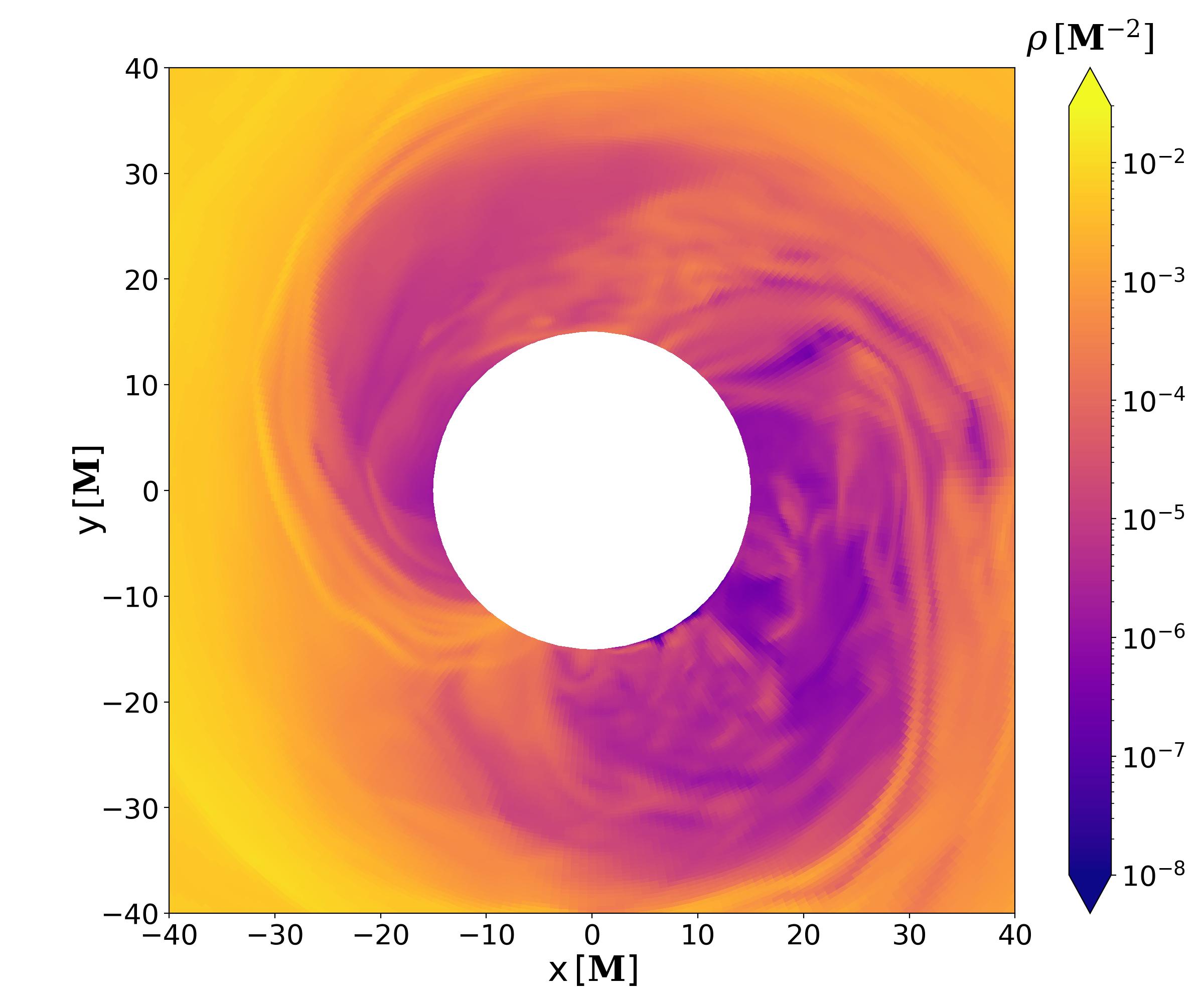}
    \end{subfigure}
    \begin{subfigure}[c]{0.49\textwidth}
        \includegraphics[width = \textwidth]{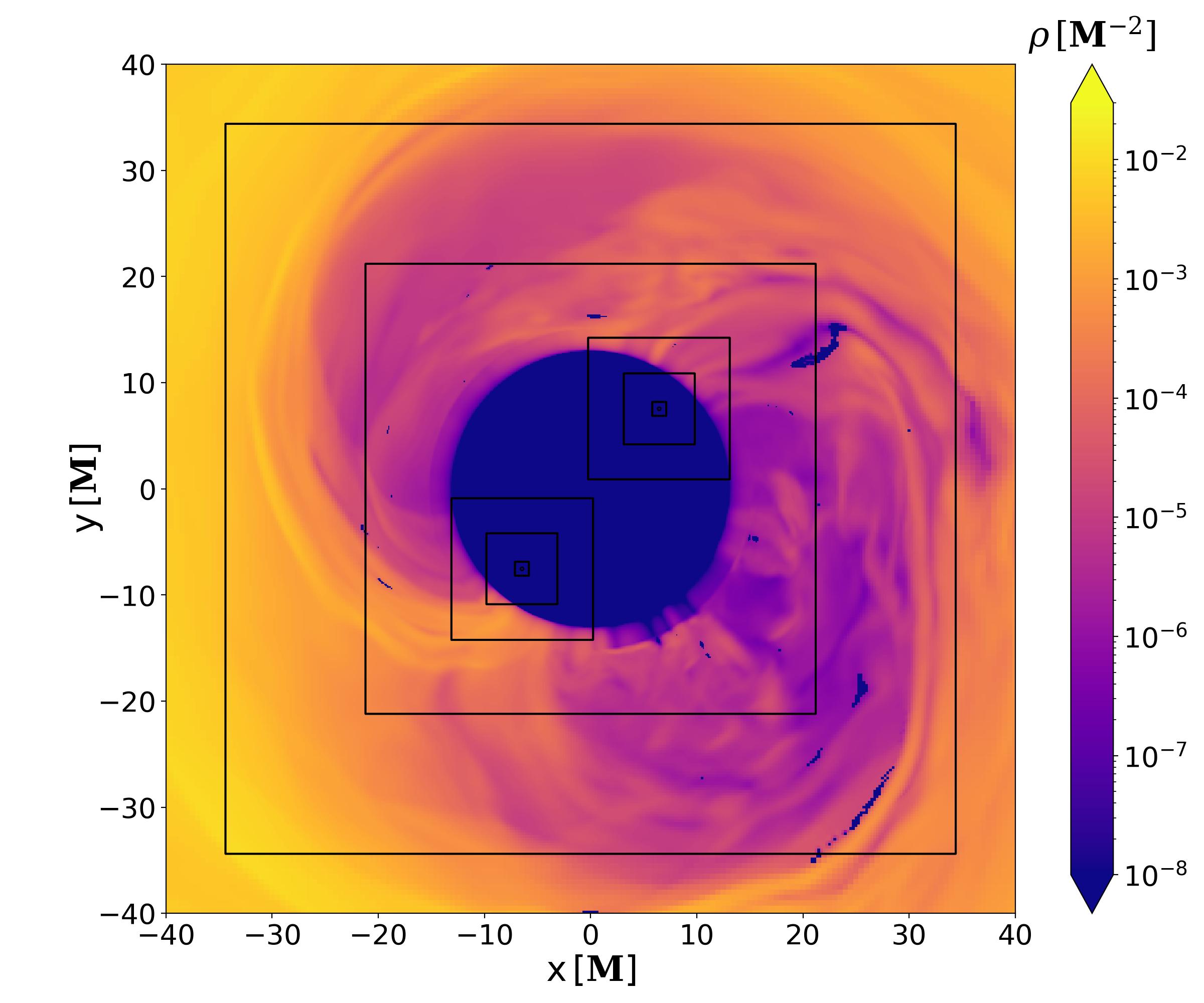}
    \end{subfigure}
    \begin{subfigure}[c]{0.49\textwidth}
        \includegraphics[width = \textwidth]{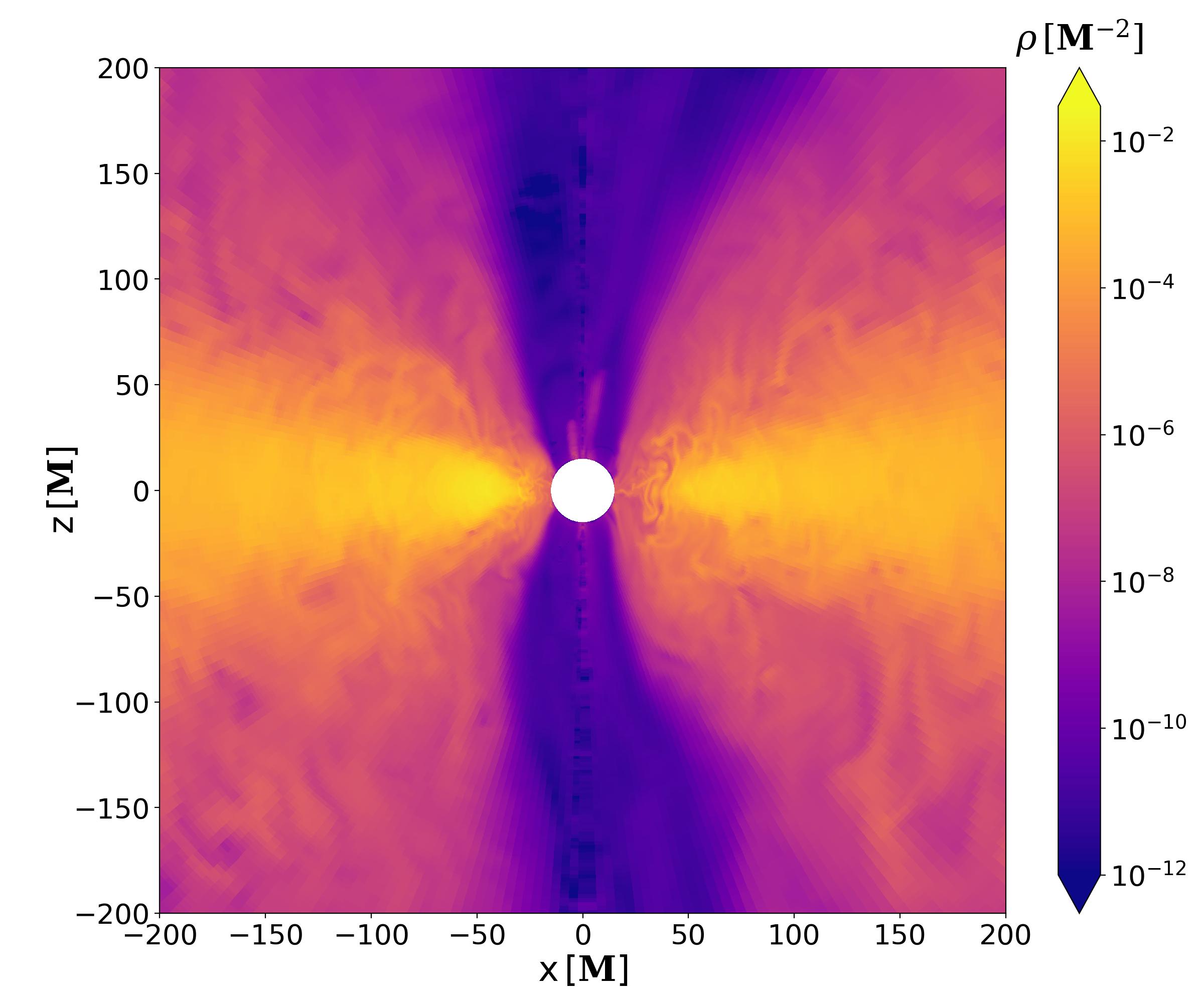}
        \captionsetup{justification = centering, format = hang}
        \caption{Before hand-off (\textsc{SphericalNR})}
    \end{subfigure}
    \begin{subfigure}[c]{0.49\textwidth}
        \includegraphics[width = \textwidth]{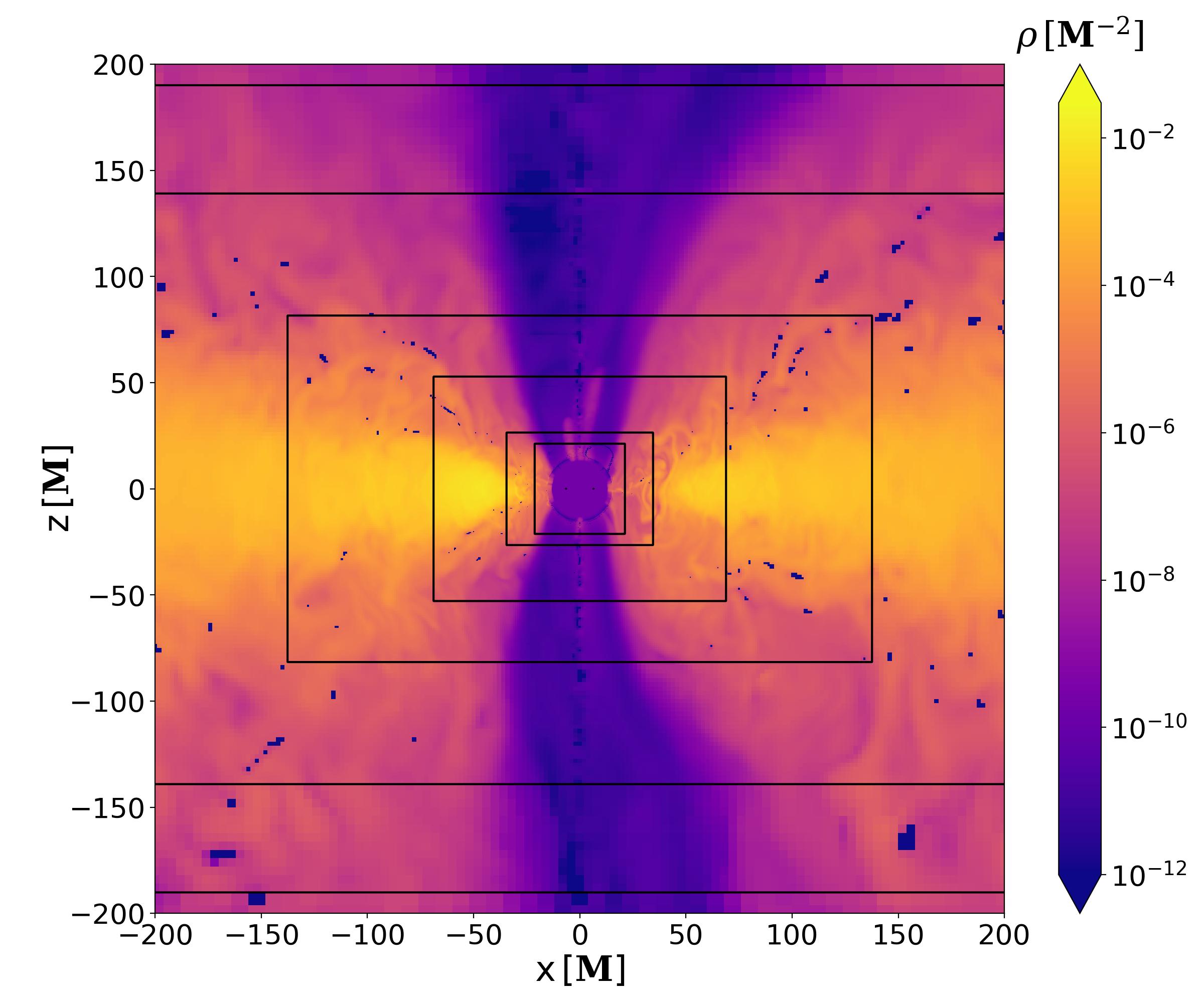}
        \captionsetup{justification = centering, format = hang}
        \caption{After hand-off (\textsc{IllinoisGRMHD})}
    \end{subfigure}
    \captionsetup{justification = raggedright}
    \caption{Mass density distribution before (left panels) and after (right panels) the hand-off step from \textsc{SphericalNR} to \textsc{IllinoisGRMHD}. The right panels show the innermost mesh refinement boundaries. A few points, especially in the low-density polar region and in the disk's magnetized corona, have been set to atmosphere.}
    \label{fig: hand-off}
\end{figure*}

Fig.~\ref{fig: hand-off} shows the mass density distribution before and after the hand-off, along with the apparent horizons of the two black holes and the boundaries of the innermost refinement levels on the Cartesian grid.

\subsection{Inspiral and merger evolution}
\label{subsec: Inspiral and merger evolution}

\subsubsection{Evolution methods}

As done for the equilibration of the CBD, during the inspiral and merger stages the gas is assumed to not self-gravitate. We implemented the same floor prescription used in the \textsc{SphericalNR} simulation and imposed by \textsc{ReadInterpolate} into \textsc{IllinoisGRMHD} starting from the code base available within the \textsc{Einstein Toolkit} release ``Karl Schwarzschild'' (label: \textsc{ET\_2023\_05}). The (vacuum) Einstein equations describing the binary black hole dynamical spacetime are solved in their BSSN formulation~\cite{ShibataNakamura1995, BaumgarteShapiro1998} by the \textsc{Cactus} thorn \textsc{ML\_BSSN}~\cite{Brown2009, McLachlanDocs, McLachlanBitBucket} adopting the moving puncture gauge~\cite{Campanelli2006, Baker2006} along with a fourth-order accurate finite difference stencil and Kreiss-Oliger dissipation applied to the evolved fields. The location of the apparent horizons is tracked via the \textsc{Cactus} thorn \textsc{AHFinderDirect}~\cite{Thornburg2004, AHFinderDirectDocs, AHFinderDirectBitBucket} and horizon diagnostics are provided by the \textsc{QuasiLocalMeasures} thorn~\cite{Dreyer2003, Schnetter2006QLM, Szabados2004, QuasiLocalMeasuresDocs, QuasiLocalMeasuresBitBucket}.

\smallskip
The mass and energy-momentum conservation equations~\eqref{eq: GRMHD equations} are solved by \textsc{IllinoisGRMHD} via a finite-volume scheme that implements the third-order accurate piecewise-parabolic (PPM) reconstruction technique described in~\cite{ColellaWoodward1984} coupled to the HLLE approximate Riemann solver and to a conservatives-to-primitives inversion method very similar to the ``2D scheme'' presented in~\cite{Noble2006}. Similarly to \textsc{SphericalNR}, \textsc{IllinoisGRMHD} solves the ideal MHD approximation to Maxwell's equations in a vector-potential based approach adopting the generalized Lorenz gauge prescription. However, for improved stability and accuracy, all EM variables in \textsc{IllinoisGRMHD} are staggered with respect to the center of the numerical cell: the vector potential sits at the edges, the scalar potential at the vertices, and the magnetic field at the faces~\cite{Etienne2015}. With this, applying Kreiss-Oliger dissipation to the EM evolved fields is typically not needed.

\smallskip
As in the CBD simulation, the time evolution of the spacetime and MHD fields is left to the \textsc{MoL} thorn. The CFL factor on the coarsest grid level is set to ${\Delta t_\text{coarse}/\Delta x_\text{coarse} = 1/32}$, where ${\Delta t_\text{coarse}}$ and ${\Delta x_\text{coarse}}$ are the timestep and spatial resolution on the same level; recalling (see Sec.~\ref{subsubsec: Destination grid setup}) that ${\Delta x_\text{coarse} = 64\,M}$\,, that means ${\Delta t_\text{coarse} = 2M}$. The same timestep is imposed on the first five refinement levels (including the coarsest one) and drops by a factor of 2 across all other refinement levels. Having a maximum timestep of ${\Delta t_\text{max} = 2M}$ allows us to set the Lorenz gauge damping factor to the value ${1.5/\Delta t_\text{max} = 0.75}$\,, thus preventing EM gauge waves from reaching the outer boundary and being reflected back into the computational domain, harming the stability of the evolution. We note that setting the Lorenz gauge damping factor to ${0.75}$ without limiting the timestep to ${2M}$ also leads to severe evolution instabilities.

\subsubsection{Radiative cooling}
\label{subsubsec: Radiative cooling (IGM)}

The radiative cooling prescription is similar to the one used in~\cite{Bowen2017, Bowen2019}, which in turn extends the method of~\cite{Noble2012} (see Eqs.~\eqref{eq: cooling rate} and~\eqref{eq: Bernoulli criterion}) from the CBD to the cavity and the minidisks; however, cooling in proximity of the black holes is handled differently here. First, two ``minidisk regions'' are defined as the spheres centered on each puncture with coordinate radius ${0.45\,a}$\,. Then, as done in~\cite{Bowen2017, Bowen2019}, the cooling timescale computed via~\eqref{eq: cooling timescale} at the inner edge of the CBD, defined as the sphere of coordinate radius ${1.5\,a}$ centered at the origin of the computational domain, is used everywhere inside the inner edge itself, except in the minidisk regions. Following once again~\cite{Bowen2017, Bowen2019}, the cooling timescale of a fluid element lying inside either of the two minidisk regions at a distance ${r}$ from the puncture is set to the Keplerian period of the circular orbit of radius ${r}$ around the latter (i.e.,~\eqref{eq: cooling timescale}, where now ${M}$ is the mass of the puncture and not the total mass of the binary).

However, in this work, the regions closest to the black holes are handled differently from~\cite{Bowen2017, Bowen2019}. Let ${i\in\left\{1, 2\right\}}$ label the two punctures and let ${r_i}$ be the distance of a given grid point from puncture ${i}$. Then, for each ${i\in\left\{1, 2\right\}}$:
\begin{itemize}
    \item ${r_i < 0.35\,M}$\,: the cooling rate is set to zero.
    \item ${0.35\,M < r_i\leq 1 M}$\,: the cooling rate is built using the cooling timescale ${t_\text{cool,\,i}^{1M}}$ computed at ${r_i = 1M}$ and is multiplied by the linear damping factor
    \begin{equation}
        f_i = \frac{r_i - 0.35\,M}{1M - 0.35\,M}\in\left[0, 1\right]\;.
    \end{equation}
    \item If, however, the two spheres centered at the punctures with radii ${1M}$ overlap (i.e., the binary is close to merging), the cooling timescale ${t_\text{cool}}$ and the damping factor ${f}$ in the overlap region are set to the averages of the respective quantities in the two spheres, weighted by the inverse distances from the punctures:
    \begin{align}
        t_\text{cool} &= \frac{\frac{t_\text{cool,\,1}^{1M}}{r_1} + \frac{t_\text{cool,\,2}^{1M}}{r_2}}{\frac{1}{r_1} + \frac{1}{r_2}} = \frac{r_2\,t_\text{cool,\,1}^{1M} + r_1\,t_\text{cool,\,2}^{1M}}{r_1 + r_2}\;,\\
        f &= \frac{\frac{f_1}{r_1} + \frac{f_2}{r_2}}{\frac{1}{r_1} + \frac{1}{r_2}} = \frac{r_2 f_1 + r_1 f_2}{r_1 + r_2}\;.
    \end{align}
\end{itemize}
The above prescription avoids cooling of material well inside the apparent horizons, whose minimum radius is ${\sim\!0.45\,M}$, and linearly damps the cooling rate to within the innermost stable circular orbit (ISCO), which is located at ${r_\text{ISCO}^\text{(num)}\simeq 1.4\,M}$ from each puncture. Because the analytical form of the coordinates on the numerical spacetime is unknown, we estimate the numerical ISCO radius ${r_\text{ISCO}^\text{(num)}}$ via the simple proportionality relation
\begin{equation}
    r_\text{ISCO}^\text{(num)} = \frac{r_\text{ISCO}\,r_\text{AH}^\text{(num)}}{r_\text{hor}}\;,
\end{equation}
where ${r_\text{AH}^\text{(num)}\simeq 0.47\,M}$ is the maximum apparent horizon radius and ${r_\text{hor}}$ and ${r_\text{ISCO}}$ denote, respectively, the maximum radius of the event horizon and the ISCO radius in Boyer-Lindquist coordinates.

\begin{figure*}[htp]
    \centering
    \begin{subfigure}[c]{0.32\textwidth}
        \includegraphics[width = \textwidth]{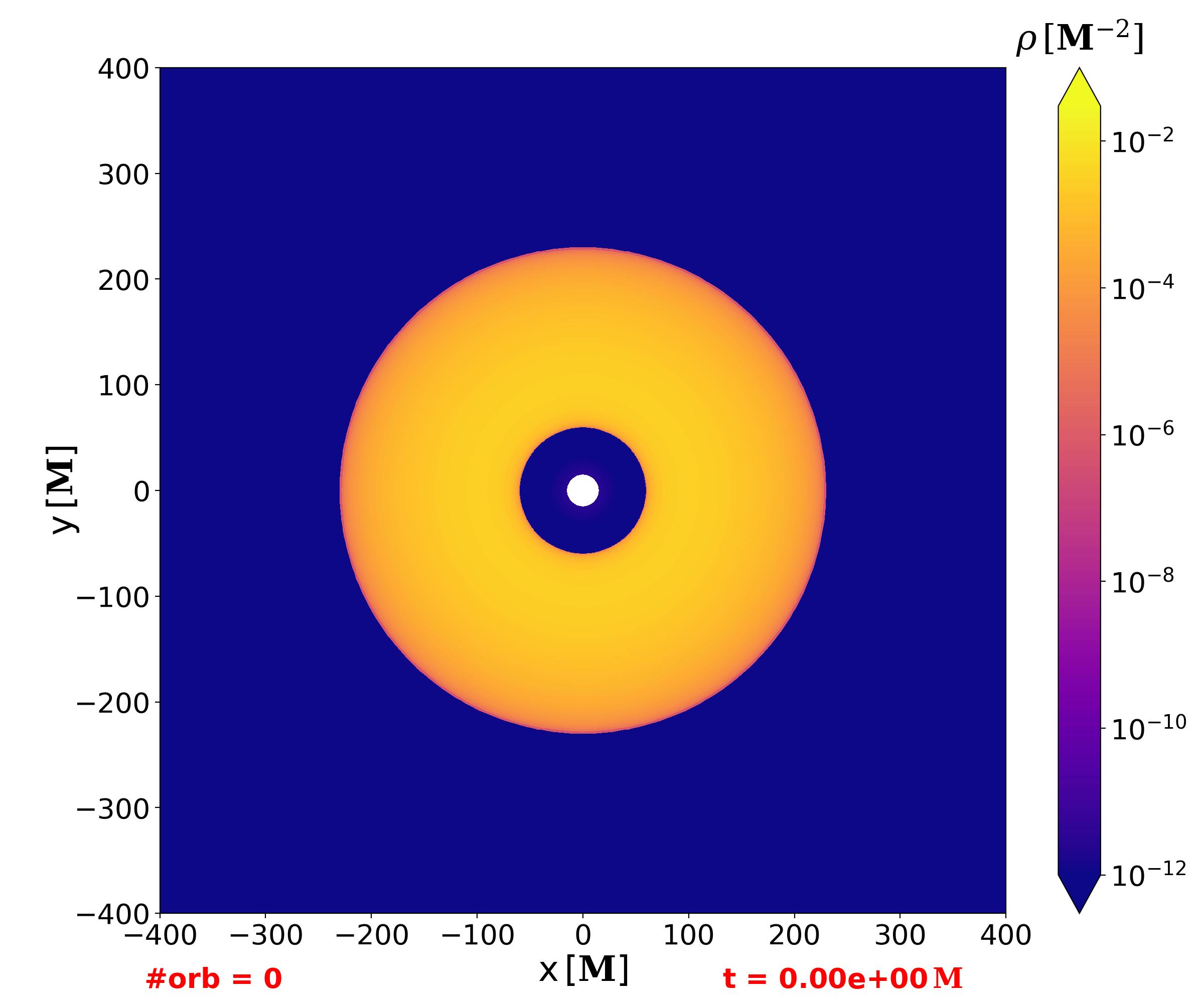}
    \end{subfigure}
    \begin{subfigure}[c]{0.32\textwidth}
        \includegraphics[width = \textwidth]{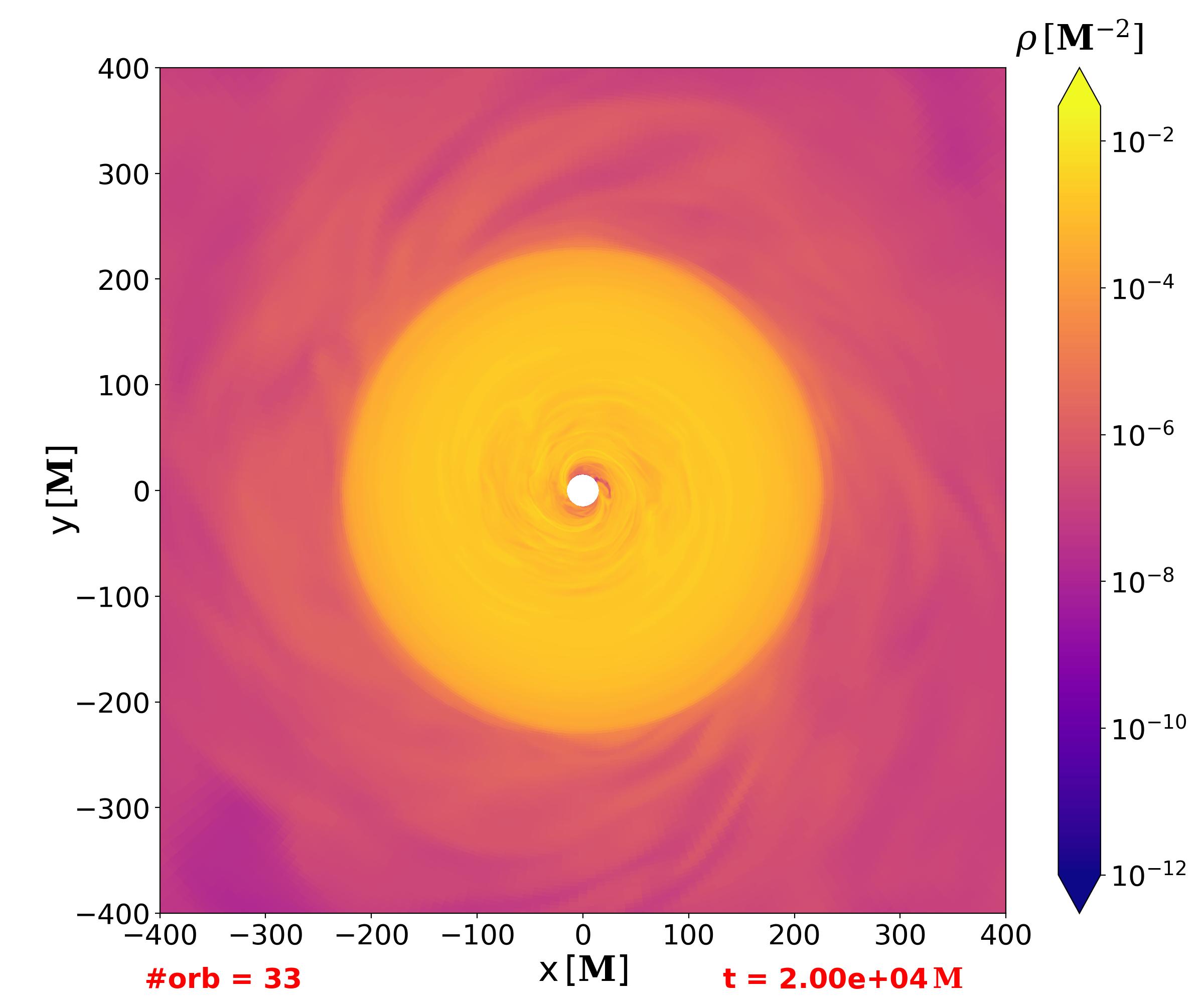}
    \end{subfigure}
    \begin{subfigure}[c]{0.32\textwidth}
        \includegraphics[width = \textwidth]{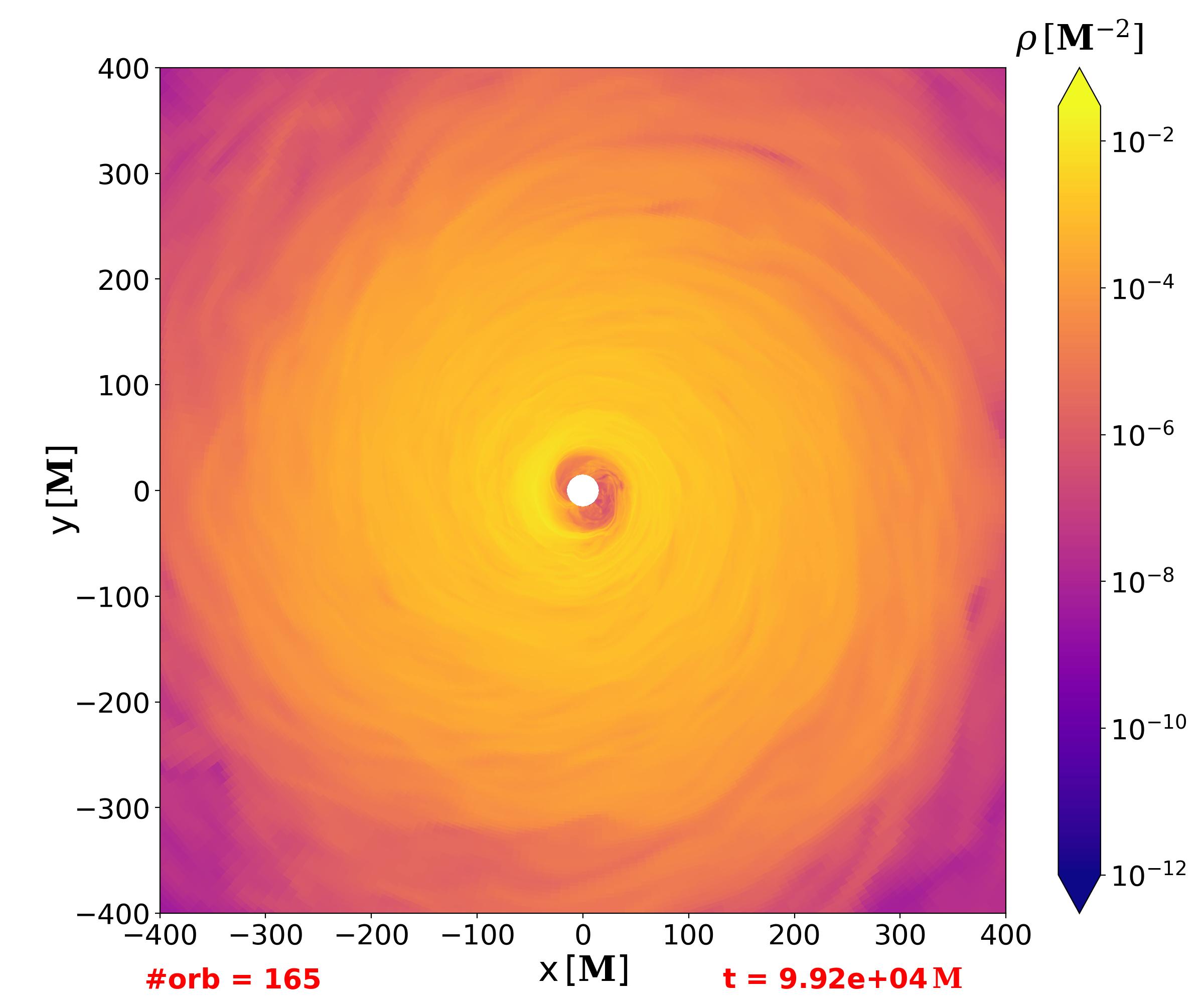}
    \end{subfigure}
    \begin{subfigure}[c]{0.32\textwidth}
        \includegraphics[width = \textwidth]{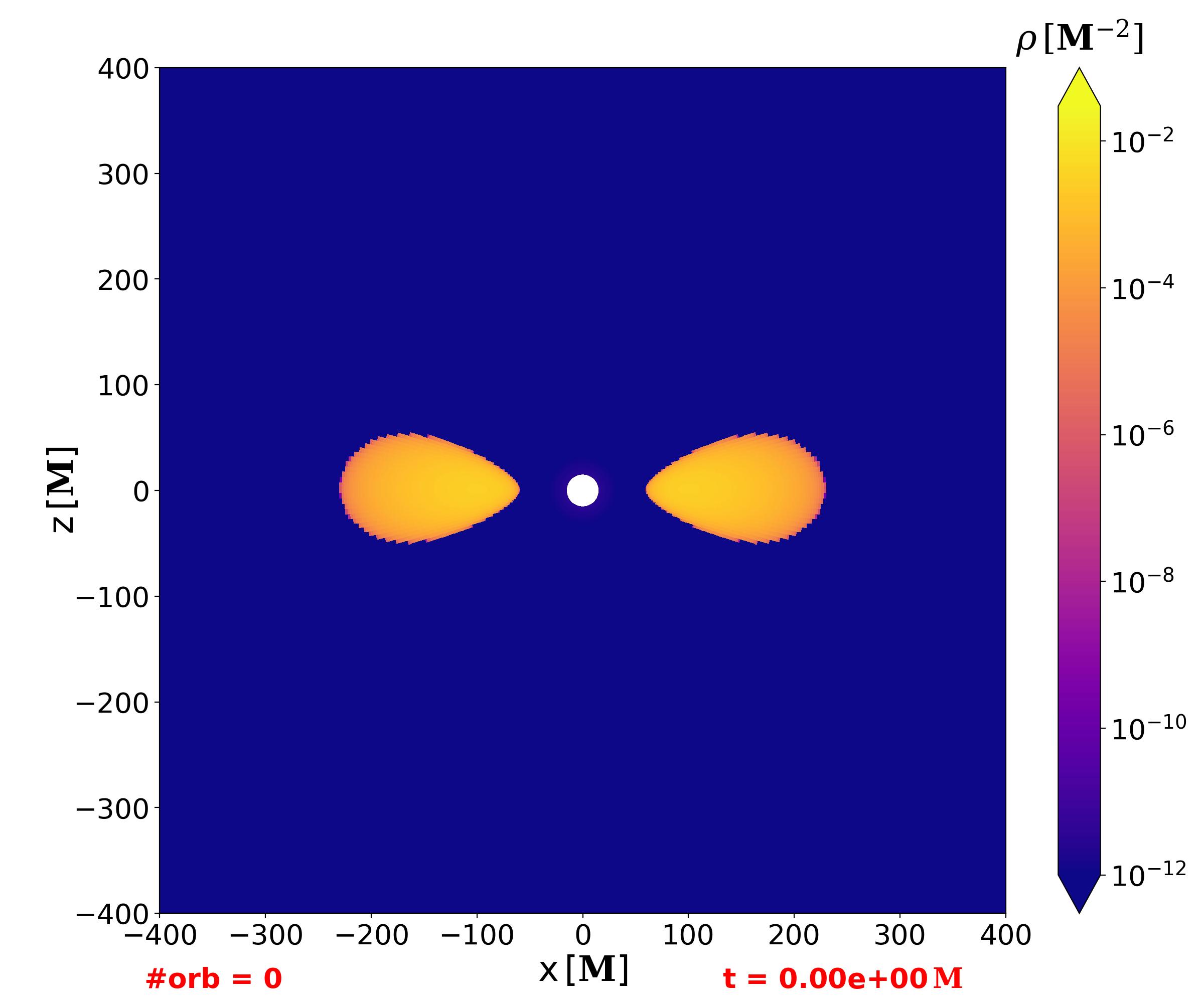}
        \caption{Initial configuration}
    \end{subfigure}
    \begin{subfigure}[c]{0.32\textwidth}
        \includegraphics[width = \textwidth]{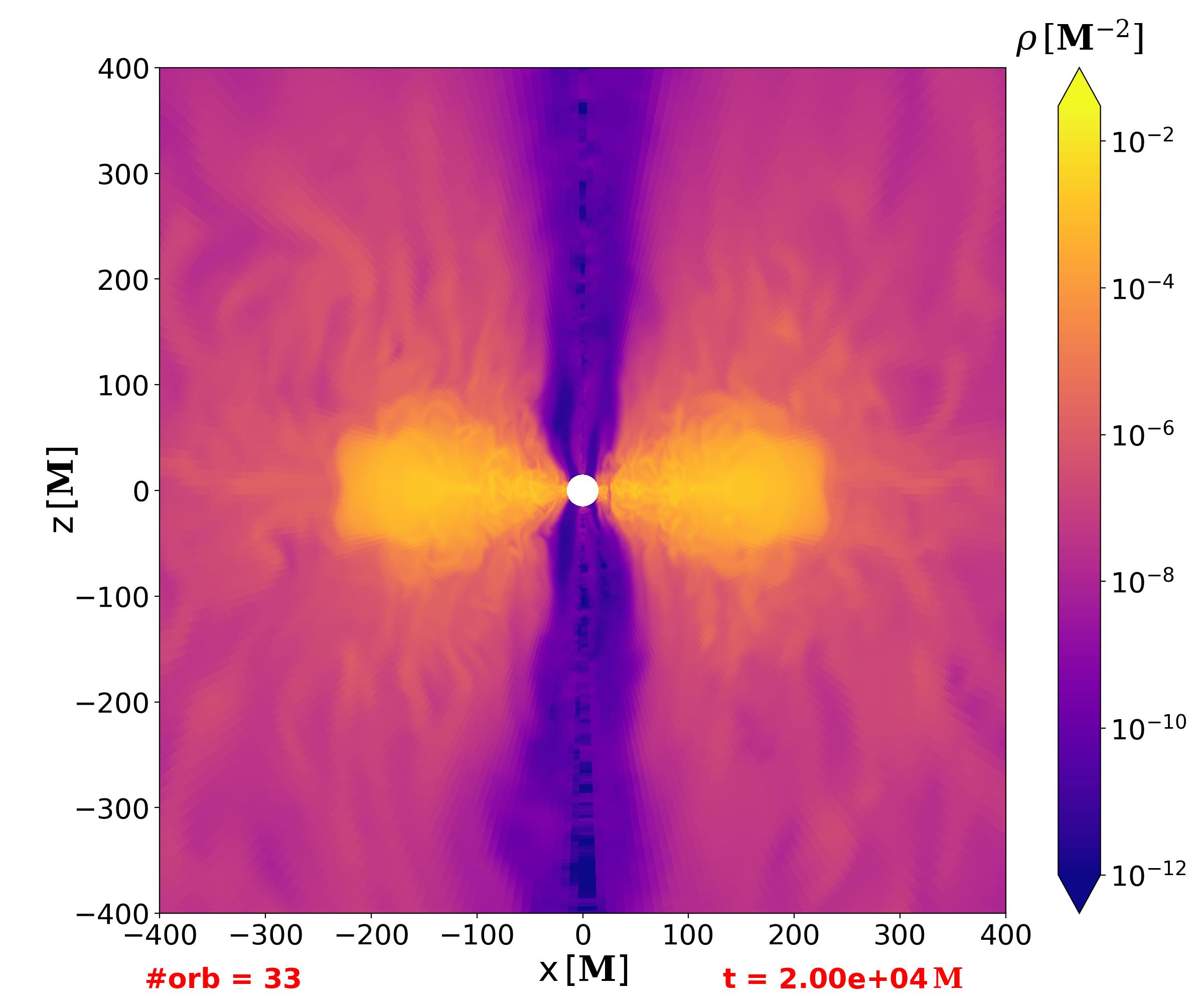}
        \caption{MRI growth}
    \end{subfigure}
    \begin{subfigure}[c]{0.32\textwidth}
        \includegraphics[width = \textwidth]{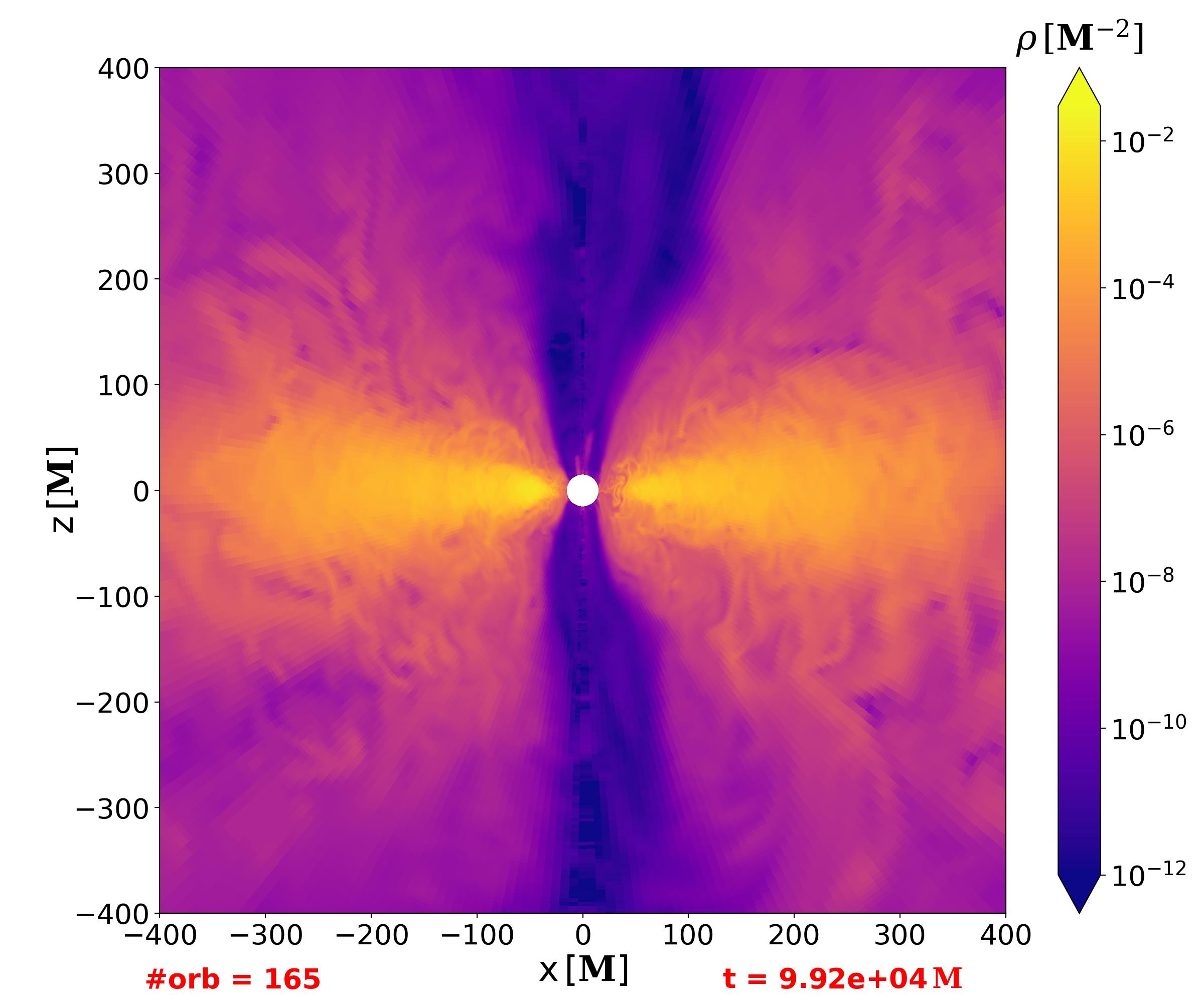}
        \caption{Quasi-steady-accretion phase}
    \end{subfigure}
    \captionsetup{justification = raggedright, format = hang}
    \caption{Face-on (top panels) and edge-on (bottom panels) views of the mass density distribution in the CBD at different stages of its evolution with \textsc{SphericalNR}. The snapshots only show the central part of the computational domain, which extends to a maximum coordinate radius ${r=20000\,M}$.}
    \label{fig: CBD evolution}
\end{figure*}

\smallskip
The method just described applies so long as the distance between the punctures is larger than a user-defined threshold, here set to ${0.9\,M}$ and corresponding to roughly twice the maximum radius of each apparent horizon. We then calculate the distance ${r_\text{CM}}$ of a given grid point from the center of mass of the binary (which is the remnant black hole after merger) and apply a prescription suitable for handling the merger:
\begin{itemize}
    \item ${r_1\leq 0.5\,M}$ or ${r_2\leq 0.5\,M}$ or ${r_\text{CM}\leq 0.5\,M}$\,: the cooling rate is set to zero.
    \item ${0.5\,M < r_1\leq 1.2\,M}$ or ${0.5\,M < r_2\leq 1.2\,M}$ or ${0.5\,M < r_\text{CM}\leq 1.2\,M}$\,: the cooling rate is built using the cooling timescale computed at a distance ${1.2\,M}$ from the center of mass of the system and is multiplied by the linear damping factor
    \begin{equation}
        g_i = \min\left(\frac{r_\text{CM} - 0.5\,M}{1.2\,M - 0.5\,M}\;,\;1\right)\in\left[0, 1\right]\;.
    \end{equation}
    \item Farther away from the merging binary or remnant black hole, the cooling timescale and rate are computed using the standard prescription of~\cite{Noble2012}, with the center of mass of the system being the center of the circular Keplerian orbit of the fluid element at hand.
\end{itemize}
Again, the above prescription produces no cooling in a region well inside the apparent horizon of the merger remnant (minimum radius ${\sim\!0.68\,M}$) and linearly damps the cooling rate to within the ISCO radius, which we estimate as ${\sim\!1.54\,M}$.

\section{Results}
\label{sec: Results}

\subsection{Circumbinary disk equilibration}
\label{subsec: Circumbinary disk equilibration}

Defining the characteristic gravitational timescale of the system as ${t_g\equiv GM/c^3\equiv M}$, the CBD simulation is carried out for a time span ${t\simeq 1.5\times 10^5 M}$, or ${\sim\!250}$ binary orbits. Similar to what was found in~\cite{LopezArmengol2021}, the magnetorotational instability (MRI) grows quickly up to ${t\simeq 30000\,M}$, creating Maxwell stress as orbital shear correlates the turbulent magnetic field; this stress boosts gas accretion onto the binary through outward angular momentum transport. As a result, gravitational potential energy is released, and the torus heats up and expands along both the equatorial and polar directions (Fig.~\ref{fig: CBD evolution}). Over time, angular momentum is transferred from the binary to the torus via gravitational torques exerted on the accretion streams, causing a portion of them to move outward, return to the CBD, and deposit their angular momentum; as a result, mass piles up at the CBD's inner edge. Eventually, a quasi-steady accretion regime is reached.

\begin{figure*}[htp]
    \centering
    \includegraphics[width = \linewidth]{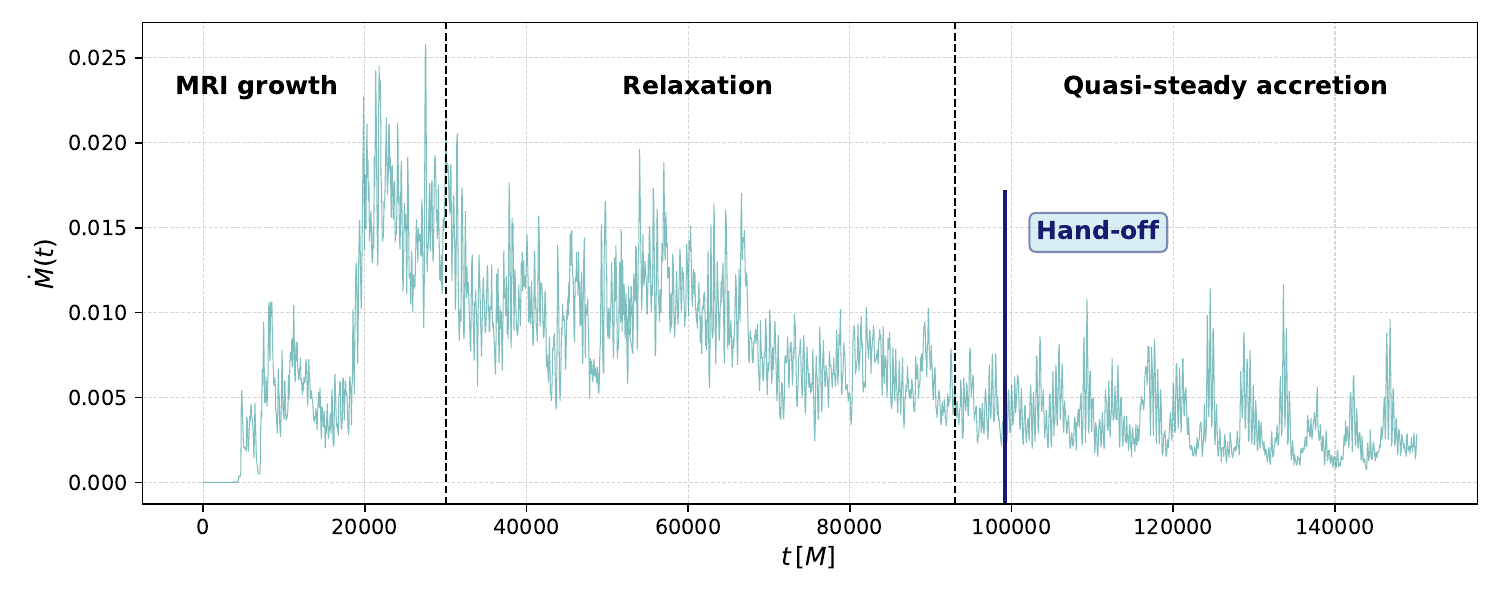}
    \captionsetup{justification = raggedright, format = hang}
    \caption{Mass accretion rate onto the inner edge of the computational domain during the whole CBD evolution with \textsc{SphericalNR}. A solid vertical blue line at ${t\simeq 99200\,M}$ marks the time of hand-off to the inspiral and merger part of the simulation.}
    \label{fig: CBD accretion rate at 15 M}
\end{figure*}

\begin{figure}[htp]
    \centering
    \includegraphics[width = \linewidth]{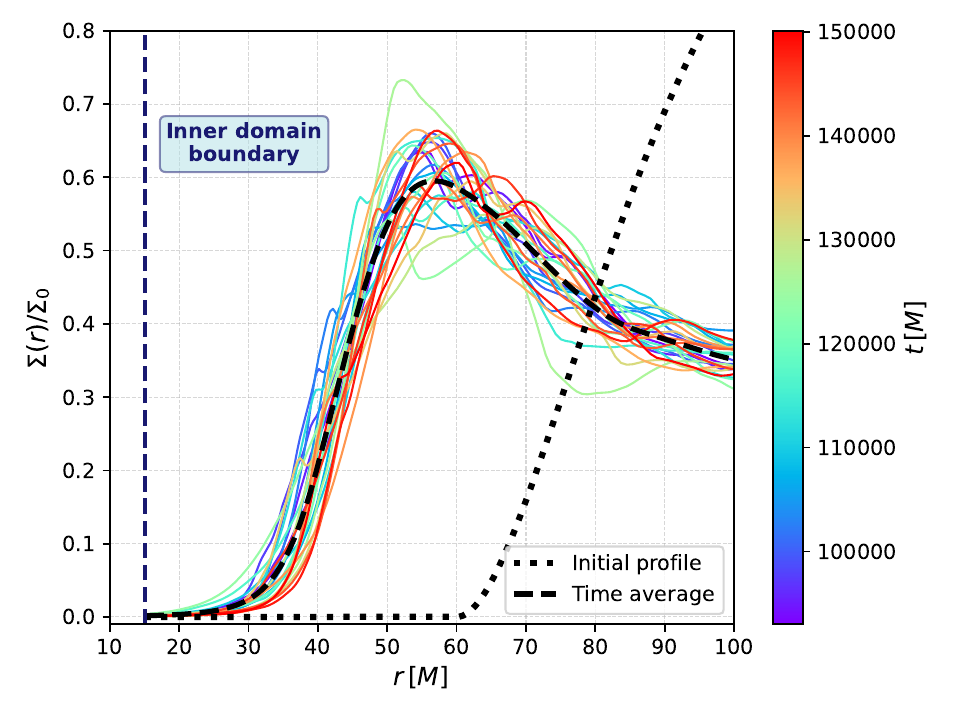}
    \captionsetup{justification = raggedright} 
    \caption{Surface density in the CBD from ${t=93000\,M}$ to ${t=150120\,M}$ during the \textsc{SphericalNR} evolution. Colored curves show radial profiles of ${\Sigma\left(r\right)\!/\,\Sigma_0}$ (with ${\Sigma_0\simeq 0.126}$ in code units) at different times with a time resolution of ${2400\,M}$. The dotted and dashed curves are the initial and time-averaged profiles, respectively.}
    \label{fig: CBD sigma vs r}
\end{figure}

These statements are quantified in Figs.~\ref{fig: CBD accretion rate at 15 M} and~\ref{fig: CBD sigma vs r}. The mass accretion rate onto the binary is defined by
\begin{equation}
    \label{eq: Mdot vs t}
    \dot{M}\!\left(t\right) = \int_\mathcal{S} d^2\sigma\,\sqrt{-g}\,\rho u^r\;,
\end{equation}
where $\mathcal{S}$ is the spherical surface centered at the origin of the computational domain with coordinate radius ${r=15\,M}$, corresponding to the inner radial grid boundary. The onset of the quasi-steady accretion stage, which occurs around ${t\simeq 93000\,M}$, is signaled by the formation of a density concentration (the ``lump'': see~\cite{Shi2012, Noble2012}) which modulates the accretion onto the binary (Fig.~\ref{fig: CBD accretion rate at 15 M}).

\smallskip
The primary goal of the CBD simulation is to establish appropriate initial conditions for the gas feeding the binary within the gap region during the inspiral and merger phases of the evolution. Therefore, a detailed analysis of the dynamics of the system at this stage is not included here; for more in-depth information, readers can refer to, e.g.,~\cite{Farris2012, Noble2021, LopezArmengol2021}. Nonetheless, in order to get a sense of how well our CBD evolution compares to the one in~\cite{Noble2012}, we calculate the surface density
\begin{equation}
    \label{eq: surface density}
    \Sigma\left(r, \phi\right) = \frac{\int_0^{\pi}d\theta\sqrt{-g}\,\rho}{\sqrt{g_{\phi\phi}\left(\theta = \frac{\pi}{2}\right)}}
\end{equation}
and plot its azimuthal average
\begin{equation}
    \label{eq: phi-averaged surface density}
    \Sigma\left(r\right) = \frac{1}{2\pi}\int_0^{2\pi}d\phi\;\Sigma\left(r, \phi\right)
\end{equation}
from the beginning of the quasi-steady accretion phase (${t=93000\,M}$) up to the end of the \textsc{SphericalNR} simulation (${t=150120\,M}$). In passing, we remark that we hand-off the CBD data to the numerical relativity portion of the simulation at ${t\simeq 99200\,M}$\,, i.e., not too long after the onset of the quasi-stationary stage of the CBD evolution (see also Sec.~\ref{subsec: Hand-off consistency}); however, we continue to evolve the CBD with \textsc{SphericalNR} until well after the time of hand-off to make sure the CBD has really reached a quasi-steady accretion regime. This means that the radial surface density profiles~\eqref{eq: phi-averaged surface density} are calculated at a fixed binary separation ${a=20\,M}$ at all times.

In Fig.~\ref{fig: CBD sigma vs r} we show ${\Sigma\left(r\right)}$ at a number of times along with its time average and initial profile; for ease of comparison with~\cite{Noble2012}, we actually plot ${\Sigma\left(r\right)/\Sigma_0}$\,, where ${\Sigma_0\simeq 0.126}$ (code units) is the initial maximum of~\eqref{eq: surface density}. Consistent with~\cite{Noble2012}, we observe a pile-up of material between ${r\simeq 50\,M}$ and ${r\simeq 60\,M}$. The overall small variation in the radial surface density profiles over the time span covered by Fig.~\ref{fig: CBD sigma vs r} provides further evidence that a quasi-stationary regime is reached around ${t\simeq 93000\,M}$.

\subsection{Hand-off consistency}
\label{subsec: Hand-off consistency}

\begin{figure*}[htp]
    \centering
    \begin{subfigure}[c]{0.49\textwidth}
        \includegraphics[width = \textwidth]{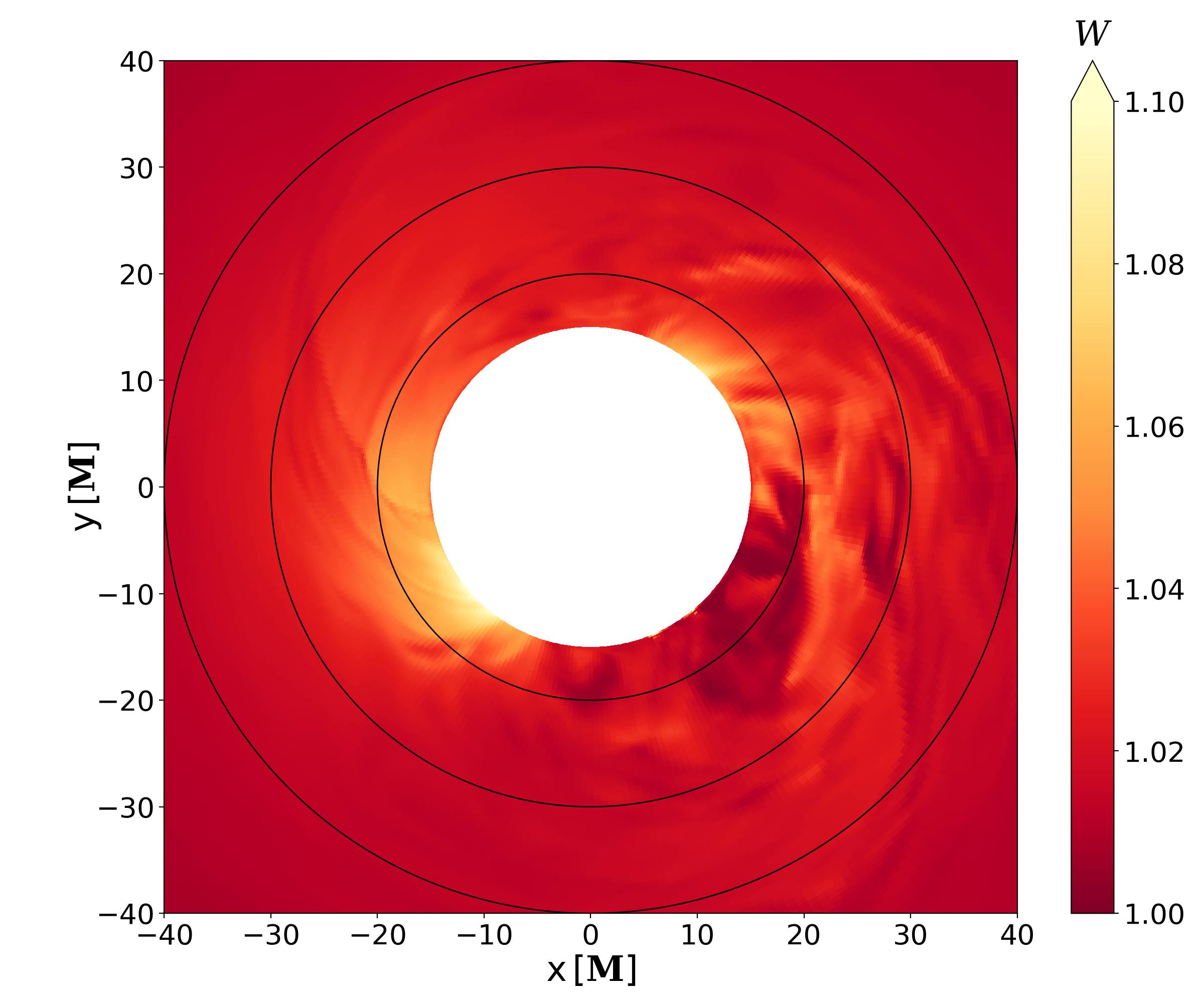}
    \end{subfigure}
    \begin{subfigure}[c]{0.49\textwidth}
        \includegraphics[width = \textwidth]{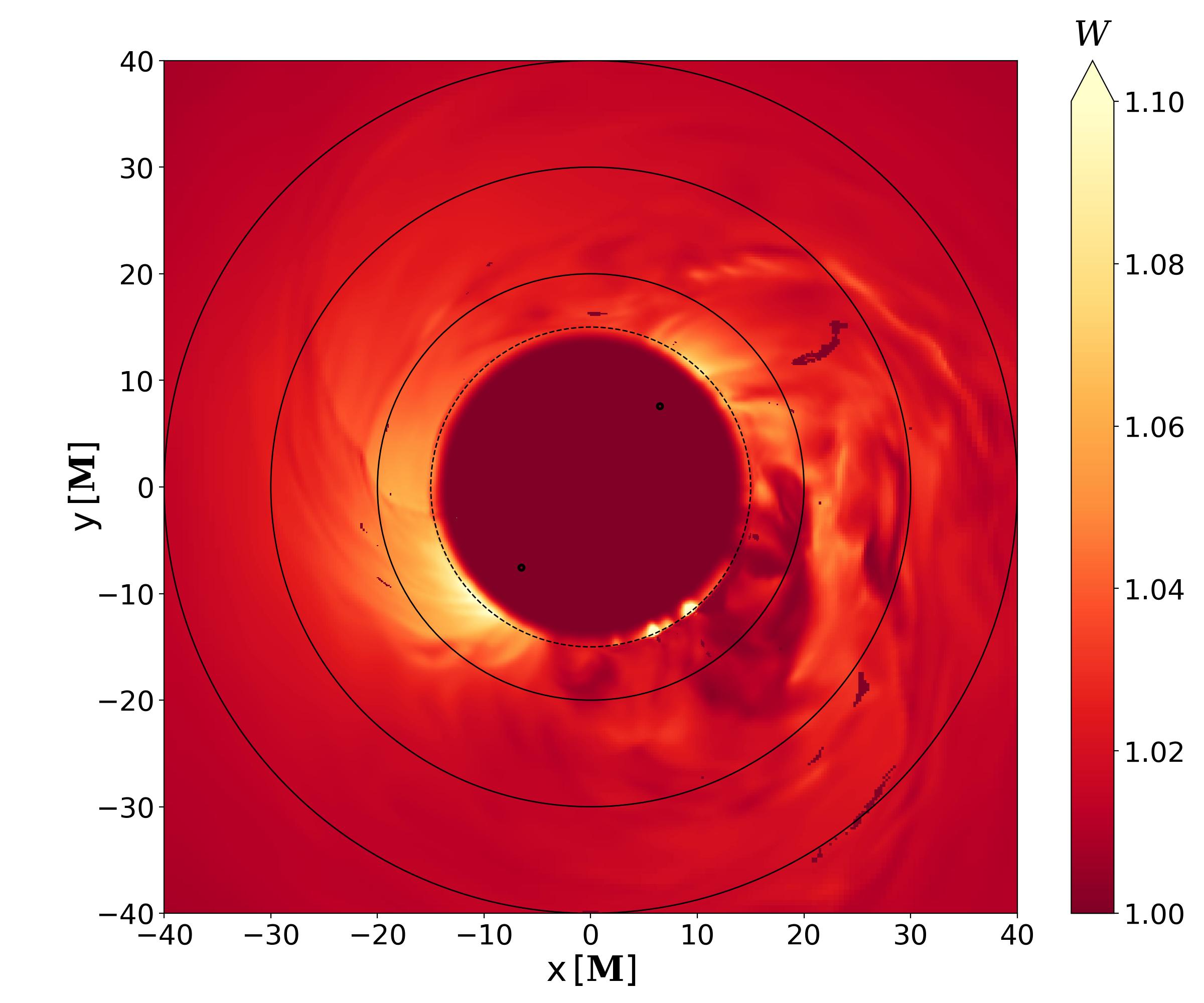}
    \end{subfigure}
    \begin{subfigure}[c]{0.49\textwidth}
        \includegraphics[width = \textwidth]{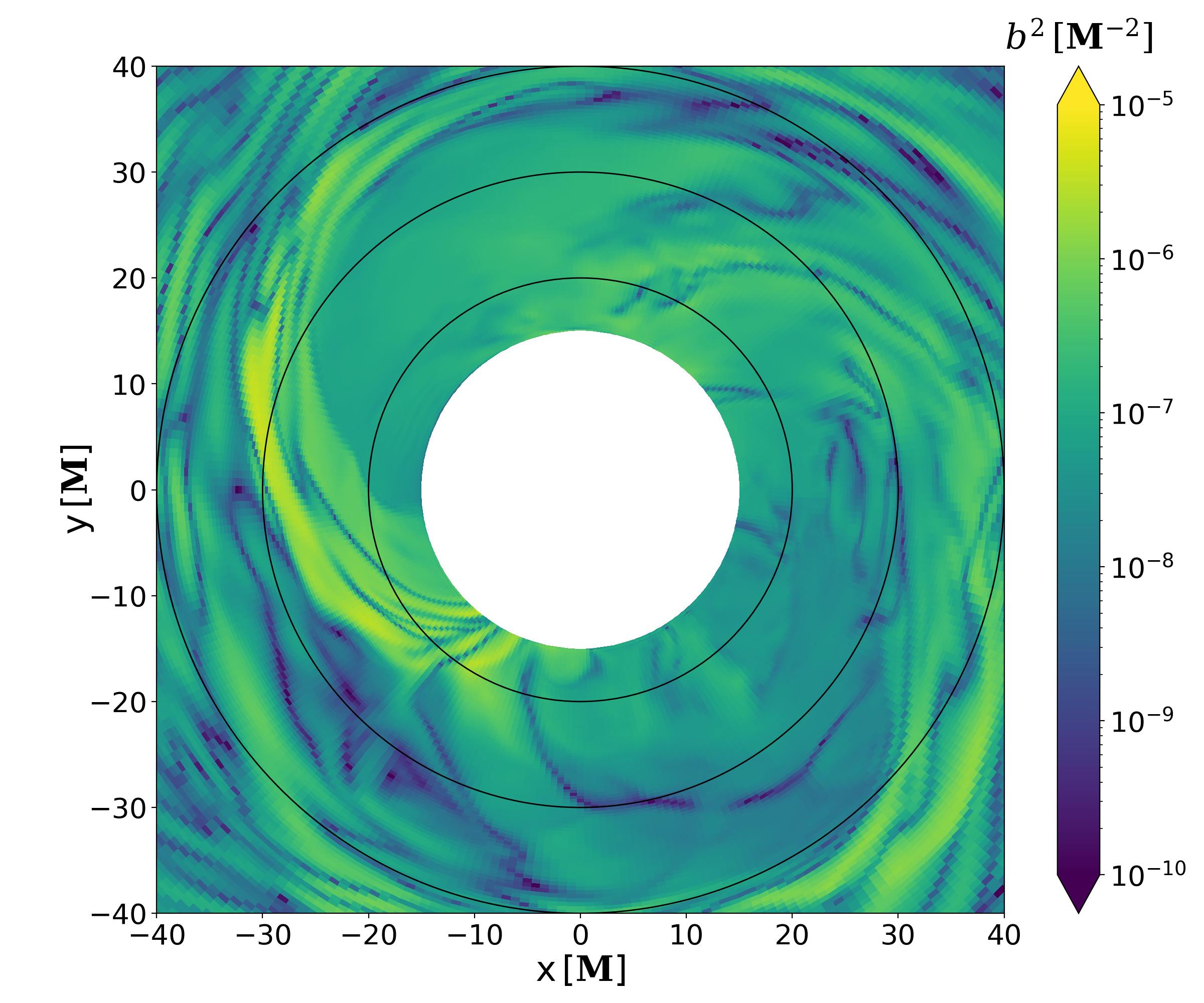}
        \caption{CBD simulation}
    \end{subfigure}
    \begin{subfigure}[c]{0.49\textwidth}
        \includegraphics[width = \textwidth]{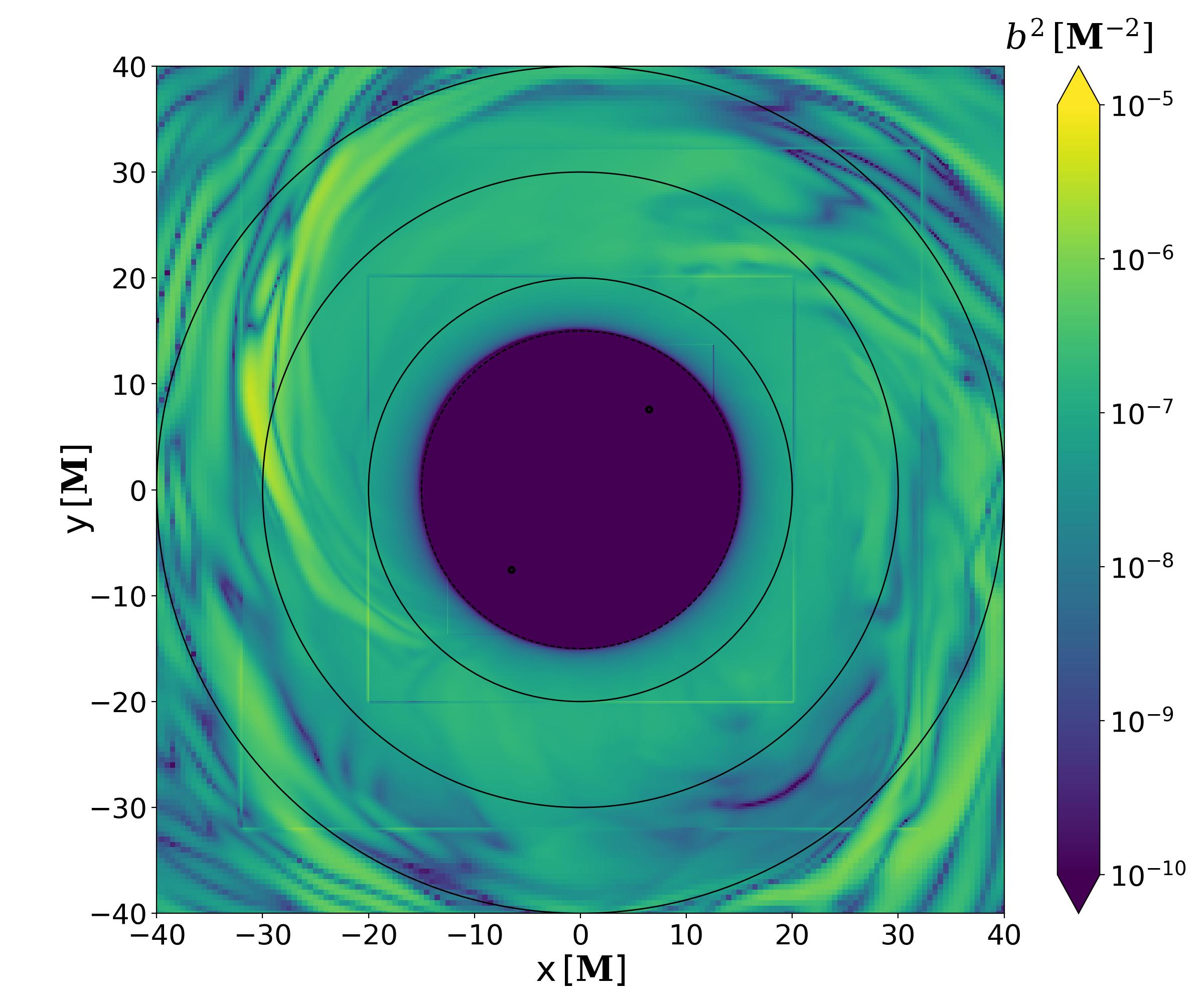}
        \caption{Inspiral+merger simulation}
    \end{subfigure}
    \captionsetup{justification = raggedright, format = hang}
    \caption{Equatorial distributions of the Lorentz factor (top panels) and squared-magnitude of the magnetic field in the fluid's frame (bottom panels) in the CBD (left panels) and inspiral+merger (right panels) simulations at the time of hand-off. The magnetic field map after the hand-off looks blurrier than the corresponding distribution before the hand-off around the cavity region as a result of the radial smoothing of the magnetic vector potential components described in Sec.~\ref{subsubsec: Interpolation of the MHD primitives}.}
    \label{fig: Lorentz factor and b2 hand-off comparison}
\end{figure*}

\begin{figure*}[htp]
    \centering
    \includegraphics[width = \textwidth]{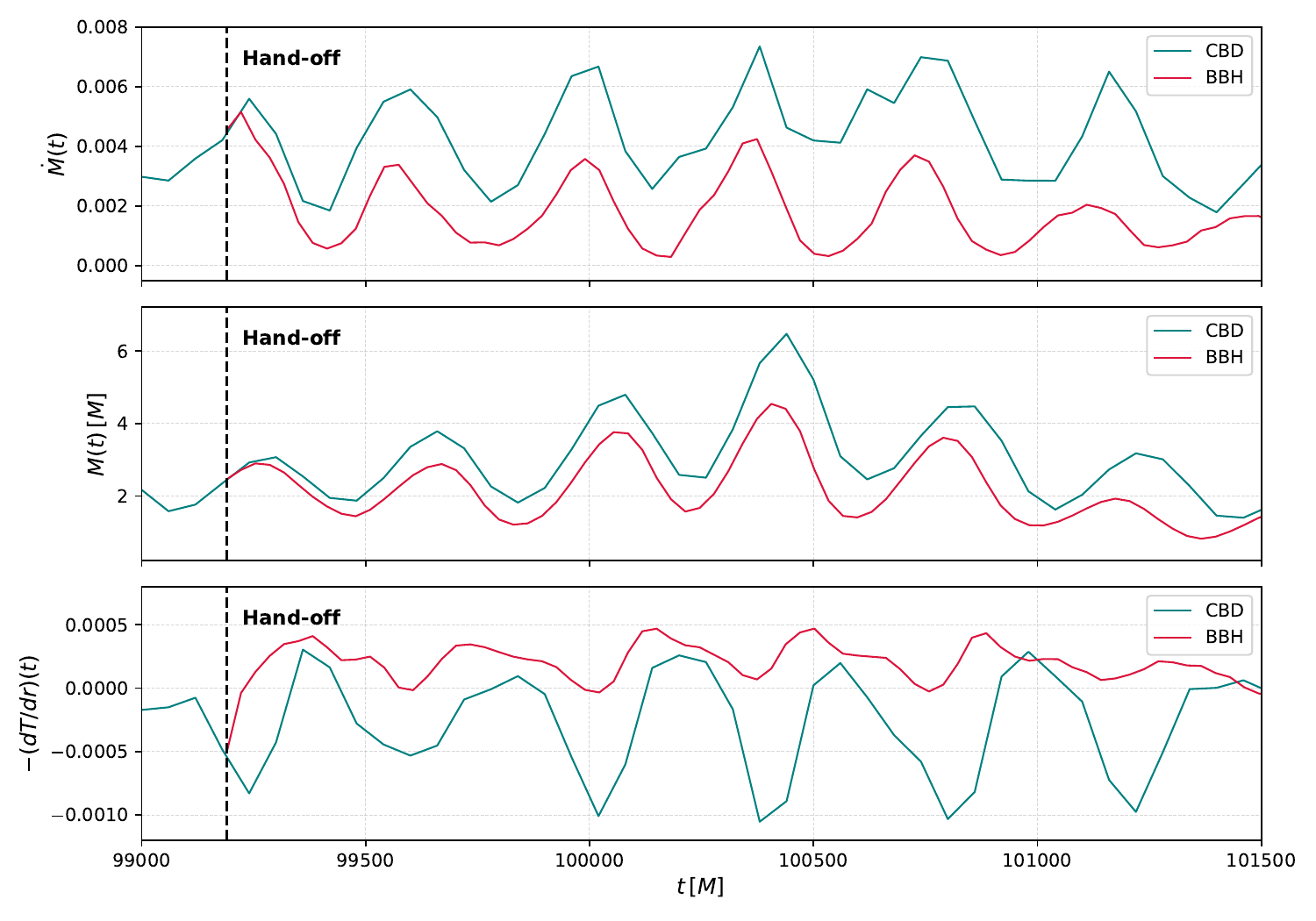}
    \captionsetup{justification = raggedright, format = hang}
    \caption{Evolution of various quantities during the first $\sim\!2300\,M$ of evolution after the hand-off for the CBD (teal) and inspiral (red) simulations. \textbf{[Top panel]} Accretion rate on a spherical surface centered of coordinate radius ${r=20\,M}$ centered around the binary's center of mass. \textbf{[Central panel]} Mass within a spherical volume of coordinate radius ${r=30\,M}$ centered at the same location. \textbf{[Bottom panel]} Radial density of gravitational torque on the same spherical surface where the accretion rate is calculated.}
    \label{fig: CBD vs BBH handoff}
\end{figure*}

\smallskip
To ensure that the hand-off step (taken at ${99200\,M}$) does not introduce any unphysical features into the system's evolution, we performed several consistency checks. Figure~\ref{fig: Lorentz factor and b2 hand-off comparison} shows that in the equatorial plane, the distributions of the Lorentz factor (upper panels) and the squared magnitude of the fluid-frame magnetic field ${b^2}$ (lower panels) remain unchanged up to interpolation errors after the hand-off process, as expected for spacetime scalars. The Lorentz factor is reset to its floor value of 1 at very few points, where the interpolation produces nonphysical or nonacceptable results (see Sec.~\ref{subsubsec: Interpolation of the MHD primitives}). Square-shaped features are visible in the post-hand-off ${b^2}$ distribution at the boundaries of the refinement levels, but they dissolve very quickly once the evolution starts. The same panel shows that the smoothing procedure applied to the magnetic vector potential outside the inner cavity (see again Sec.~\ref{subsubsec: Interpolation of the MHD primitives}) results in a radially smoothed distribution of ${b^2}$ that partially preserves the magnetic field topology, albeit introducing some artificial field between ${r=15\,M}$ and ${r=35\,M}$.

\smallskip
Next, we turn to comparing the CBD and numerical relativity simulations in a more quantitative way. For the sake of conciseness and readability, in the remainder of this section we label the former \texttt{runCBD} and the latter \texttt{runBBH}. We compare the mass accretion rate~\eqref{eq: Mdot vs t} in the two runs over a spherical surface with coordinate radius ${20\,M}$, as well as the mass contained within a sphere extending to ${r=30\,M}$, i.e.,
\begin{equation}
    M\!\left(t\right)\equiv\int_V dV\sqrt{-g}\,\rho u^t\;,
\end{equation}
where ${V}$ is the spherical volume bounded by ${r=30\,M}$. In Fig.~\ref{fig: CBD vs BBH handoff}, both quantities for both runs are displayed as functions of time for the first ${\sim\!2300\,M}$ of evolution after the hand-off. The curves agree with each other very well at the time of hand-off, confirming the accuracy of the interpolation of the MHD primitives and the spacetime setup; at later times, though, both the accretion rate and the mass of \texttt{runBBH} are lower than their equivalents in \texttt{runCBD} by a few tens of percent, as also reported by~\cite{Avara2024}.

A small part of the discrepancy is due to the slightly different spacetimes used in the two simulations, which enter the accretion rate and mass measurements through the volume element ${\sqrt{-g}}$\,; however, the observed differences early after the hand-off can be mainly ascribed to interactions between the accretion streams and the minidisks to which its cut-out makes \texttt{runCBD} blind, but are described well by \texttt{runBBH} (at later times, these differences grow even larger, but that is mostly because the binary of \texttt{runBBH} progressively accelerates its inspiral). As a result of these interactions, the distribution of gas density in the gap is altered, substantially changing the radial density of gravitational torque
\begin{equation}
    \label{eq: radial density of gravitational torque}
    -\frac{dT}{dr}\!\left(t\right)\equiv -\int_S d^2\sigma\,\sqrt{-g}\,T^{\mu\nu}\partial_\phi g_{\mu\nu}\;.
\end{equation}
In Fig.~\ref{fig: CBD vs BBH handoff}, we demonstrate this contrast by calculating~\eqref{eq: radial density of gravitational torque} on the same spherical surface where the accretion rate was computed. The torque density is nearly always negative in \texttt{runCBD}, which means fluid elements in the spherical shell around ${r=20\,M}$ overall lose angular momentum to the neighboring shells at larger radii, thereby gaining an inward radial velocity component and causing the accretion rate at ${r=20\,M}$ to increase. By contrast,~\eqref{eq: radial density of gravitational torque} is nearly always positive in \texttt{runBBH}, leading to the opposite result and invalidating the assumption that fluid flow is exclusively inward at ${r=15\,M}$. The reader is referred to, e.g., Appendix C of~\cite{Noble2012} for an in-depth analysis of the relation between~\eqref{eq: radial density of gravitational torque} and the fluid's angular momentum budget.

Given these remarks, we conclude that excising a relatively small portion of the computational domain during the CBD evolution and imposing purely inflow boundary conditions at the inner radial grid boundary (${r=15\,M}$, just outside the black holes' Hills spheres) does not fully capture the complexity of the gas dynamics in the vicinity of the binary. However, for the purposes of this paper, this limitation is not a concern because the main goal of our CBD simulation is to construct a realistic circumbinary disk whose evolved state provides the initial data for our subsequent numerical relativity simulation.

\subsection{Inspiral and merger accretion dynamics}
\label{subsec: Inspiral and merger accretion dynamics}

\subsubsection{Overview of structural evolution}
\label{subsubsec: Overview of structural evolution}

As briefly mentioned in the introduction, all previous efforts studying gas inflow in the inspiral to merger had serious limitations. Some focused on a relatively limited number of binary orbits early in the inspiral without exploring the physics of the system close to and at merger~\cite{Bowen2017, Bowen2019, Combi2022, Paschalidis2021, Bright2023}. Others evolved the system through the merger, but with an initial configuration of the matter and EM fields having little connection to prior evolutionary stages~\cite{Farris2012, Gold2014a, Gold2014b, Cattorini2021, Cattorini2022, Fedrigo2023, Cattorini2024}. Another made drastic simplifications, considering Newtonian 2D hydro with phenomenological internal stress, introducing GR only for the binary's orbital evolution~\cite{Krauth2023}. A notable exception is~\cite{Avara2024}, which employed a novel multi-patch code infrastructure in order to evolve an equal-mass, circular, nonspinning SMBBH system from an initial coordinate separation ${20\,M}$ down to a coordinate separation of ${9M}$, thus lacking only the last ${\sim\!1000\,M}$ in time before merger. The simulation in~\cite{Avara2024} could not reach merger due to its use of a post-Newtonian approximation to the actual spacetime metric; in the work we report here, this limitation is overcome by employing full NR.

\begin{figure}[htp]
    \centering
    \includegraphics[width = \linewidth]{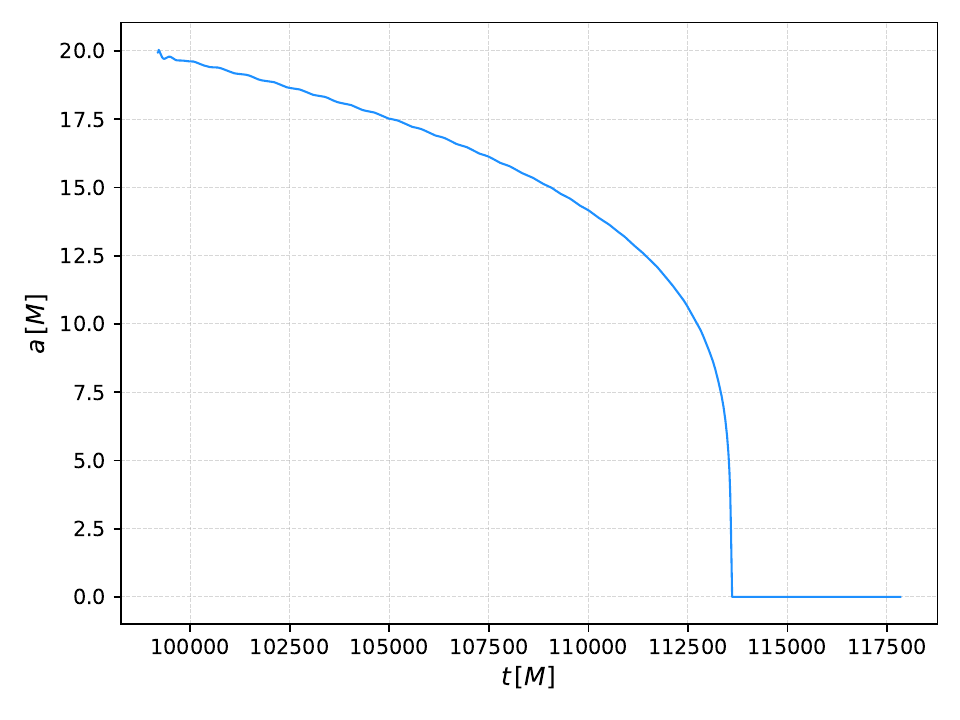}
    \captionsetup{justification = raggedright, format = hang}
    \caption{Binary separation in coordinate distance as a function of time.}
    \label{fig: binary separation}
\end{figure}

\smallskip
Throughout the period we simulate, the binary separation diminishes due to the emission of gravitational radiation; the relation between the coordinate separation and coordinate time is shown in Fig.~\ref{fig: binary separation}. As shown in this plot, the merger takes place at ${t\simeq 113600\,M}$.

\begin{figure*}[htp]
    \centering
    \begin{subfigure}[t]{0.49\textwidth}
        \includegraphics[width = \linewidth, valign = t]{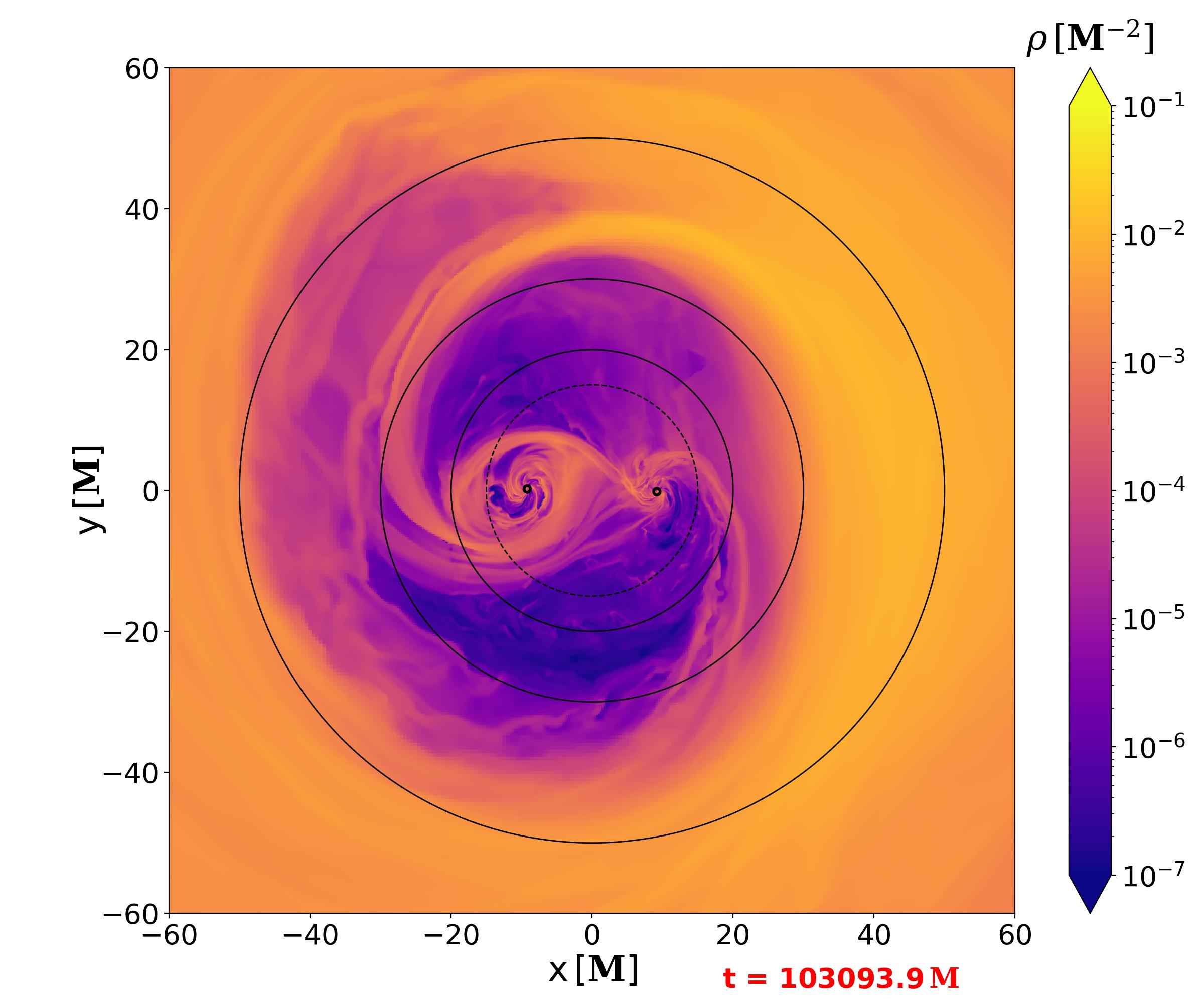}
        \captionsetup{justification = raggedright, format = hang}
        \caption{Early inspiral: the minidisk on the left (right) is in the ``disklike'' (``stream-like'') state.}
    \end{subfigure}
    \begin{subfigure}[t]{0.49\textwidth}
        \includegraphics[width = \linewidth, valign = t]{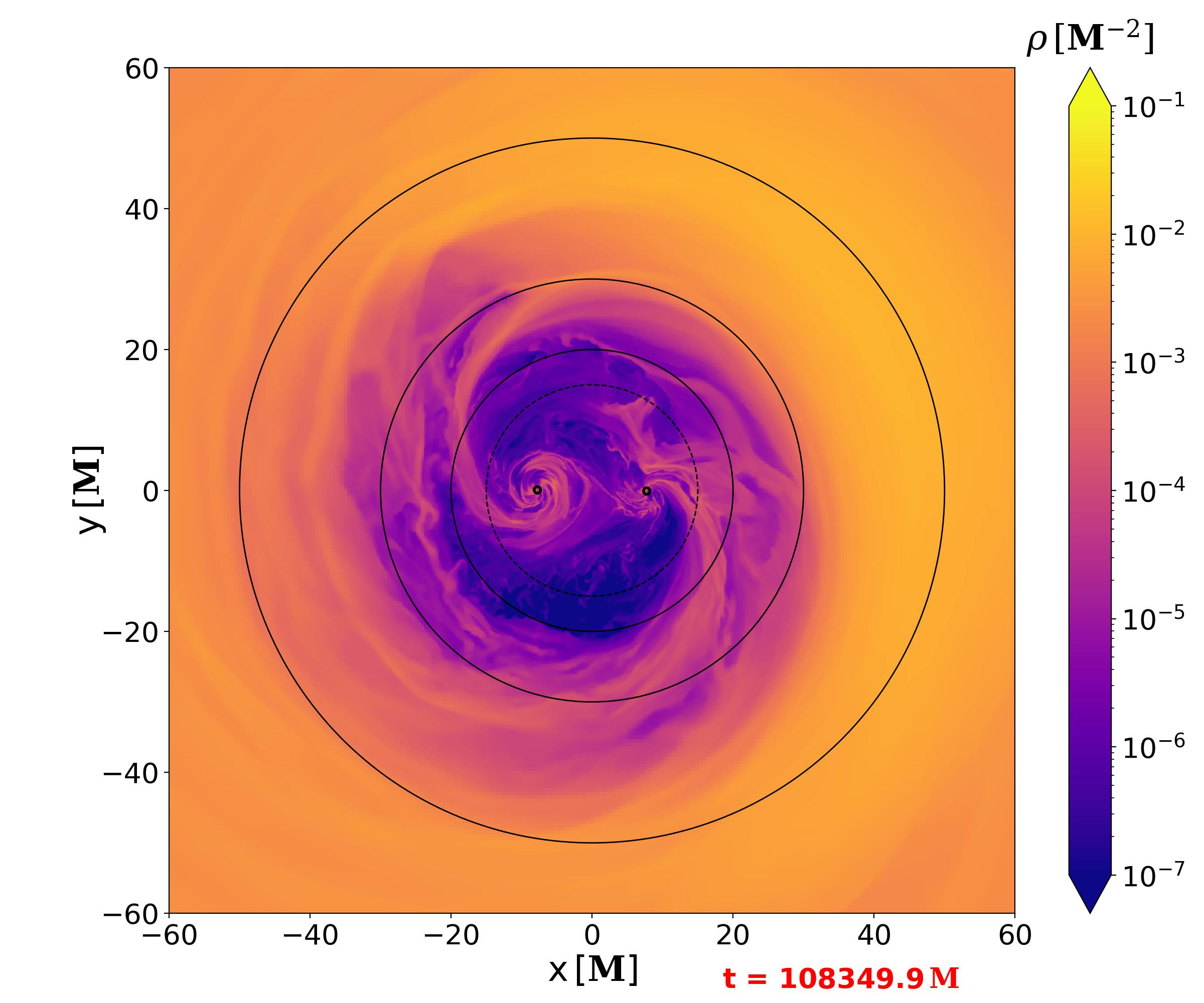}
        \captionsetup{justification = raggedright, format = hang}
        \caption{Half-way through the inspiral: the minidisk on the left is still in the ``disklike'' state, but contains relatively little mass. The inner edge of the CBD has drifted inward and a few irregular gas structures detach from it.\vspace{2\baselineskip}}
    \end{subfigure}
    \begin{subfigure}[c]{0.49\textwidth}
        \includegraphics[width = \linewidth]{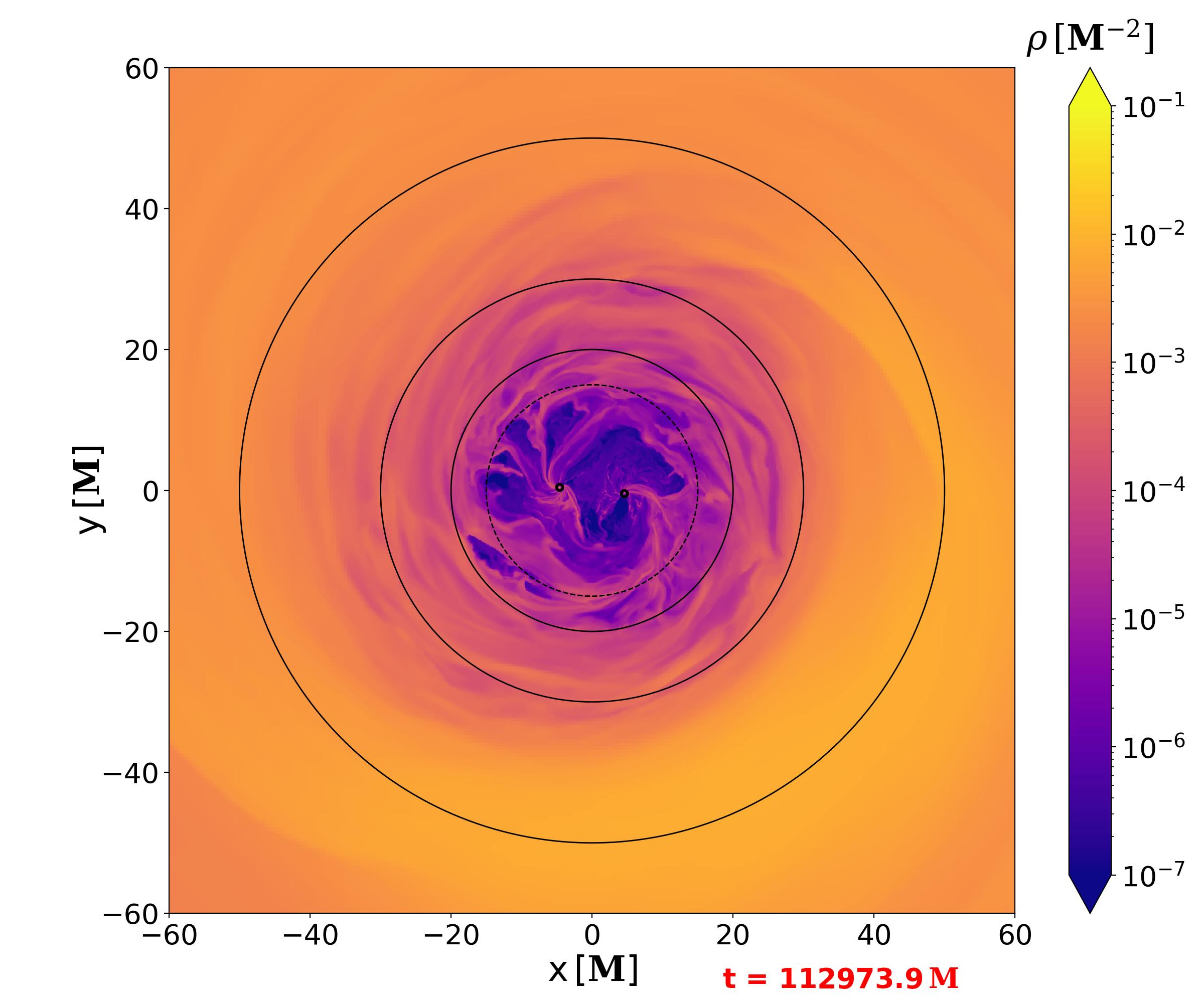}
        \captionsetup{justification = raggedright, format = hang}
        \caption{Late inspiral: the CBD's inner edge is now hard to define sharply, more irregular gas structures detach from it and wander through the cavity, and ``disklike'' minidisks cease to exist, thereby stopping sloshing from operating.}
    \end{subfigure}
    \begin{subfigure}[c]{0.49\textwidth}
        \includegraphics[width = \linewidth]{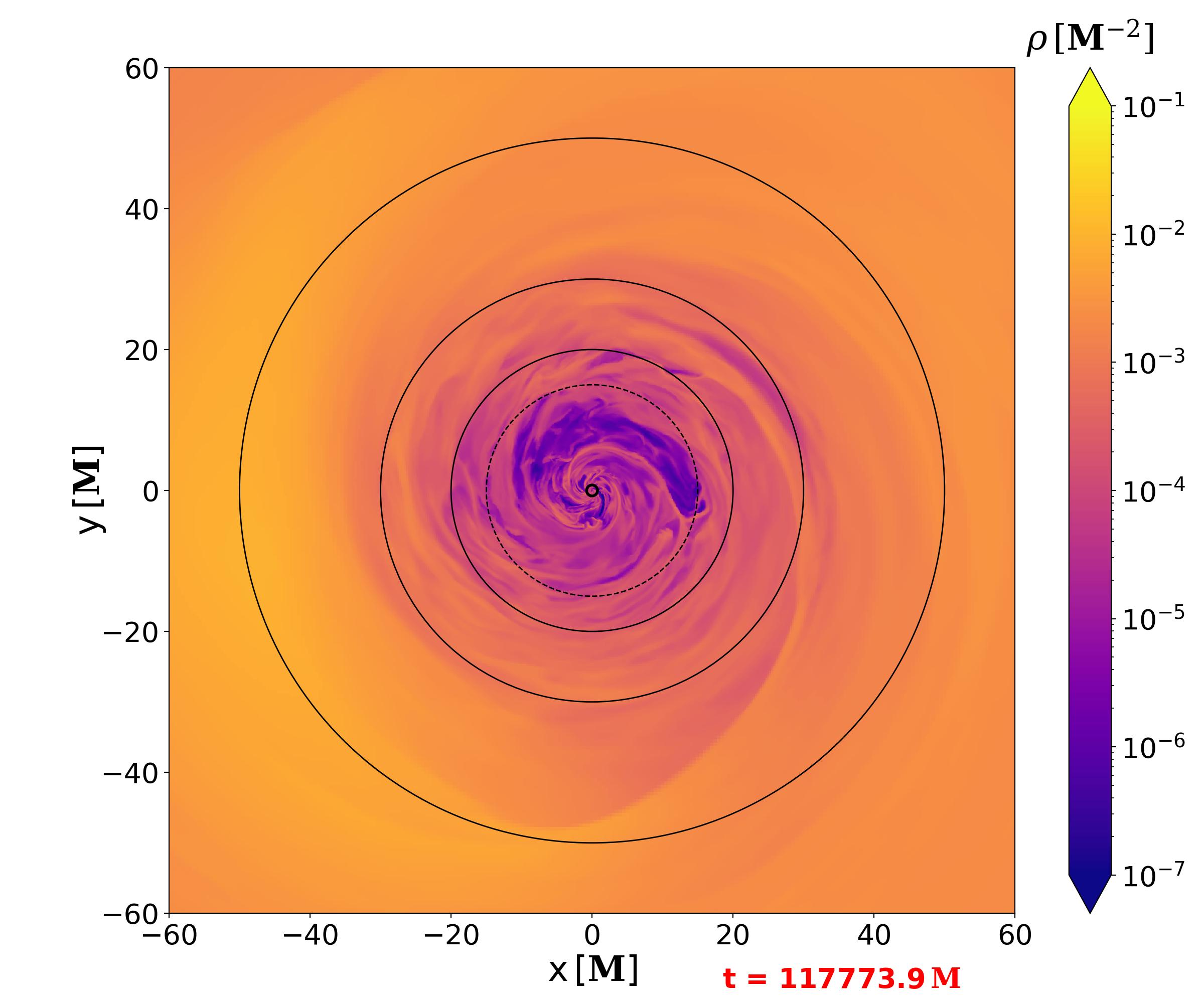}
        \captionsetup{justification = raggedright, format = hang}
        \caption{${\sim\!4200\,M}$ in time after merger, the gas hasn't yet fully repopulated the innermost region. Low-density ``bubbles'' of highly-magnetized gas are clearly visible close to the remnant.}
    \end{subfigure}
    \captionsetup{justification = raggedright, format = hang}
    \caption{Equatorial slices of the mass density distribution from the early inspiral to the postmerger stages of the NR evolution. Time advances from left to right and from top to bottom. Black circular lines mark the ${15\,M}$, ${20\,M}$, ${30\,M}$, and ${50\,M}$ radii. Note the progressive spreading of the CBD's inner edge into the gap region during the inspiral and the almost-relaxed disk structure postmerger.}
    \label{fig: evolution of mass density}
\end{figure*}

\smallskip
Figure~\ref{fig: evolution of mass density} shows face-on views of the mass density distribution around the binary for three times during the inspiral and once during the postmerger stage. Early in the evolution, each minidisk oscillates between a \textit{``disklike''} regime, with maximum disk mass and minimum accretion rate on its black hole, and a \textit{``stream-like''} regime, with only a thin thread of matter streaming toward its black hole, supplying a maximum accretion rate and supporting a minimum mass.\footnote{For further details about disk structure in these two regimes, see~\cite{Avara2024}.}
The first panel in this figure shows a moment in which the minidisk on the left is in the former regime and the one on the right is in the latter. This panel also shows how mass is exchanged through the center-of-mass region, a phenomenon called ``sloshing'' (Sec.~\ref{sec: Introduction}). The second and third panels illustrate how, as the inspiral proceeds, coherent minidisks are dissolved and the density profile of the CBD's inner edge, rather than forming a sharp boundary, evolves into an irregular but steadily more gradual decline that reaches closer and closer to the center of mass. At ${\sim\!4200\,M}$ after merger (the time illustrated in the fourth panel), gas has partially refilled the gap region, but still exhibits strong irregularities as far out as ${r\simeq 30\,M}$.

\begin{figure*}[htp]
    \centering
    \begin{subfigure}[c]{0.32\textwidth}
        \includegraphics[width = \linewidth]{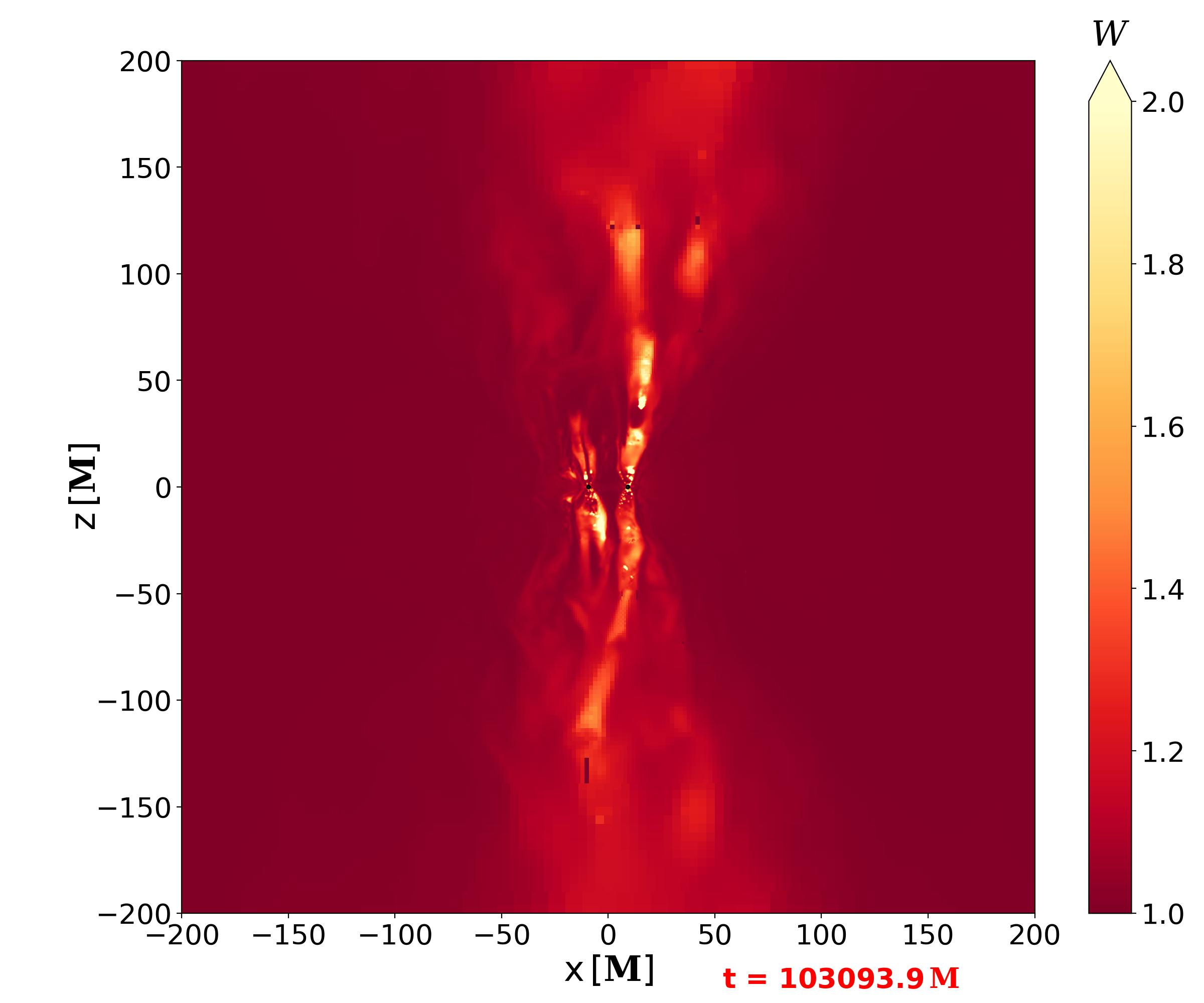}
    \end{subfigure}
    \begin{subfigure}[c]{0.32\textwidth}
        \includegraphics[width = \linewidth]{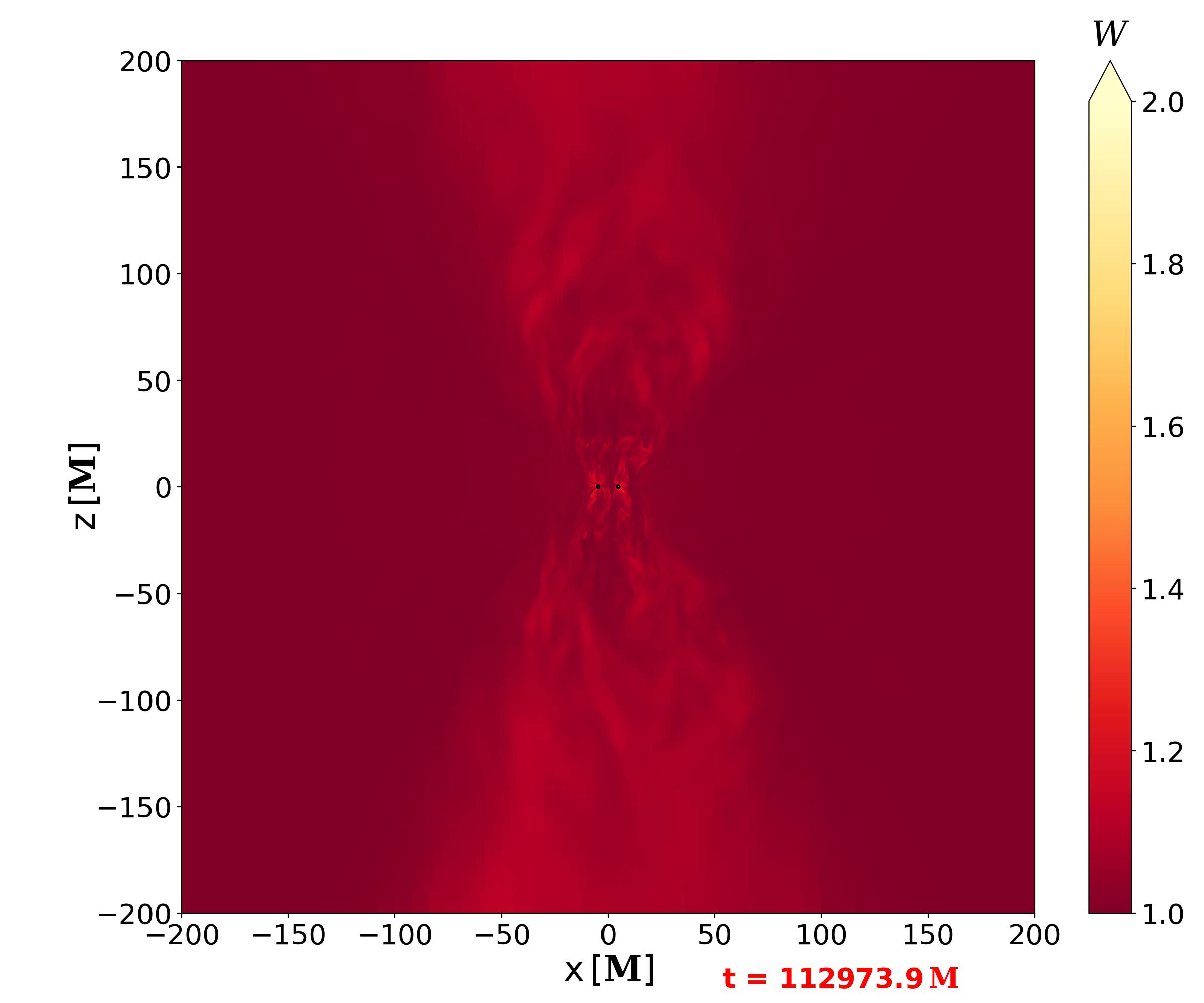}
    \end{subfigure}
    \begin{subfigure}[c]{0.32\textwidth}
        \includegraphics[width = \linewidth]{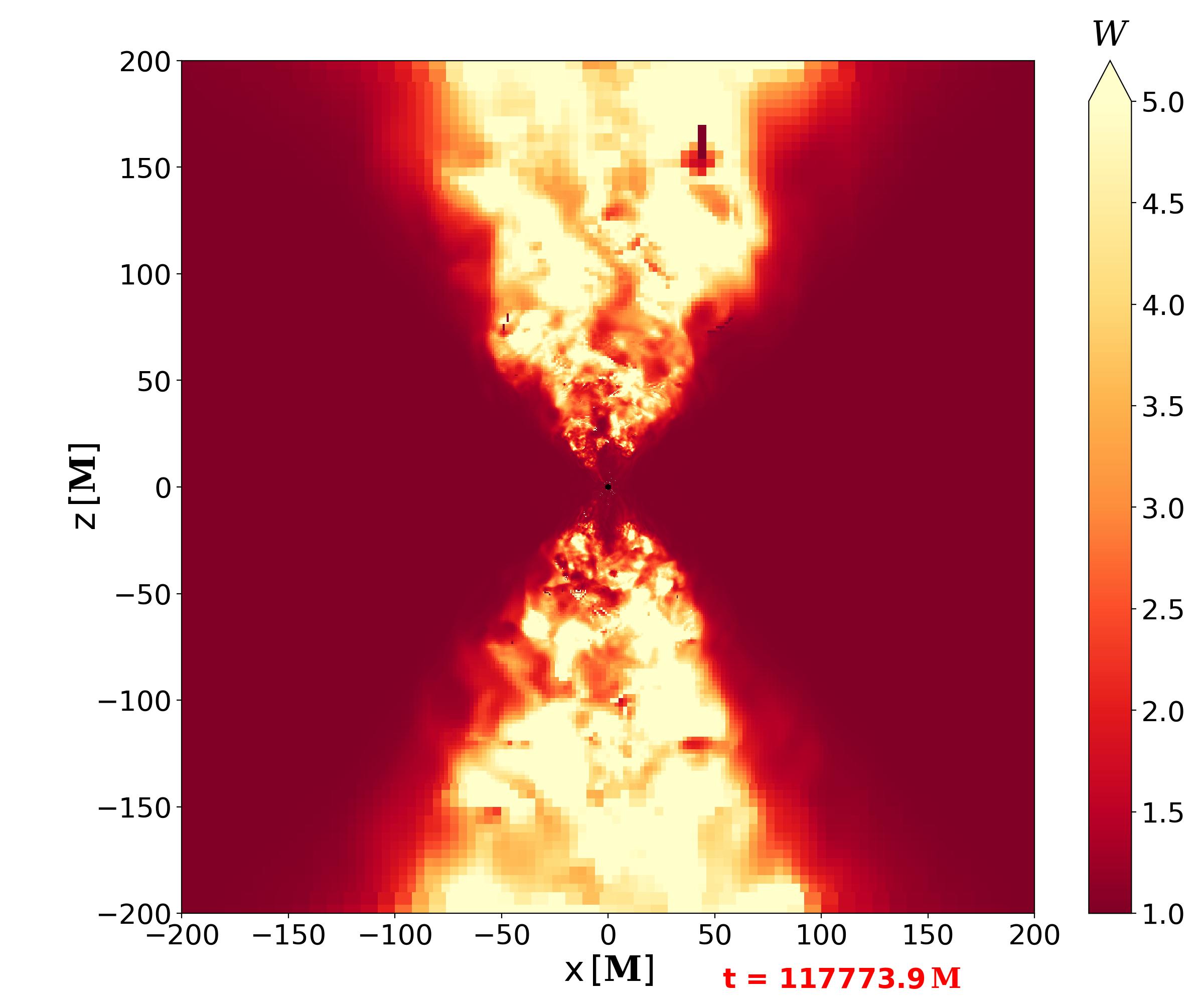}
    \end{subfigure}
    \begin{subfigure}[c]{0.32\textwidth}
        \includegraphics[width = \linewidth]{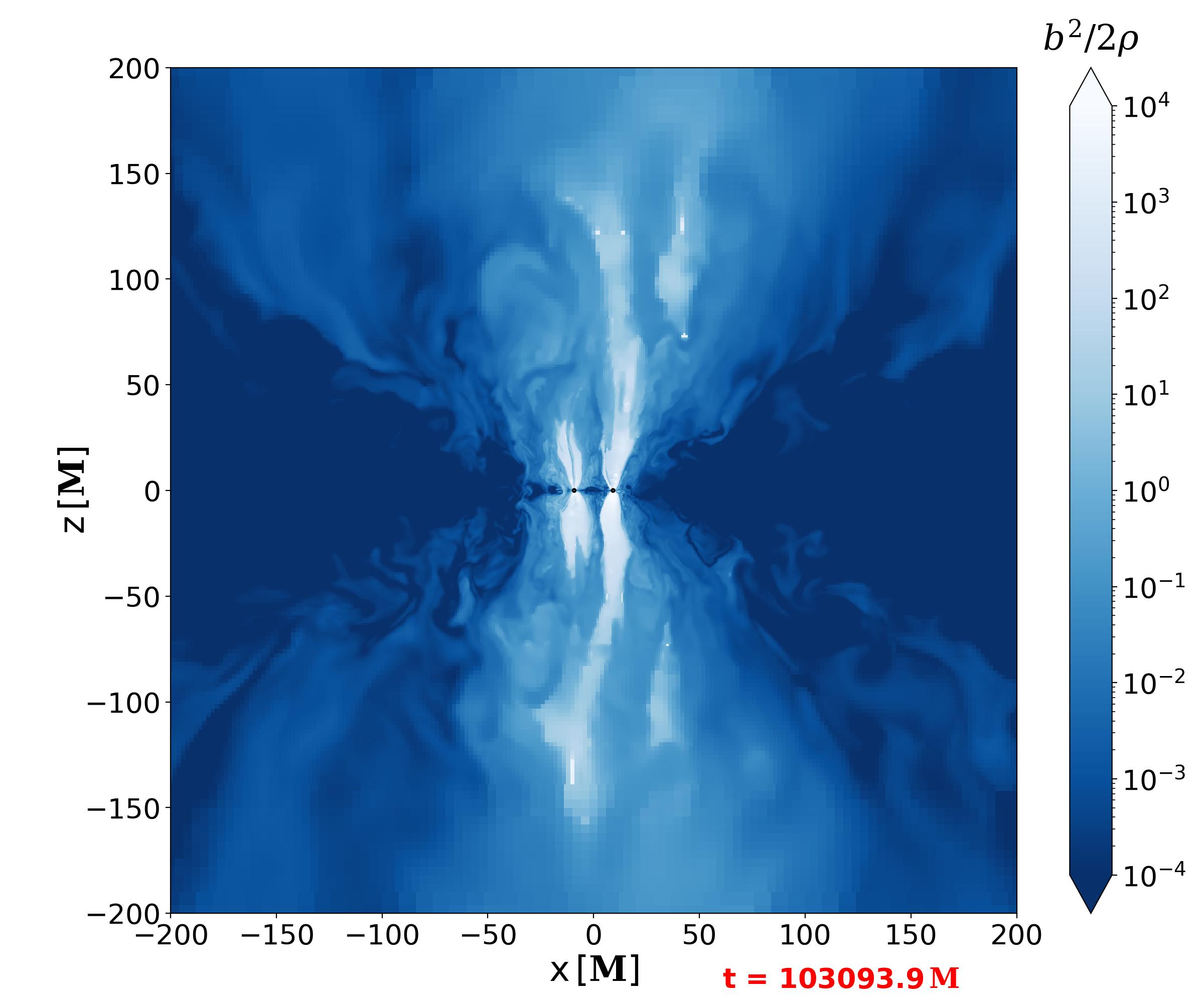}
    \end{subfigure}
    \begin{subfigure}[c]{0.32\textwidth}
        \includegraphics[width = \linewidth]{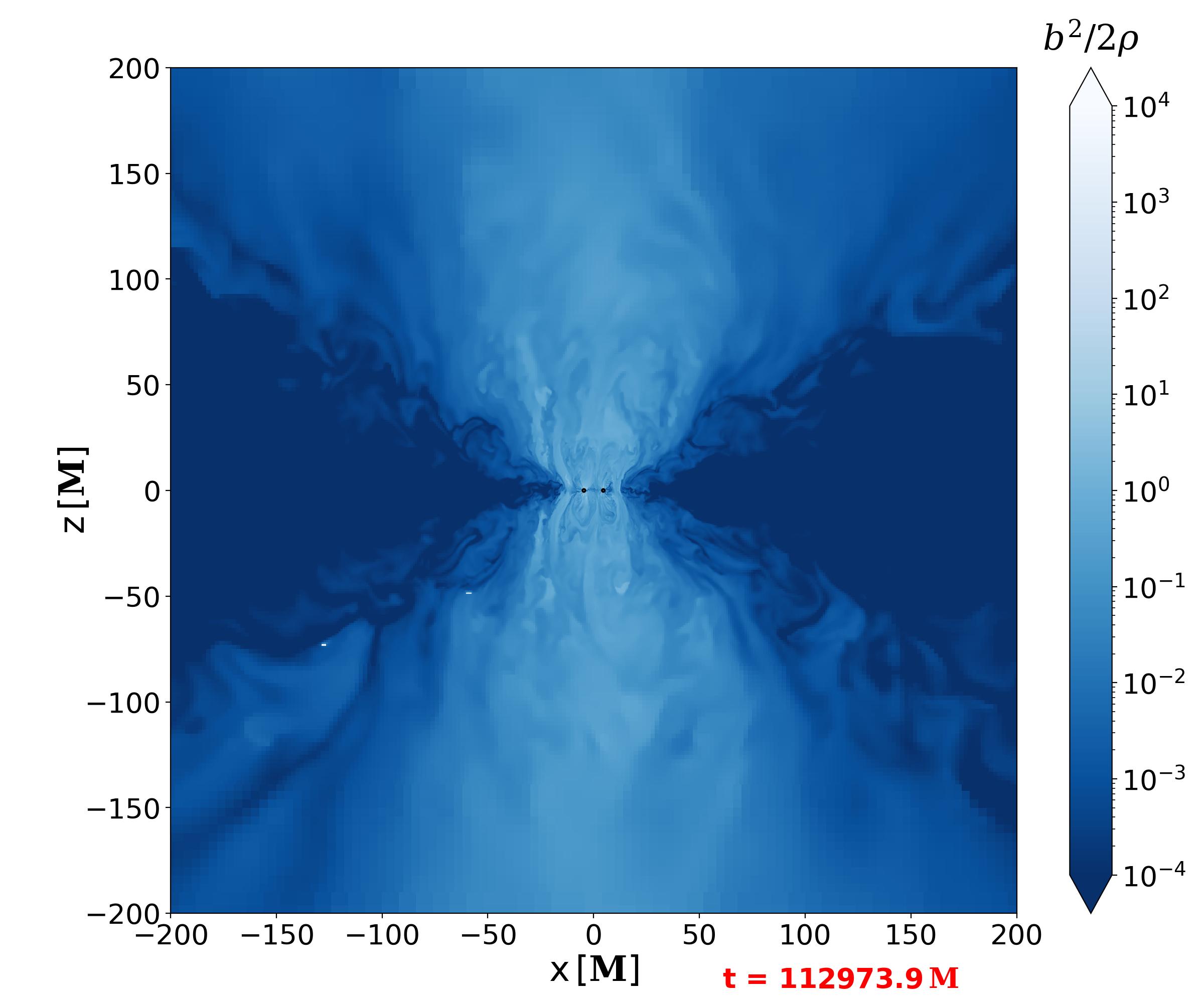}
    \end{subfigure}
    \begin{subfigure}[c]{0.32\textwidth}
        \includegraphics[width = \linewidth]{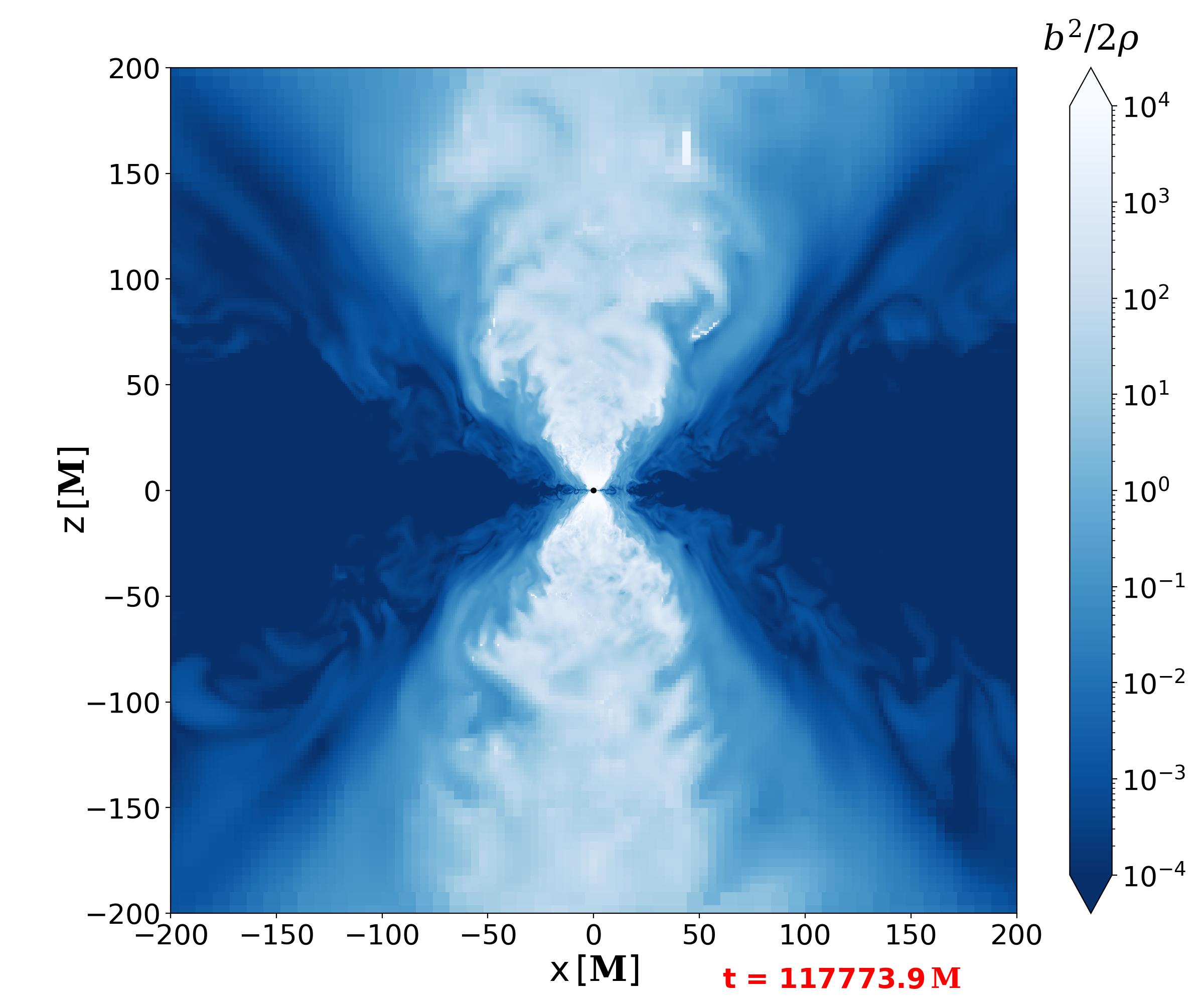}
    \end{subfigure}
    \captionsetup{justification = raggedright, format = hang}
    \caption{Polar slices of the distributions of the Lorentz factor (top panels) and magnetization ${b^2/2\rho}$ during the early inspiral (left), late inspiral (center), and postmerger (right) stages. Note that we extended the colorbar limits in the plot of the postmerger distribution of the Lorentz factor (top-right panel).}
    \label{fig: Lorentz factor and magnetization evolution}
\end{figure*}

\begin{figure*}[htp]
    \centering
    \begin{subfigure}[c]{0.32\textwidth}
        \includegraphics[width = \linewidth]{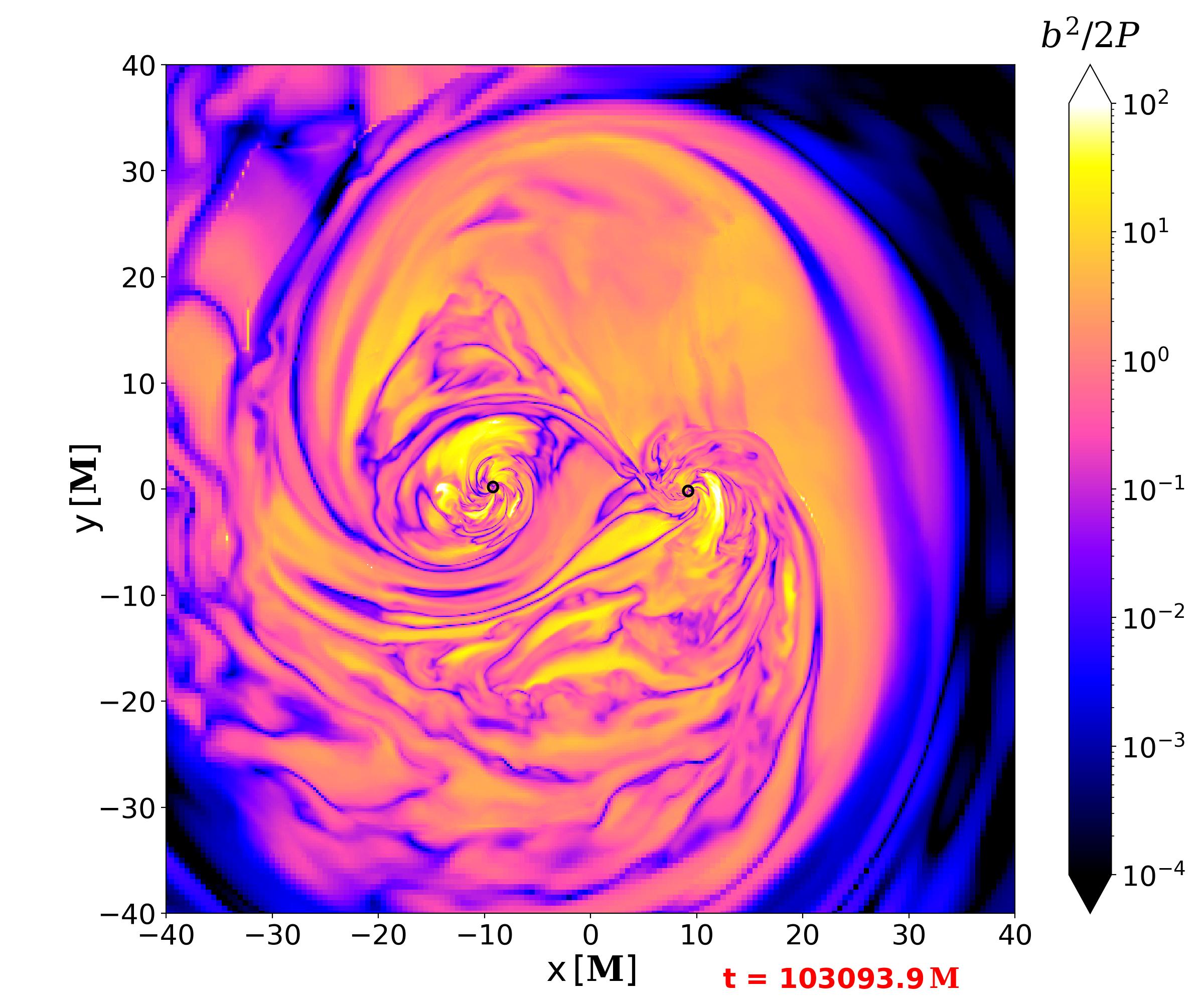}
    \end{subfigure}
    \begin{subfigure}[c]{0.32\textwidth}
        \includegraphics[width = \linewidth]{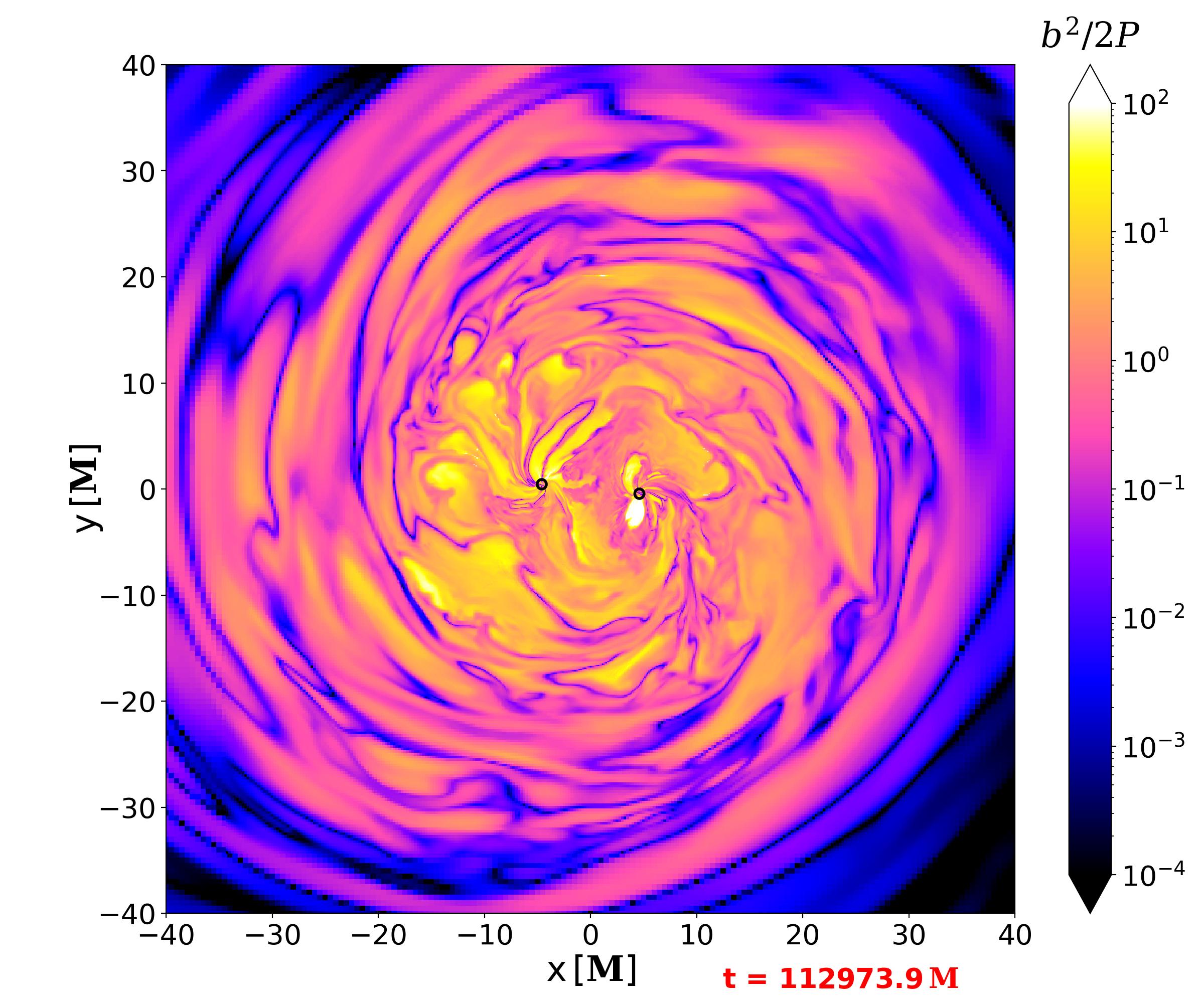}
    \end{subfigure}
    \begin{subfigure}[c]{0.32\textwidth}
        \includegraphics[width = \linewidth]{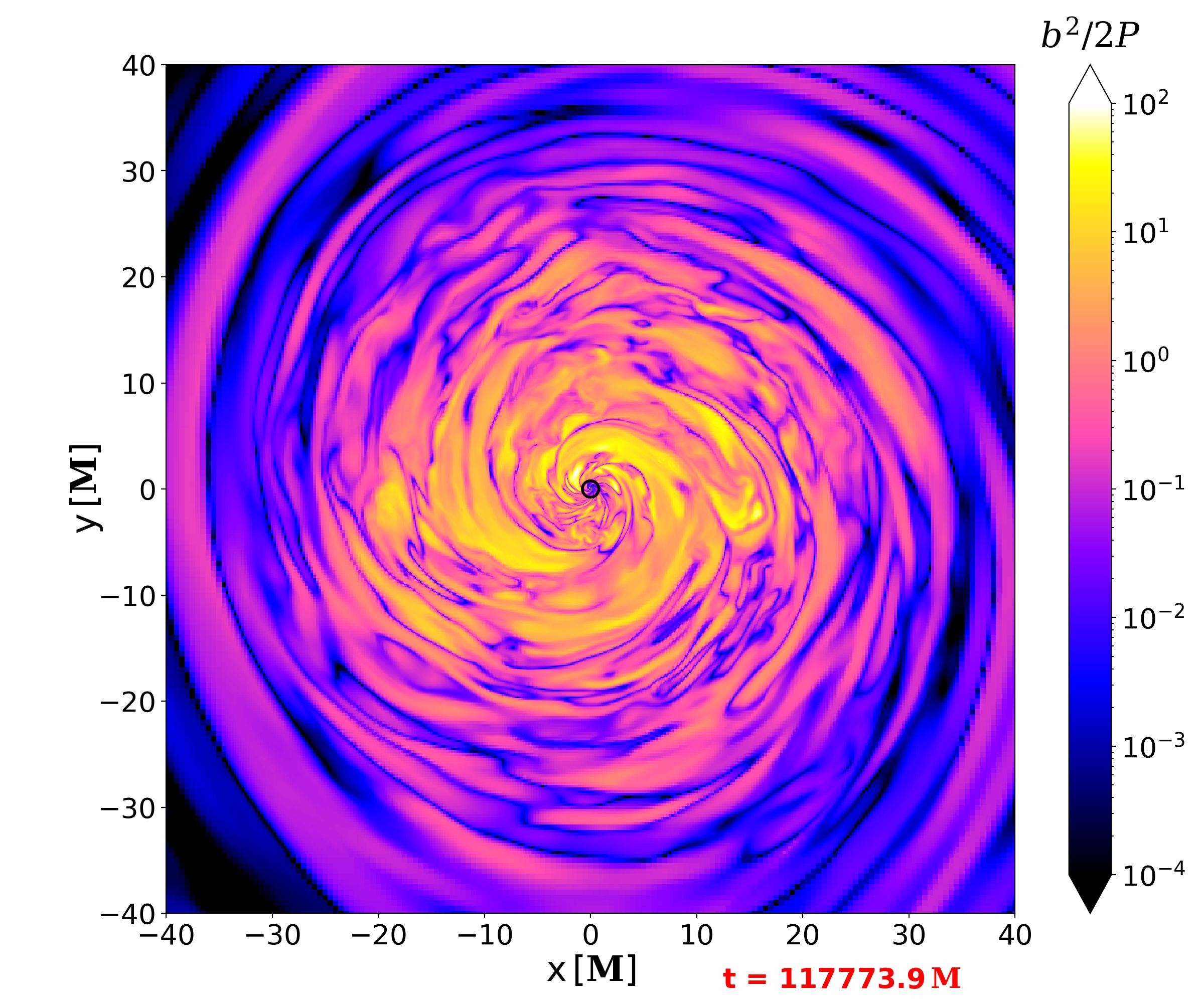}
    \end{subfigure}
    \captionsetup{justification = raggedright, format = hang}
    \caption{Equatorial slices of inverse plasma beta ${\beta^{-1}\equiv b^2/2P}$ during the early inspiral (left), late inspiral (center), and postmerger (right) stages.}
    \label{fig: evolution of plasma beta}
\end{figure*}

As matter arrives at the black holes, it carries magnetic flux that becomes attached to the black holes' horizons. In the funnel regions above and below the binary, each black hole generates a mildly relativistic outflow (top-left panel of Fig.~\ref{fig: Lorentz factor and magnetization evolution}); although magnetically dominated (bottom-left panel of Fig.~\ref{fig: Lorentz factor and magnetization evolution}), these outflows carry little Poynting flux, as they are generated by nonspinning black holes (see Sec.~\ref{subsubsec: Poynting flux} for the origin of the Poynting flux seen at this time). Throughout the cavity region, the magnetic field quickly grows strong enough to be dynamically significant; as shown in Fig.~\ref{fig: evolution of plasma beta}, in much of the cavity's volume, the inverse plasma ${\beta^{-1}\equiv b^2/2P\lesssim O\!\left(1\right)}$. The low-density regions within the cavity easily visible in Fig.~\ref{fig: evolution of mass density} are carved out by magnetic pressure.

After merger, a single, rotating black hole with dimensionless spin parameter ${\sim\!0.68}$ forms. Although the accretion rate remains close to the sum of the accretion rates onto the two black holes, the cooling luminosity increases sharply immediately after merger. Almost simultaneously, the newly acquired spin of the black hole combines with the magnetic flux it inherited from the pair of nonspinning black holes to launch a relativistic, Poynting flux-dominated jet (right panels of Fig.~\ref{fig: Lorentz factor and magnetization evolution}).

\begin{figure*}[htp]
    \centering
    \includegraphics[width = \textwidth]{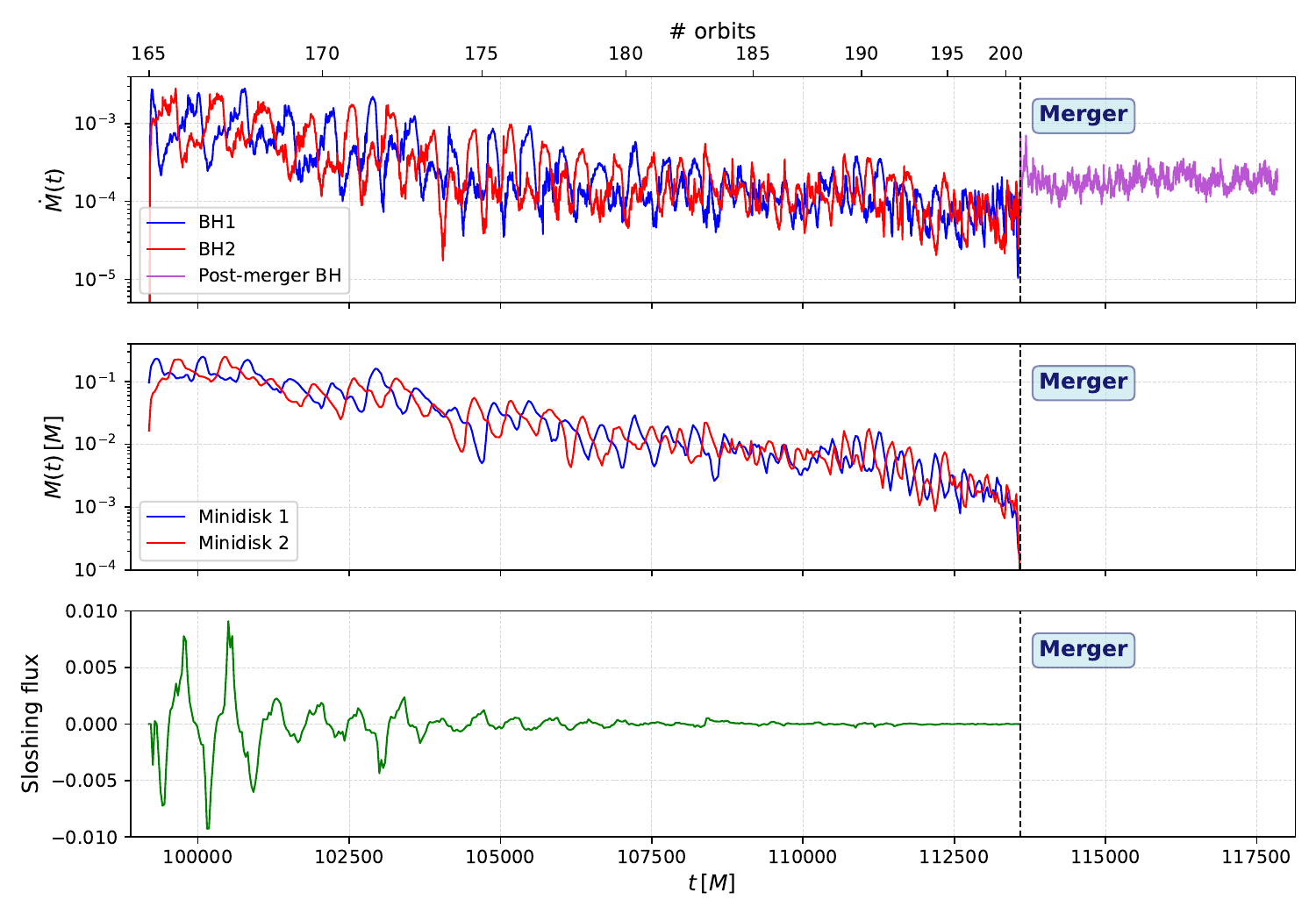}
    \captionsetup{justification = raggedright, format = hang}
    \caption{\textbf{[Top panel]} Mass accretion rate on the black holes' apparent horizons. \textbf{[Middle panel]} Minidisk masses, with minidisk regions being defined as spherical volumes around each puncture with radius ${r=0.45\,a}$, where ${a}$ is the coordinate orbital separation. The apparent horizons are excluded from the calculation. \textbf{[Bottom panel]} Mass sloshing flux through a flat, square surface with edge length ${0.8\,a}$ and orthogonal to the line connecting the two punctures. Positive flux means the flow is mainly directed from ``minidisk 2'' (red curve in the middle panel) to ``minidisk 1'' (blue curve in the middle panel). }
    \label{fig: mdot AH, minidisk mass, sloshing flux}
\end{figure*}

\subsubsection{Mass motions}
\label{subsubsec: Mass motions}

After this overview of the structural changes in the accretion flow induced by the inspiral and merger, we now turn to a more detailed account of how the flow of mass from the CBD to the black hole(s) changes during this period.

There are several noteworthy long-term trends. At the beginning of the simulation, the accretion rate onto the black holes oscillates in the range ${\sim\!2\text{--}3\times 10^{-3}}$, but beginning at ${t\simeq 101500\,M}$ it begins to decrease. From ${t\simeq 106000\,M}$ to ${t\simeq 111500\,M}$, the range of variation is again roughly constant, at ${2\text{--}4\times 10^{-4}}$, but then drops a factor ${\sim\!2}$ at ${t\simeq 112000\,M}$, where it remains for at least ${\sim\!5000\,M}$ past the merger.

As the binary shrinks, the outer edge of the very low density portion of the cavity drifts inward from ${r\simeq 50\,M}$ to ${r\simeq 20\,M}$ by ${t\simeq 112000\,M}$; by this time, it is difficult to define this edge. Outside what remains of the very low density region, there are irregular gas structures of intermediate density (Fig.~\ref{fig: evolution of mass density}). Rather than the earlier pattern comprising a pair of well-defined accretion streams, multiple streams break off from these irregular gas structures and spiral toward one or the other of the black holes, in some cases interacting with each other. It is at this point, i.e., ${\sim\!2000\,M}$ before merger, that regular minidisks disappear altogether, replaced by chaotic infalling streams; in fact, the middle panel of Fig.~\ref{fig: mdot AH, minidisk mass, sloshing flux} shows that the minidisk masses decrease by a factor ${\sim\!100}$ from the beginning of the inspiral evolution to merger. However, the total mass within a distance ${40\,M}$ of the binary (Fig.~\ref{fig: mass inside 15M, 15-30M, 30-40M}), even while oscillating across a factor of ${\sim\!2}$, remains roughly constant. In other words, the amount of mass within the gap region hardly changes during the inspiral, but begins this phase mostly concentrated in the minidisks and becomes progressively more diffuse as merger approaches.

The dynamics of minidisk destruction are complex. As we will describe in greater detail shortly, sloshing streams can begin the process by keeping the minidisks in the stream-like state. In addition, as the inner edge of the CBD stretches inward, the cavity density rises and reaches levels at which it can exert significant forces on the depleted minidisks (see panels b) and c) of Fig.~\ref{fig: evolution of mass density}). Plus, as the binary contracts, the tidal truncation radii of the minidisks, which are always $\propto a$, shrink until they are not much larger than their black holes' ISCOs. The upshot of all three of these mechanisms is that when ${q=1}$ and the black holes do not spin, the minidisks disappear ${\sim\!2000\,M}$ before the merger.

\smallskip
Due to the changes in accretion rate and the decay of the minidisks, the mass of matter close to the binary (a region we define as ${r\leq 15\,M}$) declines by roughly a factor of ${10}$ over the first ${\sim\!10500\,M}$ of the inspiral (i.e., from ${t\simeq 99200\,M}$ to ${t\simeq 109700\,M}$), but in the ${\sim\!4000\,M}$ from the end of that period to the merger recovers about half of that decline (see again Fig.~\ref{fig: mass inside 15M, 15-30M, 30-40M}). That the mass of this innermost region changes so greatly demonstrates that the accretion rate onto the black holes is significantly different from the accretion rate through a spherical surface ${15\,M}$ from the binary center of mass; the black hole accretion rate is considerably greater than the mass inflow rate during the first part of the inspiral and consistently smaller during the second part.

As shown by the swing in mass within ${30\,M}$ (Fig.~\ref{fig: mass inside 15M, 15-30M, 30-40M}), there is an even larger difference in the accretion rates through surfaces at ${r=30\,M}$ and ${r=40\,M}$. In fact, in the late inspiral, the accretion rate at ${r=30\,M}$ is ${\sim\!3\times}$ the rate at both ${r=40\,M}$ and ${r=20\,M}$. Thus, during the inspiral period the cavity region is far from a state of inflow equilibrium, and the sense of ${d|\dot{M}|/dr}$ in the cavity switches from negative to positive in the middle of the inspiral.

\begin{figure*}[htp]
    \centering
    \includegraphics[width = \textwidth]{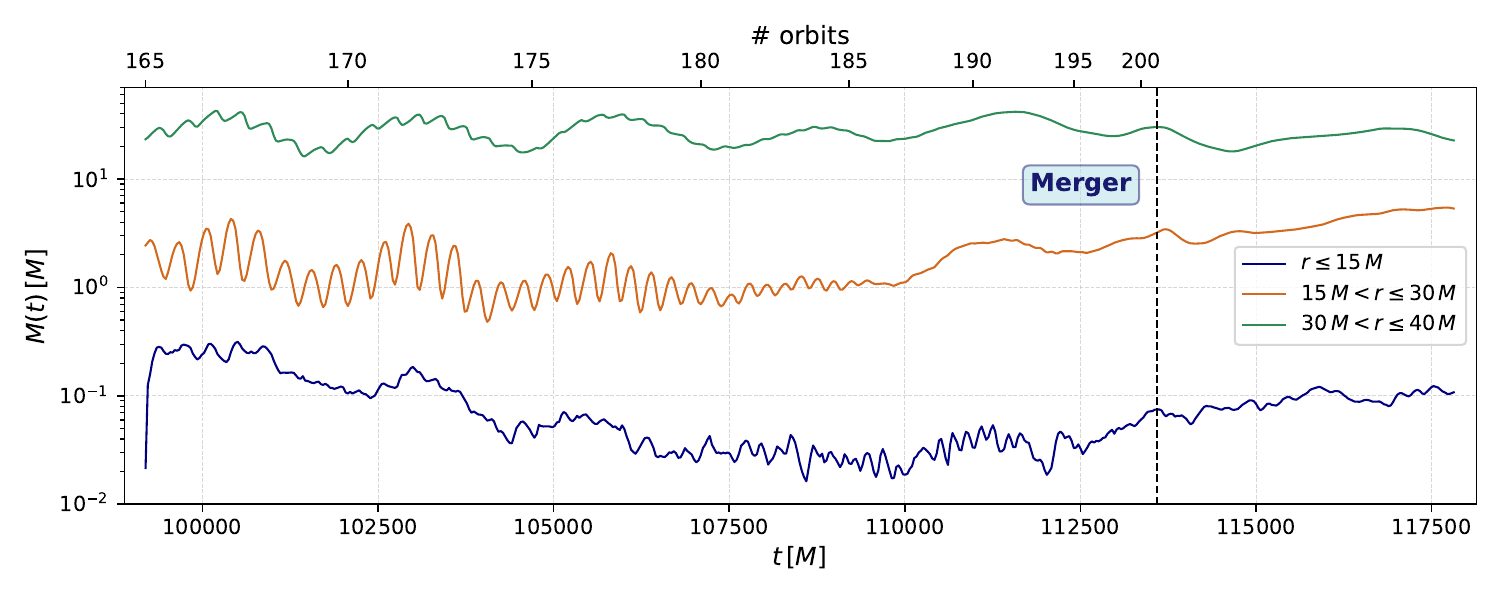}
    \captionsetup{justification = raggedright, format = hang}
    \caption{Masses in a spherical volume of coordinate radius ${r=15\,M}$ (blue) and in spherical shells ${15\,M<r\leq 30\,M}$ (orange) and ${30\,M<r\leq 40\,M}$ (green) around the binary's center of mass.}
    \label{fig: mass inside 15M, 15-30M, 30-40M}
\end{figure*}

\smallskip
The long-term trends in accretion rate and minidisk mass are subject to strong periodic modulation.
During the early to mid-inspiral stage of the evolution, peaks in the accretion rate ${\dot{M}_{AH1}}$ onto one black hole coincide with dips in the accretion rate ${\dot{M}_{AH2}}$ onto the other black hole, and vice versa (top panel of Fig.~\ref{fig: mdot AH, minidisk mass, sloshing flux}); similarly, peaks in the mass ${M_1}$ of one minidisk are simultaneous with dips in the mass ${M_2}$ of the other minidisk (middle panel of the same figure), and vice versa. Moreover, in the first ${\sim\!8000\,M}$ of the evolution, ${\dot{M}_{AH1}}$ is largest (smallest) when ${M_1}$ is smallest (largest), and the same goes for ${\dot{M}_{AH2}}$ and ${M_2}$\,: this behavior confirms the alternating ``stream-like'' and ``disklike'' nature of the minidisks during their early inspiral evolution~\cite{Avara2024}. However, later in the inspiral and up to ${t\simeq 112500\,M}$, the masses of the minidisks vary consistently with opposite phases, whereas the accretion rates onto the black holes occasionally synchronize.

Both the average binary orbital frequency ${f_\text{orb}}$ and the lump's orbital frequency ${f_\text{lump}\simeq 0.3\,f_\text{orb}}$ are imprinted on the time-dependence of ${\dot{M}_{AH1}}$\,, ${\dot{M}_{AH2}}$\,, ${M_1}$\,, and ${M_2}$ during the inspiral stage. All four of these quantities are modulated at the beat frequency ${f_\text{beat} = f_\text{orb} - f_\text{lump}}$ and at ${2 f_\text{beat}}$\,, as expected for equal mass black hole binaries~\cite{Bowen2017, Bowen2019} and confirmed by Fig.~\ref{fig: PSD Mdot, minidisk masses, sloshing flux}; in addition, we find a modulation at a frequency slightly lower than ${f_\text{lump}}$\,. Because ${\dot{M}_{AH1}}$ and ${\dot{M}_{AH2}}$\,, as well as ${M_1}$ and ${M_2}$\,, vary with opposite phases and their amplitudes are comparable, the modulation at the frequency ${f_\text{beat}}$ is suppressed in ${\dot{M}_{AH1} + \dot{M}_{AH2}}$ and ${M_1 + M_2}$ (see again Fig.~\ref{fig: PSD Mdot, minidisk masses, sloshing flux}); on the other hand, the periodic feature at the frequency ${f\lesssim f_\text{lump}}$ in both ${\dot{M}_{AH1} + \dot{M}_{AH2}}$ and ${M_1 + M_2}$ is more pronounced than in the power spectra of the individual quantities. 

\begin{figure*}[htp]
    \centering
    \includegraphics[width = \linewidth]{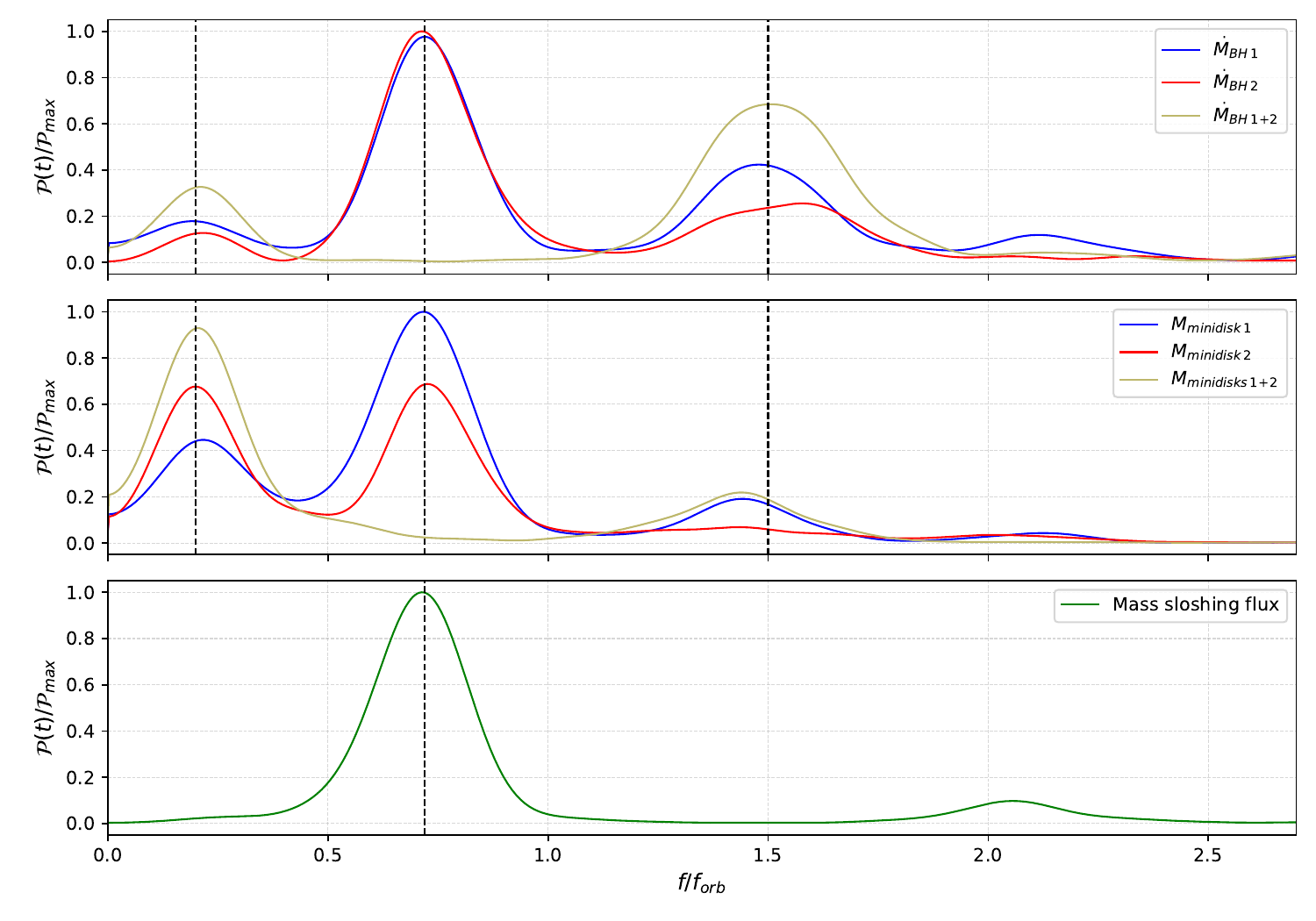}
    \captionsetup{justification = raggedright, format = hang}
    \caption{Power spectra of the quantities shown in Fig.~\ref{fig: mdot AH, minidisk mass, sloshing flux}. \textbf{[Top panel]} Power spectra of the mass accretion rates onto the individual horizons (blue and red lines) and their sum (yellow line). \textbf{[Middle panel]} Power spectra of the individual minidisk masses (blue and red lines) and their sum (yellow line). \textbf{[Bottom panel]} Power spectrum of the mass sloshing flux between the minidisks.\newline
    All of the spectra are computed via Welch's method~\cite{Welch1967} as implemented by the \textsc{scipy.signal.welch} routine from the \textsc{Python} package \textsc{SciPy}, splitting the time series into three overlapping segments and applying a Hamming window to each of them. The analysis only considers the first ${10500\,M}$ of inspiral evolution (i.e., from ${t\simeq 99200\,M}$ to ${t\simeq 109700\,M}$), where regular modulations in most of the signals occur. The frequency unit is the average orbital frequency ${f_\text{orb}}$\,, which is calculated as the peak frequency in the ${l=2\,,\;m=2}$ mode of the ${\Psi^4}$ Weyl scalar from ${t=200\,M}$ to ${t=4200\,M}$ (i.e., early) into the inspiral stage of the evolution. Vertical dashed lines located at ${0.2\,f_\text{orb}\lesssim f_\text{lump}}$\,, ${0.72\,f_\text{orb}\simeq f_\text{beat}}$\,, and ${1.5\,f_\text{orb}\gtrsim 2\,f_\text{beat}}$ mark the strongest peaks in the spectra.}
    \label{fig: PSD Mdot, minidisk masses, sloshing flux}
\end{figure*}

\smallskip
As already outlined in Sec.~\ref{subsubsec: Overview of structural evolution}, during the inspiral stage matter is transferred (``sloshes'') from one minidisk to the other when the former is in the disklike state: some of the gas from the more massive minidisk is able to escape the gravitational potential of the black hole it is orbiting and travels towards the other hole, where it contributes to filling the corresponding minidisk. The rate at which sloshing carries mass from one minidisk to the other varies at the beat frequency (bottom panel of Fig.~\ref{fig: PSD Mdot, minidisk masses, sloshing flux}). Although substantial amounts of matter are transferred by sloshing in the early inspiral, this phenomenon weakens dramatically ${\sim\!5000\,M}$ into the simulation and is negligible thereafter (see bottom panel of Fig.~\ref{fig: mdot AH, minidisk mass, sloshing flux}).

While it is still active, sloshing can disrupt the structure of the mass-receiving disk. Because the mass available for sloshing scales with the donor minidisk's mass, when the tidal truncation radius diminishes with the binary separation, such disruptions grow in importance as the binary shrinks. In fact, the disklike state disappears after ${\sim\!10500\,M}$ into the inspiral (when the binary separation is ${\lesssim 14.5\,M}$) due to sloshing streams expelling mass from the receiving minidisk; from then on, the minidisks can exist only in the stream-like state. Sloshing ends at this stage because these states have asymmetric density distributions in which the side toward the L1 point is relatively depleted.

\smallskip
The most remarkable feature of time-dependence in the mass motion is that the accretion rate is (modulo factor of 2 fluctuations) essentially constant from ${\sim\!1000\,M}$ before merger to ${\sim\!4000\,M}$ after (top panel of Fig.~\ref{fig: mdot tot, luminosity in 100M}). \textit{Despite the dramatic change in the spacetime's geometric symmetry and the large-amplitude gravitational waves generated, the change in spacetime from that of a rapidly orbiting black hole binary to a single black hole has no effect at all on the accretion rate.}

\begin{figure*}[htp]
    \centering
    \includegraphics[width = \textwidth]{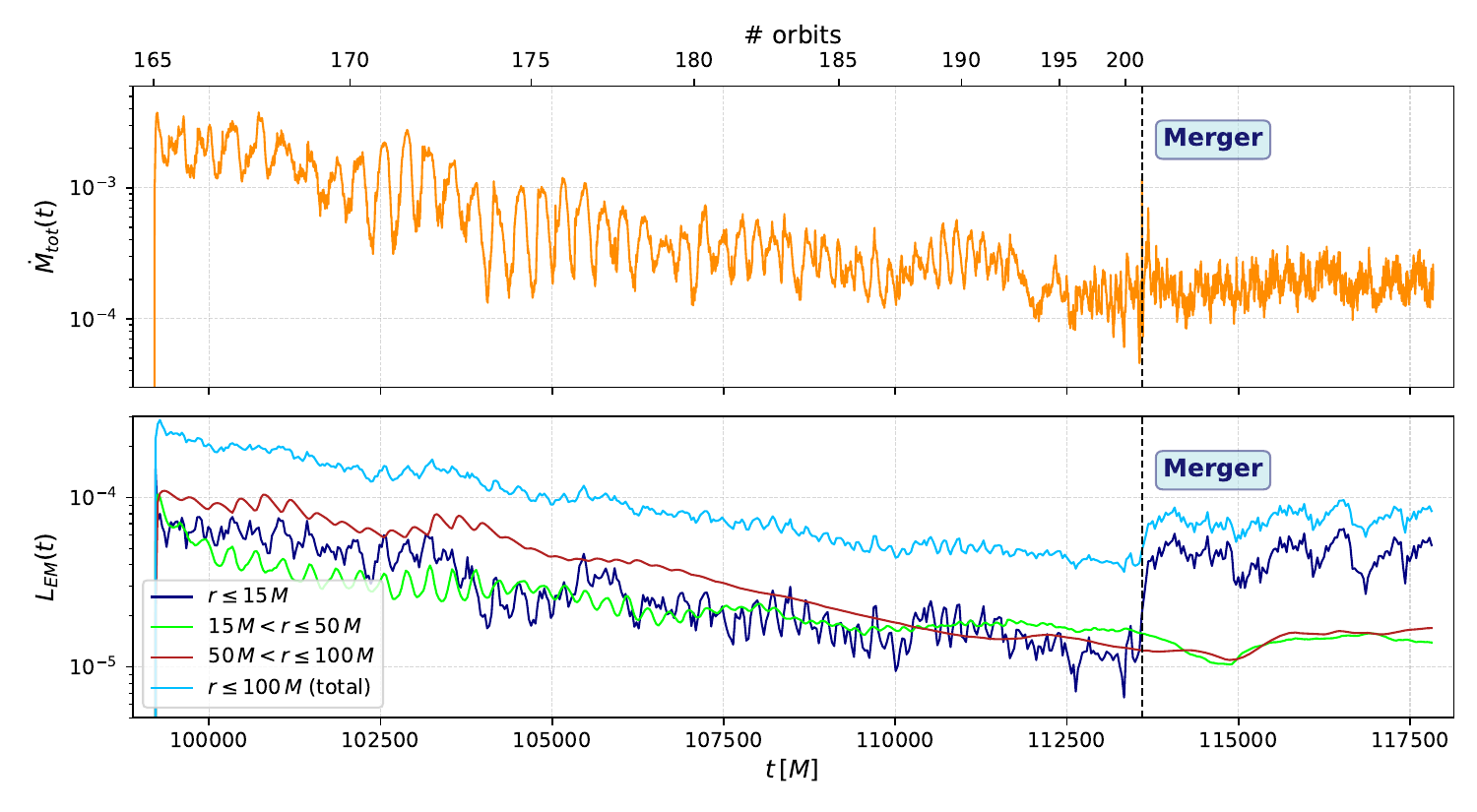}
    \captionsetup{justification = raggedright, format = hang}
    \caption{\textbf{[Top panel]} Total accretion rate onto the binary (i.e., sum of the blue and red curves in the top panel of Fig.~\ref{fig: mdot AH, minidisk mass, sloshing flux}). \textbf{[Bottom panel]} EM luminosity: total for ${r\leq 100\,M}$ (cyan); in a ${r\leq 15\,M}$ spherical volume (dark blue); and in spherical shells between ${15\,M<r\leq 50\,M}$ (green) and ${50\,M<r\leq 100\,M}$ (brown).}
    \label{fig: mdot tot, luminosity in 100M}
\end{figure*}

\begin{figure}[htp]
    \centering
    \includegraphics[width = \linewidth]{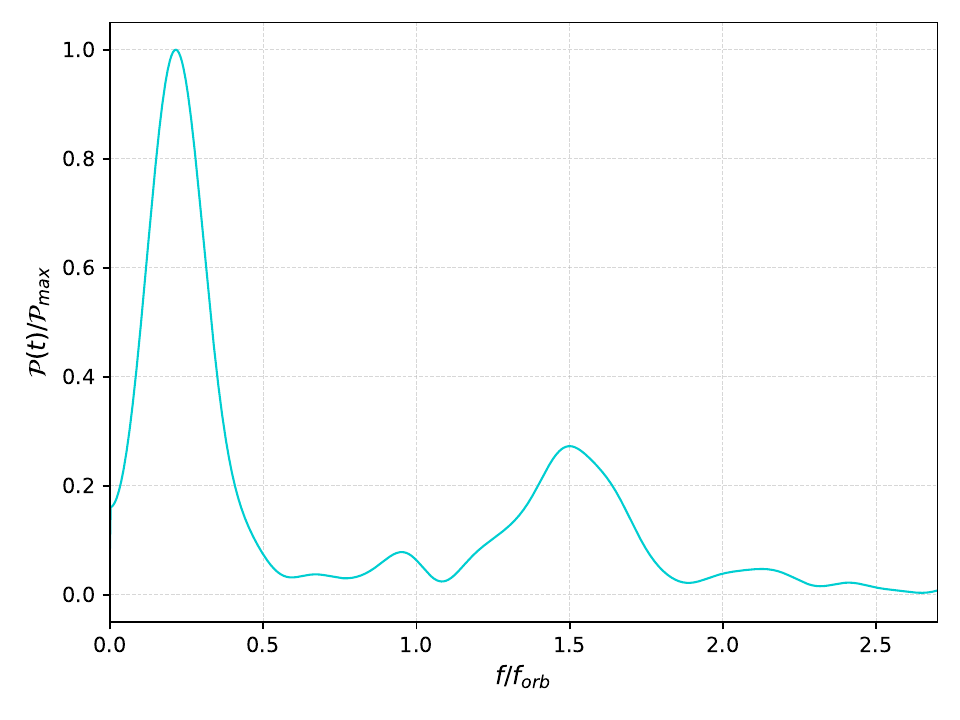}
    \captionsetup{justification = raggedright} 
    \caption{Power spectrum of ${L_\text{EM}}$ in a sphere of coordinate radius ${r=100\,M}$ around the center of mass of the binary. The spectrum is computed using the same numerical techniques described in the caption of Fig.~\ref{fig: PSD Mdot, minidisk masses, sloshing flux}.}
    \label{fig: PSD of the luminosity in r=100M}
\end{figure}

\subsubsection{Luminosity}

In this section, we report on the time evolution of the EM luminosity in the inner parts of the system. In the literature, the term ``EM luminosity'' (or simply ``luminosity'') sometimes refers to the volume-integrated cooling rate and at other times to the Poynting luminosity; in this work we follow the former convention and define the EM luminosity as
\begin{equation}
    \label{eq: EM luminosity}
    L_{EM}\equiv\int_V dV\sqrt{-g}\,\mathcal{L}_\text{cool}\;.
\end{equation}

Over the same time span in which ${\dot{M}_\text{tot}}$ falls by a factor ${\sim\!10}$, the EM luminosity of all the inner portions of the system also drops, but by a smaller factor. To illustrate this, we calculate the EM luminosity \eqref{eq: EM luminosity} in three different subregions centered at the binary's center of mass: one is a sphere with coordinate radius ${r=15\,M}$, and the other two are spherical shells, one with a radial span of ${15\,M\text{--}\,50\,M}$, the other with a span of ${50\,M\text{--}\,100\,M}$. The lightcurves for all three are shown in the bottom panel of Fig.~\ref{fig: mdot tot, luminosity in 100M} along with their sum, i.e., the total luminosity within ${r=100\,M}$. During the inspiral, the total luminosity generated within ${r=100\,M}$ falls steadily from the beginning of the simulation to immediately before the merger, reaching a level ${\sim\!4}$ smaller than at the start. The luminosities of the inner CBD (${50\,M<r\leq 100\,M}$), outer cavity (${15\,M<r\leq 50\,M}$), and the binary region (${r\leq 15\,M}$) follow almost in parallel, each accounting for ${\sim\!1/3}$ of the total. Strikingly, however, during a time of only ${\sim\!100\,M}$ surrounding the merger, the luminosity associated directly with the binary (and the merger remnant) rises by a factor ${\sim\!4}$, driving up the total luminosity by ${\sim\!50\%}$\,; both luminosities show no further trend for the ${\sim\!4200\,M}$ past merger that we studied. In sharp contrast, the luminosities of the (former) outer cavity and CBD hardly change at and after merger.

\begin{figure*}[htp]
    \centering
    \begin{subfigure}[c]{0.49\textwidth}
        \includegraphics[width = \textwidth]{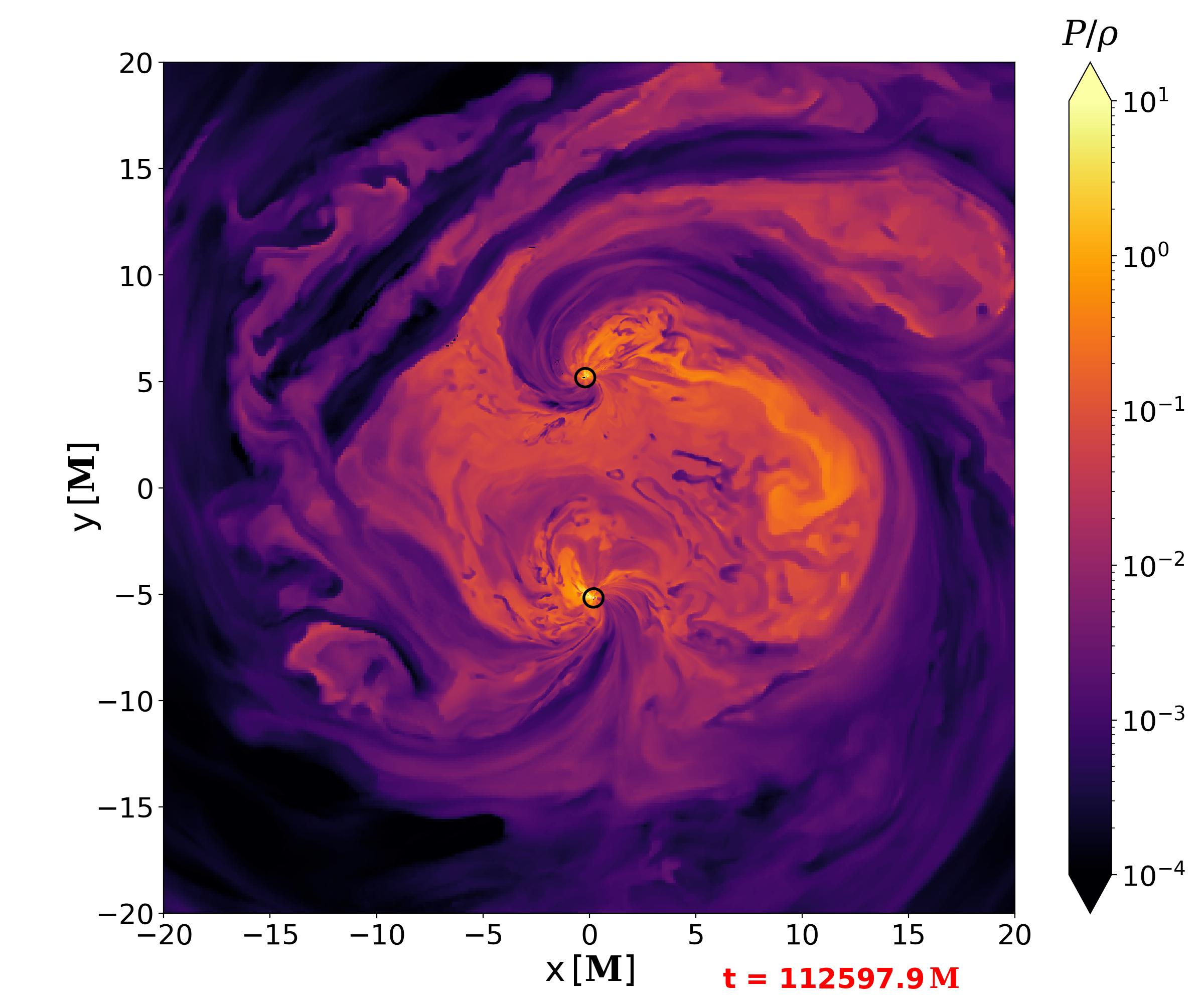}
    \end{subfigure}
    \begin{subfigure}[c]{0.49\textwidth}
        \includegraphics[width = \textwidth]{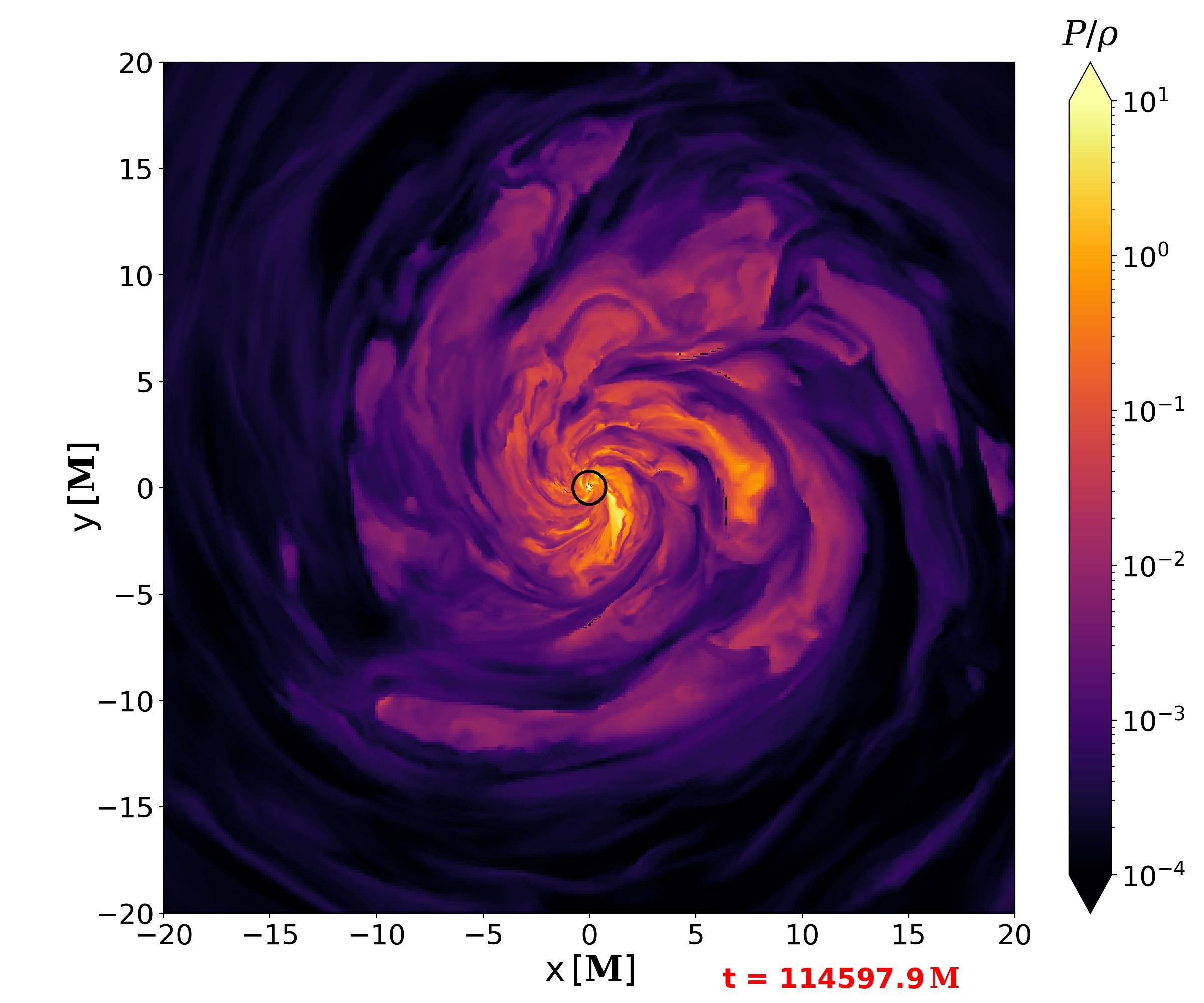}
    \end{subfigure}
    \captionsetup{justification = raggedright, format = hang}
    \caption{Equatorial slices showing the increase in temperature ${T\equiv P/\rho}$ in the vicinity of the black holes from ${\sim\!1000\,M}$ before merger (left panel) to ${\sim\!1000\,M}$ after (right panel).}
    \label{fig: temperature}
\end{figure*}

\begin{figure*}[htp]
    \centering
    \includegraphics[width = \textwidth]{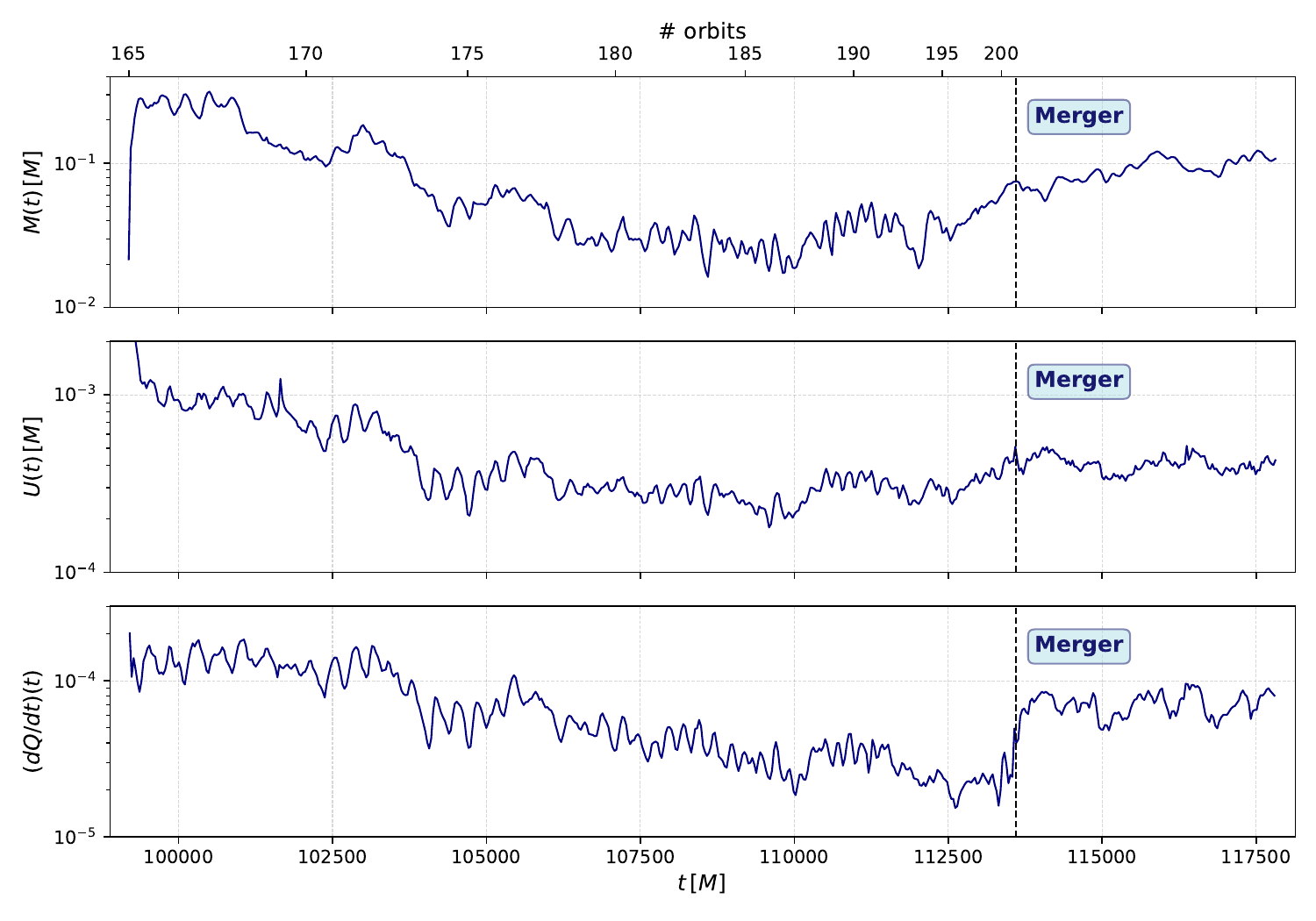}
    \captionsetup{justification = raggedright, format = hang}
    \caption{Mass (top panel), internal energy (middle panel), and heating rate (bottom panel) in a spherical volume of coordinate radius ${r=15\,M}$ centered around the binary.}
    \label{fig: M, U, dQ/dt in 15M}
\end{figure*}

\smallskip
The sudden increase in the luminosity of the innermost region despite an almost constant accretion rate means that the radiative efficiency of accretion onto the black hole(s) increases by a factor ${\sim\!4}$. Both because the radiative efficiency can change by a large factor and because the accretion rate can vary within the cavity, \textit{the mass accretion rate onto the binary is a poor proxy for the EM luminosity around the time of merger; instead, it is the rate of gas heating that drives the luminosity.}

Because our cooling rate is entirely driven by increases in specific entropy, gas heating that leads to increased radiation must be dissipative, not adiabatic. Prior to the merger, accreting matter is pulled along by the orbiting black holes, so that the gas has relatively small angular momentum with respect to the nearest black hole and its flow lines are comparatively laminar; by this means, the gas can plunge into the black holes rising in temperature by adiabatic compression, but gaining little entropy. By contrast, as a result of the merger, the time-dependent quadrupole moment of the gravitational potential disappears, leaving only its monopolar component, which is slightly weaker due to the loss of mass associated with the energy radiated in gravitational waves (${\sim\!3\%}$ of the total for equal-mass, nonspinning binaries~\cite{Campanelli2006}). The sudden change in the shape of the potential causes a concomitant swerve in the gas orbits, driving shocks in which dissipation rapidly raises the gas's entropy.

This stronger heating is illustrated in Fig.~\ref{fig: temperature}, which contrasts a measure of the gas temperature ${P/\rho \simeq 2kT/m_p c^2}$ just prior to merger with the gas temperature shortly after merger. Before merger, regions with temperatures ${kT\!\sim\!25\text{--}250\,\text{MeV}}$ are spread over distances as much as ${\sim\!10\,M}$ from the center of mass; after merger, high temperatures are confined to within a few ${M}$ around the merged black hole, but the temperature of the hottest regions is ${\sim\!5\,\text{GeV}}$.

\smallskip
These events are illustrated more quantitatively in Fig.~\ref{fig: M, U, dQ/dt in 15M}: again considering a sphere of radius ${r=15\,M}$ around the center of mass and starting from a time ${\sim\!2500\,M}$ before merger, the mass grows from ${\sim\!0.02\,M}$ to ${\sim\!0.1\,M}$ in ${\sim\!5000\,M}$ of time, and the volume-integrated internal energy ${U}$ grows from ${\sim\!3\!\times\!10^{-4}M}$ to ${\lesssim 5\!\times\!10^{-4}M}$ in ${\sim\!3000\,M}$ of time. Some of the growth in ${U}$ is due to inflow, but more is the result of heating within this volume. Denoting the internal energy density of the fluid as ${\mathfrak{u}\equiv\rho\epsilon}$ and its 4-velocity as ${u}$\,, the heating rate ${dQ/dt}$ (including both compressive and dissipative heating) can be inferred from the internal energy budget in the cavity,
\begin{equation}
    \begin{split}
        \frac{dU}{dt} =&-\int_{\partial V}d^2\sigma_i\,\sqrt{-g}\;\mathfrak{u}\,u^i +\\
                       &-\int_V dV\sqrt{-g}\;\mathcal{L}_\text{cool} + \frac{dQ}{dt}\;;
    \end{split}
\end{equation}
note that ${\partial V}$ includes both the spherical surface with radius ${r=15\,M}$ and the apparent horizons. As shown in Fig.~\ref{fig: M, U, dQ/dt in 15M}, the growth in ${U}$ is a small fraction of ${dQ/dt}$ integrated across merger; on the other hand, ${L_\text{EM}}$ follows ${dQ/dt}$ very closely (Fig.~\ref{fig: mdot tot, luminosity in 100M}). In other words, nearly all the heating at this time is radiated promptly, which in our model means the heating is nearly all dissipative.

Lastly, as vividly portrayed in Fig.~\ref{fig: mdot tot, luminosity in 100M} and demonstrated by Fig.~\ref{fig: PSD of the luminosity in r=100M}, all the luminosity radiated from the binary and cavity regions during the first half of the inspiral is modulated at the same frequencies as ${{\dot M}_\text{tot}}$\,. However, its behavior contrasts with the accretion rate in several ways. The luminosity modulation persists until merger only in the region immediately surrounding the binary (i.e., ${r\leq 15\,M}$). It is noticeably more irregular than the accretion rate modulation; and, compared to the accretion rate power spectrum, the relative strength of the two peaks in the luminosity power spectrum is reversed.

\begin{figure*}[htp]
    \centering
    \begin{subfigure}[c]{0.49\textwidth}
        \includegraphics[width = \textwidth]{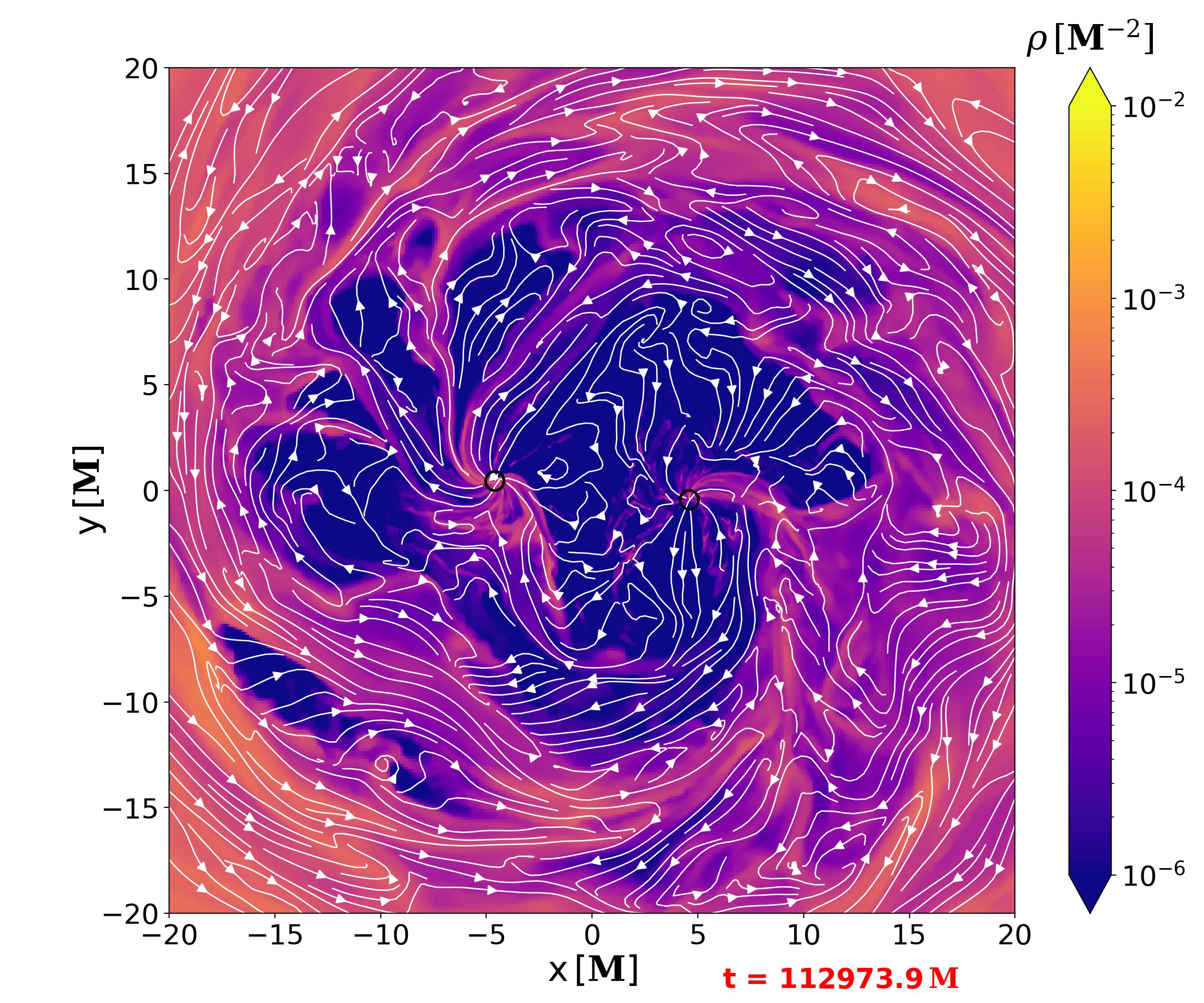}
    \end{subfigure}
    \begin{subfigure}[c]{0.49\textwidth}
        \includegraphics[width = \textwidth]{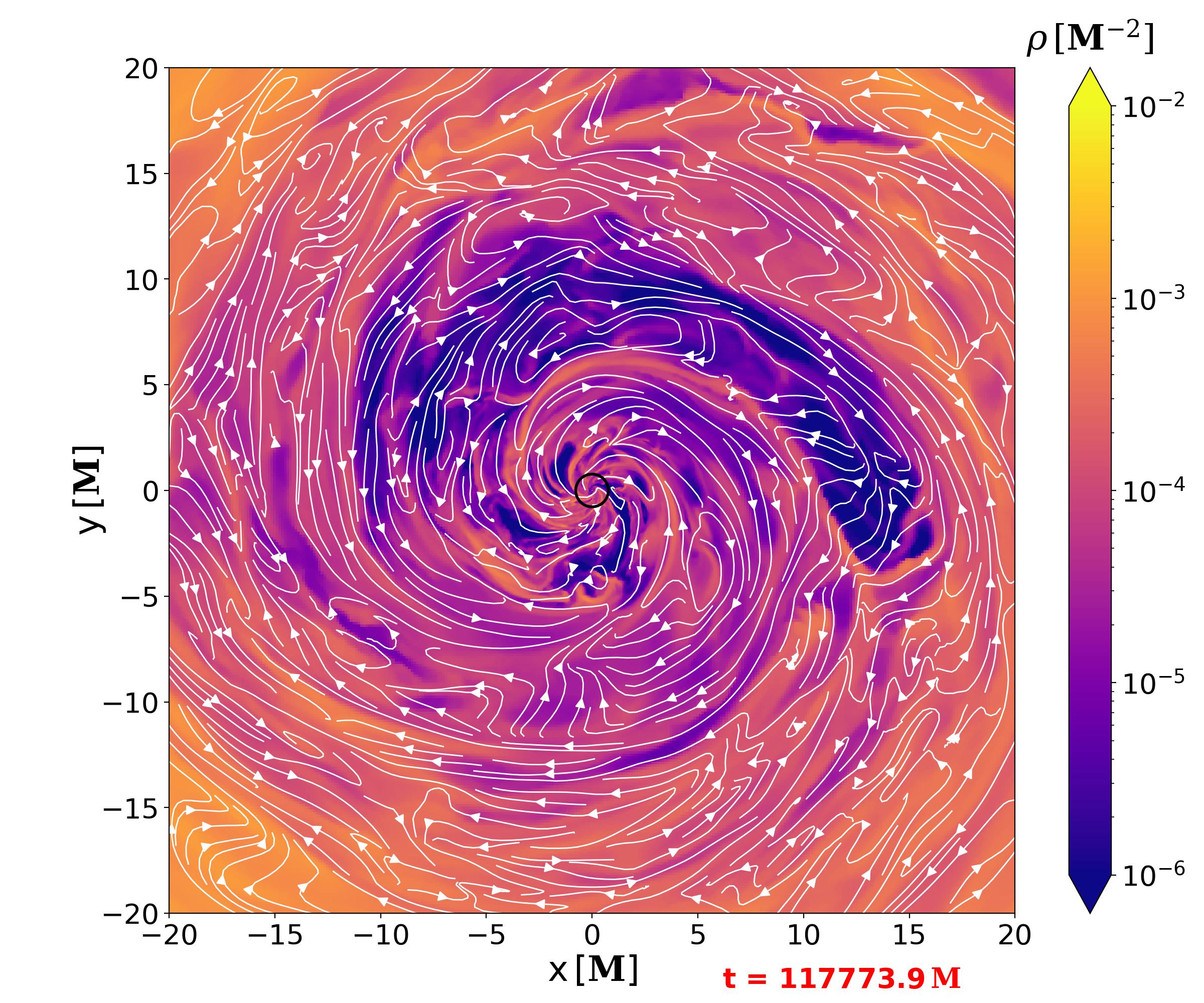}
    \end{subfigure}
    \begin{subfigure}[c]{0.49\textwidth}
        \includegraphics[width = \textwidth]{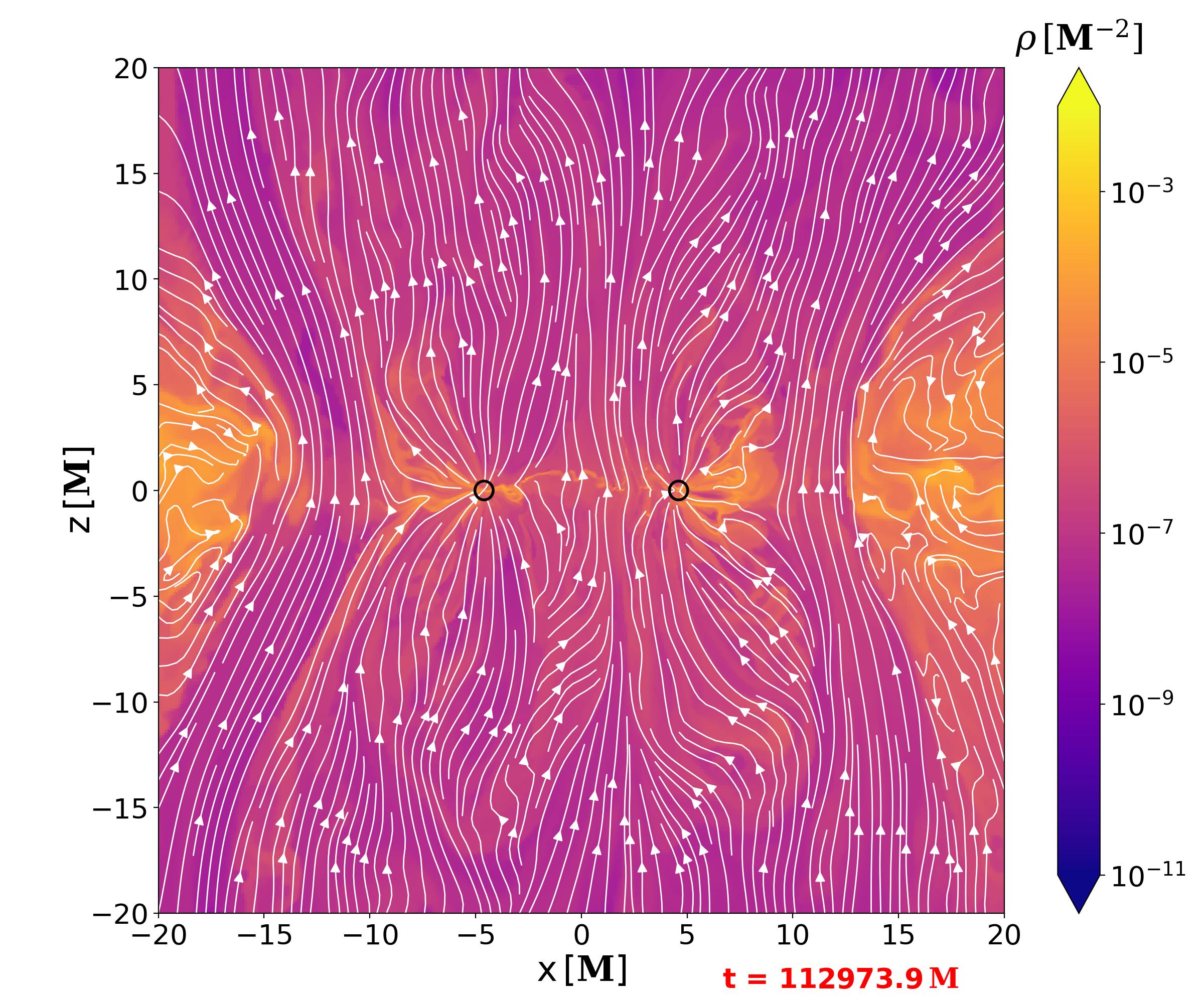}
    \end{subfigure}
    \begin{subfigure}[c]{0.49\textwidth}
        \includegraphics[width = \textwidth]{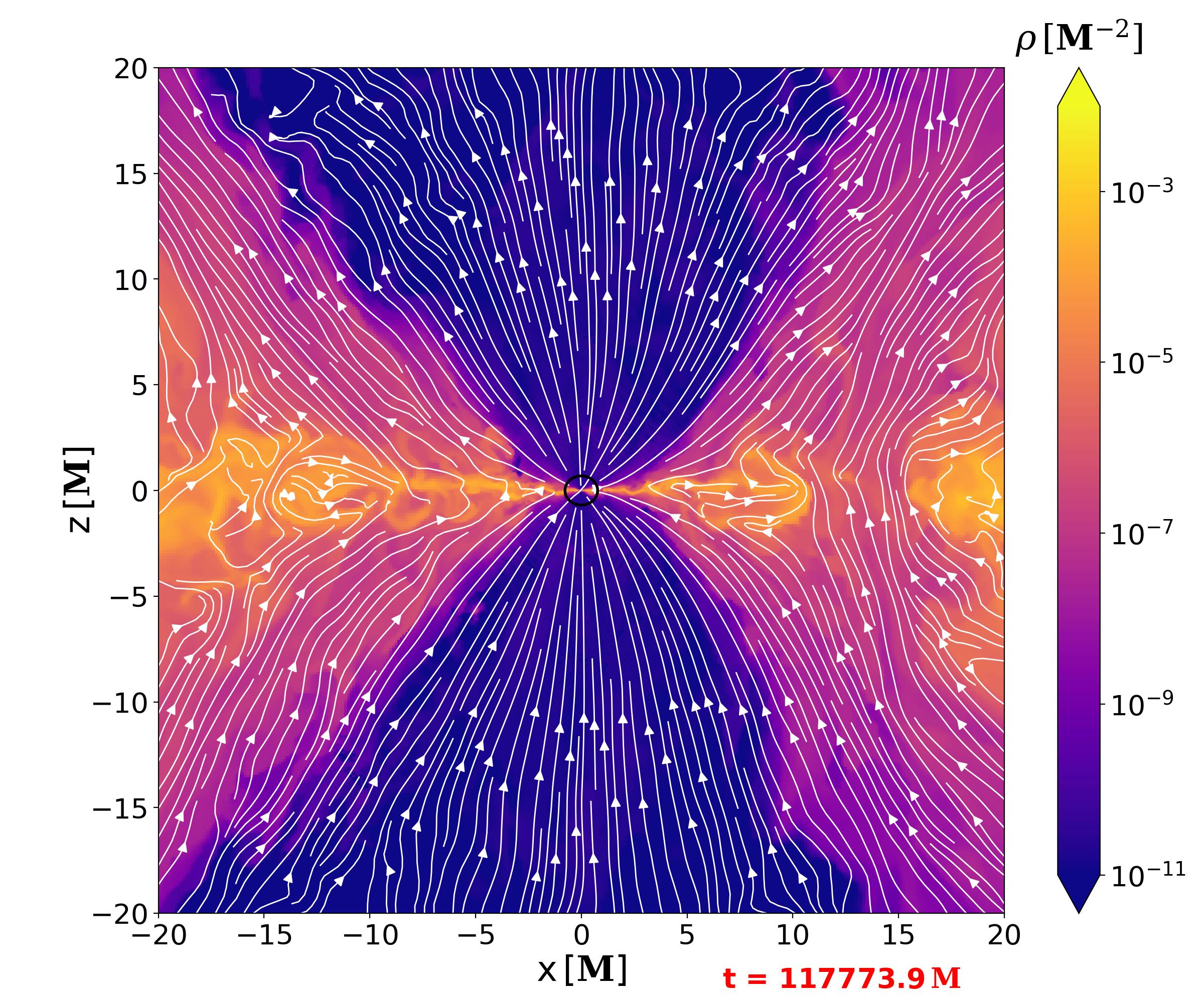}
    \end{subfigure}
    \captionsetup{justification = raggedright, format = hang}
    \caption{Projection of the magnetic field lines on the equatorial (left panels) and meridional (right panels) planes shortly before merger (upper panels) and after merger (lower panels) superimposed onto the mass density distribution. The merger remnant forces the magnetic field lines into rotation around the black hole's spin axis and deflects them so as to reduce the magnetic flux through the equatorial plane (see also Fig.~\ref{fig: Bz xy xz}).}
    \label{fig: B field lines}
\end{figure*}

\begin{figure*}[htp]
    \centering
    \includegraphics[width = \textwidth]{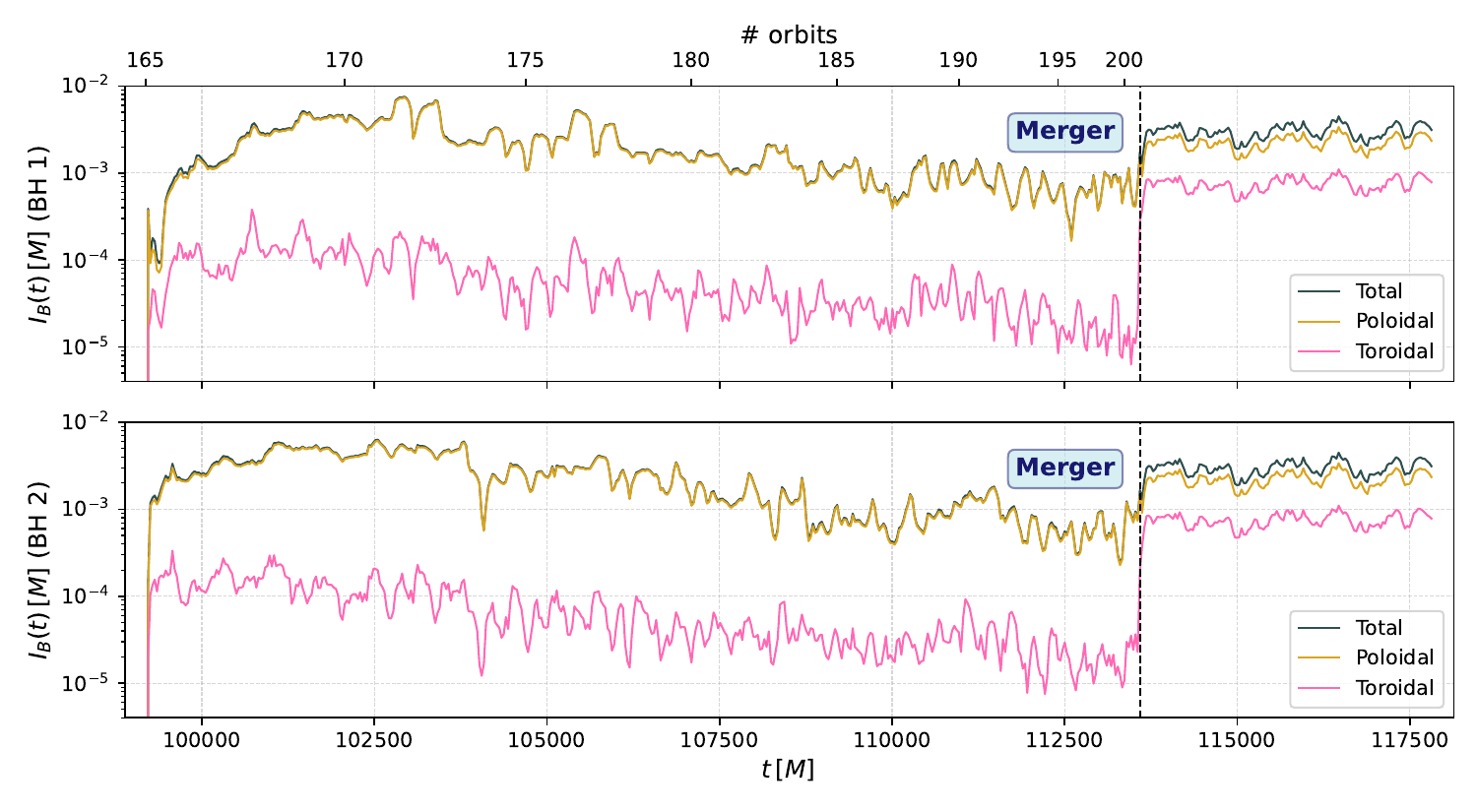}
    \captionsetup{justification = raggedright, format = hang}
    \caption{Total magnetic field energy (dark teal) and the poloidal (gold) and toroidal (magenta) contributions to the magnetic field energy in spherical volumes of coordinate radius ${r=1\,M}$ centered on the inspiralling punctures or on the merger remnant. The choice ${r=1\,M}$ ensures the ergosphere of the merged black hole is fully contained inside the integration volume.}
    \label{fig: magnetic volume integrals around the BHs}
\end{figure*}

\subsection{Electromagnetic phenomena}
\label{subsec: Electromagnetic phenomena}

\subsubsection{Character of the magnetic field}

We now direct our attention to the dynamics of the EM field. The magnetic field changes in a number of ways during inspiral and does so especially dramatically at the time of merger. For example, the relative sizes of the magnetic field's components near the black hole horizons shift sharply toward a greater contribution from toroidal field as the merger takes place; this is visually evident in the upper panels of Fig.~\ref{fig: B field lines}. To demonstrate this quantitatively, in Fig.~\ref{fig: magnetic volume integrals around the BHs} we plot the total magnetic field energy,
\begin{equation}
    \int_V\,dV\,\sqrt{-g}\,B^2\;,
\end{equation}
and the contributions to that energy from the poloidal and toroidal components,
\begin{align}
    \int_V\,dV\,\sqrt{-g}\left(B^2 - B_\phi B^\phi\right) \\
    \text{and}\;\int_V\,dV\,\sqrt{-g}\,B_\phi B^\phi\;,
\end{align}
in spheres of radius ${1M}$ around the black holes from which we excise the horizons; note that ``poloidal'' and ``toroidal'' refer to the coordinate frames centered on the punctures, not the one centered in the binary's center of mass. Although the field before merger is almost purely poloidal (in fact, mostly radial within a few ${M}$ of the black holes), the toroidal contribution makes up ${\sim\!25\%}$ of the total field fluid-frame energy density around the postmerger remnant. We attribute this sudden growth in the toroidal component to rotational shear acting on the radial component of the magnetic field, where the rotation is a combination of orbital motion around the merged black hole and frame dragging.

\begin{figure*}[htp]
    \centering
    \includegraphics[width = \textwidth]{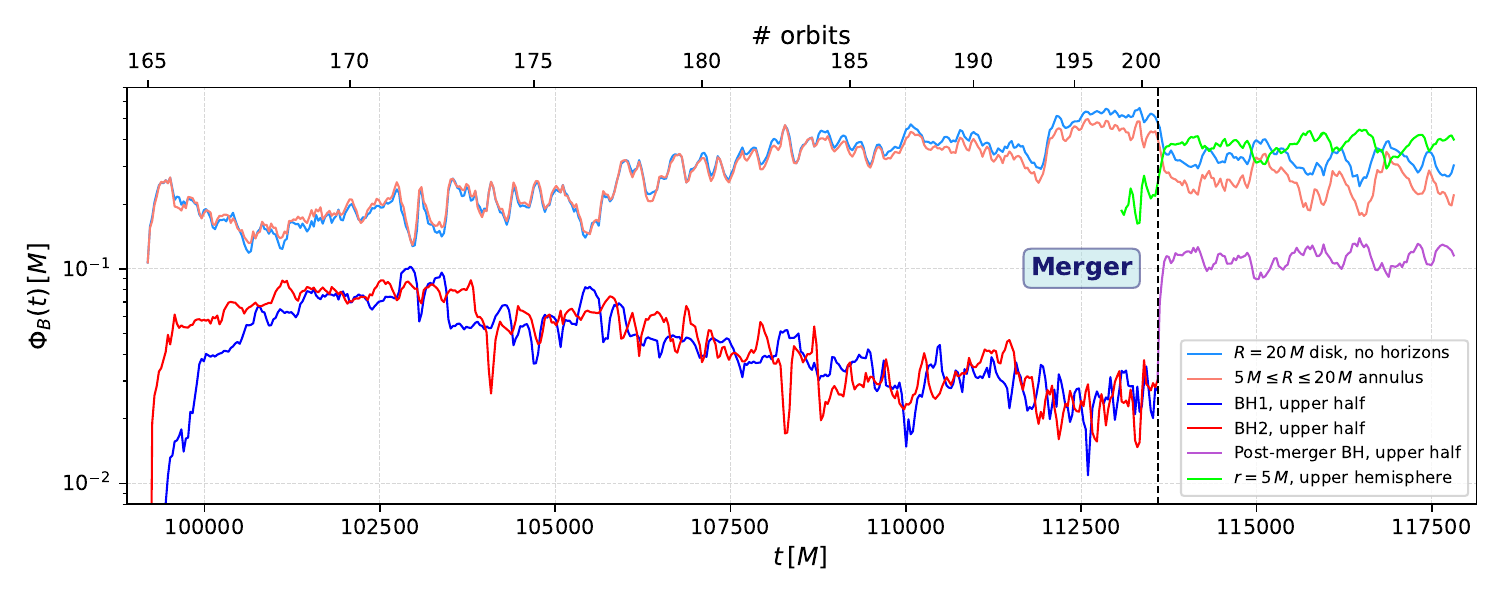}
    \captionsetup{justification = raggedright, format = hang}
    \caption{Magnetic flux through several different surfaces in the inner portion of the cavity region around the binary as a function of time: a disk of coordinate radius ${R=20\,M}$ in the equatorial plane where we excise the intersection with the black hole horizons (light blue); an annulus with ${5\,M\leq R\leq 20\,M}$ in the equatorial plane (orange); the upper halves of the two horizons during the inspiral (blue and red); the upper half of the merged black hole (violet); and the upper half of a sphere of coordinate radius ${r=5\,M}$ centered on binary's center of mass, which becomes the remnant black hole after merger (green).}
    \label{fig: magnetic flux history}
\end{figure*}

\begin{figure*}
    \centering
    \begin{subfigure}[c]{0.32\textwidth}
        \includegraphics[width = \textwidth]{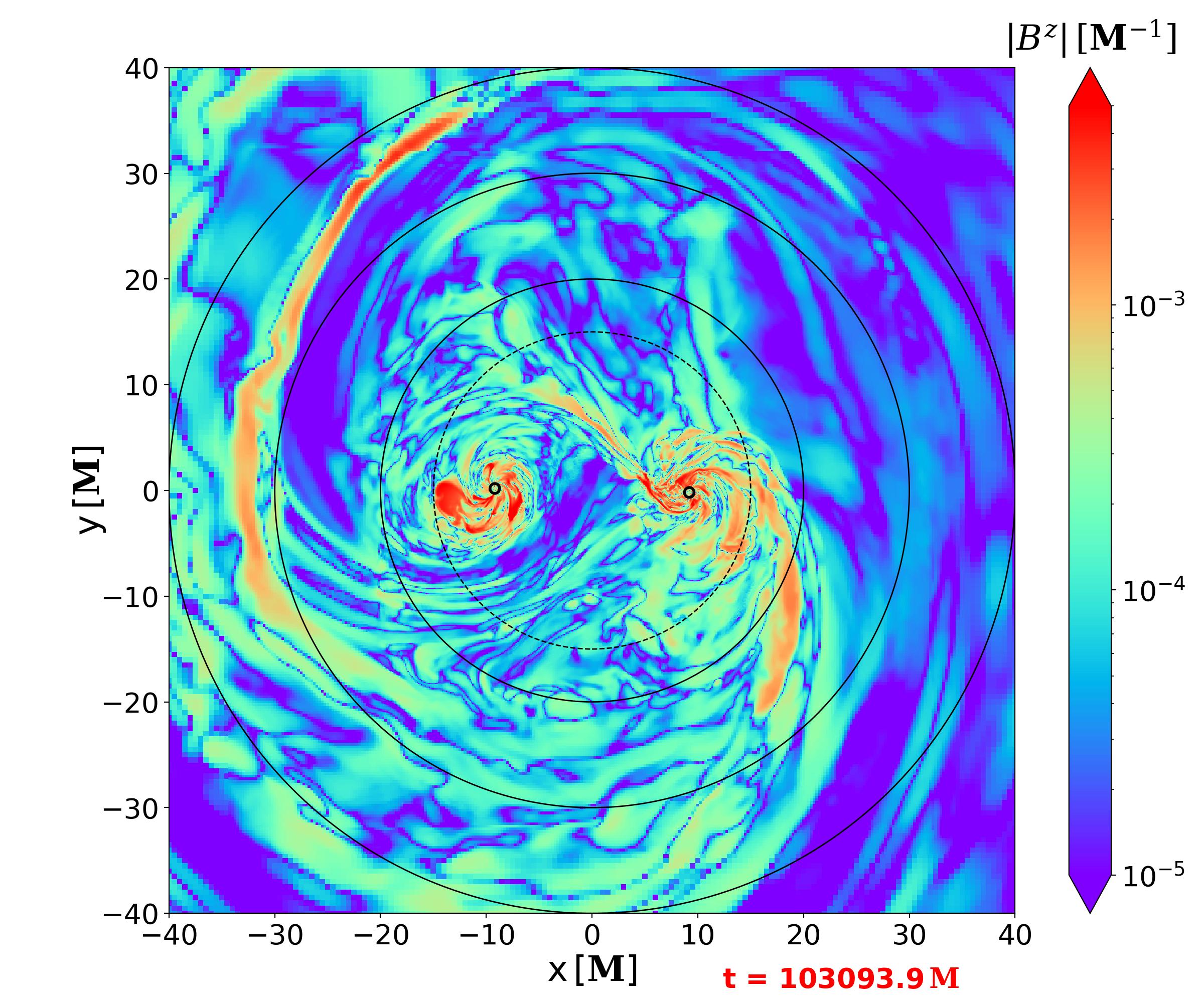}
    \end{subfigure}
    \begin{subfigure}[c]{0.32\textwidth}
        \includegraphics[width = \textwidth]{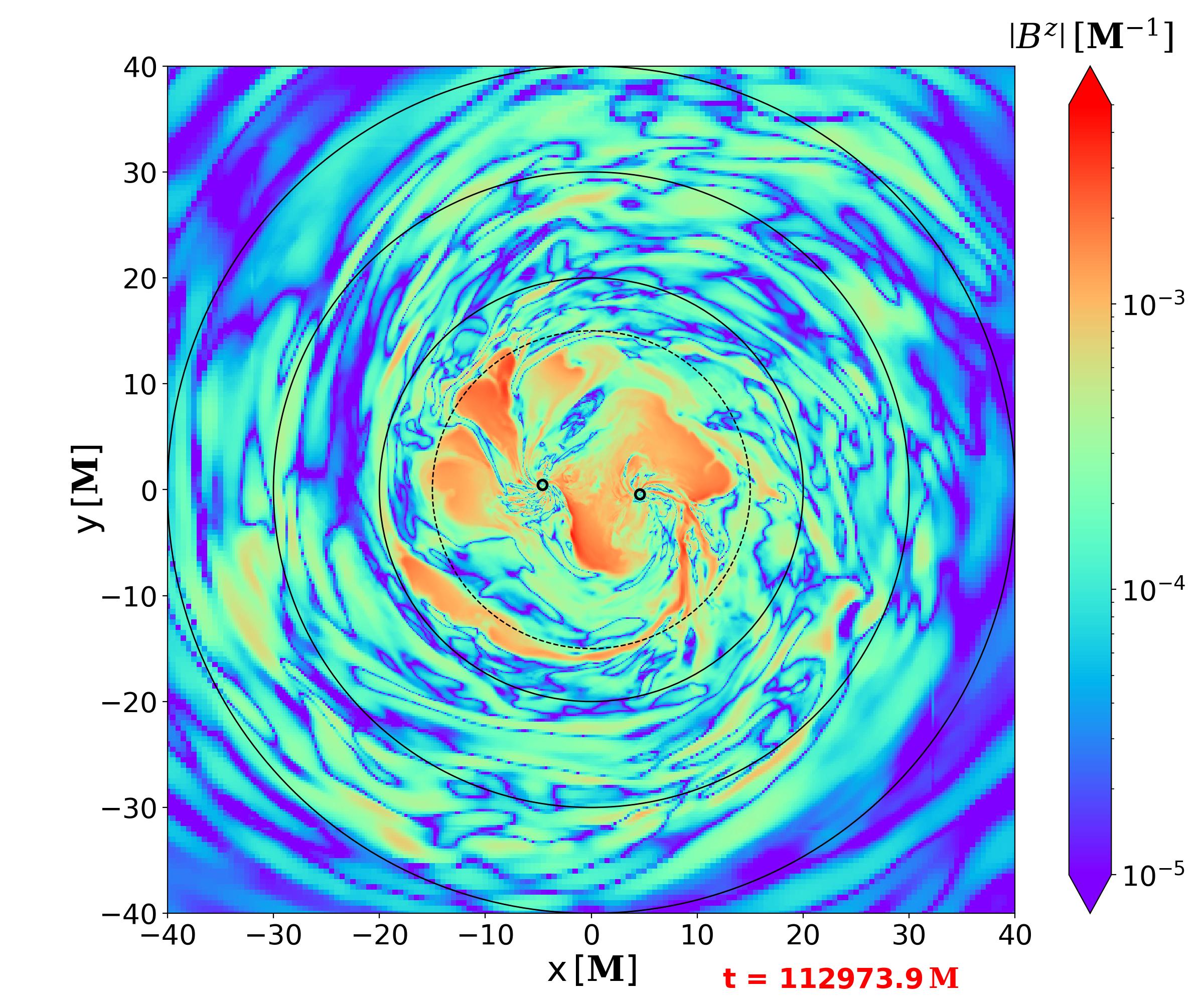}
    \end{subfigure}
    \begin{subfigure}[c]{0.32\textwidth}
        \includegraphics[width = \textwidth]{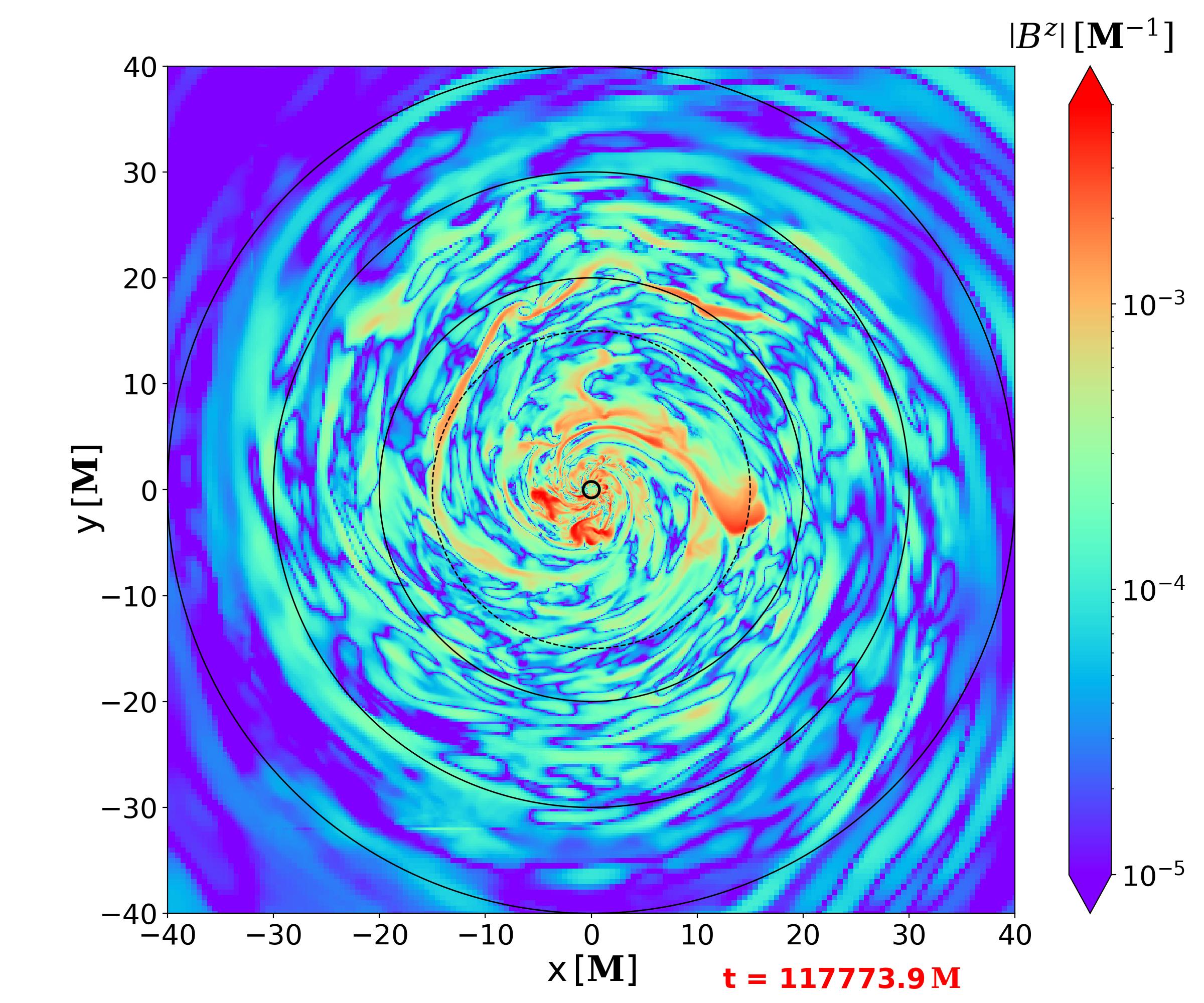}
    \end{subfigure}
    \begin{subfigure}[c]{0.32\textwidth}
        \includegraphics[width = \textwidth]{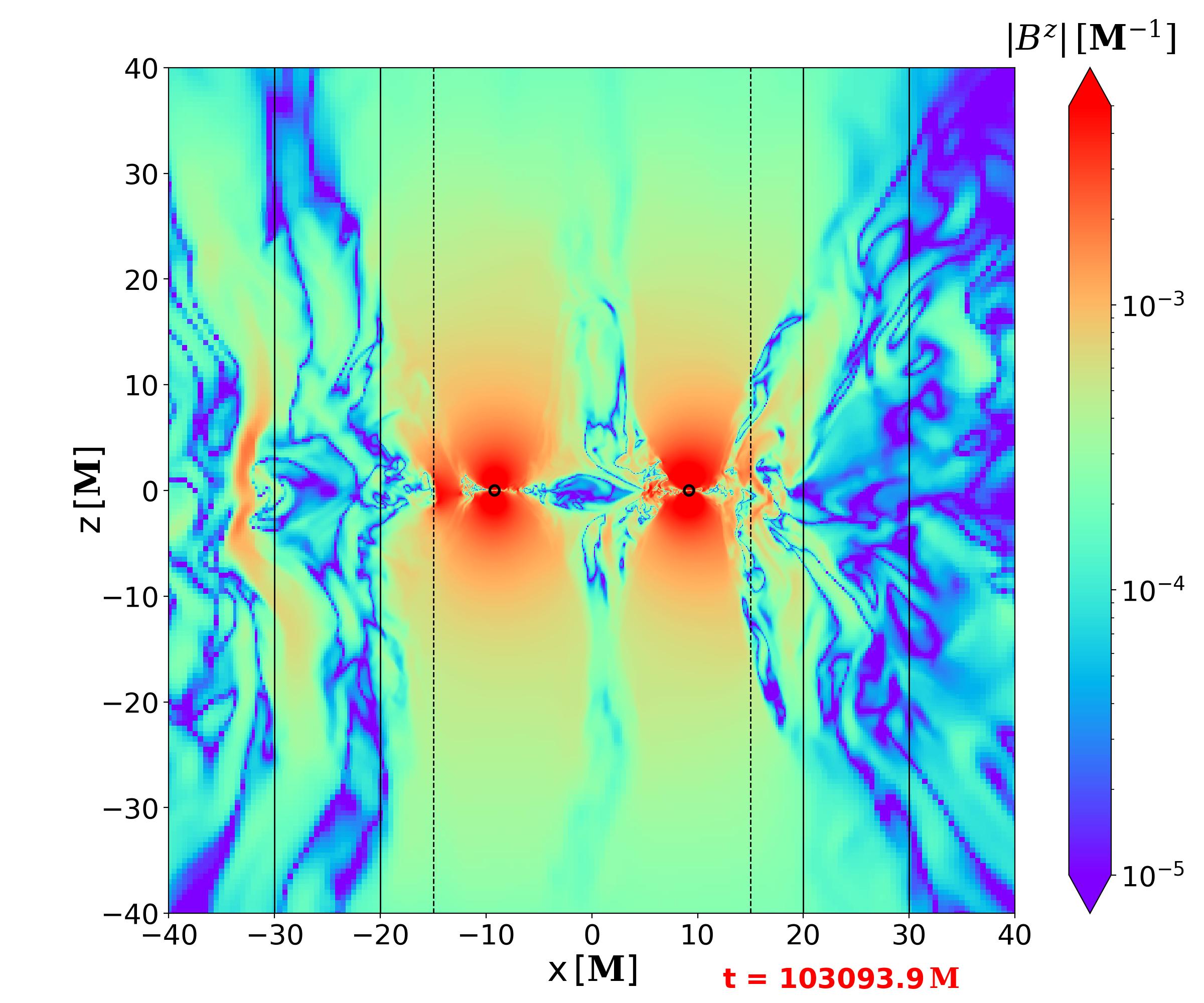}
    \end{subfigure}
    \begin{subfigure}[c]{0.32\textwidth}
        \includegraphics[width = \textwidth]{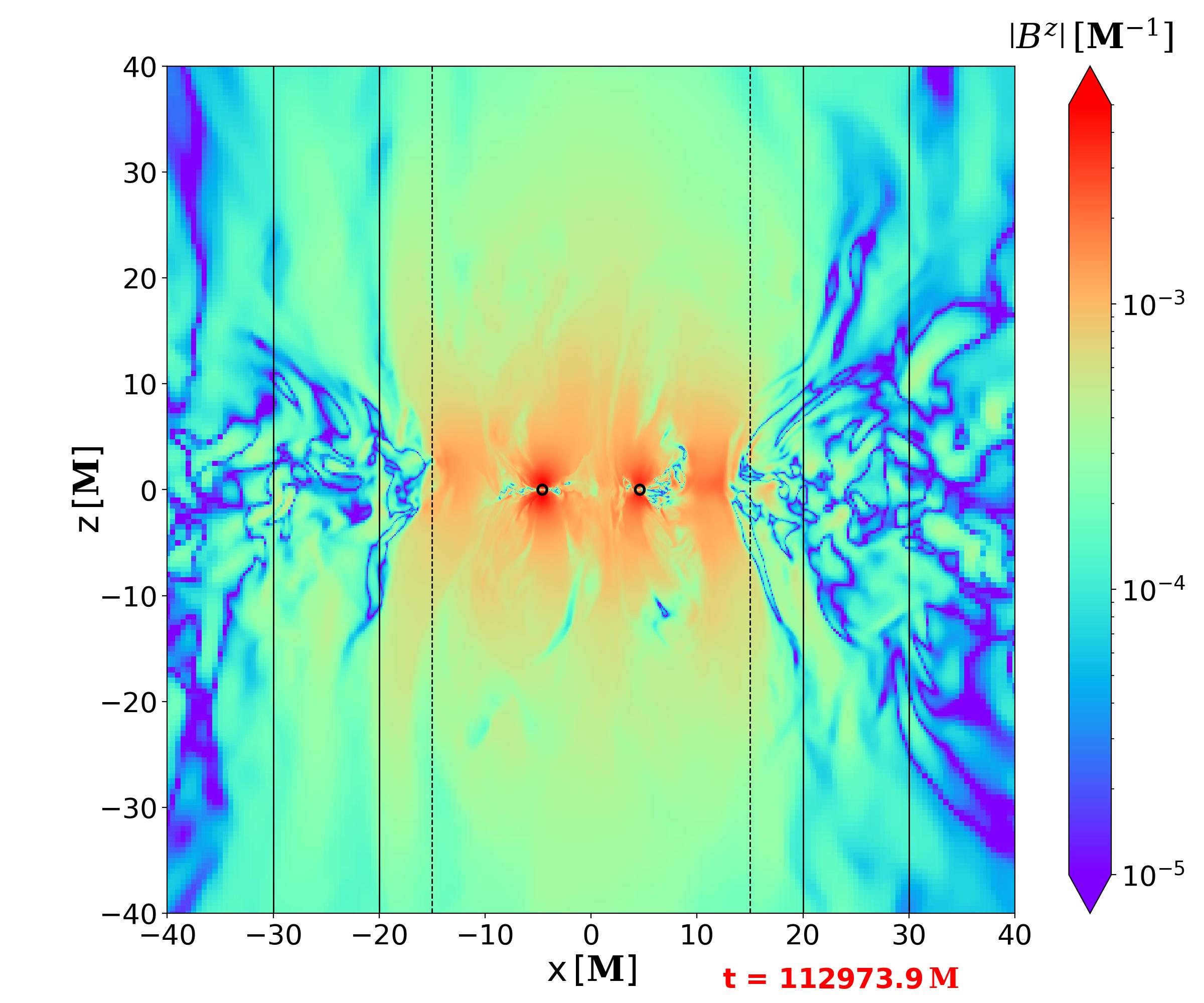}
    \end{subfigure}
    \begin{subfigure}[c]{0.32\textwidth}
        \includegraphics[width = \textwidth]{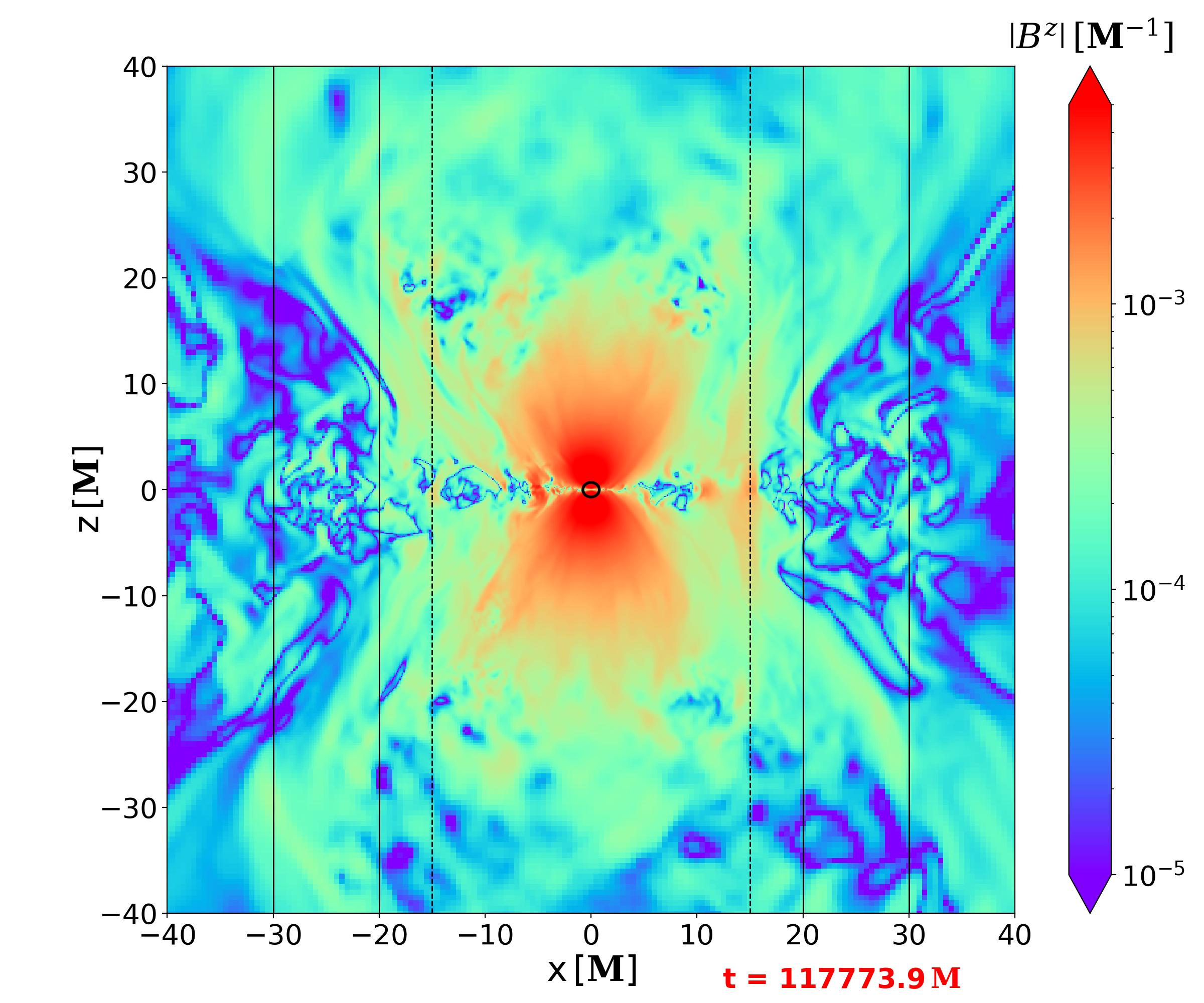}
    \end{subfigure}
    \captionsetup{justification = raggedright, format = hang}
    \caption{Equatorial (top panels) and polar (bottom panels) slices of the vertical magnetic field component ${\left\lvert B^z\right\rvert}$ in the early inspiral (left panels), shortly (${\sim\!620\,M}$) before merger (central panels), and ${\sim\!4200\,M}$ after merger (right panels). Note the progressive accretion of vertical magnetic flux during the inspiral stage and its decrease through the equatorial plane after merger, indicating deflection of magnetic field lines towards the merger remnant (see also Fig.~\ref{fig: B field lines}).}
    \label{fig: Bz xy xz}
\end{figure*}

\smallskip
To describe other field properties, we study the behavior of the magnetic flux through various open surfaces ${\mathcal{S}}$ (the no-monopole constraint makes the total magnetic flux through a closed surface identically zero) during the inspiral and merger stages of the simulation:
\begin{equation}
    \Phi_B\equiv\int_\mathcal{S}\,d^2\sigma_i\,\sqrt{-g}\,B^i\;.
\end{equation}
The evolution of the magnetic flux in the inner portion of the cavity is portrayed in Fig.~\ref{fig: magnetic flux history}. The light blue curve shows the flux through a disk of coordinate radius ${R=20\,M}$ centered around the binary's center of mass and lying on the equatorial plane; we remove the intersection of the disk with the black hole horizons from the integration region. The flux through this surface grows steadily from the beginning of the inspiral until ${\sim\!1000\,M}$ before the merger, presumably as a result of continued accretion bringing additional vertical magnetic flux into the region; Fig.~\ref{fig: Bz xy xz} appears to support this hypothesis. At the merger, this flux drops abruptly by ${\sim\!35\%}$\,, consistent with the observation that magnetic field lines are deflected from the vertical direction by the merger remnant (bottom panels of Fig.~\ref{fig: B field lines}).

The total flux inside ${20\,M}$ (combining flux on and outside the black hole horizons) rises from a minimum ${\sim\!0.3}$ at the beginning of inspiral to a maximum ${\sim\!0.6}$ immediately before merger. During this span of time, the location of the flux changes systematically. In the early inspiral, roughly half this flux threads the event horizons of the black holes, but this fraction decreases to ${\sim\!10\%}$ at the time of maximum total flux shortly before merger. This drop in the horizons' flux-share comes about because the horizons of the two black holes lose ${\sim\!65\%}$ of their total magnetic flux during this time period. However, nearly all the lost flux is given to the remaining black hole's horizon at the moment of merger, as roughly twice this amount of flux moves from the annulus ${5\,M < R < 20\,M}$ to inside ${5\,M}$.

\smallskip
These results prompt two questions. First, why does so much flux move suddenly to the merged black hole's horizon at the moment of merger, even though the total mass accretion rate onto the system's black hole(s) hardly changes? Second, what happens to the half of the magnetic flux taken from the ${5\,M < R < 20\,M}$ annulus that doesn't go to the new black hole's horizon? A partial answer to the second question may be seen from the data of Fig.~\ref{fig: magnetic flux history}: all the flux lost from the annulus moves inward, but only about half of it goes all the way to the horizon. We propose that both questions may be completely answered by considering the role of accretion ram pressure in confining magnetic flux onto black hole horizons. The ram pressure of accreting matter just outside the horizon of a black hole is ${\rho h u^r u_r}$\,, where ${h}$ is the enthalpy of the gas. At the order-of-magnitude level, this is ${\sim\!\dot{M}c/A_\text{BH}}$ because the radial component of the accreting matter's 4-velocity must be relativistic at that location; here, ${A_\text{BH}}$ denotes the horizon's surface area. If the magnetic pressure balances the ram pressure, the flux on each black hole is ${\sim\!f_\text{geom}\left(\dot{M}c\,A_\text{BH}\right)^{1/2}}$ with ${f_\text{geom}}$ a factor giving the ratio between the field component normal to the horizon and the field magnitude. As a result of the merger, ${\dot{M}}$ per black hole doubles and ${A_\text{BH}}$ per black hole increases by a factor ${\sim\!2.8}$\,, from ${\sim\!2.47}$ to ${\sim\!6.86}$\,. The result is that the flux capable of balancing ram pressure is ${\sim\!1.2\times}$ the total horizon flux pre-merger; in fact, the flux on the merged horizon is about ${2.5\times}$ the total pre-merger. This level of agreement is consistent with the fact that our argument is fundamentally order-of-magnitude. The geometric factor might contribute to the change in flux, but not through the relative increase in toroidal field we have already noted because the toroidal direction is always normal to the horizon's normal. We might also speculate that only half the flux brought within ${5\,M}$ at merger reaches the horizon because the accretion ram pressure is insufficient to confine it closer.

Ram pressure may also play a role in the decline of magnetic flux on the horizons during nearly the entire span of the inspiral. Over this time period, the accretion rate also falls, dropping roughly a factor ${\sim\!10}$ (Figs.~\ref{fig: mdot AH, minidisk mass, sloshing flux} and~\ref{fig: mdot tot, luminosity in 100M}); the ram pressure of accreting gas would then fall by essentially the same factor. As a result, the flux that the ram pressure can hold on the horizon falls by a factor ${\sim\!3}$\,; Fig.~\ref{fig: magnetic flux history} shows the actual fall is also a factor ${\sim\!3}$\,.

\subsubsection{Poynting flux}
\label{subsubsec: Poynting flux}

The EM field also generates Poynting flux, whose total amount can be found by a surface integral analogous to that used for magnetic flux:
\begin{equation}
    \label{eq: Poynting flux}
    \mathcal{F}_S\equiv\int_\mathcal{S}\,d^2\sigma_i\,\sqrt{-g}\,S^i\,.
\end{equation}
Here ${S^i}$ denotes the Poynting vector which, in the ideal MHD limit, is given by
\begin{equation}
    S^i = B^2 v^i - \left(v_j B^j\right)v^i\;.
\end{equation}
Note that surfaces appropriate to describing total Poynting flux may be closed, unlike the surfaces used for magnetic flux integrals. The Poynting flux~\eqref{eq: Poynting flux} calculated on a spherical surface of coordinate radius ${r=100\,M}$ around the binary is shown in the bottom panel of Fig.~\ref{fig: normalized magnetic flux on the AHs, Poynting flux at 100M}. This quantity remains approximately constant and relatively small (${\sim\!1\text{--}2\times 10^{-4}}$ in code units) throughout the inspiral phase, but undergoes a sharp increase by a factor ${\sim\!5}$ at the time of merger.

The observed behavior is not surprising: as already mentioned in Sec.~\ref{subsubsec: Overview of structural evolution}, the polar outflows generated by a binary comprising nonspinning black holes, albeit magnetically dominated, are able to carry Poynting flux only as a result of their orbital motion. Summing the Poynting flux within the upper and lower cones gives a total (in code-units) that begins at ${\sim\!3\text{--}4\times 10^{-5}}$ and gradually rises to double that by ${\sim\!1000\,M}$ before merger. The remainder of the Poynting flux during the inspiral (${\sim\!0.6\text{--}1.5\times 10^{-4}}$) is associated with the work done by magnetic accretion stresses inside the CBD. The jump in Poynting flux after the merger is, of course, associated with the now-spinning black hole merger remnant.

\smallskip
The balance between ram pressure and magnetic field pressure leads to a convenient nondimensional form in which to measure the magnetic flux relevant to the Poynting flux created by a magnetized and rotating black hole:
\begin{equation}
    \label{Dimensionless magnetic flux}
    \phi_B\equiv\frac{\Phi_B}{\sqrt{4\pi\dot M r_g^2 c}}\;.
\end{equation}
Here, ${\Phi_B}$ is calculated on the upper half of the black hole horizon, while ${\dot{M}}$ is integrated over the full horizon's surface. This form differs from the ram pressure criterion only in using ${4\pi r_g^2}$ in place of the horizon area; when it is adopted, the solid angle-integrated Poynting flux is a dimensionless number times ${\dot{M}c^2}$, which can be thought of as a jet efficiency. For example, if the magnetic field on the horizon of the black hole is a split monopole, one may use the analytic solution of~\cite{McKinneyGammie2004} to find that
\begin{equation}
    L_\text{BZ} = \phi_B^2\dot{M}c^2\left(\frac{4\pi r_g^2}{A_\text{BH}}\right)\left(\frac{r_g}{r_H}\right)\omega^\prime\left(1 - \omega^\prime\right)\chi^2\;,
\end{equation}
where ${r_H}$ is the radius of the horizon, ${\omega^\prime}$ is the ratio of the field-line rotation rate to the rotation rate of the horizon (typically, ${\omega^\prime\simeq 0.4}$~\cite{HawleyKrolik2006}), and ${\chi}$ is the dimensionless spin parameter; note that the field geometry can depend on the intrinsic field topology, the accreting material's equation of state, and ${\chi}$\,.

\begin{figure*}[htp]
    \centering
    \includegraphics[width = \textwidth]{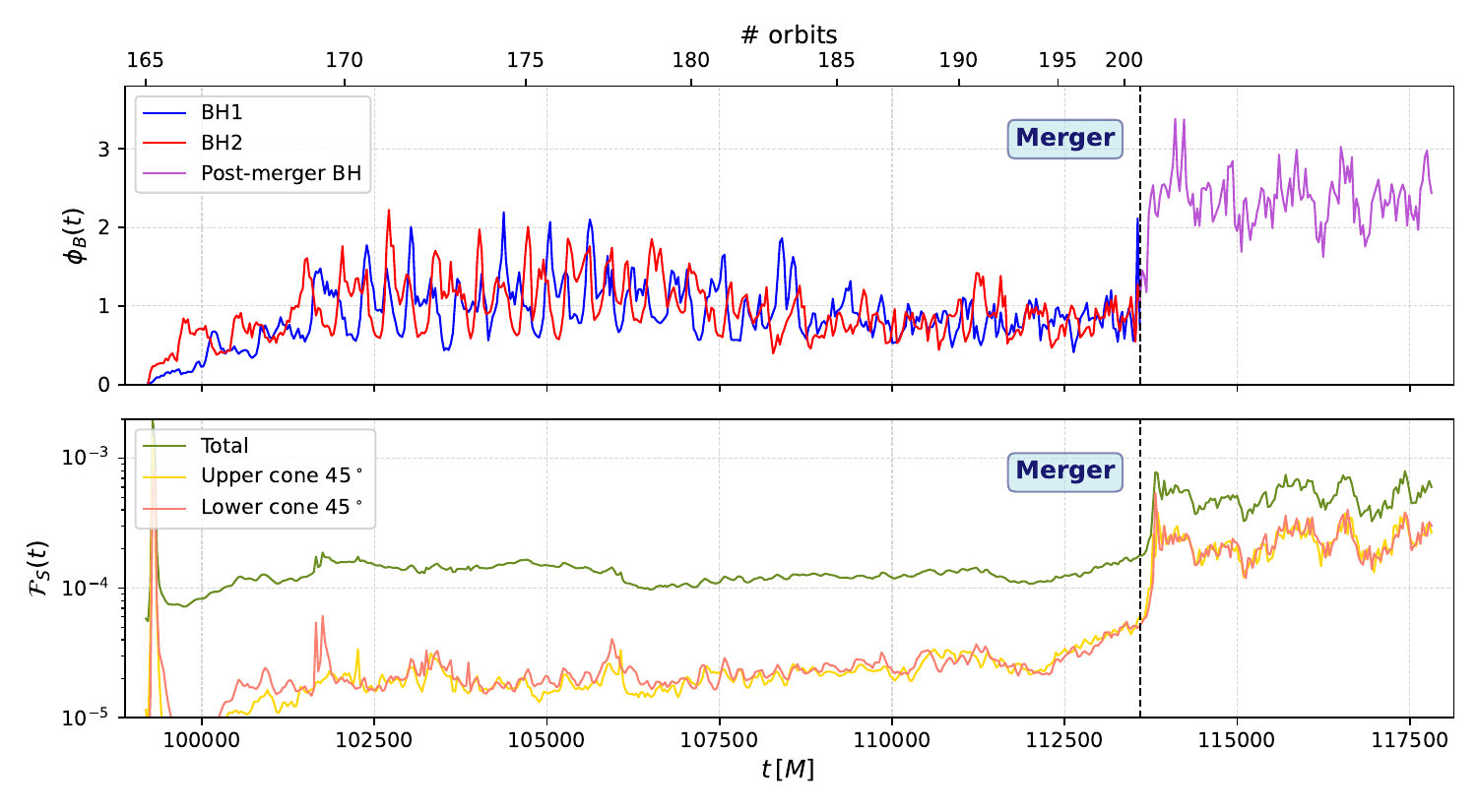}
    \captionsetup{justification = raggedright, format = hang}
    \caption{\textbf{[Upper panel]} Accretion-rate-normalized magnetic flux (see Eq.~\eqref{Dimensionless magnetic flux}) on the horizons. \textbf{[Lower panel]} Poynting flux on a spherical surface with coordinate radius ${r=100\,M}$ around the binary's center of mass. Note the sharp increase of both quantities at merger.}
    \label{fig: normalized magnetic flux on the AHs, Poynting flux at 100M}
\end{figure*}

In the upper panel of Fig.~\ref{fig: normalized magnetic flux on the AHs, Poynting flux at 100M}, we show that ${\phi_B\simeq 1\text{--}2}$ on the two individual black holes early in the inspiral, dwindling to values ${\lesssim\!1}$ in the last ${\sim\!5000\,M}$ before merger. At merger, ${\phi_B}$ jumps to ${2\text{--}3}$ on the single remaining black hole. These values of the normalized flux are small enough that the ram pressure of the accretion flow can confine the magnetic field, and the field is too weak to interrupt accretion.

\section{Discussion}
\label{sec: Discussion}

Our results have a number of implications for understanding the principal features of the final inspiral and merger of a pair of supermassive black holes.

\subsection{Gas dynamics unique to late inspiral and merger}
\label{subsec: Dynamics unique to late inspiral and merger}

\subsubsection{New features}

A very large literature now exists regarding the fluid dynamics of binary accretion. Features such as the cavity within ${\sim\!2a}$ of the center of mass and the minidisks around each of the partners in the binary have been extensively discussed. Our simulation has demonstrated that essentially all these features change dramatically when two black holes approach to ${\lesssim 20\,M}$ of one another:

\begin{itemize}
    \item When the separation is less than an order of magnitude larger than the the individual holes' ISCO radii, the masses of the minidisks rise and fall by large factors on timescales comparable to the binary orbital period. Near a minidisk's mass minimum, its density structure is better represented by a loosely-wound spiral than an axisymmetric disk.
    \item Once the separation shrinks to being only several times the ISCO radii, the minidisks never return to an axisymmetric structure.
    \item While the binary loses another factor of ${\sim\!2}$ in separation, the inner edge of the CBD, formerly quite sharply defined, spreads irregularly inward. As a result, rather than a pair of accretion streams feeding the minidisks, the flow breaks up into many irregular interacting streams. During this stage, distinct minidisks disappear. At the same time, magnetic ``bubbles'' form within the disorganized accretion flow.
    \item Immediately after the merger, when the pair of rapidly orbiting black holes is replaced by a single black hole, the gas heating rate jumps as shocks triggered by the abrupt change in the gravitational potential due to the black hole's sudden mass-loss run through nearby gas.
    \item Unlike most accreting black hole systems, the heating rate has little to do with the instantaneous accretion rate into the black hole(s) (the only other example in which this happens is tidal disruption events~\cite{Shiokawa2015, Piran2015, Ryu2023, SteinbergStone2024, Price2024}).
    \item Although inflow equilibrium is generally well-established from the inner portion of the CBD down to the black holes of the binary well after ``decoupling'' (i.e., when the binary's orbital evolution rate becomes faster than the CBD's inflow rate), in the latter inspiral it breaks down even in the cavity region surrounding the binary. In fact, as evidenced by both the strong nonmonotonicity in the cavity mass's time dependence and its uneven distribution in radius (see Figs.~\ref{fig: mass inside 15M, 15-30M, 30-40M} and~\ref{fig: M, U, dQ/dt in 15M}), ``decoupling'' does not lead to a progressive mass-starving of the gap region. Instead, it is better described as a decline in the accretion rate onto the black holes that is stabilized well before merger. Because the accretion rate declines even while mass continues to be injected from the CBD into the cavity, the gas mass in the inner portion of the cavity begins to grow ${\sim\!4000\,M}$ before the merger.
    \item The accretion rate onto the merged black hole is essentially identical to the sum of the accretion rates onto the binary black holes immediately before merger.
\end{itemize}

\subsubsection{Contrast with previous work}

The chaotic gas dynamics in the gap region that we observe close to merger contrasts sharply with some cognate previous work. The most dramatic contrast is with the work of \cite{Krauth2023}, who used 2D Newtonian hydrodynamics combined with orbital evolution given by the quadrupole gravitational wave radiation rate. In their calculation, very tightly confined streams spiral down to the black holes, where previously well-defined minidisks disappear ${\sim\!6000\,M}$ before merger without ever passing through the stream-like state. The regularity they observe in the accretion streams is likely due to the purely laminar flow in their CBD, very much unlike the MHD turbulence in disks with physical internal stresses. \cite{Bowen2017} showed that relativistic effects strengthen sloshing; use of Newtonian dynamics may therefore be the reason for extended maintenance of the disklike state, as well as the early total dissolution of the minidisks.

There are also contrasts with other work utilizing numerical relativity and 3D MHD. For example, \cite{Farris2012, Gold2014b} study a binary whose initial state contains a binary with a separation of ${10\,M}$ and a surrounding gas torus whose inner edge is, even after ${\sim\!10000\,M}$ of dynamical relaxation (with radiative cooling) in a nonevolving binary spacetime, ${\approx 15\text{--}20\,M}$ from the center of mass (see Fig.~1b and Fig.~2 of~\cite{Farris2012}). In their work, two well-defined accretion streams spiral onto the black holes' horizons until only ${\sim\!400\,M}$ before merger. These results differ qualitatively from ours, in which the gas moves much more chaotically even when the separation is larger than ${10\,M}$. Well before our binary shrinks to this separation, irregular gas structures in the inner portion of the CBD defy definition of a sharp edge, while low-density gas streams break off from these structures, occasionally interacting with each other before a fraction of their mass is captured by a black hole.

These qualitative differences can be appreciated by contrasting the bottom-left panel of Fig.~\ref{fig: evolution of mass density}---which shows the mass density distribution within the cavity when the binary separation is ${\sim\!10\,M}$---with the top panels of Fig.\,2 in~\cite{Gold2014b}. Comparing these figures also shows that the contrast between the regions of highest and lowest densities is at most ${2\text{--}3}$ orders of magnitude in~\cite{Gold2014b}, versus ${6\text{--}7}$ orders of magnitude in our case, even though the low-density (i.e., purple-ish) regions of Fig.~\ref{fig: evolution of mass density} have densities that are typically ${\sim\!3}$ orders of magnitude above the floor value. These differences in the gas dynamics within the cavity translate into differences in the mass accretion rate onto the binary. Whereas in~\cite{Farris2012, Gold2014b} this quantity drops only shortly before merger, in our simulation it settles to an approximately constant value (albeit with factor of ${2}$ oscillations) a full ${\sim\!4000\,M}$ before merger.

We believe the differences between the work of~\cite{Farris2012, Gold2014b} and ours can be most likely ascribed to the small binary separation at the initial state. At such a small separation, when orbital evolution is initiated, the timescale for binary orbital evolution is much shorter than the time for radial mass flow within the CBD; in other words, the system is already well past the start of decoupling. Consequently, the gas's spatial distribution primarily reflects the imposed initial state more than the dynamics that would have taken place if decoupling had been followed from a moment closer to its start. As we have shown, longer-term dynamical evolution places the gas in a rather different location from the one in which it is found in this earlier work.

\smallskip
A different comparison with previous work has to do with the total mass of gas in the cavity. \cite{Krolik2010} suggested that the total mass within ${\sim\!100\,M}$ might be as much as ${\sim\!1 M_8^2 M_\odot}$\,, where ${M_8}$ is the black hole mass in units of ${10^8 M_\odot}$\,. Because the code-unit of density in our simulation is arbitrary, to translate the simulation results into physical-unit numbers, it is necessary to choose specific values for both the binary black hole mass and its accretion rate at a fiducial location. Following~\cite{Schnittman2013}, we have
\begin{equation}
    \label{eq: physical mass density}
    \begin{split}
        \rho_\text{phys} &= \rho_\text{code}\frac{4\pi}{\kappa_T r_g} \frac{\dot{m}/\eta}{\dot{M}_\text{code}}\simeq \\
        &\simeq 2.14\times 10^{-12}\rho_\text{code}\,M_8^{-1}\frac{\dot{m}/\eta}{\dot M_\text{code}}\,\text{g}\,\text{cm}^{-3},
    \end{split}
\end{equation}
where ${\kappa_T}$ is the Thomson opacity, ${\dot{m}}$ is the accretion rate in Eddington units, and ${\eta}$ is a nominal radiative efficiency. In our particular case, we take as our fiducial value for the accretion rate ${\dot{M}_\text{code} = 0.002}$\,, the rate crossing the cavity at the beginning of the inspiral, and suppose it corresponds to ${\dot{m}/\eta = 1}$, so that ${\rho_\text{phys}\simeq 1.1 \times 10^{-9}\rho_\text{code}\,M_8^{-1}\text{g}\,\text{cm}^{-3}}$. The denser regions of the flow (the accretion streams during the early part of the inspiral, the midplane of the CBD afterwards) therefore have ${\rho\simeq 10^{-11} M_8^{-1}\text{g}\,\text{cm}^{-3}}$ (see Fig.~\ref{fig: evolution of mass density}). At ${t\simeq 5000\,M}$ before merger, the code-unit mass within ${r\leq 15\,M}$ is ${\sim\!0.03}$ (Fig.~\ref{fig: mass inside 15M, 15-30M, 30-40M}), so the physical mass within that zone is ${\sim\!1 \times 10^{29} M_8^2\,\text{g}\simeq 5\times 10^{-5} M_\odot}$. However, the mass in this zone begins growing ${\sim\!4000\,M}$ before merger and is ${5\times}$ larger by ${\sim\!4000\,M}$ after merger. The mass within ${40\,M}$ is ${\sim\!6\times 10^{-2} M_8^2 M_\odot}$ throughout the inspiral and immediate postmerger period. Thus, the mass near the merging binary at the time of merger is likely to be less than a solar mass, but perhaps by only ${1\text{--}3}$ orders of magnitude. Note, also, that the ratio between the nearby gas mass and the black hole mass increases ${\propto M\dot{m}/\eta}$.

\subsection{Bolometric lightcurve}
\label{subsec: Bolometric lightcurve}

Although our cooling rate prescription is ad hoc and does not account for the detailed mechanisms of EM radiation, the fact that it closely matches the heating rate suggests that it faithfully represents the bolometric cooling rate if the cooling time is short compared to the dynamical time. There may, however, be departures between our predicted lightcurve and the actual lightcurve when this assumption breaks down, as may happen when the gas is especially optically thick. In this case, the variations we predict can be sharper than would occur in reality. However, at very high temperatures (and these are often found in the cavity gas during inspiral, as in Fig.~\ref{fig: temperature}), the dominant cooling rate is almost certainly inverse Compton scattering, which is generically rapid\footnote{Its characteristic timescale in an accretion context is ${\sim\left(m_e/\mu_e\right)\dot{m}^{-1}\left(r/r_g\right)^2 r_g/c}$, where ${\mu_e}$ is the plasma's mean mass per electron, so ${m_e/\mu_e\simeq 5\times 10^{-4}}$ is common.}
and often faster than the dynamical time. As one would expect physically and is shown in the lower panel of Fig.~\ref{fig: mdot tot, luminosity in 100M}, the timescales of variation rise outward. Because the radiation variability timescale at all radii is longer than the local dynamical timescale (i.e., the timescale at which our cooling rate vents heat), we expect that these variations are reasonably accurately portrayed.

\smallskip
Our most striking prediction about the bolometric lightcurve is that during the inspiral there should be a gradual downward trend, but at the time of merger there should be a sharp rise by ${\sim\!50\%}$\,. Because this change in luminosity is associated with an even more dramatic change in the locations from which the light is radiated, it is very likely that the spectrum changes dramatically at merger; detailed calculations will be required to identify the nature of the spectral shift. The duration of the inspiral decline is ${\sim\!10^4 M\simeq 2\,M_8\,\text{months}}$; the jump and spectral change at merger may be accomplished in as little as ${\sim\!100\,M\sim 0.5 M_8\,\text{days}}$.

In physical units, the power radiated is
\begin{equation}
    L = L_\text{EM}L_E\,\frac{\dot{m}/\eta}{\dot{M}}\;,
\end{equation}
where ${L_\text{EM}}$ is the luminosity in code-units and ${L_E}$ is the Eddington luminosity. Thus, when the luminosity is ${\sim\!10^{-4}}$ in code-units, we would expect a physical luminosity ${\sim\!0.05\left(\dot{m}/\eta\right)L_E}$\,. The actual total luminosity produced within ${r=100\,M}$ is ${\sim\!3\times 10^{-4}}$ at the beginning of our simulation, drops to ${\sim\!0.5\times 10^{-4}}$ shortly before merger, and jumps back to ${\sim\!0.8\times 10^{-4}}$ at merger. In other words, if the accreting binary, when widely separated, produced a luminosity ${\dot{m}L_E}$ by conventional accretion onto the minidisks with a radiative efficiency ${\eta=0.1}$\,, its luminosity due to entirely unconventional radiation processes during its final inspiral would have gone from ${\sim\!1.5\,\dot{m}\left(\eta/0.1\right)^{-1}L_E}$ to ${\sim\!0.25\,\dot{m}\left(\eta/0.1\right)^{-1}L_E}$ to
${\sim\!0.4\,\dot{m}\left(\eta/0.1\right)^{-1}L_E}$ over the course of the inspiral and the immediate aftermath of merger. Thus, despite the diminution of the accretion rate onto the black holes by a factor ${\sim\!10}$, the actual luminosity derived from gas heating during the inspiral is depressed by no more than a factor ${\sim\!4}$ and, early in the inspiral, is actually tens of percent \textit{greater} than what might be expected from the nominal accretion rate.

\smallskip
The smoothness of the lightcurve is itself a new result. Many have speculated that when the system ``decouples'', i.e., when the binary evolves on a timescale shorter than the inflow time in the inner CBD, matter flow into the cavity ceases, leaving the immediate vicinity of the binary almost bereft of gas and suppressing photon emission from the entire cavity region until after the merger~\cite{MilosavljevicPhinney2005}. Others (e.g.,~\cite{Krolik2010}) have speculated that gas can still occupy the region around the binary during the final stages of the inspiral. In this case, it was suggested that when the GW luminosity is near its peak, the spacetime near the merging black holes is so violently distorted that, for a short time, it pushes gas streams against each other with equal violence, causing a sudden deposition of extra heat into the gas.

As Fig.~\ref{fig: M, U, dQ/dt in 15M} shows, the truth is in between. The gas mass close to the binary falls by a factor ${\sim\!10}$ during the first part of the inspiral, but as the merger approaches, rises back to ${\sim\!1/4}$ of its earlier mass. Although there is persistent heating in both the binary region and the outer cavity throughout the inspiral (Fig.~\ref{fig: mdot tot, luminosity in 100M}), it does not climax immediately before merger. The reason for this smoothness is fundamentally the same as the reason the gravitational waveform is also smooth: rather than a chaotically dynamical spacetime, there is only a coherent propagating wave, driven by the shrinking black hole binary. Heating during the inspiral can be similarly thought of as driven by the smoothly time-varying potential created by the binary. The change in the location of heating at merger is due to the fundamental change in the gravitational potential, from that of an orbiting binary to that of a single stationary mass.

Here, too, there is contrast with previous work. Over the period corresponding to the duration of our simulation, \cite{Krauth2023} found that the associated bolometric luminosity drops by a factor ${\sim\!5}$, very similar to our factor ${\sim\!4}$. However, their (2D Newtonian hydro) simulation predicts that the bolometric luminosity maintains the level reached at merger for ${\sim 10^5 M}$ afterward, whereas we see a sharp upward step at merger. They also predict that merger leads to a complete shut-off of radiation in the band associated with the innermost radii of accretion, whereas we find the innermost region rises in luminosity by a factor ${\sim\!5}$.

\subsection{Spectrum}
\label{subsec: Spectrum}

To predict the spectrum radiated, the first question that must be answered is how much of the light is thermalized. Equivalently, is the effective absorptive opacity photons face large enough that the expectation value for a photon's absorption while diffusing out of the gas is at least comparable to unity? Given the density scale we have estimated (see Eq.~\eqref{eq: physical mass density}), the electron scattering optical depth in the cavity region is
\begin{equation}
    \tau_T\sim\kappa_T\rho_\text{phys} H = 4\pi\left(H/r\right)\left(r/r_g\right)\left(\dot{m}/\eta\right)\frac{\rho_\text{code}}{\dot{M}_\text{code}}\;,
\end{equation}
where ${H}$ is the local vertical scale height at a distance ${r}$ from the center of mass of the system. The ratio ${\rho_\text{code}/\dot{M}}$ ranges from ${\sim\!10^{-4}\text{--}10}$, so if ${\left(H/r\right)\left(r/r_g\right)\sim\dot{m}/\eta\sim\!1}$, the Thomson optical depth ranges from ${\sim\!10^{-3}\text{--}10^2}$. The free-free optical depth for photons at the peak of a Planck spectrum is ${\sim\!5\times 10^{-2}\left(\rho_\text{code}/10^{-2}\right)^2\,T_5^{-7/2} M_8^{-2}\left(H/r_g\right)}$, for ${T_5}$ the gas temperature in units of ${10^5\text{K}}$; the code density is normalized to the highest density found in the cavity. Because the temperature within tens of ${M}$ of a black hole is unlikely to be much less than ${10^5\text{K}}$ and could be a good deal higher, the effective thermalization optical depth ${\left[\left(\tau_T + \tau_\text{ff}\right)\tau_\text{ff}\right]^{1/2}}$ might at most be ${\sim\!O\!\left(1\right)}$, but is generally below unity. Consequently, some fraction of the emitted luminosity may be thermalized, but it could be a small fraction. The remainder of the photon power is likely due to inverse Compton scattering of photons created in the CBD.

\smallskip
In the inner cavity (${r < 15\,M}$), the typical ratio of internal energy to rest-mass is ${\sim\!10^{-2}\simeq10\,\text{MeV}}$ (Fig.~\ref{fig: M, U, dQ/dt in 15M}). This is hot enough to drive significant ${e^\pm}$-pair creation. If most of the internal energy is converted to pair rest-mass, the optical depth could increase by a factor ${\sim\!10}$ and the temperature would fall to ${\sim\!500\,\text{keV}}$. The majority of the power from the inner region would then emerge in hard X-rays.

\subsection{Jet-launching}
\label{subsec: Jet-launching}

Even if the individual black holes do not spin, their orbital motion at the very end of the inspiral is rapid enough to drive a weak jet along the orbital axis; the acceleration in orbital motion as the binary nears merger (see also~\cite{Gold2014a}) accounts for the increase in total Poynting flux in the last ${\sim\!1000\,M}$ before merger (Fig.~\ref{fig: normalized magnetic flux on the AHs, Poynting flux at 100M}). Almost independent of the pre-merger black holes' spins, the merger product must spin reasonably fast if the binary mass-ratio is ${\sim\!1}$. Mergers in general should therefore launch jets, part of whose energy is emitted in photons beamed and boosted in the jet's direction. Sudden turn-on of jet emission would then strongly indicate two slowly spinning black holes have merged. If at least one of the merging black holes had a high spin parameter, it should have driven a jet all along, but in principle its direction might have been different from the postmerger jet's. If both black holes spun rapidly enough to drive jets, the jets could interact. When the spins are aligned with the orbit, the interaction takes place where the jet cones have widened enough for the jets to touch~\cite{Gutierrez2024Radiation}. When the spins are misaligned, relativistic spin-orbit and spin-spin interactions will cause them to precess, producing a complicated time-dependence for distant observers. We will discuss a few examples of these options in a forthcoming paper.

\section{Conclusions and outlook}
\label{sec: Conclusions and outlook}

We have presented the first 3D GRMHD plus numerical relativity simulation that, starting from an initial separation of ${20\,M}$ and prerelaxing the gas and magnetic field distributions onto the binary's spacetime, follows the inspiral, merger, and early postmerger stages of the evolution of a supermassive black hole binary (in this case, it is circular, equal-mass, and nonspinning). Throughout our simulation, we have modeled gas cooling through the emission of EM radiation in a simple though effective way. Our methods allowed us to unveil the highly chaotic nature of the gas's motion in the gap region in the late inspiral and the ultimate dissolution of the minidisks, as well as a sharp increase in the bolometric lightcurve tied to a predicted drastic change in spectrum at the time of merger.

This work establishes a foundation for future explorations of more complex SMBBH merger scenarios involving, for instance, different black hole mass ratios (e.g.,~\cite{Gold2014b}), orbital eccentricities (e.g.,~\cite{Manikantan2024}), and spin configurations (e.g.,~\cite{Paschalidis2021, Bright2023}), as well as different thermodynamic properties of accretion disks. By enhancing the physical accuracy of these models, we can deepen our understanding of the intricate processes involved in these astronomical events, ultimately enabling more precise predictions and interpretations of observational data. In particular, starting the simulations from larger binary separations is essential in order to gain deeper insight into the gas's behavior in the gap region in this regime.

\smallskip
EM observables during the late inspiral, merger, and postmerger stages are likely sensitive to the binary's mass ratio and orbital eccentricity. Specifically, if the mass ratio ${q\equiv m_1/m_2\neq 1}$, any asymmetry in the gas distribution around the binary could potentially increase accretion onto the less massive hole~\cite{Dorazio2013, Farris2014, Gold2014b}, thereby facilitating mass sloshing between the minidisks. Because the black hole orbits tend to circularize by radiating GWs, orbital eccentricity should be small when the black holes are separated by a distance ${\leq 20\,M}$~\cite{Armitage2005, ColemanMiller2005, Sesana2013}; however, eccentricity could still impact the dynamics and observable features of SMBBH systems in a significant way~\cite{Gualandris2022, Manikantan2024}.

The binary's spin configuration can have important implications for the accretion dynamics and jet power throughout the evolution of the system~\cite{LopezArmengol2021, Combi2022, Gutierrez2022, Paschalidis2021, Bright2023}. For instance, the ISCO radius shrinks with greater prograde spin; as a result, minidisks are likely more persistent when the black hole they orbit spins in a prograde sense~\cite{Paschalidis2021, Bright2023}. The spins may often be aligned with the binary's angular momentum~\cite{Bogdanovic2007, MillerKrolik2013, Sorathia2013}, but the arguments are far from airtight. When the spins are oblique, gravitomagnetic torques can deflect the accretion streams, causing the minidisks to twist and leading to time-varying, nonplanar gas orbits due to precession induced by spin-spin and spin-orbit couplings. At close separations, in-plane spins might induce ``bobbing'' of the orbital plane~\cite{Campanelli2006HangUp, Campanelli2007, Keppel2009, Hemberger2013} and possible gas shocks, along with significant interactions between the jets~\cite{Gutierrez2024Radiation, Ressler2024} and recoil of the merger remnant~\cite{Campanelli2007, LoustoZlochower2011}. Under specific conditions, phenomena such as spin flipping (``S-Flip'') or a total angular momentum reversal (``L-Flop'') could occur~\cite{Lousto2015, Lousto2016, Kesden2015, Lousto2016FlipFlop, Healy2018}. Finally, depending on the configuration, spin-orbit interactions can either increase or decrease the time to merger compared to the zero-spin case~\cite{Campanelli2006HangUp, Hemberger2013}.

\smallskip
Magnetic field strength and topology can also play a crucial role in shaping accretion flows by providing dynamical feedback that influences the behavior of infalling matter. For example, in the case of single black holes, magnetically arrested disks (MADs) have been proposed to explain the observations of M87* and Sagittarius A* by the Event Horizon Telescope (EHT) Collaboration~\cite{EHT2019}.  Although MADs have been extensively studied in the context of isolated black holes~\cite{Narayan2003, Tchekhovskoy2011, Ripperda2022}, only initial studies of their possible presence in SMBBH mergers have been conducted~\cite{Most2024}. Clearly, further exploration is required to achieve a comprehensive understanding of the effects of magnetic fields on such complex astrophysical phenomena.

\smallskip
Lastly, more accurate radiation transport models (see, e.g.,~\cite{Kinch2020, Stone2024, Tiwari2025}) than the simple cooling prescription used in this work are needed to effectively account for the intricate interactions between radiation and matter in these extreme environments. Radiation transport determines the gas's equation of state and the radiation forces on the fluid; it may, for example, affect the duration of the ``decoupled'' period. In addition, of course, a careful solution of the radiation transport problem is essential for predicting the EM spectrum seen by observers.  

These improvements, along with the availability of advanced computational infrastructure, hold the potential to yield groundbreaking insight into the astrophysical processes at the heart of supermassive black hole mergers.

\section*{Acknowledgments}

The authors would like to thank the anonymous referee for a thorough reading of the manuscript and insightful comments and suggestions. We would also like to thank Zachariah B. Etienne and Carlos Lousto for insightful physics-related discussions and Steve Fromm for a careful reading of the manuscript. We also appreciate the assistance of Roland Haas in helping us modify his \textsc{ReadInterpolate} code for our purposes.

The authors gratefully acknowledge the National Science Foundation (NSF) for financial support from grants AST-2009260, AST-2009330, PHY-2409706, PHY-2110338, PHY-2110339, and OAC-2004044, as well as the National Aeronautics and Space Administration (NASA) for financial support from TCAN Grant No. 80NSSC24K0100. Work at Oak Ridge National Laboratory is supported under contract DE-AC05-00OR22725 with the U.S. Department of Energy (DOE). VM was also supported by the Exascale Computing Project (17-SC-20-SC), a collaborative effort of the DOE Office of Science and the National Nuclear Security Administration. 

This research used resources from the Texas Advanced Computing Center's (TACC) Frontera and Vista supercomputer allocations (award PHY20010). Additional resources were provided by the BlueSky, Green Prairies, and Lagoon clusters of the Rochester Institute of Technology (RIT) acquired with NSF grants PHY-2018420, PHY-0722703, PHY-1229173, and PHY-1726215.

\section*{Data availability}
The data that support the findings of this article are not publicly available upon publication because it is not technically feasible and/or the cost of preparing, depositing, and hosting the data would be prohibitive within the terms of this research project. The data are available from the authors upon reasonable request.

\appendix
\section{Analysis tools}

The primary physical insights from the inspiral and merger simulation presented in this work stem from integrating the MHD quantities over appropriate surfaces and volumes. Although some analysis was conducted during the simulation, most of it had to be performed \textit{post hoc}, as identifying the key physical quantities in advance was not entirely feasible. This Appendix provides some detail about two codes---\textsc{CactusSurfaceIntegrals} and \textsc{CactusVolumeIntegrals}---that we developed in order to calculate most of the surface and volume integrals presented here. The codes are version-controlled in a private GitHub
repository, but we plan to make them available to the public in the near future.

\subsection{Surface integrals}

We developed \textsc{CactusSurfaceIntegrals}, an MPI-parallel \textsc{C++} code capable of calculating the integrals of various quantities of physical interest over planar, spherical, or cylindrical surfaces embedded in \textsc{Cactus} grids with Cartesian topology and box-in-box mesh refinement. Since the main application of the code is (for now) the analysis of binary black hole simulations, the integration surface can be fixed in time or centered around the location of either black hole or centered in the center of mass of the binary system. Furthermore, the extent of the surface can be made proportional to the binary separation.

In a nutshell, the code's workflow is the following:
\begin{enumerate}
    \item Each MPI task is assigned a chunk of the integration surface.
    \item Each point on the surface chunk is assigned the \textsc{Cactus} grid component with the finest available resolution that overlaps that point.
    \item All the needed fields are interpolated on each point of the surface chunk using the previously assigned grid patch.
    \item The integral(s) is (are) computed by each MPI rank on the corresponding surface chunk, then the local contributions are summed and written to file.
\end{enumerate}

A typical binary black hole simulation involves  multiple MPI processes, so that the GRMHD fields are scattered across multiple files (possibly one file per process). Each of those files---in HDF5 format~\cite{HDF5}---typically contains field data on multiple \textsc{Cactus} grid components. A given \textsc{Cactus} grid component may overlap multiple chunks of the integration surface, corresponding to as many MPI processes in \textsc{CactusSurfaceIntegrals}, so that multiple MPI processes may need to open each file at once. For this reason, \textsc{CactusSurfaceIntegrals} builds MPI communicators listing which MPI processes should open any given file, and then opens each file collectively across that communicator using the appropriate routines from the HDF5 library.

\smallskip
Interpolations of the fields and their spatial derivatives are carried out using a three-dimensional Newton scheme (divided differences algorithm) which has been successfully tested against known linear and quadratic functions. The user can choose the interpolation order, which is set to 4 in this work. The interpolation may occasionally fail to provide physically meaningful results: for example, the interpolated fluid's mass density and pressure may be slightly below their floor values, or the fluid's velocity may be slightly superluminal. First, the number of incorrect interpolations is typically reduced by interpolating the fields' logarithm, possibly adding a constant to the field values in order to avoid negative arguments in the logarithmic function. Second, if incorrect interpolations occur at a given point, the user can choose whether to abort execution or replace the interpolated value at that point with the averaged field values interpolated at the four nearest neighboring points; note that this may require exchanging data among neighboring MPI processes. If that still fails, the interpolated value can be replaced by a user-specified default value.

\smallskip
Finally, the code can calculate the coordinate area of the integration surface and compare it with the corresponding exact area to provide an upper bound to the quality of the integration---measuring the surface area does not involve any field interpolations, which naturally come with their own truncation errors. We always set the number of points on the integration surface in such a way that the surface resolution at any given point is no coarser than resolution of the \textsc{Cactus} grid around the same point. This way, we typically find that \textsc{CactusSurfaceIntegrals} underestimates the integration area by at most a fraction ${10^{-8}}$.

\subsection{Volume integrals}

We also developed \textsc{CactusVolumeIntegrals}, an MPI-parallel \textsc{C++} code to compute spherical volume integrals over \textsc{Cactus} grids with Cartesian topology and box-in-box mesh refinement. As in \textsc{CactusSurfaceIntegrals}, the center of the sphere can either be fixed in time, coincide with the location of either puncture, or be located at the center of mass of the binary; also, its radius can be made proportional to the binary separation. In addition to that, apparent horizons or spherical volumes of arbitrary radii centered on the punctures can be excluded from the integration domain.

The workflow of the code is essentially as follows:
\begin{enumerate}
    \item Each MPI process reads one and only one \textsc{Cactus} HDF5 file containing the information about how the GRMHD fields are laid out in the simulation domain. Note that the code must run as many MPI processes as there are files to read.
    \item All grid components in each file are checked for overlap against the integration sphere by means of Arvo's algorithm~\cite{Arvo1990}.
    \item All MPI processes must know which grid components overlap the integration sphere (no matter which files those components belong to), so that only the components with the finest available resolution are used in the integration. This step involves collective MPI operations.
    \item Each MPI process calculates the integral(s) over all grid components (or part of them) with the finest available resolution that overlap the sphere and belong to the corresponding HDF5 file. The contribution from each grid point is weighed by a factor ${1/2^{rl}}$, where ${rl\in\left\{0, 1, 2, \dots\,\right\}}$ denotes the refinement level the given grid point belongs to and ${rl = 0}$ corresponds to the coarsest level.
\end{enumerate}

Similar to \textsc{CactusSurfaceIntegrals}, this code can calculate the integration volume numerically and compare it to the exact value to provide an estimate of the integration accuracy. In the applications involved in the work presented here, \textsc{CactusVolumeIntegrals} typically underestimates the volume of the integration region by only ${0.3\text{--}0.8\%}$\,.

\bibliographystyle{apsrev4-2}
\bibliography{PRDBibliography}  

\begin{thebibliography}{158}%
\makeatletter
\providecommand \@ifxundefined [1]{%
 \@ifx{#1\undefined}
}%
\providecommand \@ifnum [1]{%
 \ifnum #1\expandafter \@firstoftwo
 \else \expandafter \@secondoftwo
 \fi
}%
\providecommand \@ifx [1]{%
 \ifx #1\expandafter \@firstoftwo
 \else \expandafter \@secondoftwo
 \fi
}%
\providecommand \natexlab [1]{#1}%
\providecommand \enquote  [1]{``#1''}%
\providecommand \bibnamefont  [1]{#1}%
\providecommand \bibfnamefont [1]{#1}%
\providecommand \citenamefont [1]{#1}%
\providecommand \href@noop [0]{\@secondoftwo}%
\providecommand \href [0]{\begingroup \@sanitize@url \@href}%
\providecommand \@href[1]{\@@startlink{#1}\@@href}%
\providecommand \@@href[1]{\endgroup#1\@@endlink}%
\providecommand \@sanitize@url [0]{\catcode `\\12\catcode `\$12\catcode
  `\&12\catcode `\#12\catcode `\^12\catcode `\_12\catcode `\%12\relax}%
\providecommand \@@startlink[1]{}%
\providecommand \@@endlink[0]{}%
\providecommand \url  [0]{\begingroup\@sanitize@url \@url }%
\providecommand \@url [1]{\endgroup\@href {#1}{\urlprefix }}%
\providecommand \urlprefix  [0]{URL }%
\providecommand \Eprint [0]{\href }%
\providecommand \doibase [0]{https://doi.org/}%
\providecommand \selectlanguage [0]{\@gobble}%
\providecommand \bibinfo  [0]{\@secondoftwo}%
\providecommand \bibfield  [0]{\@secondoftwo}%
\providecommand \translation [1]{[#1]}%
\providecommand \BibitemOpen [0]{}%
\providecommand \bibitemStop [0]{}%
\providecommand \bibitemNoStop [0]{.\EOS\space}%
\providecommand \EOS [0]{\spacefactor3000\relax}%
\providecommand \BibitemShut  [1]{\csname bibitem#1\endcsname}%
\let\auto@bib@innerbib\@empty
\bibitem [{\citenamefont {{Schramm}}(1992)}]{Schramm1992}%
  \BibitemOpen
  \bibfield  {author} {\bibinfo {author} {\bibfnamefont {D.~N.}\ \bibnamefont
  {{Schramm}}},\ }\href {https://doi.org/10.1016/0920-5632(92)90180-Z}
  {\bibfield  {journal} {\bibinfo  {journal} {Nuclear Physics B Proceedings
  Supplements}\ }\textbf {\bibinfo {volume} {28}},\ \bibinfo {pages} {243}
  (\bibinfo {year} {1992})}\BibitemShut {NoStop}%
\bibitem [{\citenamefont {{Peebles}}(1993)}]{Peebles1993}%
  \BibitemOpen
  \bibfield  {author} {\bibinfo {author} {\bibfnamefont {P.~J.~E.}\
  \bibnamefont {{Peebles}}},\ }\href {https://doi.org/10.1515/9780691206721}
  {\emph {\bibinfo {title} {Principles of Physical Cosmology}}}\ (\bibinfo
  {publisher} {Princeton University Press},\ \bibinfo {year}
  {1993})\BibitemShut {NoStop}%
\bibitem [{\citenamefont {{Schneider}}(2006)}]{Schneider2006}%
  \BibitemOpen
  \bibfield  {author} {\bibinfo {author} {\bibfnamefont {P.}~\bibnamefont
  {{Schneider}}},\ }\href@noop {} {\emph {\bibinfo {title} {Extragalactic
  Astronomy and Cosmology}}}\ (\bibinfo  {publisher} {Springer},\ \bibinfo
  {year} {2006})\BibitemShut {NoStop}%
\bibitem [{\citenamefont {{Kormendy}}\ and\ \citenamefont
  {{Ho}}(2013)}]{KormendyHo2013}%
  \BibitemOpen
  \bibfield  {author} {\bibinfo {author} {\bibfnamefont {J.}~\bibnamefont
  {{Kormendy}}}\ and\ \bibinfo {author} {\bibfnamefont {L.~C.}\ \bibnamefont
  {{Ho}}},\ }\href {https://doi.org/10.1146/annurev-astro-082708-101811}
  {\bibfield  {journal} {\bibinfo  {journal} {\araa}\ }\textbf {\bibinfo
  {volume} {51}},\ \bibinfo {pages} {511} (\bibinfo {year} {2013})},\ \Eprint
  {https://arxiv.org/abs/1304.7762} {arXiv:1304.7762 [astro-ph.CO]}
  \BibitemShut {NoStop}%
\bibitem [{\citenamefont {{Bogdanovi{\'c}}}\ \emph {et~al.}(2022)\citenamefont
  {{Bogdanovi{\'c}}}, \citenamefont {{Miller}},\ and\ \citenamefont
  {{Blecha}}}]{Bogdanovic2022}%
  \BibitemOpen
  \bibfield  {author} {\bibinfo {author} {\bibfnamefont {T.}~\bibnamefont
  {{Bogdanovi{\'c}}}}, \bibinfo {author} {\bibfnamefont {M.~C.}\ \bibnamefont
  {{Miller}}},\ and\ \bibinfo {author} {\bibfnamefont {L.}~\bibnamefont
  {{Blecha}}},\ }\href {https://doi.org/10.1007/s41114-022-00037-8} {\bibfield
  {journal} {\bibinfo  {journal} {Living Reviews in Relativity}\ }\textbf
  {\bibinfo {volume} {25}},\ \bibinfo {eid} {3} (\bibinfo {year} {2022})},\
  \Eprint {https://arxiv.org/abs/2109.03262} {arXiv:2109.03262 [astro-ph.HE]}
  \BibitemShut {NoStop}%
\bibitem [{\citenamefont {{Volonteri}}\ \emph {et~al.}(2003)\citenamefont
  {{Volonteri}}, \citenamefont {{Haardt}},\ and\ \citenamefont
  {{Madau}}}]{Volonteri2003}%
  \BibitemOpen
  \bibfield  {author} {\bibinfo {author} {\bibfnamefont {M.}~\bibnamefont
  {{Volonteri}}}, \bibinfo {author} {\bibfnamefont {F.}~\bibnamefont
  {{Haardt}}},\ and\ \bibinfo {author} {\bibfnamefont {P.}~\bibnamefont
  {{Madau}}},\ }\href {https://doi.org/10.1086/344675} {\bibfield  {journal}
  {\bibinfo  {journal} {\apj}\ }\textbf {\bibinfo {volume} {582}},\ \bibinfo
  {pages} {559} (\bibinfo {year} {2003})},\ \Eprint
  {https://arxiv.org/abs/astro-ph/0207276} {arXiv:astro-ph/0207276 [astro-ph]}
  \BibitemShut {NoStop}%
\bibitem [{\citenamefont {{Campanelli}}\ \emph {et~al.}(2010)\citenamefont
  {{Campanelli}}, \citenamefont {{Lousto}}, \citenamefont {{Mundim}},
  \citenamefont {{Nakano}}, \citenamefont {{Zlochower}},\ and\ \citenamefont
  {{Bischof}}}]{Campanelli2010}%
  \BibitemOpen
  \bibfield  {author} {\bibinfo {author} {\bibfnamefont {M.}~\bibnamefont
  {{Campanelli}}}, \bibinfo {author} {\bibfnamefont {C.~O.}\ \bibnamefont
  {{Lousto}}}, \bibinfo {author} {\bibfnamefont {B.~C.}\ \bibnamefont
  {{Mundim}}}, \bibinfo {author} {\bibfnamefont {H.}~\bibnamefont {{Nakano}}},
  \bibinfo {author} {\bibfnamefont {Y.}~\bibnamefont {{Zlochower}}},\ and\
  \bibinfo {author} {\bibfnamefont {H.-P.}\ \bibnamefont {{Bischof}}},\ }\href
  {https://doi.org/10.1088/0264-9381/27/8/084034} {\bibfield  {journal}
  {\bibinfo  {journal} {Classical and Quantum Gravity}\ }\textbf {\bibinfo
  {volume} {27}},\ \bibinfo {eid} {084034} (\bibinfo {year} {2010})},\ \Eprint
  {https://arxiv.org/abs/1001.3834} {arXiv:1001.3834 [gr-qc]} \BibitemShut
  {NoStop}%
\bibitem [{\citenamefont {{J. Antoniadis \textit{et al.} (EPTA Collaboration
  and InPTA Collaboration)}}(2023)}]{Antoniadis2023}%
  \BibitemOpen
  \bibfield  {author} {\bibinfo {author} {\bibnamefont {{J. Antoniadis
  \textit{et al.} (EPTA Collaboration and InPTA Collaboration)}}},\ }\href
  {https://doi.org/10.1051/0004-6361/202346844} {\bibfield  {journal} {\bibinfo
   {journal} {Astronomy \& Astrosphysics}\ }\textbf {\bibinfo {volume} {678}},\
  \bibinfo {eid} {A50} (\bibinfo {year} {2023})},\ \Eprint
  {https://arxiv.org/abs/2306.16214} {arXiv:2306.16214 [astro-ph.HE]}
  \BibitemShut {NoStop}%
\bibitem [{\citenamefont {{M. Colpi \textit{et al.}}}(2024)}]{Colpi2024}%
  \BibitemOpen
  \bibfield  {author} {\bibinfo {author} {\bibnamefont {{M. Colpi \textit{et
  al.}}}},\ }\href {https://doi.org/10.48550/arXiv.2402.07571} {\bibfield
  {journal} {\bibinfo  {journal} {arXiv e-prints}\ ,\ \bibinfo {eid}
  {arXiv:2402.07571}} (\bibinfo {year} {2024})},\ \Eprint
  {https://arxiv.org/abs/2402.07571} {arXiv:2402.07571 [astro-ph.CO]}
  \BibitemShut {NoStop}%
\bibitem [{\citenamefont {{P. Amaro-Seoane \textit{et
  al.}}}(2023)}]{AmaroSeoane2023}%
  \BibitemOpen
  \bibfield  {author} {\bibinfo {author} {\bibnamefont {{P. Amaro-Seoane
  \textit{et al.}}}},\ }\href {https://doi.org/10.1007/s41114-022-00041-y}
  {\bibfield  {journal} {\bibinfo  {journal} {Living Reviews in Relativity}\
  }\textbf {\bibinfo {volume} {26}},\ \bibinfo {eid} {2} (\bibinfo {year}
  {2023})},\ \Eprint {https://arxiv.org/abs/2203.06016} {arXiv:2203.06016
  [gr-qc]} \BibitemShut {NoStop}%
\bibitem [{\citenamefont {{Dal Canton}}\ \emph {et~al.}(2019)\citenamefont
  {{Dal Canton}}, \citenamefont {{Mangiagli}}, \citenamefont {{Noble}},
  \citenamefont {{Schnittman}}, \citenamefont {{Ptak}}, \citenamefont
  {{Klein}}, \citenamefont {{Sesana}},\ and\ \citenamefont
  {{Camp}}}]{DalCanton2019}%
  \BibitemOpen
  \bibfield  {author} {\bibinfo {author} {\bibfnamefont {T.}~\bibnamefont {{Dal
  Canton}}}, \bibinfo {author} {\bibfnamefont {A.}~\bibnamefont {{Mangiagli}}},
  \bibinfo {author} {\bibfnamefont {S.~C.}\ \bibnamefont {{Noble}}}, \bibinfo
  {author} {\bibfnamefont {J.}~\bibnamefont {{Schnittman}}}, \bibinfo {author}
  {\bibfnamefont {A.}~\bibnamefont {{Ptak}}}, \bibinfo {author} {\bibfnamefont
  {A.}~\bibnamefont {{Klein}}}, \bibinfo {author} {\bibfnamefont
  {A.}~\bibnamefont {{Sesana}}},\ and\ \bibinfo {author} {\bibfnamefont
  {J.}~\bibnamefont {{Camp}}},\ }\href
  {https://doi.org/10.3847/1538-4357/ab505a} {\bibfield  {journal} {\bibinfo
  {journal} {\apj}\ }\textbf {\bibinfo {volume} {886}},\ \bibinfo {eid} {146}
  (\bibinfo {year} {2019})},\ \Eprint {https://arxiv.org/abs/1902.01538}
  {arXiv:1902.01538 [astro-ph.HE]} \BibitemShut {NoStop}%
\bibitem [{\citenamefont {{Duffell}}\ \emph {et~al.}(2024)\citenamefont
  {{Duffell}}, \citenamefont {{Dittmann}}, \citenamefont {{D'Orazio}},
  \citenamefont {{Franchini}}, \citenamefont {{Kratter}}, \citenamefont
  {{Penzlin}}, \citenamefont {{Ragusa}}, \citenamefont {{Siwek}}, \citenamefont
  {{Tiede}}, \citenamefont {{Wang}}, \citenamefont {{Zrake}}, \citenamefont
  {{Dempsey}}, \citenamefont {{Haiman}}, \citenamefont {{Lupi}}, \citenamefont
  {{Pirog}},\ and\ \citenamefont {{Ryan}}}]{Duffell2024}%
  \BibitemOpen
  \bibfield  {author} {\bibinfo {author} {\bibfnamefont {P.~C.}\ \bibnamefont
  {{Duffell}}}, \bibinfo {author} {\bibfnamefont {A.~J.}\ \bibnamefont
  {{Dittmann}}}, \bibinfo {author} {\bibfnamefont {D.~J.}\ \bibnamefont
  {{D'Orazio}}}, \bibinfo {author} {\bibfnamefont {A.}~\bibnamefont
  {{Franchini}}}, \bibinfo {author} {\bibfnamefont {K.~M.}\ \bibnamefont
  {{Kratter}}}, \bibinfo {author} {\bibfnamefont {A.~B.~T.}\ \bibnamefont
  {{Penzlin}}}, \bibinfo {author} {\bibfnamefont {E.}~\bibnamefont {{Ragusa}}},
  \bibinfo {author} {\bibfnamefont {M.}~\bibnamefont {{Siwek}}}, \bibinfo
  {author} {\bibfnamefont {C.}~\bibnamefont {{Tiede}}}, \bibinfo {author}
  {\bibfnamefont {H.}~\bibnamefont {{Wang}}}, \bibinfo {author} {\bibfnamefont
  {J.}~\bibnamefont {{Zrake}}}, \bibinfo {author} {\bibfnamefont {A.~M.}\
  \bibnamefont {{Dempsey}}}, \bibinfo {author} {\bibfnamefont {Z.}~\bibnamefont
  {{Haiman}}}, \bibinfo {author} {\bibfnamefont {A.}~\bibnamefont {{Lupi}}},
  \bibinfo {author} {\bibfnamefont {M.}~\bibnamefont {{Pirog}}},\ and\ \bibinfo
  {author} {\bibfnamefont {G.}~\bibnamefont {{Ryan}}},\ }\href
  {https://doi.org/10.3847/1538-4357/ad5a7e} {\bibfield  {journal} {\bibinfo
  {journal} {\apj}\ }\textbf {\bibinfo {volume} {970}},\ \bibinfo {eid} {156}
  (\bibinfo {year} {2024})},\ \Eprint {https://arxiv.org/abs/2402.13039}
  {arXiv:2402.13039 [astro-ph.SR]} \BibitemShut {NoStop}%
\bibitem [{\citenamefont {{Guti{\'e}rrez}}\ \emph
  {et~al.}(2024{\natexlab{a}})\citenamefont {{Guti{\'e}rrez}}, \citenamefont
  {{Combi}},\ and\ \citenamefont {{Ryan}}}]{Gutierrez2024}%
  \BibitemOpen
  \bibfield  {author} {\bibinfo {author} {\bibfnamefont {E.~M.}\ \bibnamefont
  {{Guti{\'e}rrez}}}, \bibinfo {author} {\bibfnamefont {L.}~\bibnamefont
  {{Combi}}},\ and\ \bibinfo {author} {\bibfnamefont {G.}~\bibnamefont
  {{Ryan}}},\ }\href {https://doi.org/10.48550/arXiv.2405.14843} {\bibfield
  {journal} {\bibinfo  {journal} {arXiv e-prints}\ ,\ \bibinfo {eid}
  {arXiv:2405.14843}} (\bibinfo {year} {2024}{\natexlab{a}})},\ \Eprint
  {https://arxiv.org/abs/2405.14843} {arXiv:2405.14843 [astro-ph.HE]}
  \BibitemShut {NoStop}%
\bibitem [{\citenamefont {Noble}\ and\ \citenamefont
  {Krolik}()}]{NobleKrolik2025}%
  \BibitemOpen
  \bibfield  {author} {\bibinfo {author} {\bibfnamefont {S.~C.}\ \bibnamefont
  {Noble}}\ and\ \bibinfo {author} {\bibfnamefont {J.~H.}\ \bibnamefont
  {Krolik}},\ }\href@noop {} {\bibinfo  {journal} {in preparation}\
  }\BibitemShut {NoStop}%
\bibitem [{\citenamefont {{Pringle}}(1991)}]{Pringle1991}%
  \BibitemOpen
\bibfield  {journal} {  }\bibfield  {author} {\bibinfo {author} {\bibfnamefont
  {J.~E.}\ \bibnamefont {{Pringle}}},\ }\href
  {https://doi.org/10.1093/mnras/248.4.754} {\bibfield  {journal} {\bibinfo
  {journal} {\mnras}\ }\textbf {\bibinfo {volume} {248}},\ \bibinfo {pages}
  {754} (\bibinfo {year} {1991})}\BibitemShut {NoStop}%
\bibitem [{\citenamefont {{Artymowicz}}\ and\ \citenamefont
  {{Lubow}}(1994)}]{ArtymowiczLubow1994}%
  \BibitemOpen
  \bibfield  {author} {\bibinfo {author} {\bibfnamefont {P.}~\bibnamefont
  {{Artymowicz}}}\ and\ \bibinfo {author} {\bibfnamefont {S.~H.}\ \bibnamefont
  {{Lubow}}},\ }\href {https://doi.org/10.1086/173679} {\bibfield  {journal}
  {\bibinfo  {journal} {\apj}\ }\textbf {\bibinfo {volume} {421}},\ \bibinfo
  {pages} {651} (\bibinfo {year} {1994})}\BibitemShut {NoStop}%
\bibitem [{\citenamefont {{MacFadyen}}\ and\ \citenamefont
  {{Milosavljevi{\'c}}}(2008)}]{MacFadyenMilosavljevic2008}%
  \BibitemOpen
  \bibfield  {author} {\bibinfo {author} {\bibfnamefont {A.~I.}\ \bibnamefont
  {{MacFadyen}}}\ and\ \bibinfo {author} {\bibfnamefont {M.}~\bibnamefont
  {{Milosavljevi{\'c}}}},\ }\href {https://doi.org/10.1086/523869} {\bibfield
  {journal} {\bibinfo  {journal} {\apj}\ }\textbf {\bibinfo {volume} {672}},\
  \bibinfo {pages} {83} (\bibinfo {year} {2008})},\ \Eprint
  {https://arxiv.org/abs/astro-ph/0607467} {arXiv:astro-ph/0607467 [astro-ph]}
  \BibitemShut {NoStop}%
\bibitem [{\citenamefont {{Shi}}\ \emph {et~al.}(2012)\citenamefont {{Shi}},
  \citenamefont {{Krolik}}, \citenamefont {{Lubow}},\ and\ \citenamefont
  {{Hawley}}}]{Shi2012}%
  \BibitemOpen
  \bibfield  {author} {\bibinfo {author} {\bibfnamefont {J.-M.}\ \bibnamefont
  {{Shi}}}, \bibinfo {author} {\bibfnamefont {J.~H.}\ \bibnamefont {{Krolik}}},
  \bibinfo {author} {\bibfnamefont {S.~H.}\ \bibnamefont {{Lubow}}},\ and\
  \bibinfo {author} {\bibfnamefont {J.~F.}\ \bibnamefont {{Hawley}}},\ }\href
  {https://doi.org/10.1088/0004-637X/749/2/118} {\bibfield  {journal} {\bibinfo
   {journal} {\apj}\ }\textbf {\bibinfo {volume} {749}},\ \bibinfo {eid} {118}
  (\bibinfo {year} {2012})},\ \Eprint {https://arxiv.org/abs/1110.4866}
  {arXiv:1110.4866 [astro-ph.HE]} \BibitemShut {NoStop}%
\bibitem [{\citenamefont {{Noble}}\ \emph {et~al.}(2012)\citenamefont
  {{Noble}}, \citenamefont {{Mundim}}, \citenamefont {{Nakano}}, \citenamefont
  {{Krolik}}, \citenamefont {{Campanelli}}, \citenamefont {{Zlochower}},\ and\
  \citenamefont {{Yunes}}}]{Noble2012}%
  \BibitemOpen
  \bibfield  {author} {\bibinfo {author} {\bibfnamefont {S.~C.}\ \bibnamefont
  {{Noble}}}, \bibinfo {author} {\bibfnamefont {B.~C.}\ \bibnamefont
  {{Mundim}}}, \bibinfo {author} {\bibfnamefont {H.}~\bibnamefont {{Nakano}}},
  \bibinfo {author} {\bibfnamefont {J.~H.}\ \bibnamefont {{Krolik}}}, \bibinfo
  {author} {\bibfnamefont {M.}~\bibnamefont {{Campanelli}}}, \bibinfo {author}
  {\bibfnamefont {Y.}~\bibnamefont {{Zlochower}}},\ and\ \bibinfo {author}
  {\bibfnamefont {N.}~\bibnamefont {{Yunes}}},\ }\href
  {https://doi.org/10.1088/0004-637X/755/1/51} {\bibfield  {journal} {\bibinfo
  {journal} {\apj}\ }\textbf {\bibinfo {volume} {755}},\ \bibinfo {eid} {51}
  (\bibinfo {year} {2012})},\ \Eprint {https://arxiv.org/abs/1204.1073}
  {arXiv:1204.1073 [astro-ph.HE]} \BibitemShut {NoStop}%
\bibitem [{\citenamefont {{Farris}}\ \emph {et~al.}(2014)\citenamefont
  {{Farris}}, \citenamefont {{Duffell}}, \citenamefont {{MacFadyen}},\ and\
  \citenamefont {{Haiman}}}]{Farris2014}%
  \BibitemOpen
  \bibfield  {author} {\bibinfo {author} {\bibfnamefont {B.~D.}\ \bibnamefont
  {{Farris}}}, \bibinfo {author} {\bibfnamefont {P.}~\bibnamefont {{Duffell}}},
  \bibinfo {author} {\bibfnamefont {A.~I.}\ \bibnamefont {{MacFadyen}}},\ and\
  \bibinfo {author} {\bibfnamefont {Z.}~\bibnamefont {{Haiman}}},\ }\href
  {https://doi.org/10.1088/0004-637X/783/2/134} {\bibfield  {journal} {\bibinfo
   {journal} {\apj}\ }\textbf {\bibinfo {volume} {783}},\ \bibinfo {eid} {134}
  (\bibinfo {year} {2014})},\ \Eprint {https://arxiv.org/abs/1310.0492}
  {arXiv:1310.0492 [astro-ph.HE]} \BibitemShut {NoStop}%
\bibitem [{\citenamefont {{Shi}}\ and\ \citenamefont
  {{Krolik}}(2015)}]{ShiKrolik2015}%
  \BibitemOpen
  \bibfield  {author} {\bibinfo {author} {\bibfnamefont {J.-M.}\ \bibnamefont
  {{Shi}}}\ and\ \bibinfo {author} {\bibfnamefont {J.~H.}\ \bibnamefont
  {{Krolik}}},\ }\href {https://doi.org/10.1088/0004-637X/807/2/131} {\bibfield
   {journal} {\bibinfo  {journal} {\apj}\ }\textbf {\bibinfo {volume} {807}},\
  \bibinfo {eid} {131} (\bibinfo {year} {2015})},\ \Eprint
  {https://arxiv.org/abs/1503.05561} {arXiv:1503.05561 [astro-ph.HE]}
  \BibitemShut {NoStop}%
\bibitem [{\citenamefont {{Bowen}}\ \emph {et~al.}(2017)\citenamefont
  {{Bowen}}, \citenamefont {{Campanelli}}, \citenamefont {{Krolik}},
  \citenamefont {{Mewes}},\ and\ \citenamefont {{Noble}}}]{Bowen2017}%
  \BibitemOpen
  \bibfield  {author} {\bibinfo {author} {\bibfnamefont {D.~B.}\ \bibnamefont
  {{Bowen}}}, \bibinfo {author} {\bibfnamefont {M.}~\bibnamefont
  {{Campanelli}}}, \bibinfo {author} {\bibfnamefont {J.~H.}\ \bibnamefont
  {{Krolik}}}, \bibinfo {author} {\bibfnamefont {V.}~\bibnamefont {{Mewes}}},\
  and\ \bibinfo {author} {\bibfnamefont {S.~C.}\ \bibnamefont {{Noble}}},\
  }\href {https://doi.org/10.3847/1538-4357/aa63f3} {\bibfield  {journal}
  {\bibinfo  {journal} {\apj}\ }\textbf {\bibinfo {volume} {838}},\ \bibinfo
  {eid} {42} (\bibinfo {year} {2017})},\ \Eprint
  {https://arxiv.org/abs/1612.02373} {arXiv:1612.02373 [astro-ph.HE]}
  \BibitemShut {NoStop}%
\bibitem [{\citenamefont {{Bowen}}\ \emph {et~al.}(2019)\citenamefont
  {{Bowen}}, \citenamefont {{Mewes}}, \citenamefont {{Noble}}, \citenamefont
  {{Avara}}, \citenamefont {{Campanelli}},\ and\ \citenamefont
  {{Krolik}}}]{Bowen2019}%
  \BibitemOpen
  \bibfield  {author} {\bibinfo {author} {\bibfnamefont {D.~B.}\ \bibnamefont
  {{Bowen}}}, \bibinfo {author} {\bibfnamefont {V.}~\bibnamefont {{Mewes}}},
  \bibinfo {author} {\bibfnamefont {S.~C.}\ \bibnamefont {{Noble}}}, \bibinfo
  {author} {\bibfnamefont {M.}~\bibnamefont {{Avara}}}, \bibinfo {author}
  {\bibfnamefont {M.}~\bibnamefont {{Campanelli}}},\ and\ \bibinfo {author}
  {\bibfnamefont {J.~H.}\ \bibnamefont {{Krolik}}},\ }\href
  {https://doi.org/10.3847/1538-4357/ab2453} {\bibfield  {journal} {\bibinfo
  {journal} {\apj}\ }\textbf {\bibinfo {volume} {879}},\ \bibinfo {eid} {76}
  (\bibinfo {year} {2019})},\ \Eprint {https://arxiv.org/abs/1904.12048}
  {arXiv:1904.12048 [astro-ph.HE]} \BibitemShut {NoStop}%
\bibitem [{\citenamefont {{Artymowicz}}\ and\ \citenamefont
  {{Lubow}}(1996)}]{ArtymowiczLubow1996}%
  \BibitemOpen
  \bibfield  {author} {\bibinfo {author} {\bibfnamefont {P.}~\bibnamefont
  {{Artymowicz}}}\ and\ \bibinfo {author} {\bibfnamefont {S.~H.}\ \bibnamefont
  {{Lubow}}},\ }\href {https://doi.org/10.1086/310200} {\bibfield  {journal}
  {\bibinfo  {journal} {\apjl}\ }\textbf {\bibinfo {volume} {467}},\ \bibinfo
  {pages} {L77} (\bibinfo {year} {1996})}\BibitemShut {NoStop}%
\bibitem [{\citenamefont {{Avara}}\ \emph {et~al.}(2024)\citenamefont
  {{Avara}}, \citenamefont {{Krolik}}, \citenamefont {{Campanelli}},
  \citenamefont {{Noble}}, \citenamefont {{Bowen}},\ and\ \citenamefont
  {{Ryu}}}]{Avara2024}%
  \BibitemOpen
  \bibfield  {author} {\bibinfo {author} {\bibfnamefont {M.~J.}\ \bibnamefont
  {{Avara}}}, \bibinfo {author} {\bibfnamefont {J.~H.}\ \bibnamefont
  {{Krolik}}}, \bibinfo {author} {\bibfnamefont {M.}~\bibnamefont
  {{Campanelli}}}, \bibinfo {author} {\bibfnamefont {S.~C.}\ \bibnamefont
  {{Noble}}}, \bibinfo {author} {\bibfnamefont {D.}~\bibnamefont {{Bowen}}},\
  and\ \bibinfo {author} {\bibfnamefont {T.}~\bibnamefont {{Ryu}}},\ }\href
  {https://doi.org/10.3847/1538-4357/ad5bda} {\bibfield  {journal} {\bibinfo
  {journal} {\apj}\ }\textbf {\bibinfo {volume} {974}},\ \bibinfo {eid} {242}
  (\bibinfo {year} {2024})},\ \Eprint {https://arxiv.org/abs/2305.18538}
  {arXiv:2305.18538 [astro-ph.HE]} \BibitemShut {NoStop}%
\bibitem [{\citenamefont {{Westernacher-Schneider}}\ \emph
  {et~al.}(2024)\citenamefont {{Westernacher-Schneider}}, \citenamefont
  {{Zrake}}, \citenamefont {{MacFadyen}},\ and\ \citenamefont
  {{Haiman}}}]{Westernacher-Schneider2024}%
  \BibitemOpen
  \bibfield  {author} {\bibinfo {author} {\bibfnamefont {J.~R.}\ \bibnamefont
  {{Westernacher-Schneider}}}, \bibinfo {author} {\bibfnamefont
  {J.}~\bibnamefont {{Zrake}}}, \bibinfo {author} {\bibfnamefont
  {A.}~\bibnamefont {{MacFadyen}}},\ and\ \bibinfo {author} {\bibfnamefont
  {Z.}~\bibnamefont {{Haiman}}},\ }\href
  {https://doi.org/10.3847/1538-4357/ad1a17} {\bibfield  {journal} {\bibinfo
  {journal} {\apj}\ }\textbf {\bibinfo {volume} {962}},\ \bibinfo {eid} {76}
  (\bibinfo {year} {2024})},\ \Eprint {https://arxiv.org/abs/2307.01154}
  {arXiv:2307.01154 [astro-ph.HE]} \BibitemShut {NoStop}%
\bibitem [{\citenamefont {{Roedig}}\ \emph {et~al.}(2014)\citenamefont
  {{Roedig}}, \citenamefont {{Krolik}},\ and\ \citenamefont
  {{Miller}}}]{Roedig2014}%
  \BibitemOpen
  \bibfield  {author} {\bibinfo {author} {\bibfnamefont {C.}~\bibnamefont
  {{Roedig}}}, \bibinfo {author} {\bibfnamefont {J.~H.}\ \bibnamefont
  {{Krolik}}},\ and\ \bibinfo {author} {\bibfnamefont {M.~C.}\ \bibnamefont
  {{Miller}}},\ }\href {https://doi.org/10.1088/0004-637X/785/2/115} {\bibfield
   {journal} {\bibinfo  {journal} {\apj}\ }\textbf {\bibinfo {volume} {785}},\
  \bibinfo {eid} {115} (\bibinfo {year} {2014})},\ \Eprint
  {https://arxiv.org/abs/1402.7098} {arXiv:1402.7098 [astro-ph.HE]}
  \BibitemShut {NoStop}%
\bibitem [{\citenamefont {{Farris}}\ \emph {et~al.}(2015)\citenamefont
  {{Farris}}, \citenamefont {{Duffell}}, \citenamefont {{MacFadyen}},\ and\
  \citenamefont {{Haiman}}}]{Farris2015}%
  \BibitemOpen
  \bibfield  {author} {\bibinfo {author} {\bibfnamefont {B.~D.}\ \bibnamefont
  {{Farris}}}, \bibinfo {author} {\bibfnamefont {P.}~\bibnamefont {{Duffell}}},
  \bibinfo {author} {\bibfnamefont {A.~I.}\ \bibnamefont {{MacFadyen}}},\ and\
  \bibinfo {author} {\bibfnamefont {Z.}~\bibnamefont {{Haiman}}},\ }\href
  {https://doi.org/10.1093/mnrasl/slu160} {\bibfield  {journal} {\bibinfo
  {journal} {\mnras}\ }\textbf {\bibinfo {volume} {446}},\ \bibinfo {pages}
  {L36} (\bibinfo {year} {2015})},\ \Eprint {https://arxiv.org/abs/1406.0007}
  {arXiv:1406.0007 [astro-ph.HE]} \BibitemShut {NoStop}%
\bibitem [{\citenamefont {{d'Ascoli}}\ \emph {et~al.}(2018)\citenamefont
  {{d'Ascoli}}, \citenamefont {{Noble}}, \citenamefont {{Bowen}}, \citenamefont
  {{Campanelli}}, \citenamefont {{Krolik}},\ and\ \citenamefont
  {{Mewes}}}]{dAscoli2018}%
  \BibitemOpen
  \bibfield  {author} {\bibinfo {author} {\bibfnamefont {S.}~\bibnamefont
  {{d'Ascoli}}}, \bibinfo {author} {\bibfnamefont {S.~C.}\ \bibnamefont
  {{Noble}}}, \bibinfo {author} {\bibfnamefont {D.~B.}\ \bibnamefont
  {{Bowen}}}, \bibinfo {author} {\bibfnamefont {M.}~\bibnamefont
  {{Campanelli}}}, \bibinfo {author} {\bibfnamefont {J.~H.}\ \bibnamefont
  {{Krolik}}},\ and\ \bibinfo {author} {\bibfnamefont {V.}~\bibnamefont
  {{Mewes}}},\ }\href {https://doi.org/10.3847/1538-4357/aad8b4} {\bibfield
  {journal} {\bibinfo  {journal} {\apj}\ }\textbf {\bibinfo {volume} {865}},\
  \bibinfo {eid} {140} (\bibinfo {year} {2018})},\ \Eprint
  {https://arxiv.org/abs/1806.05697} {arXiv:1806.05697 [astro-ph.HE]}
  \BibitemShut {NoStop}%
\bibitem [{\citenamefont {{Guti{\'e}rrez}}\ \emph {et~al.}(2022)\citenamefont
  {{Guti{\'e}rrez}}, \citenamefont {{Combi}}, \citenamefont {{Noble}},
  \citenamefont {{Campanelli}}, \citenamefont {{Krolik}}, \citenamefont
  {{L{\'o}pez Armengol}},\ and\ \citenamefont {{Garc{\'\i}a}}}]{Gutierrez2022}%
  \BibitemOpen
  \bibfield  {author} {\bibinfo {author} {\bibfnamefont {E.~M.}\ \bibnamefont
  {{Guti{\'e}rrez}}}, \bibinfo {author} {\bibfnamefont {L.}~\bibnamefont
  {{Combi}}}, \bibinfo {author} {\bibfnamefont {S.~C.}\ \bibnamefont
  {{Noble}}}, \bibinfo {author} {\bibfnamefont {M.}~\bibnamefont
  {{Campanelli}}}, \bibinfo {author} {\bibfnamefont {J.~H.}\ \bibnamefont
  {{Krolik}}}, \bibinfo {author} {\bibfnamefont {F.}~\bibnamefont {{L{\'o}pez
  Armengol}}},\ and\ \bibinfo {author} {\bibfnamefont {F.}~\bibnamefont
  {{Garc{\'\i}a}}},\ }\href {https://doi.org/10.3847/1538-4357/ac56de}
  {\bibfield  {journal} {\bibinfo  {journal} {\apj}\ }\textbf {\bibinfo
  {volume} {928}},\ \bibinfo {eid} {137} (\bibinfo {year} {2022})},\ \Eprint
  {https://arxiv.org/abs/2112.09773} {arXiv:2112.09773 [astro-ph.HE]}
  \BibitemShut {NoStop}%
\bibitem [{\citenamefont {{Nagele}}\ \emph {et~al.}()\citenamefont {{Nagele}},
  \citenamefont {{Krolik}},\ and\ \citenamefont {{Noble}}}]{Nagele2025}%
  \BibitemOpen
  \bibfield  {author} {\bibinfo {author} {\bibfnamefont {C.}~\bibnamefont
  {{Nagele}}}, \bibinfo {author} {\bibfnamefont {J.}~\bibnamefont {{Krolik}}},\
  and\ \bibinfo {author} {\bibfnamefont {S.}~\bibnamefont {{Noble}}},\
  }\href@noop {} {\bibinfo  {journal} {in preparation}\ }\BibitemShut {NoStop}%
\bibitem [{\citenamefont {{Kinch}}\ \emph {et~al.}(2020)\citenamefont
  {{Kinch}}, \citenamefont {{Noble}}, \citenamefont {{Schnittman}},\ and\
  \citenamefont {{Krolik}}}]{Kinch2020}%
  \BibitemOpen
\bibfield  {journal} {  }\bibfield  {author} {\bibinfo {author} {\bibfnamefont
  {B.~E.}\ \bibnamefont {{Kinch}}}, \bibinfo {author} {\bibfnamefont {S.~C.}\
  \bibnamefont {{Noble}}}, \bibinfo {author} {\bibfnamefont {J.~D.}\
  \bibnamefont {{Schnittman}}},\ and\ \bibinfo {author} {\bibfnamefont {J.~H.}\
  \bibnamefont {{Krolik}}},\ }\href {https://doi.org/10.3847/1538-4357/abc176}
  {\bibfield  {journal} {\bibinfo  {journal} {\apj}\ }\textbf {\bibinfo
  {volume} {904}},\ \bibinfo {eid} {117} (\bibinfo {year} {2020})},\ \Eprint
  {https://arxiv.org/abs/2009.01914} {arXiv:2009.01914 [astro-ph.HE]}
  \BibitemShut {NoStop}%
\bibitem [{\citenamefont {{Farris}}\ \emph {et~al.}(2012)\citenamefont
  {{Farris}}, \citenamefont {{Gold}}, \citenamefont {{Paschalidis}},
  \citenamefont {{Etienne}},\ and\ \citenamefont {{Shapiro}}}]{Farris2012}%
  \BibitemOpen
  \bibfield  {author} {\bibinfo {author} {\bibfnamefont {B.~D.}\ \bibnamefont
  {{Farris}}}, \bibinfo {author} {\bibfnamefont {R.}~\bibnamefont {{Gold}}},
  \bibinfo {author} {\bibfnamefont {V.}~\bibnamefont {{Paschalidis}}}, \bibinfo
  {author} {\bibfnamefont {Z.~B.}\ \bibnamefont {{Etienne}}},\ and\ \bibinfo
  {author} {\bibfnamefont {S.~L.}\ \bibnamefont {{Shapiro}}},\ }\href
  {https://doi.org/10.1103/PhysRevLett.109.221102} {\bibfield  {journal}
  {\bibinfo  {journal} {\prl}\ }\textbf {\bibinfo {volume} {109}},\ \bibinfo
  {eid} {221102} (\bibinfo {year} {2012})},\ \Eprint
  {https://arxiv.org/abs/1207.3354} {arXiv:1207.3354 [astro-ph.HE]}
  \BibitemShut {NoStop}%
\bibitem [{\citenamefont {{Gold}}\ \emph
  {et~al.}(2014{\natexlab{a}})\citenamefont {{Gold}}, \citenamefont
  {{Paschalidis}}, \citenamefont {{Etienne}}, \citenamefont {{Shapiro}},\ and\
  \citenamefont {{Pfeiffer}}}]{Gold2014a}%
  \BibitemOpen
  \bibfield  {author} {\bibinfo {author} {\bibfnamefont {R.}~\bibnamefont
  {{Gold}}}, \bibinfo {author} {\bibfnamefont {V.}~\bibnamefont
  {{Paschalidis}}}, \bibinfo {author} {\bibfnamefont {Z.~B.}\ \bibnamefont
  {{Etienne}}}, \bibinfo {author} {\bibfnamefont {S.~L.}\ \bibnamefont
  {{Shapiro}}},\ and\ \bibinfo {author} {\bibfnamefont {H.~P.}\ \bibnamefont
  {{Pfeiffer}}},\ }\href {https://doi.org/10.1103/PhysRevD.89.064060}
  {\bibfield  {journal} {\bibinfo  {journal} {\prd}\ }\textbf {\bibinfo
  {volume} {89}},\ \bibinfo {eid} {064060} (\bibinfo {year}
  {2014}{\natexlab{a}})},\ \Eprint {https://arxiv.org/abs/1312.0600}
  {arXiv:1312.0600 [astro-ph.HE]} \BibitemShut {NoStop}%
\bibitem [{\citenamefont {{Gold}}\ \emph
  {et~al.}(2014{\natexlab{b}})\citenamefont {{Gold}}, \citenamefont
  {{Paschalidis}}, \citenamefont {{Ruiz}}, \citenamefont {{Shapiro}},
  \citenamefont {{Etienne}},\ and\ \citenamefont {{Pfeiffer}}}]{Gold2014b}%
  \BibitemOpen
  \bibfield  {author} {\bibinfo {author} {\bibfnamefont {R.}~\bibnamefont
  {{Gold}}}, \bibinfo {author} {\bibfnamefont {V.}~\bibnamefont
  {{Paschalidis}}}, \bibinfo {author} {\bibfnamefont {M.}~\bibnamefont
  {{Ruiz}}}, \bibinfo {author} {\bibfnamefont {S.~L.}\ \bibnamefont
  {{Shapiro}}}, \bibinfo {author} {\bibfnamefont {Z.~B.}\ \bibnamefont
  {{Etienne}}},\ and\ \bibinfo {author} {\bibfnamefont {H.~P.}\ \bibnamefont
  {{Pfeiffer}}},\ }\href {https://doi.org/10.1103/PhysRevD.90.104030}
  {\bibfield  {journal} {\bibinfo  {journal} {\prd}\ }\textbf {\bibinfo
  {volume} {90}},\ \bibinfo {eid} {104030} (\bibinfo {year}
  {2014}{\natexlab{b}})},\ \Eprint {https://arxiv.org/abs/1410.1543}
  {arXiv:1410.1543 [astro-ph.GA]} \BibitemShut {NoStop}%
\bibitem [{\citenamefont {{Paschalidis}}\ \emph {et~al.}(2021)\citenamefont
  {{Paschalidis}}, \citenamefont {{Bright}}, \citenamefont {{Ruiz}},\ and\
  \citenamefont {{Gold}}}]{Paschalidis2021}%
  \BibitemOpen
  \bibfield  {author} {\bibinfo {author} {\bibfnamefont {V.}~\bibnamefont
  {{Paschalidis}}}, \bibinfo {author} {\bibfnamefont {J.}~\bibnamefont
  {{Bright}}}, \bibinfo {author} {\bibfnamefont {M.}~\bibnamefont {{Ruiz}}},\
  and\ \bibinfo {author} {\bibfnamefont {R.}~\bibnamefont {{Gold}}},\ }\href
  {https://doi.org/10.3847/2041-8213/abee21} {\bibfield  {journal} {\bibinfo
  {journal} {\apjl}\ }\textbf {\bibinfo {volume} {910}},\ \bibinfo {eid} {L26}
  (\bibinfo {year} {2021})},\ \Eprint {https://arxiv.org/abs/2102.06712}
  {arXiv:2102.06712 [astro-ph.HE]} \BibitemShut {NoStop}%
\bibitem [{\citenamefont {{Bright}}\ and\ \citenamefont
  {{Paschalidis}}(2023)}]{Bright2023}%
  \BibitemOpen
  \bibfield  {author} {\bibinfo {author} {\bibfnamefont {J.~C.}\ \bibnamefont
  {{Bright}}}\ and\ \bibinfo {author} {\bibfnamefont {V.}~\bibnamefont
  {{Paschalidis}}},\ }\href {https://doi.org/10.1093/mnras/stad091} {\bibfield
  {journal} {\bibinfo  {journal} {\mnras}\ }\textbf {\bibinfo {volume} {520}},\
  \bibinfo {pages} {392} (\bibinfo {year} {2023})},\ \Eprint
  {https://arxiv.org/abs/2210.15686} {arXiv:2210.15686 [astro-ph.HE]}
  \BibitemShut {NoStop}%
\bibitem [{\citenamefont {{Cattorini}}\ \emph {et~al.}(2021)\citenamefont
  {{Cattorini}}, \citenamefont {{Giacomazzo}}, \citenamefont {{Haardt}},\ and\
  \citenamefont {{Colpi}}}]{Cattorini2021}%
  \BibitemOpen
  \bibfield  {author} {\bibinfo {author} {\bibfnamefont {F.}~\bibnamefont
  {{Cattorini}}}, \bibinfo {author} {\bibfnamefont {B.}~\bibnamefont
  {{Giacomazzo}}}, \bibinfo {author} {\bibfnamefont {F.}~\bibnamefont
  {{Haardt}}},\ and\ \bibinfo {author} {\bibfnamefont {M.}~\bibnamefont
  {{Colpi}}},\ }\href {https://doi.org/10.1103/PhysRevD.103.103022} {\bibfield
  {journal} {\bibinfo  {journal} {\prd}\ }\textbf {\bibinfo {volume} {103}},\
  \bibinfo {eid} {103022} (\bibinfo {year} {2021})},\ \Eprint
  {https://arxiv.org/abs/2102.13166} {arXiv:2102.13166 [astro-ph.HE]}
  \BibitemShut {NoStop}%
\bibitem [{\citenamefont {{Cattorini}}\ \emph {et~al.}(2022)\citenamefont
  {{Cattorini}}, \citenamefont {{Maggioni}}, \citenamefont {{Giacomazzo}},
  \citenamefont {{Haardt}}, \citenamefont {{Colpi}},\ and\ \citenamefont
  {{Covino}}}]{Cattorini2022}%
  \BibitemOpen
  \bibfield  {author} {\bibinfo {author} {\bibfnamefont {F.}~\bibnamefont
  {{Cattorini}}}, \bibinfo {author} {\bibfnamefont {S.}~\bibnamefont
  {{Maggioni}}}, \bibinfo {author} {\bibfnamefont {B.}~\bibnamefont
  {{Giacomazzo}}}, \bibinfo {author} {\bibfnamefont {F.}~\bibnamefont
  {{Haardt}}}, \bibinfo {author} {\bibfnamefont {M.}~\bibnamefont {{Colpi}}},\
  and\ \bibinfo {author} {\bibfnamefont {S.}~\bibnamefont {{Covino}}},\ }\href
  {https://doi.org/10.3847/2041-8213/ac6755} {\bibfield  {journal} {\bibinfo
  {journal} {\apjl}\ }\textbf {\bibinfo {volume} {930}},\ \bibinfo {eid} {L1}
  (\bibinfo {year} {2022})},\ \Eprint {https://arxiv.org/abs/2202.08282}
  {arXiv:2202.08282 [astro-ph.HE]} \BibitemShut {NoStop}%
\bibitem [{\citenamefont {{Fedrigo}}\ \emph {et~al.}(2023)\citenamefont
  {{Fedrigo}}, \citenamefont {{Cattorini}}, \citenamefont {{Giacomazzo}},\ and\
  \citenamefont {{Colpi}}}]{Fedrigo2023}%
  \BibitemOpen
  \bibfield  {author} {\bibinfo {author} {\bibfnamefont {G.}~\bibnamefont
  {{Fedrigo}}}, \bibinfo {author} {\bibfnamefont {F.}~\bibnamefont
  {{Cattorini}}}, \bibinfo {author} {\bibfnamefont {B.}~\bibnamefont
  {{Giacomazzo}}},\ and\ \bibinfo {author} {\bibfnamefont {M.}~\bibnamefont
  {{Colpi}}},\ }\href {https://doi.org/10.48550/arXiv.2309.03949} {\bibfield
  {journal} {\bibinfo  {journal} {arXiv e-prints}\ ,\ \bibinfo {eid}
  {arXiv:2309.03949}} (\bibinfo {year} {2023})},\ \Eprint
  {https://arxiv.org/abs/2309.03949} {arXiv:2309.03949 [astro-ph.HE]}
  \BibitemShut {NoStop}%
\bibitem [{\citenamefont {{Cattorini}}\ \emph {et~al.}(2024)\citenamefont
  {{Cattorini}}, \citenamefont {{Giacomazzo}}, \citenamefont {{Colpi}},\ and\
  \citenamefont {{Haardt}}}]{Cattorini2024}%
  \BibitemOpen
  \bibfield  {author} {\bibinfo {author} {\bibfnamefont {F.}~\bibnamefont
  {{Cattorini}}}, \bibinfo {author} {\bibfnamefont {B.}~\bibnamefont
  {{Giacomazzo}}}, \bibinfo {author} {\bibfnamefont {M.}~\bibnamefont
  {{Colpi}}},\ and\ \bibinfo {author} {\bibfnamefont {F.}~\bibnamefont
  {{Haardt}}},\ }\href {https://doi.org/10.1103/PhysRevD.109.043004} {\bibfield
   {journal} {\bibinfo  {journal} {\prd}\ }\textbf {\bibinfo {volume} {109}},\
  \bibinfo {eid} {043004} (\bibinfo {year} {2024})},\ \Eprint
  {https://arxiv.org/abs/2309.05738} {arXiv:2309.05738 [astro-ph.HE]}
  \BibitemShut {NoStop}%
\bibitem [{\citenamefont {{Manikantan}}\ \emph {et~al.}(2024)\citenamefont
  {{Manikantan}}, \citenamefont {{Paschalidis}},\ and\ \citenamefont
  {{Bozzola}}}]{Manikantan2024}%
  \BibitemOpen
  \bibfield  {author} {\bibinfo {author} {\bibfnamefont {V.}~\bibnamefont
  {{Manikantan}}}, \bibinfo {author} {\bibfnamefont {V.}~\bibnamefont
  {{Paschalidis}}},\ and\ \bibinfo {author} {\bibfnamefont {G.}~\bibnamefont
  {{Bozzola}}},\ }\href {https://doi.org/10.48550/arXiv.2411.11955} {\bibfield
  {journal} {\bibinfo  {journal} {arXiv e-prints}\ ,\ \bibinfo {eid}
  {arXiv:2411.11955}} (\bibinfo {year} {2024})},\ \Eprint
  {https://arxiv.org/abs/2411.11955} {arXiv:2411.11955 [astro-ph.HE]}
  \BibitemShut {NoStop}%
\bibitem [{\citenamefont {{Noble}}\ \emph {et~al.}(2021)\citenamefont
  {{Noble}}, \citenamefont {{Krolik}}, \citenamefont {{Campanelli}},
  \citenamefont {{Zlochower}}, \citenamefont {{Mundim}}, \citenamefont
  {{Nakano}},\ and\ \citenamefont {{Zilh{\~a}o}}}]{Noble2021}%
  \BibitemOpen
  \bibfield  {author} {\bibinfo {author} {\bibfnamefont {S.~C.}\ \bibnamefont
  {{Noble}}}, \bibinfo {author} {\bibfnamefont {J.~H.}\ \bibnamefont
  {{Krolik}}}, \bibinfo {author} {\bibfnamefont {M.}~\bibnamefont
  {{Campanelli}}}, \bibinfo {author} {\bibfnamefont {Y.}~\bibnamefont
  {{Zlochower}}}, \bibinfo {author} {\bibfnamefont {B.~C.}\ \bibnamefont
  {{Mundim}}}, \bibinfo {author} {\bibfnamefont {H.}~\bibnamefont {{Nakano}}},\
  and\ \bibinfo {author} {\bibfnamefont {M.}~\bibnamefont {{Zilh{\~a}o}}},\
  }\href {https://doi.org/10.3847/1538-4357/ac2229} {\bibfield  {journal}
  {\bibinfo  {journal} {\apj}\ }\textbf {\bibinfo {volume} {922}},\ \bibinfo
  {eid} {175} (\bibinfo {year} {2021})},\ \Eprint
  {https://arxiv.org/abs/2103.12100} {arXiv:2103.12100 [astro-ph.HE]}
  \BibitemShut {NoStop}%
\bibitem [{\citenamefont {{Lopez Armengol}}\ \emph {et~al.}(2021)\citenamefont
  {{Lopez Armengol}}, \citenamefont {{Combi}}, \citenamefont {{Campanelli}},
  \citenamefont {{Noble}}, \citenamefont {{Krolik}}, \citenamefont {{Bowen}},
  \citenamefont {{Avara}}, \citenamefont {{Mewes}},\ and\ \citenamefont
  {{Nakano}}}]{LopezArmengol2021}%
  \BibitemOpen
  \bibfield  {author} {\bibinfo {author} {\bibfnamefont {F.~G.}\ \bibnamefont
  {{Lopez Armengol}}}, \bibinfo {author} {\bibfnamefont {L.}~\bibnamefont
  {{Combi}}}, \bibinfo {author} {\bibfnamefont {M.}~\bibnamefont
  {{Campanelli}}}, \bibinfo {author} {\bibfnamefont {S.~C.}\ \bibnamefont
  {{Noble}}}, \bibinfo {author} {\bibfnamefont {J.~H.}\ \bibnamefont
  {{Krolik}}}, \bibinfo {author} {\bibfnamefont {D.~B.}\ \bibnamefont
  {{Bowen}}}, \bibinfo {author} {\bibfnamefont {M.~J.}\ \bibnamefont
  {{Avara}}}, \bibinfo {author} {\bibfnamefont {V.}~\bibnamefont {{Mewes}}},\
  and\ \bibinfo {author} {\bibfnamefont {H.}~\bibnamefont {{Nakano}}},\ }\href
  {https://doi.org/10.3847/1538-4357/abf0af} {\bibfield  {journal} {\bibinfo
  {journal} {\apj}\ }\textbf {\bibinfo {volume} {913}},\ \bibinfo {eid} {16}
  (\bibinfo {year} {2021})},\ \Eprint {https://arxiv.org/abs/2102.00243}
  {arXiv:2102.00243 [astro-ph.HE]} \BibitemShut {NoStop}%
\bibitem [{\citenamefont {{Combi}}\ \emph {et~al.}(2022)\citenamefont
  {{Combi}}, \citenamefont {{Lopez Armengol}}, \citenamefont {{Campanelli}},
  \citenamefont {{Noble}}, \citenamefont {{Avara}}, \citenamefont {{Krolik}},\
  and\ \citenamefont {{Bowen}}}]{Combi2022}%
  \BibitemOpen
  \bibfield  {author} {\bibinfo {author} {\bibfnamefont {L.}~\bibnamefont
  {{Combi}}}, \bibinfo {author} {\bibfnamefont {F.~G.}\ \bibnamefont {{Lopez
  Armengol}}}, \bibinfo {author} {\bibfnamefont {M.}~\bibnamefont
  {{Campanelli}}}, \bibinfo {author} {\bibfnamefont {S.~C.}\ \bibnamefont
  {{Noble}}}, \bibinfo {author} {\bibfnamefont {M.}~\bibnamefont {{Avara}}},
  \bibinfo {author} {\bibfnamefont {J.~H.}\ \bibnamefont {{Krolik}}},\ and\
  \bibinfo {author} {\bibfnamefont {D.}~\bibnamefont {{Bowen}}},\ }\href
  {https://doi.org/10.3847/1538-4357/ac532a} {\bibfield  {journal} {\bibinfo
  {journal} {\apj}\ }\textbf {\bibinfo {volume} {928}},\ \bibinfo {eid} {187}
  (\bibinfo {year} {2022})},\ \Eprint {https://arxiv.org/abs/2109.01307}
  {arXiv:2109.01307 [astro-ph.HE]} \BibitemShut {NoStop}%
\bibitem [{\citenamefont {{Mundim}}\ \emph {et~al.}(2014)\citenamefont
  {{Mundim}}, \citenamefont {{Nakano}}, \citenamefont {{Yunes}}, \citenamefont
  {{Campanelli}}, \citenamefont {{Noble}},\ and\ \citenamefont
  {{Zlochower}}}]{Mundim2014}%
  \BibitemOpen
  \bibfield  {author} {\bibinfo {author} {\bibfnamefont {B.~C.}\ \bibnamefont
  {{Mundim}}}, \bibinfo {author} {\bibfnamefont {H.}~\bibnamefont {{Nakano}}},
  \bibinfo {author} {\bibfnamefont {N.}~\bibnamefont {{Yunes}}}, \bibinfo
  {author} {\bibfnamefont {M.}~\bibnamefont {{Campanelli}}}, \bibinfo {author}
  {\bibfnamefont {S.~C.}\ \bibnamefont {{Noble}}},\ and\ \bibinfo {author}
  {\bibfnamefont {Y.}~\bibnamefont {{Zlochower}}},\ }\href
  {https://doi.org/10.1103/PhysRevD.89.084008} {\bibfield  {journal} {\bibinfo
  {journal} {\prd}\ }\textbf {\bibinfo {volume} {89}},\ \bibinfo {eid} {084008}
  (\bibinfo {year} {2014})},\ \Eprint {https://arxiv.org/abs/1312.6731}
  {arXiv:1312.6731 [gr-qc]} \BibitemShut {NoStop}%
\bibitem [{\citenamefont {{Ireland}}\ \emph {et~al.}(2016)\citenamefont
  {{Ireland}}, \citenamefont {{Mundim}}, \citenamefont {{Nakano}},\ and\
  \citenamefont {{Campanelli}}}]{Ireland2016}%
  \BibitemOpen
  \bibfield  {author} {\bibinfo {author} {\bibfnamefont {B.}~\bibnamefont
  {{Ireland}}}, \bibinfo {author} {\bibfnamefont {B.~C.}\ \bibnamefont
  {{Mundim}}}, \bibinfo {author} {\bibfnamefont {H.}~\bibnamefont {{Nakano}}},\
  and\ \bibinfo {author} {\bibfnamefont {M.}~\bibnamefont {{Campanelli}}},\
  }\href {https://doi.org/10.1103/PhysRevD.93.104057} {\bibfield  {journal}
  {\bibinfo  {journal} {\prd}\ }\textbf {\bibinfo {volume} {93}},\ \bibinfo
  {eid} {104057} (\bibinfo {year} {2016})}\BibitemShut {NoStop}%
\bibitem [{\citenamefont {{Nakano}}\ \emph {et~al.}(2016)\citenamefont
  {{Nakano}}, \citenamefont {{Ireland}}, \citenamefont {{Campanelli}},\ and\
  \citenamefont {{West}}}]{Nakano2016}%
  \BibitemOpen
  \bibfield  {author} {\bibinfo {author} {\bibfnamefont {H.}~\bibnamefont
  {{Nakano}}}, \bibinfo {author} {\bibfnamefont {B.}~\bibnamefont {{Ireland}}},
  \bibinfo {author} {\bibfnamefont {M.}~\bibnamefont {{Campanelli}}},\ and\
  \bibinfo {author} {\bibfnamefont {E.~J.}\ \bibnamefont {{West}}},\ }\href
  {https://doi.org/10.1088/0264-9381/33/24/247001} {\bibfield  {journal}
  {\bibinfo  {journal} {Classical and Quantum Gravity}\ }\textbf {\bibinfo
  {volume} {33}},\ \bibinfo {eid} {247001} (\bibinfo {year} {2016})},\ \Eprint
  {https://arxiv.org/abs/1608.01033} {arXiv:1608.01033 [gr-qc]} \BibitemShut
  {NoStop}%
\bibitem [{\citenamefont {{Combi}}\ \emph {et~al.}(2021)\citenamefont
  {{Combi}}, \citenamefont {{Armengol}}, \citenamefont {{Campanelli}},
  \citenamefont {{Ireland}}, \citenamefont {{Noble}}, \citenamefont
  {{Nakano}},\ and\ \citenamefont {{Bowen}}}]{Combi2021}%
  \BibitemOpen
  \bibfield  {author} {\bibinfo {author} {\bibfnamefont {L.}~\bibnamefont
  {{Combi}}}, \bibinfo {author} {\bibfnamefont {F.~G.~L.}\ \bibnamefont
  {{Armengol}}}, \bibinfo {author} {\bibfnamefont {M.}~\bibnamefont
  {{Campanelli}}}, \bibinfo {author} {\bibfnamefont {B.}~\bibnamefont
  {{Ireland}}}, \bibinfo {author} {\bibfnamefont {S.~C.}\ \bibnamefont
  {{Noble}}}, \bibinfo {author} {\bibfnamefont {H.}~\bibnamefont {{Nakano}}},\
  and\ \bibinfo {author} {\bibfnamefont {D.}~\bibnamefont {{Bowen}}},\ }\href
  {https://doi.org/10.1103/PhysRevD.104.044041} {\bibfield  {journal} {\bibinfo
   {journal} {\prd}\ }\textbf {\bibinfo {volume} {104}},\ \bibinfo {eid}
  {044041} (\bibinfo {year} {2021})},\ \Eprint
  {https://arxiv.org/abs/2103.15707} {arXiv:2103.15707 [gr-qc]} \BibitemShut
  {NoStop}%
\bibitem [{\citenamefont {{Combi}}\ and\ \citenamefont
  {{Ressler}}(2024)}]{CombiRessler2024}%
  \BibitemOpen
  \bibfield  {author} {\bibinfo {author} {\bibfnamefont {L.}~\bibnamefont
  {{Combi}}}\ and\ \bibinfo {author} {\bibfnamefont {S.~M.}\ \bibnamefont
  {{Ressler}}},\ }\href {https://doi.org/10.48550/arXiv.2403.13308} {\bibfield
  {journal} {\bibinfo  {journal} {arXiv e-prints}\ ,\ \bibinfo {eid}
  {arXiv:2403.13308}} (\bibinfo {year} {2024})},\ \Eprint
  {https://arxiv.org/abs/2403.13308} {arXiv:2403.13308 [gr-qc]} \BibitemShut
  {NoStop}%
\bibitem [{\citenamefont {{Mewes}}\ \emph {et~al.}(2018)\citenamefont
  {{Mewes}}, \citenamefont {{Zlochower}}, \citenamefont {{Campanelli}},
  \citenamefont {{Ruchlin}}, \citenamefont {{Etienne}},\ and\ \citenamefont
  {{Baumgarte}}}]{Mewes2018}%
  \BibitemOpen
  \bibfield  {author} {\bibinfo {author} {\bibfnamefont {V.}~\bibnamefont
  {{Mewes}}}, \bibinfo {author} {\bibfnamefont {Y.}~\bibnamefont
  {{Zlochower}}}, \bibinfo {author} {\bibfnamefont {M.}~\bibnamefont
  {{Campanelli}}}, \bibinfo {author} {\bibfnamefont {I.}~\bibnamefont
  {{Ruchlin}}}, \bibinfo {author} {\bibfnamefont {Z.~B.}\ \bibnamefont
  {{Etienne}}},\ and\ \bibinfo {author} {\bibfnamefont {T.~W.}\ \bibnamefont
  {{Baumgarte}}},\ }\href {https://doi.org/10.1103/PhysRevD.97.084059}
  {\bibfield  {journal} {\bibinfo  {journal} {\prd}\ }\textbf {\bibinfo
  {volume} {97}},\ \bibinfo {eid} {084059} (\bibinfo {year} {2018})},\ \Eprint
  {https://arxiv.org/abs/1802.09625} {arXiv:1802.09625 [gr-qc]} \BibitemShut
  {NoStop}%
\bibitem [{\citenamefont {{Mewes}}\ \emph {et~al.}(2020)\citenamefont
  {{Mewes}}, \citenamefont {{Zlochower}}, \citenamefont {{Campanelli}},
  \citenamefont {{Baumgarte}}, \citenamefont {{Etienne}}, \citenamefont
  {{Armengol}},\ and\ \citenamefont {{Cipolletta}}}]{Mewes2020}%
  \BibitemOpen
  \bibfield  {author} {\bibinfo {author} {\bibfnamefont {V.}~\bibnamefont
  {{Mewes}}}, \bibinfo {author} {\bibfnamefont {Y.}~\bibnamefont
  {{Zlochower}}}, \bibinfo {author} {\bibfnamefont {M.}~\bibnamefont
  {{Campanelli}}}, \bibinfo {author} {\bibfnamefont {T.~W.}\ \bibnamefont
  {{Baumgarte}}}, \bibinfo {author} {\bibfnamefont {Z.~B.}\ \bibnamefont
  {{Etienne}}}, \bibinfo {author} {\bibfnamefont {F.~G.~L.}\ \bibnamefont
  {{Armengol}}},\ and\ \bibinfo {author} {\bibfnamefont {F.}~\bibnamefont
  {{Cipolletta}}},\ }\href {https://doi.org/10.1103/PhysRevD.101.104007}
  {\bibfield  {journal} {\bibinfo  {journal} {\prd}\ }\textbf {\bibinfo
  {volume} {101}},\ \bibinfo {eid} {104007} (\bibinfo {year} {2020})},\ \Eprint
  {https://arxiv.org/abs/2002.06225} {arXiv:2002.06225 [gr-qc]} \BibitemShut
  {NoStop}%
\bibitem [{\citenamefont {{Ji}}\ \emph {et~al.}(2023)\citenamefont {{Ji}},
  \citenamefont {{Mewes}}, \citenamefont {{Zlochower}}, \citenamefont
  {{Ennoggi}}, \citenamefont {{Armengol}}, \citenamefont {{Campanelli}},
  \citenamefont {{Cipolletta}},\ and\ \citenamefont {{Etienne}}}]{Ji2023}%
  \BibitemOpen
  \bibfield  {author} {\bibinfo {author} {\bibfnamefont {L.}~\bibnamefont
  {{Ji}}}, \bibinfo {author} {\bibfnamefont {V.}~\bibnamefont {{Mewes}}},
  \bibinfo {author} {\bibfnamefont {Y.}~\bibnamefont {{Zlochower}}}, \bibinfo
  {author} {\bibfnamefont {L.}~\bibnamefont {{Ennoggi}}}, \bibinfo {author}
  {\bibfnamefont {F.~G.~L.}\ \bibnamefont {{Armengol}}}, \bibinfo {author}
  {\bibfnamefont {M.}~\bibnamefont {{Campanelli}}}, \bibinfo {author}
  {\bibfnamefont {F.}~\bibnamefont {{Cipolletta}}},\ and\ \bibinfo {author}
  {\bibfnamefont {Z.~B.}\ \bibnamefont {{Etienne}}},\ }\href
  {https://doi.org/10.1103/PhysRevD.108.104005} {\bibfield  {journal} {\bibinfo
   {journal} {\prd}\ }\textbf {\bibinfo {volume} {108}},\ \bibinfo {eid}
  {104005} (\bibinfo {year} {2023})},\ \Eprint
  {https://arxiv.org/abs/2305.01537} {arXiv:2305.01537 [gr-qc]} \BibitemShut
  {NoStop}%
\bibitem [{\citenamefont {{Etienne}}\ \emph {et~al.}(2015)\citenamefont
  {{Etienne}}, \citenamefont {{Paschalidis}}, \citenamefont {{Haas}},
  \citenamefont {{M{\"o}sta}},\ and\ \citenamefont {{Shapiro}}}]{Etienne2015}%
  \BibitemOpen
  \bibfield  {author} {\bibinfo {author} {\bibfnamefont {Z.~B.}\ \bibnamefont
  {{Etienne}}}, \bibinfo {author} {\bibfnamefont {V.}~\bibnamefont
  {{Paschalidis}}}, \bibinfo {author} {\bibfnamefont {R.}~\bibnamefont
  {{Haas}}}, \bibinfo {author} {\bibfnamefont {P.}~\bibnamefont
  {{M{\"o}sta}}},\ and\ \bibinfo {author} {\bibfnamefont {S.~L.}\ \bibnamefont
  {{Shapiro}}},\ }\href {https://doi.org/10.1088/0264-9381/32/17/175009}
  {\bibfield  {journal} {\bibinfo  {journal} {Classical and Quantum Gravity}\
  }\textbf {\bibinfo {volume} {32}},\ \bibinfo {eid} {175009} (\bibinfo {year}
  {2015})},\ \Eprint {https://arxiv.org/abs/1501.07276} {arXiv:1501.07276
  [astro-ph.HE]} \BibitemShut {NoStop}%
\bibitem [{\citenamefont {{Werneck}}\ \emph {et~al.}(2023)\citenamefont
  {{Werneck}}, \citenamefont {{Etienne}}, \citenamefont {{Murguia-Berthier}},
  \citenamefont {{Haas}}, \citenamefont {{Cipolletta}}, \citenamefont
  {{Noble}}, \citenamefont {{Ennoggi}}, \citenamefont {{Lopez Armengol}},
  \citenamefont {{Giacomazzo}}, \citenamefont {{Assump{\c{c}}{\~a}o}},
  \citenamefont {{Faber}}, \citenamefont {{Gupte}}, \citenamefont {{Kelly}},\
  and\ \citenamefont {{Krolik}}}]{Werneck2023}%
  \BibitemOpen
  \bibfield  {author} {\bibinfo {author} {\bibfnamefont {L.~R.}\ \bibnamefont
  {{Werneck}}}, \bibinfo {author} {\bibfnamefont {Z.~B.}\ \bibnamefont
  {{Etienne}}}, \bibinfo {author} {\bibfnamefont {A.}~\bibnamefont
  {{Murguia-Berthier}}}, \bibinfo {author} {\bibfnamefont {R.}~\bibnamefont
  {{Haas}}}, \bibinfo {author} {\bibfnamefont {F.}~\bibnamefont
  {{Cipolletta}}}, \bibinfo {author} {\bibfnamefont {S.~C.}\ \bibnamefont
  {{Noble}}}, \bibinfo {author} {\bibfnamefont {L.}~\bibnamefont {{Ennoggi}}},
  \bibinfo {author} {\bibfnamefont {F.~G.}\ \bibnamefont {{Lopez Armengol}}},
  \bibinfo {author} {\bibfnamefont {B.}~\bibnamefont {{Giacomazzo}}}, \bibinfo
  {author} {\bibfnamefont {T.}~\bibnamefont {{Assump{\c{c}}{\~a}o}}}, \bibinfo
  {author} {\bibfnamefont {J.}~\bibnamefont {{Faber}}}, \bibinfo {author}
  {\bibfnamefont {T.}~\bibnamefont {{Gupte}}}, \bibinfo {author} {\bibfnamefont
  {B.~J.}\ \bibnamefont {{Kelly}}},\ and\ \bibinfo {author} {\bibfnamefont
  {J.~H.}\ \bibnamefont {{Krolik}}},\ }\href
  {https://doi.org/10.1103/PhysRevD.107.044037} {\bibfield  {journal} {\bibinfo
   {journal} {\prd}\ }\textbf {\bibinfo {volume} {107}},\ \bibinfo {eid}
  {044037} (\bibinfo {year} {2023})},\ \Eprint
  {https://arxiv.org/abs/2208.14487} {arXiv:2208.14487 [gr-qc]} \BibitemShut
  {NoStop}%
\bibitem [{\citenamefont {Darmois}(1927)}]{Darmois1927}%
  \BibitemOpen
  \bibfield  {author} {\bibinfo {author} {\bibfnamefont {G.}~\bibnamefont
  {Darmois}},\ }\href {http://www.numdam.org/item/MSM\_1927\_\_25\_\_1\_0}
  {\emph {\bibinfo {title} {Les \'equations de la gravitation
  einsteinienne}}},\ \bibinfo {series} {M\'emorial des sciences
  math\'ematiques}\ No.~\bibinfo {number} {25}\ (\bibinfo  {publisher}
  {Gauthier-Villars},\ \bibinfo {year} {1927})\BibitemShut {NoStop}%
\bibitem [{\citenamefont {{Four{\`e}s-Bruhat}}(1952)}]{FouresBruhat1952}%
  \BibitemOpen
  \bibfield  {author} {\bibinfo {author} {\bibfnamefont {Y.}~\bibnamefont
  {{Four{\`e}s-Bruhat}}},\ }\href {https://doi.org/10.1007/BF02392131}
  {\bibfield  {journal} {\bibinfo  {journal} {Acta Mathematica}\ }\textbf
  {\bibinfo {volume} {88}},\ \bibinfo {pages} {141} (\bibinfo {year}
  {1952})}\BibitemShut {NoStop}%
\bibitem [{\citenamefont {Arnowitt}\ \emph {et~al.}(2008)\citenamefont
  {Arnowitt}, \citenamefont {Deser},\ and\ \citenamefont
  {Misner}}]{Arnowitt2008}%
  \BibitemOpen
  \bibfield  {author} {\bibinfo {author} {\bibfnamefont {R.}~\bibnamefont
  {Arnowitt}}, \bibinfo {author} {\bibfnamefont {S.}~\bibnamefont {Deser}},\
  and\ \bibinfo {author} {\bibfnamefont {C.~W.}\ \bibnamefont {Misner}},\
  }\href {https://doi.org/10.1007/s10714-008-0661-1} {\bibfield  {journal}
  {\bibinfo  {journal} {General Relativity and Gravitation}\ }\textbf {\bibinfo
  {volume} {40}},\ \bibinfo {pages} {1997} (\bibinfo {year} {2008})},\ \Eprint
  {https://arxiv.org/abs/gr-qc/0405109} {arXiv:gr-qc/0405109 [gr-qc]}
  \BibitemShut {NoStop}%
\bibitem [{\citenamefont {Gourgoulhon}(2012)}]{Gourgoulhon2012}%
  \BibitemOpen
  \bibfield  {author} {\bibinfo {author} {\bibfnamefont {{\'E}.}~\bibnamefont
  {Gourgoulhon}},\ }\href {https://books.google.it/books?id=HKcMBwAAQBAJ}
  {\emph {\bibinfo {title} {3+1 Formalism in General Relativity: Bases of
  Numerical Relativity}}},\ Lecture Notes in Physics\ (\bibinfo  {publisher}
  {Springer Berlin Heidelberg},\ \bibinfo {year} {2012})\BibitemShut {NoStop}%
\bibitem [{\citenamefont {{Baumgarte}}\ and\ \citenamefont
  {{Shapiro}}(1998)}]{BaumgarteShapiro1998}%
  \BibitemOpen
  \bibfield  {author} {\bibinfo {author} {\bibfnamefont {T.~W.}\ \bibnamefont
  {{Baumgarte}}}\ and\ \bibinfo {author} {\bibfnamefont {S.~L.}\ \bibnamefont
  {{Shapiro}}},\ }\href {https://doi.org/10.1103/PhysRevD.59.024007} {\bibfield
   {journal} {\bibinfo  {journal} {\prd}\ }\textbf {\bibinfo {volume} {59}},\
  \bibinfo {eid} {024007} (\bibinfo {year} {1998})},\ \Eprint
  {https://arxiv.org/abs/gr-qc/9810065} {arXiv:gr-qc/9810065 [gr-qc]}
  \BibitemShut {NoStop}%
\bibitem [{\citenamefont {{Shibata}}\ and\ \citenamefont
  {{Nakamura}}(1995)}]{ShibataNakamura1995}%
  \BibitemOpen
  \bibfield  {author} {\bibinfo {author} {\bibfnamefont {M.}~\bibnamefont
  {{Shibata}}}\ and\ \bibinfo {author} {\bibfnamefont {T.}~\bibnamefont
  {{Nakamura}}},\ }\href {https://doi.org/10.1103/PhysRevD.52.5428} {\bibfield
  {journal} {\bibinfo  {journal} {\prd}\ }\textbf {\bibinfo {volume} {52}},\
  \bibinfo {pages} {5428} (\bibinfo {year} {1995})}\BibitemShut {NoStop}%
\bibitem [{\citenamefont {{Campanelli}}\ \emph
  {et~al.}(2006{\natexlab{a}})\citenamefont {{Campanelli}}, \citenamefont
  {{Lousto}}, \citenamefont {{Marronetti}},\ and\ \citenamefont
  {{Zlochower}}}]{Campanelli2006}%
  \BibitemOpen
  \bibfield  {author} {\bibinfo {author} {\bibfnamefont {M.}~\bibnamefont
  {{Campanelli}}}, \bibinfo {author} {\bibfnamefont {C.~O.}\ \bibnamefont
  {{Lousto}}}, \bibinfo {author} {\bibfnamefont {P.}~\bibnamefont
  {{Marronetti}}},\ and\ \bibinfo {author} {\bibfnamefont {Y.}~\bibnamefont
  {{Zlochower}}},\ }\href {https://doi.org/10.1103/PhysRevLett.96.111101}
  {\bibfield  {journal} {\bibinfo  {journal} {\prl}\ }\textbf {\bibinfo
  {volume} {96}},\ \bibinfo {eid} {111101} (\bibinfo {year}
  {2006}{\natexlab{a}})},\ \Eprint {https://arxiv.org/abs/gr-qc/0511048}
  {arXiv:gr-qc/0511048 [gr-qc]} \BibitemShut {NoStop}%
\bibitem [{\citenamefont {{Baker}}\ \emph {et~al.}(2006)\citenamefont
  {{Baker}}, \citenamefont {{Centrella}}, \citenamefont {{Choi}}, \citenamefont
  {{Koppitz}},\ and\ \citenamefont {{van Meter}}}]{Baker2006}%
  \BibitemOpen
  \bibfield  {author} {\bibinfo {author} {\bibfnamefont {J.~G.}\ \bibnamefont
  {{Baker}}}, \bibinfo {author} {\bibfnamefont {J.}~\bibnamefont
  {{Centrella}}}, \bibinfo {author} {\bibfnamefont {D.-I.}\ \bibnamefont
  {{Choi}}}, \bibinfo {author} {\bibfnamefont {M.}~\bibnamefont {{Koppitz}}},\
  and\ \bibinfo {author} {\bibfnamefont {J.}~\bibnamefont {{van Meter}}},\
  }\href {https://doi.org/10.1103/PhysRevLett.96.111102} {\bibfield  {journal}
  {\bibinfo  {journal} {\prl}\ }\textbf {\bibinfo {volume} {96}},\ \bibinfo
  {eid} {111102} (\bibinfo {year} {2006})},\ \Eprint
  {https://arxiv.org/abs/gr-qc/0511103} {arXiv:gr-qc/0511103 [gr-qc]}
  \BibitemShut {NoStop}%
\bibitem [{\citenamefont {{Banyuls}}\ \emph {et~al.}(1997)\citenamefont
  {{Banyuls}}, \citenamefont {{Font}}, \citenamefont {{Ib{\'a}{\~n}ez}},
  \citenamefont {{Mart{\'\i}}},\ and\ \citenamefont
  {{Miralles}}}]{Banyuls1997}%
  \BibitemOpen
  \bibfield  {author} {\bibinfo {author} {\bibfnamefont {F.}~\bibnamefont
  {{Banyuls}}}, \bibinfo {author} {\bibfnamefont {J.~A.}\ \bibnamefont
  {{Font}}}, \bibinfo {author} {\bibfnamefont {J.~M.}\ \bibnamefont
  {{Ib{\'a}{\~n}ez}}}, \bibinfo {author} {\bibfnamefont {J.~M.}\ \bibnamefont
  {{Mart{\'\i}}}},\ and\ \bibinfo {author} {\bibfnamefont {J.~A.}\ \bibnamefont
  {{Miralles}}},\ }\href {https://doi.org/10.1086/303604} {\bibfield  {journal}
  {\bibinfo  {journal} {\apj}\ }\textbf {\bibinfo {volume} {476}},\ \bibinfo
  {pages} {221} (\bibinfo {year} {1997})}\BibitemShut {NoStop}%
\bibitem [{\citenamefont {{Bonazzola}}\ \emph {et~al.}(2004)\citenamefont
  {{Bonazzola}}, \citenamefont {{Gourgoulhon}}, \citenamefont
  {{Grandcl{\'e}ment}},\ and\ \citenamefont {{Novak}}}]{Bonazzola2004}%
  \BibitemOpen
  \bibfield  {author} {\bibinfo {author} {\bibfnamefont {S.}~\bibnamefont
  {{Bonazzola}}}, \bibinfo {author} {\bibfnamefont {E.}~\bibnamefont
  {{Gourgoulhon}}}, \bibinfo {author} {\bibfnamefont {P.}~\bibnamefont
  {{Grandcl{\'e}ment}}},\ and\ \bibinfo {author} {\bibfnamefont
  {J.}~\bibnamefont {{Novak}}},\ }\href
  {https://doi.org/10.1103/PhysRevD.70.104007} {\bibfield  {journal} {\bibinfo
  {journal} {\prd}\ }\textbf {\bibinfo {volume} {70}},\ \bibinfo {eid} {104007}
  (\bibinfo {year} {2004})},\ \Eprint {https://arxiv.org/abs/gr-qc/0307082}
  {arXiv:gr-qc/0307082 [gr-qc]} \BibitemShut {NoStop}%
\bibitem [{\citenamefont {{Shibata}}\ \emph {et~al.}(2004)\citenamefont
  {{Shibata}}, \citenamefont {{Ury{\= u}}},\ and\ \citenamefont
  {{Friedman}}}]{Shibata2004}%
  \BibitemOpen
  \bibfield  {author} {\bibinfo {author} {\bibfnamefont {M.}~\bibnamefont
  {{Shibata}}}, \bibinfo {author} {\bibfnamefont {K.}~\bibnamefont {{Ury{\=
  u}}}},\ and\ \bibinfo {author} {\bibfnamefont {J.~L.}\ \bibnamefont
  {{Friedman}}},\ }\href {https://doi.org/10.1103/PhysRevD.70.044044}
  {\bibfield  {journal} {\bibinfo  {journal} {\prd}\ }\textbf {\bibinfo
  {volume} {70}},\ \bibinfo {eid} {044044} (\bibinfo {year} {2004})},\ \Eprint
  {https://arxiv.org/abs/gr-qc/0407036} {arXiv:gr-qc/0407036 [gr-qc]}
  \BibitemShut {NoStop}%
\bibitem [{\citenamefont {{Brown}}(2009)}]{BrownCovariant2009}%
  \BibitemOpen
  \bibfield  {author} {\bibinfo {author} {\bibfnamefont {J.~D.}\ \bibnamefont
  {{Brown}}},\ }\href {https://doi.org/10.1103/PhysRevD.79.104029} {\bibfield
  {journal} {\bibinfo  {journal} {\prd}\ }\textbf {\bibinfo {volume} {79}},\
  \bibinfo {eid} {104029} (\bibinfo {year} {2009})},\ \Eprint
  {https://arxiv.org/abs/0902.3652} {arXiv:0902.3652 [gr-qc]} \BibitemShut
  {NoStop}%
\bibitem [{\citenamefont {{Montero}}\ and\ \citenamefont
  {{Cordero-Carri{\'o}n}}(2012)}]{Montero2012}%
  \BibitemOpen
  \bibfield  {author} {\bibinfo {author} {\bibfnamefont {P.~J.}\ \bibnamefont
  {{Montero}}}\ and\ \bibinfo {author} {\bibfnamefont {I.}~\bibnamefont
  {{Cordero-Carri{\'o}n}}},\ }\href
  {https://doi.org/10.1103/PhysRevD.85.124037} {\bibfield  {journal} {\bibinfo
  {journal} {\prd}\ }\textbf {\bibinfo {volume} {85}},\ \bibinfo {eid} {124037}
  (\bibinfo {year} {2012})},\ \Eprint {https://arxiv.org/abs/1204.5377}
  {arXiv:1204.5377 [gr-qc]} \BibitemShut {NoStop}%
\bibitem [{\citenamefont {{Baumgarte}}\ \emph {et~al.}(2013)\citenamefont
  {{Baumgarte}}, \citenamefont {{Montero}}, \citenamefont
  {{Cordero-Carri{\'o}n}},\ and\ \citenamefont {{M{\"u}ller}}}]{Baumgarte2013}%
  \BibitemOpen
  \bibfield  {author} {\bibinfo {author} {\bibfnamefont {T.~W.}\ \bibnamefont
  {{Baumgarte}}}, \bibinfo {author} {\bibfnamefont {P.~J.}\ \bibnamefont
  {{Montero}}}, \bibinfo {author} {\bibfnamefont {I.}~\bibnamefont
  {{Cordero-Carri{\'o}n}}},\ and\ \bibinfo {author} {\bibfnamefont
  {E.}~\bibnamefont {{M{\"u}ller}}},\ }\href
  {https://doi.org/10.1103/PhysRevD.87.044026} {\bibfield  {journal} {\bibinfo
  {journal} {\prd}\ }\textbf {\bibinfo {volume} {87}},\ \bibinfo {eid} {044026}
  (\bibinfo {year} {2013})},\ \Eprint {https://arxiv.org/abs/1211.6632}
  {arXiv:1211.6632 [gr-qc]} \BibitemShut {NoStop}%
\bibitem [{\citenamefont {{Montero}}\ \emph {et~al.}(2014)\citenamefont
  {{Montero}}, \citenamefont {{Baumgarte}},\ and\ \citenamefont
  {{M{\"u}ller}}}]{Montero2014}%
  \BibitemOpen
  \bibfield  {author} {\bibinfo {author} {\bibfnamefont {P.~J.}\ \bibnamefont
  {{Montero}}}, \bibinfo {author} {\bibfnamefont {T.~W.}\ \bibnamefont
  {{Baumgarte}}},\ and\ \bibinfo {author} {\bibfnamefont {E.}~\bibnamefont
  {{M{\"u}ller}}},\ }\href {https://doi.org/10.1103/PhysRevD.89.084043}
  {\bibfield  {journal} {\bibinfo  {journal} {\prd}\ }\textbf {\bibinfo
  {volume} {89}},\ \bibinfo {eid} {084043} (\bibinfo {year} {2014})},\ \Eprint
  {https://arxiv.org/abs/1309.7808} {arXiv:1309.7808 [gr-qc]} \BibitemShut
  {NoStop}%
\bibitem [{\citenamefont {{Sanchis-Gual}}\ \emph {et~al.}(2014)\citenamefont
  {{Sanchis-Gual}}, \citenamefont {{Montero}}, \citenamefont {{Font}},
  \citenamefont {{M{\"u}ller}},\ and\ \citenamefont
  {{Baumgarte}}}]{SanchisGual2014}%
  \BibitemOpen
  \bibfield  {author} {\bibinfo {author} {\bibfnamefont {N.}~\bibnamefont
  {{Sanchis-Gual}}}, \bibinfo {author} {\bibfnamefont {P.~J.}\ \bibnamefont
  {{Montero}}}, \bibinfo {author} {\bibfnamefont {J.~A.}\ \bibnamefont
  {{Font}}}, \bibinfo {author} {\bibfnamefont {E.}~\bibnamefont
  {{M{\"u}ller}}},\ and\ \bibinfo {author} {\bibfnamefont {T.~W.}\ \bibnamefont
  {{Baumgarte}}},\ }\href {https://doi.org/10.1103/PhysRevD.89.104033}
  {\bibfield  {journal} {\bibinfo  {journal} {\prd}\ }\textbf {\bibinfo
  {volume} {89}},\ \bibinfo {eid} {104033} (\bibinfo {year} {2014})},\ \Eprint
  {https://arxiv.org/abs/1403.3653} {arXiv:1403.3653 [gr-qc]} \BibitemShut
  {NoStop}%
\bibitem [{\citenamefont {{Baumgarte}}\ \emph {et~al.}(2015)\citenamefont
  {{Baumgarte}}, \citenamefont {{Montero}},\ and\ \citenamefont
  {{M{\"u}ller}}}]{Baumgarte2015}%
  \BibitemOpen
  \bibfield  {author} {\bibinfo {author} {\bibfnamefont {T.~W.}\ \bibnamefont
  {{Baumgarte}}}, \bibinfo {author} {\bibfnamefont {P.~J.}\ \bibnamefont
  {{Montero}}},\ and\ \bibinfo {author} {\bibfnamefont {E.}~\bibnamefont
  {{M{\"u}ller}}},\ }\href {https://doi.org/10.1103/PhysRevD.91.064035}
  {\bibfield  {journal} {\bibinfo  {journal} {\prd}\ }\textbf {\bibinfo
  {volume} {91}},\ \bibinfo {eid} {064035} (\bibinfo {year} {2015})},\ \Eprint
  {https://arxiv.org/abs/1501.05259} {arXiv:1501.05259 [gr-qc]} \BibitemShut
  {NoStop}%
\bibitem [{\citenamefont {{Zilh{\~a}o}}\ \emph {et~al.}(2015)\citenamefont
  {{Zilh{\~a}o}}, \citenamefont {{Noble}}, \citenamefont {{Campanelli}},\ and\
  \citenamefont {{Zlochower}}}]{Zilhao2015}%
  \BibitemOpen
  \bibfield  {author} {\bibinfo {author} {\bibfnamefont {M.}~\bibnamefont
  {{Zilh{\~a}o}}}, \bibinfo {author} {\bibfnamefont {S.~C.}\ \bibnamefont
  {{Noble}}}, \bibinfo {author} {\bibfnamefont {M.}~\bibnamefont
  {{Campanelli}}},\ and\ \bibinfo {author} {\bibfnamefont {Y.}~\bibnamefont
  {{Zlochower}}},\ }\href {https://doi.org/10.1103/PhysRevD.91.024034}
  {\bibfield  {journal} {\bibinfo  {journal} {\prd}\ }\textbf {\bibinfo
  {volume} {91}},\ \bibinfo {eid} {024034} (\bibinfo {year} {2015})},\ \Eprint
  {https://arxiv.org/abs/1409.4787} {arXiv:1409.4787 [gr-qc]} \BibitemShut
  {NoStop}%
\bibitem [{\citenamefont {Goodale}\ \emph {et~al.}(2003)\citenamefont
  {Goodale}, \citenamefont {Allen}, \citenamefont {Lanfermann}, \citenamefont
  {Mass{\'o}}, \citenamefont {Radke}, \citenamefont {Seidel},\ and\
  \citenamefont {Shalf}}]{GoodaleCactus2003}%
  \BibitemOpen
  \bibfield  {author} {\bibinfo {author} {\bibfnamefont {T.}~\bibnamefont
  {Goodale}}, \bibinfo {author} {\bibfnamefont {G.}~\bibnamefont {Allen}},
  \bibinfo {author} {\bibfnamefont {G.}~\bibnamefont {Lanfermann}}, \bibinfo
  {author} {\bibfnamefont {J.}~\bibnamefont {Mass{\'o}}}, \bibinfo {author}
  {\bibfnamefont {T.}~\bibnamefont {Radke}}, \bibinfo {author} {\bibfnamefont
  {E.}~\bibnamefont {Seidel}},\ and\ \bibinfo {author} {\bibfnamefont
  {J.}~\bibnamefont {Shalf}},\ }in\ \href@noop {} {\emph {\bibinfo {booktitle}
  {High Performance Computing for Computational Science --- VECPAR 2002}}},\
  \bibinfo {editor} {edited by\ \bibinfo {editor} {\bibfnamefont {J.~M. L.~M.}\
  \bibnamefont {Palma}}, \bibinfo {editor} {\bibfnamefont {A.~A.}\ \bibnamefont
  {Sousa}}, \bibinfo {editor} {\bibfnamefont {J.}~\bibnamefont {Dongarra}},\
  and\ \bibinfo {editor} {\bibfnamefont {V.}~\bibnamefont {Hern{\'a}ndez}}}\
  (\bibinfo  {publisher} {Springer Berlin Heidelberg},\ \bibinfo {address}
  {Berlin, Heidelberg},\ \bibinfo {year} {2003})\ pp.\ \bibinfo {pages}
  {197--227}\BibitemShut {NoStop}%
\bibitem [{\citenamefont {{L{\"o}ffler}}\ \emph {et~al.}(2012)\citenamefont
  {{L{\"o}ffler}}, \citenamefont {{Faber}}, \citenamefont {{Bentivegna}},
  \citenamefont {{Bode}}, \citenamefont {{Diener}}, \citenamefont {{Haas}},
  \citenamefont {{Hinder}}, \citenamefont {{Mundim}}, \citenamefont {{Ott}},
  \citenamefont {{Schnetter}}, \citenamefont {{Allen}}, \citenamefont
  {{Campanelli}},\ and\ \citenamefont {{Laguna}}}]{Loffler2012}%
  \BibitemOpen
  \bibfield  {author} {\bibinfo {author} {\bibfnamefont {F.}~\bibnamefont
  {{L{\"o}ffler}}}, \bibinfo {author} {\bibfnamefont {J.}~\bibnamefont
  {{Faber}}}, \bibinfo {author} {\bibfnamefont {E.}~\bibnamefont
  {{Bentivegna}}}, \bibinfo {author} {\bibfnamefont {T.}~\bibnamefont
  {{Bode}}}, \bibinfo {author} {\bibfnamefont {P.}~\bibnamefont {{Diener}}},
  \bibinfo {author} {\bibfnamefont {R.}~\bibnamefont {{Haas}}}, \bibinfo
  {author} {\bibfnamefont {I.}~\bibnamefont {{Hinder}}}, \bibinfo {author}
  {\bibfnamefont {B.~C.}\ \bibnamefont {{Mundim}}}, \bibinfo {author}
  {\bibfnamefont {C.~D.}\ \bibnamefont {{Ott}}}, \bibinfo {author}
  {\bibfnamefont {E.}~\bibnamefont {{Schnetter}}}, \bibinfo {author}
  {\bibfnamefont {G.}~\bibnamefont {{Allen}}}, \bibinfo {author} {\bibfnamefont
  {M.}~\bibnamefont {{Campanelli}}},\ and\ \bibinfo {author} {\bibfnamefont
  {P.}~\bibnamefont {{Laguna}}},\ }\href
  {https://doi.org/10.1088/0264-9381/29/11/115001} {\bibfield  {journal}
  {\bibinfo  {journal} {Classical and Quantum Gravity}\ }\textbf {\bibinfo
  {volume} {29}},\ \bibinfo {eid} {115001} (\bibinfo {year} {2012})},\ \Eprint
  {https://arxiv.org/abs/1111.3344} {arXiv:1111.3344 [gr-qc]} \BibitemShut
  {NoStop}%
\bibitem [{\citenamefont {{Zilh{\~a}o}}\ and\ \citenamefont
  {{L{\"o}ffler}}(2013)}]{Zilhao2013}%
  \BibitemOpen
  \bibfield  {author} {\bibinfo {author} {\bibfnamefont {M.}~\bibnamefont
  {{Zilh{\~a}o}}}\ and\ \bibinfo {author} {\bibfnamefont {F.}~\bibnamefont
  {{L{\"o}ffler}}},\ }\href {https://doi.org/10.1142/S0217751X13400149}
  {\bibfield  {journal} {\bibinfo  {journal} {International Journal of Modern
  Physics A}\ }\textbf {\bibinfo {volume} {28}},\ \bibinfo {eid} {1340014-126}
  (\bibinfo {year} {2013})},\ \Eprint {https://arxiv.org/abs/1305.5299}
  {arXiv:1305.5299 [gr-qc]} \BibitemShut {NoStop}%
\bibitem [{\citenamefont {{Schnetter}}\ \emph {et~al.}(2004)\citenamefont
  {{Schnetter}}, \citenamefont {{Hawley}},\ and\ \citenamefont
  {{Hawke}}}]{Schnetter2004}%
  \BibitemOpen
  \bibfield  {author} {\bibinfo {author} {\bibfnamefont {E.}~\bibnamefont
  {{Schnetter}}}, \bibinfo {author} {\bibfnamefont {S.~H.}\ \bibnamefont
  {{Hawley}}},\ and\ \bibinfo {author} {\bibfnamefont {I.}~\bibnamefont
  {{Hawke}}},\ }\href {https://doi.org/10.1088/0264-9381/21/6/014} {\bibfield
  {journal} {\bibinfo  {journal} {Classical and Quantum Gravity}\ }\textbf
  {\bibinfo {volume} {21}},\ \bibinfo {pages} {1465} (\bibinfo {year}
  {2004})},\ \Eprint {https://arxiv.org/abs/gr-qc/0310042} {arXiv:gr-qc/0310042
  [gr-qc]} \BibitemShut {NoStop}%
\bibitem [{\citenamefont {{Schnetter}}\ \emph
  {et~al.}(2006{\natexlab{a}})\citenamefont {{Schnetter}}, \citenamefont
  {{Diener}}, \citenamefont {{Dorband}},\ and\ \citenamefont
  {{Tiglio}}}]{Schnetter2006}%
  \BibitemOpen
  \bibfield  {author} {\bibinfo {author} {\bibfnamefont {E.}~\bibnamefont
  {{Schnetter}}}, \bibinfo {author} {\bibfnamefont {P.}~\bibnamefont
  {{Diener}}}, \bibinfo {author} {\bibfnamefont {E.~N.}\ \bibnamefont
  {{Dorband}}},\ and\ \bibinfo {author} {\bibfnamefont {M.}~\bibnamefont
  {{Tiglio}}},\ }\href {https://doi.org/10.1088/0264-9381/23/16/S14} {\bibfield
   {journal} {\bibinfo  {journal} {Classical and Quantum Gravity}\ }\textbf
  {\bibinfo {volume} {23}},\ \bibinfo {pages} {S553} (\bibinfo {year}
  {2006}{\natexlab{a}})},\ \Eprint {https://arxiv.org/abs/gr-qc/0602104}
  {arXiv:gr-qc/0602104 [gr-qc]} \BibitemShut {NoStop}%
\bibitem [{\citenamefont {{Schnetter}}(2003)}]{CarpetDocs}%
  \BibitemOpen
  \bibfield  {author} {\bibinfo {author} {\bibfnamefont {E.}~\bibnamefont
  {{Schnetter}}},\ }\href@noop {} {\bibinfo {title} {\texttt{Carpet}
  documentation}},\ \bibinfo {howpublished}
  {\url{https://einsteintoolkit.org/arrangementguide/Carpet/documentation.html}}
  (\bibinfo {year} {2003})\BibitemShut {NoStop}%
\bibitem [{\citenamefont {{Schnetter}}()}]{CarpetBitBucket}%
  \BibitemOpen
  \bibfield  {author} {\bibinfo {author} {\bibfnamefont {E.}~\bibnamefont
  {{Schnetter}}},\ }\href@noop {} {\bibinfo {title} {\texttt{Carpet} bitbucket
  repository}},\ \bibinfo {howpublished}
  {\url{https://bitbucket.org/eschnett/carpet/src/master}}\BibitemShut
  {NoStop}%
\bibitem [{\citenamefont {{Balsara}}\ \emph {et~al.}(2016)\citenamefont
  {{Balsara}}, \citenamefont {{Garain}},\ and\ \citenamefont
  {{Shu}}}]{Balsara2016}%
  \BibitemOpen
  \bibfield  {author} {\bibinfo {author} {\bibfnamefont {D.~S.}\ \bibnamefont
  {{Balsara}}}, \bibinfo {author} {\bibfnamefont {S.}~\bibnamefont
  {{Garain}}},\ and\ \bibinfo {author} {\bibfnamefont {C.-W.}\ \bibnamefont
  {{Shu}}},\ }\href {https://doi.org/10.1016/j.jcp.2016.09.009} {\bibfield
  {journal} {\bibinfo  {journal} {Journal of Computational Physics}\ }\textbf
  {\bibinfo {volume} {326}},\ \bibinfo {pages} {780} (\bibinfo {year}
  {2016})}\BibitemShut {NoStop}%
\bibitem [{\citenamefont {{Castro}}\ \emph {et~al.}(2011)\citenamefont
  {{Castro}}, \citenamefont {{Costa}},\ and\ \citenamefont
  {{Don}}}]{Castro2011}%
  \BibitemOpen
  \bibfield  {author} {\bibinfo {author} {\bibfnamefont {M.}~\bibnamefont
  {{Castro}}}, \bibinfo {author} {\bibfnamefont {B.}~\bibnamefont {{Costa}}},\
  and\ \bibinfo {author} {\bibfnamefont {W.~S.}\ \bibnamefont {{Don}}},\ }\href
  {https://doi.org/10.1016/j.jcp.2010.11.028} {\bibfield  {journal} {\bibinfo
  {journal} {Journal of Computational Physics}\ }\textbf {\bibinfo {volume}
  {230}},\ \bibinfo {pages} {1766} (\bibinfo {year} {2011})}\BibitemShut
  {NoStop}%
\bibitem [{\citenamefont {{Tchekhovskoy}}\ \emph {et~al.}(2007)\citenamefont
  {{Tchekhovskoy}}, \citenamefont {{McKinney}},\ and\ \citenamefont
  {{Narayan}}}]{Tchekhovskoy2007}%
  \BibitemOpen
  \bibfield  {author} {\bibinfo {author} {\bibfnamefont {A.}~\bibnamefont
  {{Tchekhovskoy}}}, \bibinfo {author} {\bibfnamefont {J.~C.}\ \bibnamefont
  {{McKinney}}},\ and\ \bibinfo {author} {\bibfnamefont {R.}~\bibnamefont
  {{Narayan}}},\ }\href {https://doi.org/10.1111/j.1365-2966.2007.11876.x}
  {\bibfield  {journal} {\bibinfo  {journal} {\mnras}\ }\textbf {\bibinfo
  {volume} {379}},\ \bibinfo {pages} {469} (\bibinfo {year} {2007})},\ \Eprint
  {https://arxiv.org/abs/0704.2608} {arXiv:0704.2608 [astro-ph]} \BibitemShut
  {NoStop}%
\bibitem [{\citenamefont {{Rembiasz}}\ \emph {et~al.}(2016)\citenamefont
  {{Rembiasz}}, \citenamefont {{Obergaulinger}}, \citenamefont
  {{Cerd{\'a}-Dur{\'a}n}}, \citenamefont {{M{\"u}ller}},\ and\ \citenamefont
  {{Aloy}}}]{Rembiasz2016}%
  \BibitemOpen
  \bibfield  {author} {\bibinfo {author} {\bibfnamefont {T.}~\bibnamefont
  {{Rembiasz}}}, \bibinfo {author} {\bibfnamefont {M.}~\bibnamefont
  {{Obergaulinger}}}, \bibinfo {author} {\bibfnamefont {P.}~\bibnamefont
  {{Cerd{\'a}-Dur{\'a}n}}}, \bibinfo {author} {\bibfnamefont {E.}~\bibnamefont
  {{M{\"u}ller}}},\ and\ \bibinfo {author} {\bibfnamefont {M.~A.}\ \bibnamefont
  {{Aloy}}},\ }\href {https://doi.org/10.1093/mnras/stv2917} {\bibfield
  {journal} {\bibinfo  {journal} {\mnras}\ }\textbf {\bibinfo {volume} {456}},\
  \bibinfo {pages} {3782} (\bibinfo {year} {2016})},\ \Eprint
  {https://arxiv.org/abs/1508.04799} {arXiv:1508.04799 [astro-ph.SR]}
  \BibitemShut {NoStop}%
\bibitem [{\citenamefont {{Rembiasz}}\ \emph {et~al.}(2017)\citenamefont
  {{Rembiasz}}, \citenamefont {{Obergaulinger}}, \citenamefont
  {{Cerd{\'a}-Dur{\'a}n}}, \citenamefont {{Aloy}},\ and\ \citenamefont
  {{M{\"u}ller}}}]{Rembiasz2017}%
  \BibitemOpen
  \bibfield  {author} {\bibinfo {author} {\bibfnamefont {T.}~\bibnamefont
  {{Rembiasz}}}, \bibinfo {author} {\bibfnamefont {M.}~\bibnamefont
  {{Obergaulinger}}}, \bibinfo {author} {\bibfnamefont {P.}~\bibnamefont
  {{Cerd{\'a}-Dur{\'a}n}}}, \bibinfo {author} {\bibfnamefont {M.-{\'A}.}\
  \bibnamefont {{Aloy}}},\ and\ \bibinfo {author} {\bibfnamefont
  {E.}~\bibnamefont {{M{\"u}ller}}},\ }\href
  {https://doi.org/10.3847/1538-4365/aa6254} {\bibfield  {journal} {\bibinfo
  {journal} {\apjs}\ }\textbf {\bibinfo {volume} {230}},\ \bibinfo {eid} {18}
  (\bibinfo {year} {2017})},\ \Eprint {https://arxiv.org/abs/1611.05858}
  {arXiv:1611.05858 [astro-ph.IM]} \BibitemShut {NoStop}%
\bibitem [{\citenamefont {{Mahlmann}}\ \emph {et~al.}(2021)\citenamefont
  {{Mahlmann}}, \citenamefont {{Aloy}}, \citenamefont {{Mewes}},\ and\
  \citenamefont {{Cerd{\'a}-Dur{\'a}n}}}]{Mahlmann2021}%
  \BibitemOpen
  \bibfield  {author} {\bibinfo {author} {\bibfnamefont {J.~F.}\ \bibnamefont
  {{Mahlmann}}}, \bibinfo {author} {\bibfnamefont {M.~A.}\ \bibnamefont
  {{Aloy}}}, \bibinfo {author} {\bibfnamefont {V.}~\bibnamefont {{Mewes}}},\
  and\ \bibinfo {author} {\bibfnamefont {P.}~\bibnamefont
  {{Cerd{\'a}-Dur{\'a}n}}},\ }\href
  {https://doi.org/10.1051/0004-6361/202038908} {\bibfield  {journal} {\bibinfo
   {journal} {\aap}\ }\textbf {\bibinfo {volume} {647}},\ \bibinfo {eid} {A58}
  (\bibinfo {year} {2021})},\ \Eprint {https://arxiv.org/abs/2007.06599}
  {arXiv:2007.06599 [physics.comp-ph]} \BibitemShut {NoStop}%
\bibitem [{\citenamefont {{Harten}}\ \emph {et~al.}(1983)\citenamefont
  {{Harten}}, \citenamefont {{Lax}},\ and\ \citenamefont {{van
  Leer}}}]{HLLE1983}%
  \BibitemOpen
  \bibfield  {author} {\bibinfo {author} {\bibfnamefont {A.}~\bibnamefont
  {{Harten}}}, \bibinfo {author} {\bibfnamefont {P.~D.}\ \bibnamefont
  {{Lax}}},\ and\ \bibinfo {author} {\bibfnamefont {B.}~\bibnamefont {{van
  Leer}}},\ }\href {https://doi.org/10.1137/1025002} {\bibfield  {journal}
  {\bibinfo  {journal} {SIAM Review}\ }\textbf {\bibinfo {volume} {25}},\
  \bibinfo {pages} {35} (\bibinfo {year} {1983})},\ \Eprint
  {https://arxiv.org/abs/https://doi.org/10.1137/1025002}
  {https://doi.org/10.1137/1025002} \BibitemShut {NoStop}%
\bibitem [{\citenamefont {{Kreiss}}\ and\ \citenamefont
  {{Oliger}}(1973)}]{KreissOligerBook}%
  \BibitemOpen
  \bibfield  {author} {\bibinfo {author} {\bibfnamefont {H.}~\bibnamefont
  {{Kreiss}}}\ and\ \bibinfo {author} {\bibfnamefont {J.}~\bibnamefont
  {{Oliger}}},\ }\href {https://books.google.com/books?id=wj9szwEACAAJ} {\emph
  {\bibinfo {title} {Methods for the approximate solution of time dependent
  problems}}},\ GARP Publications Series\ (\bibinfo  {publisher} {SELBSTVERL.)
  FEBR},\ \bibinfo {year} {1973})\BibitemShut {NoStop}%
\bibitem [{\citenamefont {{Schnetter}}\ and\ \citenamefont
  {{Kelly}}()}]{DissipationDocs}%
  \BibitemOpen
  \bibfield  {author} {\bibinfo {author} {\bibfnamefont {E.}~\bibnamefont
  {{Schnetter}}}\ and\ \bibinfo {author} {\bibfnamefont {B.~J.}\ \bibnamefont
  {{Kelly}}},\ }\href@noop {} {\bibinfo {title} {\texttt{Dissipation}
  documentation}},\ \bibinfo {howpublished}
  {\url{https://einsteintoolkit.org/thornguide/CactusNumerical/Dissipation/documentation.html}}\BibitemShut
  {NoStop}%
\bibitem [{\citenamefont {{Hawke}}({\natexlab{a}})}]{MoLBitBucket}%
  \BibitemOpen
  \bibfield  {author} {\bibinfo {author} {\bibfnamefont {I.}~\bibnamefont
  {{Hawke}}},\ }\href@noop {} {\bibinfo {title} {\texttt{Method of Lines (MoL)}
  bitbucket repository}},\ \bibinfo {howpublished}
  {\url{https://bitbucket.org/cactuscode/cactusnumerical/src/master/MoL/}}
  ({\natexlab{a}})\BibitemShut {NoStop}%
\bibitem [{\citenamefont {{Hawke}}({\natexlab{b}})}]{MoLDocs}%
  \BibitemOpen
  \bibfield  {author} {\bibinfo {author} {\bibfnamefont {I.}~\bibnamefont
  {{Hawke}}},\ }\href@noop {} {\bibinfo {title} {\texttt{Method of Lines (MoL)}
  documentation}},\ \bibinfo {howpublished}
  {\url{https://einsteintoolkit.org/thornguide/CactusNumerical/MoL/documentation.html}}
  ({\natexlab{b}})\BibitemShut {NoStop}%
\bibitem [{\citenamefont {Spiteri}\ and\ \citenamefont
  {Ruuth}(2002)}]{SpiteriRuuth2002}%
  \BibitemOpen
  \bibfield  {author} {\bibinfo {author} {\bibfnamefont {R.~J.}\ \bibnamefont
  {Spiteri}}\ and\ \bibinfo {author} {\bibfnamefont {S.~J.}\ \bibnamefont
  {Ruuth}},\ }\href {https://doi.org/10.1137/S0036142901389025} {\bibfield
  {journal} {\bibinfo  {journal} {SIAM Journal on Numerical Analysis}\ }\textbf
  {\bibinfo {volume} {40}},\ \bibinfo {pages} {469} (\bibinfo {year} {2002})},\
  \Eprint {https://arxiv.org/abs/https://doi.org/10.1137/S0036142901389025}
  {https://doi.org/10.1137/S0036142901389025} \BibitemShut {NoStop}%
\bibitem [{\citenamefont {{Zlochower}}\ \emph {et~al.}(2016)\citenamefont
  {{Zlochower}}, \citenamefont {{Nakano}}, \citenamefont {{Mundim}},
  \citenamefont {{Campanelli}}, \citenamefont {{Noble}},\ and\ \citenamefont
  {{Zilh{\~a}o}}}]{Zlochower2016}%
  \BibitemOpen
  \bibfield  {author} {\bibinfo {author} {\bibfnamefont {Y.}~\bibnamefont
  {{Zlochower}}}, \bibinfo {author} {\bibfnamefont {H.}~\bibnamefont
  {{Nakano}}}, \bibinfo {author} {\bibfnamefont {B.~C.}\ \bibnamefont
  {{Mundim}}}, \bibinfo {author} {\bibfnamefont {M.}~\bibnamefont
  {{Campanelli}}}, \bibinfo {author} {\bibfnamefont {S.}~\bibnamefont
  {{Noble}}},\ and\ \bibinfo {author} {\bibfnamefont {M.}~\bibnamefont
  {{Zilh{\~a}o}}},\ }\href {https://doi.org/10.1103/PhysRevD.93.124072}
  {\bibfield  {journal} {\bibinfo  {journal} {\prd}\ }\textbf {\bibinfo
  {volume} {93}},\ \bibinfo {eid} {124072} (\bibinfo {year}
  {2016})}\BibitemShut {NoStop}%
\bibitem [{\citenamefont {{Ireland}}\ \emph {et~al.}(2019)\citenamefont
  {{Ireland}}, \citenamefont {{Birnholtz}}, \citenamefont {{Nakano}},
  \citenamefont {{West}},\ and\ \citenamefont {{Campanelli}}}]{Ireland2019}%
  \BibitemOpen
  \bibfield  {author} {\bibinfo {author} {\bibfnamefont {B.}~\bibnamefont
  {{Ireland}}}, \bibinfo {author} {\bibfnamefont {O.}~\bibnamefont
  {{Birnholtz}}}, \bibinfo {author} {\bibfnamefont {H.}~\bibnamefont
  {{Nakano}}}, \bibinfo {author} {\bibfnamefont {E.}~\bibnamefont {{West}}},\
  and\ \bibinfo {author} {\bibfnamefont {M.}~\bibnamefont {{Campanelli}}},\
  }\href {https://doi.org/10.1103/PhysRevD.100.024015} {\bibfield  {journal}
  {\bibinfo  {journal} {\prd}\ }\textbf {\bibinfo {volume} {100}},\ \bibinfo
  {eid} {024015} (\bibinfo {year} {2019})},\ \Eprint
  {https://arxiv.org/abs/1904.03443} {arXiv:1904.03443 [gr-qc]} \BibitemShut
  {NoStop}%
\bibitem [{\citenamefont {{Kastaun}}\ and\ \citenamefont
  {{Galeazzi}}(2015)}]{KastaunGaleazzi2015}%
  \BibitemOpen
  \bibfield  {author} {\bibinfo {author} {\bibfnamefont {W.}~\bibnamefont
  {{Kastaun}}}\ and\ \bibinfo {author} {\bibfnamefont {F.}~\bibnamefont
  {{Galeazzi}}},\ }\href {https://doi.org/10.1103/PhysRevD.91.064027}
  {\bibfield  {journal} {\bibinfo  {journal} {\prd}\ }\textbf {\bibinfo
  {volume} {91}},\ \bibinfo {eid} {064027} (\bibinfo {year} {2015})},\ \Eprint
  {https://arxiv.org/abs/1411.7975} {arXiv:1411.7975 [gr-qc]} \BibitemShut
  {NoStop}%
\bibitem [{\citenamefont {{Foucart}}\ \emph {et~al.}(2021)\citenamefont
  {{Foucart}}, \citenamefont {{M{\"o}sta}}, \citenamefont {{Ramirez}},
  \citenamefont {{Wright}}, \citenamefont {{Darbha}},\ and\ \citenamefont
  {{Kasen}}}]{Foucart2021}%
  \BibitemOpen
  \bibfield  {author} {\bibinfo {author} {\bibfnamefont {F.}~\bibnamefont
  {{Foucart}}}, \bibinfo {author} {\bibfnamefont {P.}~\bibnamefont
  {{M{\"o}sta}}}, \bibinfo {author} {\bibfnamefont {T.}~\bibnamefont
  {{Ramirez}}}, \bibinfo {author} {\bibfnamefont {A.~J.}\ \bibnamefont
  {{Wright}}}, \bibinfo {author} {\bibfnamefont {S.}~\bibnamefont {{Darbha}}},\
  and\ \bibinfo {author} {\bibfnamefont {D.}~\bibnamefont {{Kasen}}},\ }\href
  {https://doi.org/10.1103/PhysRevD.104.123010} {\bibfield  {journal} {\bibinfo
   {journal} {\prd}\ }\textbf {\bibinfo {volume} {104}},\ \bibinfo {eid}
  {123010} (\bibinfo {year} {2021})},\ \Eprint
  {https://arxiv.org/abs/2109.00565} {arXiv:2109.00565 [astro-ph.HE]}
  \BibitemShut {NoStop}%
\bibitem [{\citenamefont {Bowen}\ and\ \citenamefont {York}(1980)}]{Bowen1980}%
  \BibitemOpen
  \bibfield  {author} {\bibinfo {author} {\bibfnamefont {J.~M.}\ \bibnamefont
  {Bowen}}\ and\ \bibinfo {author} {\bibfnamefont {J.~W.}\ \bibnamefont
  {York}},\ }\href {https://doi.org/10.1103/PhysRevD.21.2047} {\bibfield
  {journal} {\bibinfo  {journal} {\prd}\ }\textbf {\bibinfo {volume} {21}},\
  \bibinfo {pages} {2047} (\bibinfo {year} {1980})}\BibitemShut {NoStop}%
\bibitem [{\citenamefont {{Brandt}}\ and\ \citenamefont
  {{Br{\"u}gmann}}(1997)}]{Brandt1997}%
  \BibitemOpen
  \bibfield  {author} {\bibinfo {author} {\bibfnamefont {S.}~\bibnamefont
  {{Brandt}}}\ and\ \bibinfo {author} {\bibfnamefont {B.}~\bibnamefont
  {{Br{\"u}gmann}}},\ }\href {https://doi.org/10.1103/PhysRevLett.78.3606}
  {\bibfield  {journal} {\bibinfo  {journal} {\prl}\ }\textbf {\bibinfo
  {volume} {78}},\ \bibinfo {pages} {3606} (\bibinfo {year} {1997})},\ \Eprint
  {https://arxiv.org/abs/gr-qc/9703066} {arXiv:gr-qc/9703066 [gr-qc]}
  \BibitemShut {NoStop}%
\bibitem [{\citenamefont {{Cook}}(2000)}]{Cook2000}%
  \BibitemOpen
  \bibfield  {author} {\bibinfo {author} {\bibfnamefont {G.~B.}\ \bibnamefont
  {{Cook}}},\ }\href {https://doi.org/10.12942/lrr-2000-5} {\bibfield
  {journal} {\bibinfo  {journal} {Living Reviews in Relativity}\ }\textbf
  {\bibinfo {volume} {3}},\ \bibinfo {eid} {5} (\bibinfo {year} {2000})},\
  \Eprint {https://arxiv.org/abs/gr-qc/0007085} {arXiv:gr-qc/0007085 [gr-qc]}
  \BibitemShut {NoStop}%
\bibitem [{\citenamefont {{Ansorg}}\ \emph {et~al.}(2004)\citenamefont
  {{Ansorg}}, \citenamefont {{Br{\"u}gmann}},\ and\ \citenamefont
  {{Tichy}}}]{Ansorg2004}%
  \BibitemOpen
  \bibfield  {author} {\bibinfo {author} {\bibfnamefont {M.}~\bibnamefont
  {{Ansorg}}}, \bibinfo {author} {\bibfnamefont {B.}~\bibnamefont
  {{Br{\"u}gmann}}},\ and\ \bibinfo {author} {\bibfnamefont {W.}~\bibnamefont
  {{Tichy}}},\ }\href {https://doi.org/10.1103/PhysRevD.70.064011} {\bibfield
  {journal} {\bibinfo  {journal} {\prd}\ }\textbf {\bibinfo {volume} {70}},\
  \bibinfo {eid} {064011} (\bibinfo {year} {2004})},\ \Eprint
  {https://arxiv.org/abs/gr-qc/0404056} {arXiv:gr-qc/0404056 [gr-qc]}
  \BibitemShut {NoStop}%
\bibitem [{\citenamefont {{Haas}}(2020)}]{ReadInterpolateDocs}%
  \BibitemOpen
  \bibfield  {author} {\bibinfo {author} {\bibfnamefont {R.}~\bibnamefont
  {{Haas}}},\ }\href@noop {} {\bibinfo {title} {\texttt{ReadInterpolate}
  documentation}},\ \bibinfo {howpublished}
  {\url{https://einsteintoolkit.org/thornguide/EinsteinInitialData/ReadInterpolate/documentation.html}}
  (\bibinfo {year} {2020})\BibitemShut {NoStop}%
\bibitem [{\citenamefont {{Haas}}()}]{ReadInterpolateBitBucket}%
  \BibitemOpen
  \bibfield  {author} {\bibinfo {author} {\bibfnamefont {R.}~\bibnamefont
  {{Haas}}},\ }\href@noop {} {\bibinfo {title} {\texttt{ReadInterpolate}
  bitbucket repository}},\ \bibinfo {howpublished}
  {\url{https://bitbucket.org/einsteintoolkit/einsteininitialdata/src/master/ReadInterpolate}}\BibitemShut
  {NoStop}%
\bibitem [{\citenamefont {{The HDF Group}}()}]{HDF5}%
  \BibitemOpen
  \bibfield  {author} {\bibinfo {author} {\bibnamefont {{The HDF Group}}},\
  }\href@noop {} {\bibinfo {title} {{Hierarchical Data Format, version 5}}},\
  \bibinfo {howpublished} {\url{https://github.com/HDFGroup/hdf5}}\BibitemShut
  {NoStop}%
\bibitem [{\citenamefont {{Prasad}}\ \emph {et~al.}(2022)\citenamefont
  {{Prasad}}, \citenamefont {{Gupta}}, \citenamefont {{Bose}},\ and\
  \citenamefont {{Krishnan}}}]{Prasad2022}%
  \BibitemOpen
  \bibfield  {author} {\bibinfo {author} {\bibfnamefont {V.}~\bibnamefont
  {{Prasad}}}, \bibinfo {author} {\bibfnamefont {A.}~\bibnamefont {{Gupta}}},
  \bibinfo {author} {\bibfnamefont {S.}~\bibnamefont {{Bose}}},\ and\ \bibinfo
  {author} {\bibfnamefont {B.}~\bibnamefont {{Krishnan}}},\ }\href
  {https://doi.org/10.1103/PhysRevD.105.044019} {\bibfield  {journal} {\bibinfo
   {journal} {\prd}\ }\textbf {\bibinfo {volume} {105}},\ \bibinfo {eid}
  {044019} (\bibinfo {year} {2022})},\ \Eprint
  {https://arxiv.org/abs/2106.02595} {arXiv:2106.02595 [gr-qc]} \BibitemShut
  {NoStop}%
\bibitem [{\citenamefont {{Prasad}}(2024)}]{Prasad2024}%
  \BibitemOpen
  \bibfield  {author} {\bibinfo {author} {\bibfnamefont {V.}~\bibnamefont
  {{Prasad}}},\ }\href {https://doi.org/10.1103/PhysRevD.109.044033} {\bibfield
   {journal} {\bibinfo  {journal} {\prd}\ }\textbf {\bibinfo {volume} {109}},\
  \bibinfo {eid} {044033} (\bibinfo {year} {2024})}\BibitemShut {NoStop}%
\bibitem [{\citenamefont {{Brown}}\ \emph {et~al.}(2009)\citenamefont
  {{Brown}}, \citenamefont {{Diener}}, \citenamefont {{Sarbach}}, \citenamefont
  {{Schnetter}},\ and\ \citenamefont {{Tiglio}}}]{Brown2009}%
  \BibitemOpen
  \bibfield  {author} {\bibinfo {author} {\bibfnamefont {D.}~\bibnamefont
  {{Brown}}}, \bibinfo {author} {\bibfnamefont {P.}~\bibnamefont {{Diener}}},
  \bibinfo {author} {\bibfnamefont {O.}~\bibnamefont {{Sarbach}}}, \bibinfo
  {author} {\bibfnamefont {E.}~\bibnamefont {{Schnetter}}},\ and\ \bibinfo
  {author} {\bibfnamefont {M.}~\bibnamefont {{Tiglio}}},\ }\href
  {https://doi.org/10.1103/PhysRevD.79.044023} {\bibfield  {journal} {\bibinfo
  {journal} {\prd}\ }\textbf {\bibinfo {volume} {79}},\ \bibinfo {eid} {044023}
  (\bibinfo {year} {2009})},\ \Eprint {https://arxiv.org/abs/0809.3533}
  {arXiv:0809.3533 [gr-qc]} \BibitemShut {NoStop}%
\bibitem [{\citenamefont {{Schnetter}}(2009)}]{McLachlanDocs}%
  \BibitemOpen
  \bibfield  {author} {\bibinfo {author} {\bibfnamefont {E.}~\bibnamefont
  {{Schnetter}}},\ }\href@noop {} {\bibinfo {title} {\texttt{McLachlan}
  documentation}},\ \bibinfo {howpublished}
  {\url{https://einsteintoolkit.org/arrangementguide/McLachlan/documentation.html}}
  (\bibinfo {year} {2009})\BibitemShut {NoStop}%
\bibitem [{\citenamefont {{Schnetter}}\ \emph {et~al.}()\citenamefont
  {{Schnetter}}, \citenamefont {{Diener}}, \citenamefont {{Tao}},\ and\
  \citenamefont {{Hinder}}}]{McLachlanBitBucket}%
  \BibitemOpen
  \bibfield  {author} {\bibinfo {author} {\bibfnamefont {E.}~\bibnamefont
  {{Schnetter}}}, \bibinfo {author} {\bibfnamefont {P.}~\bibnamefont
  {{Diener}}}, \bibinfo {author} {\bibfnamefont {J.}~\bibnamefont {{Tao}}},\
  and\ \bibinfo {author} {\bibfnamefont {I.}~\bibnamefont {{Hinder}}},\
  }\href@noop {} {\bibinfo {title} {\texttt{McLachlan} bitbucket repository}},\
  \bibinfo {howpublished}
  {\url{https://bitbucket.org/einsteintoolkit/mclachlan/src/master}}\BibitemShut
  {NoStop}%
\bibitem [{\citenamefont {{Thornburg}}(2004)}]{Thornburg2004}%
  \BibitemOpen
  \bibfield  {author} {\bibinfo {author} {\bibfnamefont {J.}~\bibnamefont
  {{Thornburg}}},\ }\href {https://doi.org/10.1088/0264-9381/21/2/026}
  {\bibfield  {journal} {\bibinfo  {journal} {Classical and Quantum Gravity}\
  }\textbf {\bibinfo {volume} {21}},\ \bibinfo {pages} {743} (\bibinfo {year}
  {2004})},\ \Eprint {https://arxiv.org/abs/gr-qc/0306056} {arXiv:gr-qc/0306056
  [gr-qc]} \BibitemShut {NoStop}%
\bibitem [{\citenamefont {{Thornburg}}({\natexlab{a}})}]{AHFinderDirectDocs}%
  \BibitemOpen
  \bibfield  {author} {\bibinfo {author} {\bibfnamefont {J.}~\bibnamefont
  {{Thornburg}}},\ }\href@noop {} {\bibinfo {title} {\texttt{AHFinderDirect}
  documentation}},\ \bibinfo {howpublished}
  {\url{https://einsteintoolkit.org/thornguide/EinsteinAnalysis/AHFinderDirect/documentation.html}}
  ({\natexlab{a}})\BibitemShut {NoStop}%
\bibitem [{\citenamefont
  {{Thornburg}}({\natexlab{b}})}]{AHFinderDirectBitBucket}%
  \BibitemOpen
  \bibfield  {author} {\bibinfo {author} {\bibfnamefont {J.}~\bibnamefont
  {{Thornburg}}},\ }\href@noop {} {\bibinfo {title} {\texttt{AHFinderDirect}
  bitbucket repository}},\ \bibinfo {howpublished}
  {\url{https://bitbucket.org/einsteintoolkit/einsteinanalysis/src/master/AHFinderDirect}}
  ({\natexlab{b}})\BibitemShut {NoStop}%
\bibitem [{\citenamefont {{Dreyer}}\ \emph {et~al.}(2003)\citenamefont
  {{Dreyer}}, \citenamefont {{Krishnan}}, \citenamefont {{Shoemaker}},\ and\
  \citenamefont {{Schnetter}}}]{Dreyer2003}%
  \BibitemOpen
  \bibfield  {author} {\bibinfo {author} {\bibfnamefont {O.}~\bibnamefont
  {{Dreyer}}}, \bibinfo {author} {\bibfnamefont {B.}~\bibnamefont
  {{Krishnan}}}, \bibinfo {author} {\bibfnamefont {D.}~\bibnamefont
  {{Shoemaker}}},\ and\ \bibinfo {author} {\bibfnamefont {E.}~\bibnamefont
  {{Schnetter}}},\ }\href {https://doi.org/10.1103/PhysRevD.67.024018}
  {\bibfield  {journal} {\bibinfo  {journal} {\prd}\ }\textbf {\bibinfo
  {volume} {67}},\ \bibinfo {eid} {024018} (\bibinfo {year} {2003})},\ \Eprint
  {https://arxiv.org/abs/gr-qc/0206008} {arXiv:gr-qc/0206008 [gr-qc]}
  \BibitemShut {NoStop}%
\bibitem [{\citenamefont {{Schnetter}}\ \emph
  {et~al.}(2006{\natexlab{b}})\citenamefont {{Schnetter}}, \citenamefont
  {{Krishnan}},\ and\ \citenamefont {{Beyer}}}]{Schnetter2006QLM}%
  \BibitemOpen
  \bibfield  {author} {\bibinfo {author} {\bibfnamefont {E.}~\bibnamefont
  {{Schnetter}}}, \bibinfo {author} {\bibfnamefont {B.}~\bibnamefont
  {{Krishnan}}},\ and\ \bibinfo {author} {\bibfnamefont {F.}~\bibnamefont
  {{Beyer}}},\ }\href {https://doi.org/10.1103/PhysRevD.74.024028} {\bibfield
  {journal} {\bibinfo  {journal} {\prd}\ }\textbf {\bibinfo {volume} {74}},\
  \bibinfo {eid} {024028} (\bibinfo {year} {2006}{\natexlab{b}})},\ \Eprint
  {https://arxiv.org/abs/gr-qc/0604015} {arXiv:gr-qc/0604015 [gr-qc]}
  \BibitemShut {NoStop}%
\bibitem [{\citenamefont {{Szabados}}(2004)}]{Szabados2004}%
  \BibitemOpen
  \bibfield  {author} {\bibinfo {author} {\bibfnamefont {L.~B.}\ \bibnamefont
  {{Szabados}}},\ }\href {https://doi.org/10.12942/lrr-2004-4} {\bibfield
  {journal} {\bibinfo  {journal} {Living Reviews in Relativity}\ }\textbf
  {\bibinfo {volume} {7}},\ \bibinfo {eid} {4} (\bibinfo {year}
  {2004})}\BibitemShut {NoStop}%
\bibitem [{\citenamefont {{Schnetter}}\ and\ \citenamefont
  {{Haas}}(2017)}]{QuasiLocalMeasuresDocs}%
  \BibitemOpen
  \bibfield  {author} {\bibinfo {author} {\bibfnamefont {E.}~\bibnamefont
  {{Schnetter}}}\ and\ \bibinfo {author} {\bibfnamefont {R.}~\bibnamefont
  {{Haas}}},\ }\href@noop {} {\bibinfo {title} {\texttt{QuasiLocalMeasures}
  documentation}},\ \bibinfo {howpublished}
  {\url{https://einsteintoolkit.org/thornguide/EinsteinAnalysis/QuasiLocalMeasures/documentation.html}}
  (\bibinfo {year} {2017})\BibitemShut {NoStop}%
\bibitem [{\citenamefont {{Schnetter}}\ and\ \citenamefont
  {{Haas}}()}]{QuasiLocalMeasuresBitBucket}%
  \BibitemOpen
  \bibfield  {author} {\bibinfo {author} {\bibfnamefont {E.}~\bibnamefont
  {{Schnetter}}}\ and\ \bibinfo {author} {\bibfnamefont {R.}~\bibnamefont
  {{Haas}}},\ }\href@noop {} {\bibinfo {title} {\texttt{QuasiLocalMeasures}
  bitbucket repository}},\ \bibinfo {howpublished}
  {\url{https://bitbucket.org/einsteintoolkit/einsteinanalysis/src/master/QuasiLocalMeasures}}\BibitemShut
  {NoStop}%
\bibitem [{\citenamefont {{Colella}}\ and\ \citenamefont
  {{Woodward}}(1984)}]{ColellaWoodward1984}%
  \BibitemOpen
  \bibfield  {author} {\bibinfo {author} {\bibfnamefont {P.}~\bibnamefont
  {{Colella}}}\ and\ \bibinfo {author} {\bibfnamefont {P.~R.}\ \bibnamefont
  {{Woodward}}},\ }\href {https://doi.org/10.1016/0021-9991(84)90143-8}
  {\bibfield  {journal} {\bibinfo  {journal} {Journal of Computational
  Physics}\ }\textbf {\bibinfo {volume} {54}},\ \bibinfo {pages} {174}
  (\bibinfo {year} {1984})}\BibitemShut {NoStop}%
\bibitem [{\citenamefont {{Noble}}\ \emph {et~al.}(2006)\citenamefont
  {{Noble}}, \citenamefont {{Gammie}}, \citenamefont {{McKinney}},\ and\
  \citenamefont {{Del Zanna}}}]{Noble2006}%
  \BibitemOpen
  \bibfield  {author} {\bibinfo {author} {\bibfnamefont {S.~C.}\ \bibnamefont
  {{Noble}}}, \bibinfo {author} {\bibfnamefont {C.~F.}\ \bibnamefont
  {{Gammie}}}, \bibinfo {author} {\bibfnamefont {J.~C.}\ \bibnamefont
  {{McKinney}}},\ and\ \bibinfo {author} {\bibfnamefont {L.}~\bibnamefont {{Del
  Zanna}}},\ }\href {https://doi.org/10.1086/500349} {\bibfield  {journal}
  {\bibinfo  {journal} {\apj}\ }\textbf {\bibinfo {volume} {641}},\ \bibinfo
  {pages} {626} (\bibinfo {year} {2006})},\ \Eprint
  {https://arxiv.org/abs/astro-ph/0512420} {arXiv:astro-ph/0512420 [astro-ph]}
  \BibitemShut {NoStop}%
\bibitem [{\citenamefont {{Krauth}}\ \emph {et~al.}(2023)\citenamefont
  {{Krauth}}, \citenamefont {{Davelaar}}, \citenamefont {{Haiman}},
  \citenamefont {{Westernacher-Schneider}}, \citenamefont {{Zrake}},\ and\
  \citenamefont {{MacFadyen}}}]{Krauth2023}%
  \BibitemOpen
  \bibfield  {author} {\bibinfo {author} {\bibfnamefont {L.~M.}\ \bibnamefont
  {{Krauth}}}, \bibinfo {author} {\bibfnamefont {J.}~\bibnamefont
  {{Davelaar}}}, \bibinfo {author} {\bibfnamefont {Z.}~\bibnamefont
  {{Haiman}}}, \bibinfo {author} {\bibfnamefont {J.~R.}\ \bibnamefont
  {{Westernacher-Schneider}}}, \bibinfo {author} {\bibfnamefont
  {J.}~\bibnamefont {{Zrake}}},\ and\ \bibinfo {author} {\bibfnamefont
  {A.}~\bibnamefont {{MacFadyen}}},\ }\href
  {https://doi.org/10.1093/mnras/stad3095} {\bibfield  {journal} {\bibinfo
  {journal} {\mnras}\ }\textbf {\bibinfo {volume} {526}},\ \bibinfo {pages}
  {5441} (\bibinfo {year} {2023})},\ \Eprint {https://arxiv.org/abs/2304.02575}
  {arXiv:2304.02575 [astro-ph.HE]} \BibitemShut {NoStop}%
\bibitem [{\citenamefont {Welch}(1967)}]{Welch1967}%
  \BibitemOpen
  \bibfield  {author} {\bibinfo {author} {\bibfnamefont {P.}~\bibnamefont
  {Welch}},\ }\href {https://doi.org/10.1109/TAU.1967.1161901} {\bibfield
  {journal} {\bibinfo  {journal} {IEEE Transactions on Audio and
  Electroacoustics}\ }\textbf {\bibinfo {volume} {15}},\ \bibinfo {pages} {70}
  (\bibinfo {year} {1967})}\BibitemShut {NoStop}%
\bibitem [{\citenamefont {{McKinney}}\ and\ \citenamefont
  {{Gammie}}(2004)}]{McKinneyGammie2004}%
  \BibitemOpen
  \bibfield  {author} {\bibinfo {author} {\bibfnamefont {J.~C.}\ \bibnamefont
  {{McKinney}}}\ and\ \bibinfo {author} {\bibfnamefont {C.~F.}\ \bibnamefont
  {{Gammie}}},\ }\href {https://doi.org/10.1086/422244} {\bibfield  {journal}
  {\bibinfo  {journal} {\apj}\ }\textbf {\bibinfo {volume} {611}},\ \bibinfo
  {pages} {977} (\bibinfo {year} {2004})},\ \Eprint
  {https://arxiv.org/abs/astro-ph/0404512} {arXiv:astro-ph/0404512 [astro-ph]}
  \BibitemShut {NoStop}%
\bibitem [{\citenamefont {{Hawley}}\ and\ \citenamefont
  {{Krolik}}(2006)}]{HawleyKrolik2006}%
  \BibitemOpen
  \bibfield  {author} {\bibinfo {author} {\bibfnamefont {J.~F.}\ \bibnamefont
  {{Hawley}}}\ and\ \bibinfo {author} {\bibfnamefont {J.~H.}\ \bibnamefont
  {{Krolik}}},\ }\href {https://doi.org/10.1086/500385} {\bibfield  {journal}
  {\bibinfo  {journal} {\apj}\ }\textbf {\bibinfo {volume} {641}},\ \bibinfo
  {pages} {103} (\bibinfo {year} {2006})},\ \Eprint
  {https://arxiv.org/abs/astro-ph/0512227} {arXiv:astro-ph/0512227 [astro-ph]}
  \BibitemShut {NoStop}%
\bibitem [{\citenamefont {{Shiokawa}}\ \emph {et~al.}(2015)\citenamefont
  {{Shiokawa}}, \citenamefont {{Krolik}}, \citenamefont {{Cheng}},
  \citenamefont {{Piran}},\ and\ \citenamefont {{Noble}}}]{Shiokawa2015}%
  \BibitemOpen
  \bibfield  {author} {\bibinfo {author} {\bibfnamefont {H.}~\bibnamefont
  {{Shiokawa}}}, \bibinfo {author} {\bibfnamefont {J.~H.}\ \bibnamefont
  {{Krolik}}}, \bibinfo {author} {\bibfnamefont {R.~M.}\ \bibnamefont
  {{Cheng}}}, \bibinfo {author} {\bibfnamefont {T.}~\bibnamefont {{Piran}}},\
  and\ \bibinfo {author} {\bibfnamefont {S.~C.}\ \bibnamefont {{Noble}}},\
  }\href {https://doi.org/10.1088/0004-637X/804/2/85} {\bibfield  {journal}
  {\bibinfo  {journal} {\apj}\ }\textbf {\bibinfo {volume} {804}},\ \bibinfo
  {eid} {85} (\bibinfo {year} {2015})},\ \Eprint
  {https://arxiv.org/abs/1501.04365} {arXiv:1501.04365 [astro-ph.HE]}
  \BibitemShut {NoStop}%
\bibitem [{\citenamefont {{Piran}}\ \emph {et~al.}(2015)\citenamefont
  {{Piran}}, \citenamefont {{Svirski}}, \citenamefont {{Krolik}}, \citenamefont
  {{Cheng}},\ and\ \citenamefont {{Shiokawa}}}]{Piran2015}%
  \BibitemOpen
  \bibfield  {author} {\bibinfo {author} {\bibfnamefont {T.}~\bibnamefont
  {{Piran}}}, \bibinfo {author} {\bibfnamefont {G.}~\bibnamefont {{Svirski}}},
  \bibinfo {author} {\bibfnamefont {J.}~\bibnamefont {{Krolik}}}, \bibinfo
  {author} {\bibfnamefont {R.~M.}\ \bibnamefont {{Cheng}}},\ and\ \bibinfo
  {author} {\bibfnamefont {H.}~\bibnamefont {{Shiokawa}}},\ }\href
  {https://doi.org/10.1088/0004-637X/806/2/164} {\bibfield  {journal} {\bibinfo
   {journal} {\apj}\ }\textbf {\bibinfo {volume} {806}},\ \bibinfo {eid} {164}
  (\bibinfo {year} {2015})},\ \Eprint {https://arxiv.org/abs/1502.05792}
  {arXiv:1502.05792 [astro-ph.HE]} \BibitemShut {NoStop}%
\bibitem [{\citenamefont {{Ryu}}\ \emph {et~al.}(2023)\citenamefont {{Ryu}},
  \citenamefont {{Krolik}}, \citenamefont {{Piran}}, \citenamefont {{Noble}},\
  and\ \citenamefont {{Avara}}}]{Ryu2023}%
  \BibitemOpen
  \bibfield  {author} {\bibinfo {author} {\bibfnamefont {T.}~\bibnamefont
  {{Ryu}}}, \bibinfo {author} {\bibfnamefont {J.}~\bibnamefont {{Krolik}}},
  \bibinfo {author} {\bibfnamefont {T.}~\bibnamefont {{Piran}}}, \bibinfo
  {author} {\bibfnamefont {S.~C.}\ \bibnamefont {{Noble}}},\ and\ \bibinfo
  {author} {\bibfnamefont {M.}~\bibnamefont {{Avara}}},\ }\href
  {https://doi.org/10.3847/1538-4357/acf5de} {\bibfield  {journal} {\bibinfo
  {journal} {\apj}\ }\textbf {\bibinfo {volume} {957}},\ \bibinfo {eid} {12}
  (\bibinfo {year} {2023})},\ \Eprint {https://arxiv.org/abs/2305.05333}
  {arXiv:2305.05333 [astro-ph.HE]} \BibitemShut {NoStop}%
\bibitem [{\citenamefont {{Steinberg}}\ and\ \citenamefont
  {{Stone}}(2024)}]{SteinbergStone2024}%
  \BibitemOpen
  \bibfield  {author} {\bibinfo {author} {\bibfnamefont {E.}~\bibnamefont
  {{Steinberg}}}\ and\ \bibinfo {author} {\bibfnamefont {N.~C.}\ \bibnamefont
  {{Stone}}},\ }\href {https://doi.org/10.1038/s41586-023-06875-y} {\bibfield
  {journal} {\bibinfo  {journal} {\nat}\ }\textbf {\bibinfo {volume} {625}},\
  \bibinfo {pages} {463} (\bibinfo {year} {2024})},\ \Eprint
  {https://arxiv.org/abs/2206.10641} {arXiv:2206.10641 [astro-ph.HE]}
  \BibitemShut {NoStop}%
\bibitem [{\citenamefont {{Price}}\ \emph {et~al.}(2024)\citenamefont
  {{Price}}, \citenamefont {{Liptai}}, \citenamefont {{Mandel}}, \citenamefont
  {{Shepherd}}, \citenamefont {{Lodato}},\ and\ \citenamefont
  {{Levin}}}]{Price2024}%
  \BibitemOpen
  \bibfield  {author} {\bibinfo {author} {\bibfnamefont {D.~J.}\ \bibnamefont
  {{Price}}}, \bibinfo {author} {\bibfnamefont {D.}~\bibnamefont {{Liptai}}},
  \bibinfo {author} {\bibfnamefont {I.}~\bibnamefont {{Mandel}}}, \bibinfo
  {author} {\bibfnamefont {J.}~\bibnamefont {{Shepherd}}}, \bibinfo {author}
  {\bibfnamefont {G.}~\bibnamefont {{Lodato}}},\ and\ \bibinfo {author}
  {\bibfnamefont {Y.}~\bibnamefont {{Levin}}},\ }\href
  {https://doi.org/10.48550/arXiv.2404.09381} {\bibfield  {journal} {\bibinfo
  {journal} {arXiv e-prints}\ ,\ \bibinfo {eid} {arXiv:2404.09381}} (\bibinfo
  {year} {2024})},\ \Eprint {https://arxiv.org/abs/2404.09381}
  {arXiv:2404.09381 [astro-ph.HE]} \BibitemShut {NoStop}%
\bibitem [{\citenamefont {{Krolik}}(2010)}]{Krolik2010}%
  \BibitemOpen
  \bibfield  {author} {\bibinfo {author} {\bibfnamefont {J.~H.}\ \bibnamefont
  {{Krolik}}},\ }\href {https://doi.org/10.1088/0004-637X/709/2/774} {\bibfield
   {journal} {\bibinfo  {journal} {\apj}\ }\textbf {\bibinfo {volume} {709}},\
  \bibinfo {pages} {774} (\bibinfo {year} {2010})},\ \Eprint
  {https://arxiv.org/abs/0911.5711} {arXiv:0911.5711 [astro-ph.CO]}
  \BibitemShut {NoStop}%
\bibitem [{\citenamefont {{Schnittman}}\ \emph {et~al.}(2013)\citenamefont
  {{Schnittman}}, \citenamefont {{Krolik}},\ and\ \citenamefont
  {{Noble}}}]{Schnittman2013}%
  \BibitemOpen
  \bibfield  {author} {\bibinfo {author} {\bibfnamefont {J.~D.}\ \bibnamefont
  {{Schnittman}}}, \bibinfo {author} {\bibfnamefont {J.~H.}\ \bibnamefont
  {{Krolik}}},\ and\ \bibinfo {author} {\bibfnamefont {S.~C.}\ \bibnamefont
  {{Noble}}},\ }\href {https://doi.org/10.1088/0004-637X/769/2/156} {\bibfield
  {journal} {\bibinfo  {journal} {\apj}\ }\textbf {\bibinfo {volume} {769}},\
  \bibinfo {eid} {156} (\bibinfo {year} {2013})},\ \Eprint
  {https://arxiv.org/abs/1207.2693} {arXiv:1207.2693 [astro-ph.HE]}
  \BibitemShut {NoStop}%
\bibitem [{\citenamefont {{Milosavljevi{\'c}}}\ and\ \citenamefont
  {{Phinney}}(2005)}]{MilosavljevicPhinney2005}%
  \BibitemOpen
  \bibfield  {author} {\bibinfo {author} {\bibfnamefont {M.}~\bibnamefont
  {{Milosavljevi{\'c}}}}\ and\ \bibinfo {author} {\bibfnamefont {E.~S.}\
  \bibnamefont {{Phinney}}},\ }\href {https://doi.org/10.1086/429618}
  {\bibfield  {journal} {\bibinfo  {journal} {\apjl}\ }\textbf {\bibinfo
  {volume} {622}},\ \bibinfo {pages} {L93} (\bibinfo {year} {2005})},\ \Eprint
  {https://arxiv.org/abs/astro-ph/0410343} {arXiv:astro-ph/0410343 [astro-ph]}
  \BibitemShut {NoStop}%
\bibitem [{\citenamefont {{Guti{\'e}rrez}}\ \emph
  {et~al.}(2024{\natexlab{b}})\citenamefont {{Guti{\'e}rrez}}, \citenamefont
  {{Combi}}, \citenamefont {{Romero}},\ and\ \citenamefont
  {{Campanelli}}}]{Gutierrez2024Radiation}%
  \BibitemOpen
  \bibfield  {author} {\bibinfo {author} {\bibfnamefont {E.~M.}\ \bibnamefont
  {{Guti{\'e}rrez}}}, \bibinfo {author} {\bibfnamefont {L.}~\bibnamefont
  {{Combi}}}, \bibinfo {author} {\bibfnamefont {G.~E.}\ \bibnamefont
  {{Romero}}},\ and\ \bibinfo {author} {\bibfnamefont {M.}~\bibnamefont
  {{Campanelli}}},\ }\href {https://doi.org/10.1093/mnras/stae1473} {\bibfield
  {journal} {\bibinfo  {journal} {\mnras}\ }\textbf {\bibinfo {volume} {532}},\
  \bibinfo {pages} {506} (\bibinfo {year} {2024}{\natexlab{b}})},\ \Eprint
  {https://arxiv.org/abs/2301.04280} {arXiv:2301.04280 [astro-ph.HE]}
  \BibitemShut {NoStop}%
\bibitem [{\citenamefont {{D'Orazio}}\ \emph {et~al.}(2013)\citenamefont
  {{D'Orazio}}, \citenamefont {{Haiman}},\ and\ \citenamefont
  {{MacFadyen}}}]{Dorazio2013}%
  \BibitemOpen
  \bibfield  {author} {\bibinfo {author} {\bibfnamefont {D.~J.}\ \bibnamefont
  {{D'Orazio}}}, \bibinfo {author} {\bibfnamefont {Z.}~\bibnamefont
  {{Haiman}}},\ and\ \bibinfo {author} {\bibfnamefont {A.}~\bibnamefont
  {{MacFadyen}}},\ }\href {https://doi.org/10.1093/mnras/stt1787} {\bibfield
  {journal} {\bibinfo  {journal} {\mnras}\ }\textbf {\bibinfo {volume} {436}},\
  \bibinfo {pages} {2997} (\bibinfo {year} {2013})},\ \Eprint
  {https://arxiv.org/abs/1210.0536} {arXiv:1210.0536 [astro-ph.GA]}
  \BibitemShut {NoStop}%
\bibitem [{\citenamefont {{Armitage}}\ and\ \citenamefont
  {{Natarajan}}(2005)}]{Armitage2005}%
  \BibitemOpen
  \bibfield  {author} {\bibinfo {author} {\bibfnamefont {P.~J.}\ \bibnamefont
  {{Armitage}}}\ and\ \bibinfo {author} {\bibfnamefont {P.}~\bibnamefont
  {{Natarajan}}},\ }\href {https://doi.org/10.1086/497108} {\bibfield
  {journal} {\bibinfo  {journal} {\apj}\ }\textbf {\bibinfo {volume} {634}},\
  \bibinfo {pages} {921} (\bibinfo {year} {2005})},\ \Eprint
  {https://arxiv.org/abs/astro-ph/0508493} {arXiv:astro-ph/0508493 [astro-ph]}
  \BibitemShut {NoStop}%
\bibitem [{\citenamefont {{Miller}}\ \emph {et~al.}(2005)\citenamefont
  {{Miller}}, \citenamefont {{Freitag}}, \citenamefont {{Hamilton}},\ and\
  \citenamefont {{Lauburg}}}]{ColemanMiller2005}%
  \BibitemOpen
  \bibfield  {author} {\bibinfo {author} {\bibfnamefont {M.~C.}\ \bibnamefont
  {{Miller}}}, \bibinfo {author} {\bibfnamefont {M.}~\bibnamefont {{Freitag}}},
  \bibinfo {author} {\bibfnamefont {D.~P.}\ \bibnamefont {{Hamilton}}},\ and\
  \bibinfo {author} {\bibfnamefont {V.~M.}\ \bibnamefont {{Lauburg}}},\ }\href
  {https://doi.org/10.1086/497335} {\bibfield  {journal} {\bibinfo  {journal}
  {\apjl}\ }\textbf {\bibinfo {volume} {631}},\ \bibinfo {pages} {L117}
  (\bibinfo {year} {2005})},\ \Eprint {https://arxiv.org/abs/astro-ph/0507133}
  {arXiv:astro-ph/0507133 [astro-ph]} \BibitemShut {NoStop}%
\bibitem [{\citenamefont {{Sesana}}(2013)}]{Sesana2013}%
  \BibitemOpen
  \bibfield  {author} {\bibinfo {author} {\bibfnamefont {A.}~\bibnamefont
  {{Sesana}}},\ }\href {https://doi.org/10.1088/0264-9381/30/22/224014}
  {\bibfield  {journal} {\bibinfo  {journal} {Classical and Quantum Gravity}\
  }\textbf {\bibinfo {volume} {30}},\ \bibinfo {eid} {224014} (\bibinfo {year}
  {2013})},\ \Eprint {https://arxiv.org/abs/1307.2600} {arXiv:1307.2600
  [astro-ph.CO]} \BibitemShut {NoStop}%
\bibitem [{\citenamefont {{Gualandris}}\ \emph {et~al.}(2022)\citenamefont
  {{Gualandris}}, \citenamefont {{Khan}}, \citenamefont {{Bortolas}},
  \citenamefont {{Bonetti}}, \citenamefont {{Sesana}}, \citenamefont
  {{Berczik}},\ and\ \citenamefont {{Holley-Bockelmann}}}]{Gualandris2022}%
  \BibitemOpen
  \bibfield  {author} {\bibinfo {author} {\bibfnamefont {A.}~\bibnamefont
  {{Gualandris}}}, \bibinfo {author} {\bibfnamefont {F.~M.}\ \bibnamefont
  {{Khan}}}, \bibinfo {author} {\bibfnamefont {E.}~\bibnamefont {{Bortolas}}},
  \bibinfo {author} {\bibfnamefont {M.}~\bibnamefont {{Bonetti}}}, \bibinfo
  {author} {\bibfnamefont {A.}~\bibnamefont {{Sesana}}}, \bibinfo {author}
  {\bibfnamefont {P.}~\bibnamefont {{Berczik}}},\ and\ \bibinfo {author}
  {\bibfnamefont {K.}~\bibnamefont {{Holley-Bockelmann}}},\ }\href
  {https://doi.org/10.1093/mnras/stac241} {\bibfield  {journal} {\bibinfo
  {journal} {\mnras}\ }\textbf {\bibinfo {volume} {511}},\ \bibinfo {pages}
  {4753} (\bibinfo {year} {2022})},\ \Eprint {https://arxiv.org/abs/2201.08646}
  {arXiv:2201.08646 [astro-ph.GA]} \BibitemShut {NoStop}%
\bibitem [{\citenamefont {{Bogdanovi{\'c}}}\ \emph {et~al.}(2007)\citenamefont
  {{Bogdanovi{\'c}}}, \citenamefont {{Reynolds}},\ and\ \citenamefont
  {{Miller}}}]{Bogdanovic2007}%
  \BibitemOpen
  \bibfield  {author} {\bibinfo {author} {\bibfnamefont {T.}~\bibnamefont
  {{Bogdanovi{\'c}}}}, \bibinfo {author} {\bibfnamefont {C.~S.}\ \bibnamefont
  {{Reynolds}}},\ and\ \bibinfo {author} {\bibfnamefont {M.~C.}\ \bibnamefont
  {{Miller}}},\ }\href {https://doi.org/10.1086/518769} {\bibfield  {journal}
  {\bibinfo  {journal} {\apjl}\ }\textbf {\bibinfo {volume} {661}},\ \bibinfo
  {pages} {L147} (\bibinfo {year} {2007})},\ \Eprint
  {https://arxiv.org/abs/astro-ph/0703054} {arXiv:astro-ph/0703054 [astro-ph]}
  \BibitemShut {NoStop}%
\bibitem [{\citenamefont {{Miller}}\ and\ \citenamefont
  {{Krolik}}(2013)}]{MillerKrolik2013}%
  \BibitemOpen
  \bibfield  {author} {\bibinfo {author} {\bibfnamefont {M.~C.}\ \bibnamefont
  {{Miller}}}\ and\ \bibinfo {author} {\bibfnamefont {J.~H.}\ \bibnamefont
  {{Krolik}}},\ }\href {https://doi.org/10.1088/0004-637X/774/1/43} {\bibfield
  {journal} {\bibinfo  {journal} {\apj}\ }\textbf {\bibinfo {volume} {774}},\
  \bibinfo {eid} {43} (\bibinfo {year} {2013})},\ \Eprint
  {https://arxiv.org/abs/1307.6569} {arXiv:1307.6569 [astro-ph.HE]}
  \BibitemShut {NoStop}%
\bibitem [{\citenamefont {{Sorathia}}\ \emph {et~al.}(2013)\citenamefont
  {{Sorathia}}, \citenamefont {{Krolik}},\ and\ \citenamefont
  {{Hawley}}}]{Sorathia2013}%
  \BibitemOpen
  \bibfield  {author} {\bibinfo {author} {\bibfnamefont {K.~A.}\ \bibnamefont
  {{Sorathia}}}, \bibinfo {author} {\bibfnamefont {J.~H.}\ \bibnamefont
  {{Krolik}}},\ and\ \bibinfo {author} {\bibfnamefont {J.~F.}\ \bibnamefont
  {{Hawley}}},\ }\href {https://doi.org/10.1088/0004-637X/777/1/21} {\bibfield
  {journal} {\bibinfo  {journal} {\apj}\ }\textbf {\bibinfo {volume} {777}},\
  \bibinfo {eid} {21} (\bibinfo {year} {2013})},\ \Eprint
  {https://arxiv.org/abs/1309.0290} {arXiv:1309.0290 [astro-ph.HE]}
  \BibitemShut {NoStop}%
\bibitem [{\citenamefont {{Campanelli}}\ \emph
  {et~al.}(2006{\natexlab{b}})\citenamefont {{Campanelli}}, \citenamefont
  {{Lousto}},\ and\ \citenamefont {{Zlochower}}}]{Campanelli2006HangUp}%
  \BibitemOpen
  \bibfield  {author} {\bibinfo {author} {\bibfnamefont {M.}~\bibnamefont
  {{Campanelli}}}, \bibinfo {author} {\bibfnamefont {C.~O.}\ \bibnamefont
  {{Lousto}}},\ and\ \bibinfo {author} {\bibfnamefont {Y.}~\bibnamefont
  {{Zlochower}}},\ }\href {https://doi.org/10.1103/PhysRevD.74.041501}
  {\bibfield  {journal} {\bibinfo  {journal} {\prd}\ }\textbf {\bibinfo
  {volume} {74}},\ \bibinfo {eid} {041501} (\bibinfo {year}
  {2006}{\natexlab{b}})},\ \Eprint {https://arxiv.org/abs/gr-qc/0604012}
  {arXiv:gr-qc/0604012 [gr-qc]} \BibitemShut {NoStop}%
\bibitem [{\citenamefont {{Campanelli}}\ \emph {et~al.}(2007)\citenamefont
  {{Campanelli}}, \citenamefont {{Lousto}}, \citenamefont {{Zlochower}},\ and\
  \citenamefont {{Merritt}}}]{Campanelli2007}%
  \BibitemOpen
  \bibfield  {author} {\bibinfo {author} {\bibfnamefont {M.}~\bibnamefont
  {{Campanelli}}}, \bibinfo {author} {\bibfnamefont {C.}~\bibnamefont
  {{Lousto}}}, \bibinfo {author} {\bibfnamefont {Y.}~\bibnamefont
  {{Zlochower}}},\ and\ \bibinfo {author} {\bibfnamefont {D.}~\bibnamefont
  {{Merritt}}},\ }\href {https://doi.org/10.1086/516712} {\bibfield  {journal}
  {\bibinfo  {journal} {\apjl}\ }\textbf {\bibinfo {volume} {659}},\ \bibinfo
  {pages} {L5} (\bibinfo {year} {2007})},\ \Eprint
  {https://arxiv.org/abs/gr-qc/0701164} {arXiv:gr-qc/0701164 [gr-qc]}
  \BibitemShut {NoStop}%
\bibitem [{\citenamefont {{Keppel}}\ \emph {et~al.}(2009)\citenamefont
  {{Keppel}}, \citenamefont {{Nichols}}, \citenamefont {{Chen}},\ and\
  \citenamefont {{Thorne}}}]{Keppel2009}%
  \BibitemOpen
  \bibfield  {author} {\bibinfo {author} {\bibfnamefont {D.}~\bibnamefont
  {{Keppel}}}, \bibinfo {author} {\bibfnamefont {D.~A.}\ \bibnamefont
  {{Nichols}}}, \bibinfo {author} {\bibfnamefont {Y.}~\bibnamefont {{Chen}}},\
  and\ \bibinfo {author} {\bibfnamefont {K.~S.}\ \bibnamefont {{Thorne}}},\
  }\href {https://doi.org/10.1103/PhysRevD.80.124015} {\bibfield  {journal}
  {\bibinfo  {journal} {\prd}\ }\textbf {\bibinfo {volume} {80}},\ \bibinfo
  {eid} {124015} (\bibinfo {year} {2009})},\ \Eprint
  {https://arxiv.org/abs/0902.4077} {arXiv:0902.4077 [gr-qc]} \BibitemShut
  {NoStop}%
\bibitem [{\citenamefont {{Hemberger}}\ \emph {et~al.}(2013)\citenamefont
  {{Hemberger}}, \citenamefont {{Lovelace}}, \citenamefont {{Loredo}},
  \citenamefont {{Kidder}}, \citenamefont {{Scheel}}, \citenamefont
  {{Szil{\'a}gyi}}, \citenamefont {{Taylor}},\ and\ \citenamefont
  {{Teukolsky}}}]{Hemberger2013}%
  \BibitemOpen
  \bibfield  {author} {\bibinfo {author} {\bibfnamefont {D.~A.}\ \bibnamefont
  {{Hemberger}}}, \bibinfo {author} {\bibfnamefont {G.}~\bibnamefont
  {{Lovelace}}}, \bibinfo {author} {\bibfnamefont {T.~J.}\ \bibnamefont
  {{Loredo}}}, \bibinfo {author} {\bibfnamefont {L.~E.}\ \bibnamefont
  {{Kidder}}}, \bibinfo {author} {\bibfnamefont {M.~A.}\ \bibnamefont
  {{Scheel}}}, \bibinfo {author} {\bibfnamefont {B.}~\bibnamefont
  {{Szil{\'a}gyi}}}, \bibinfo {author} {\bibfnamefont {N.~W.}\ \bibnamefont
  {{Taylor}}},\ and\ \bibinfo {author} {\bibfnamefont {S.~A.}\ \bibnamefont
  {{Teukolsky}}},\ }\href {https://doi.org/10.1103/PhysRevD.88.064014}
  {\bibfield  {journal} {\bibinfo  {journal} {\prd}\ }\textbf {\bibinfo
  {volume} {88}},\ \bibinfo {eid} {064014} (\bibinfo {year} {2013})},\ \Eprint
  {https://arxiv.org/abs/1305.5991} {arXiv:1305.5991 [gr-qc]} \BibitemShut
  {NoStop}%
\bibitem [{\citenamefont {{Ressler}}\ \emph {et~al.}(2024)\citenamefont
  {{Ressler}}, \citenamefont {{Combi}}, \citenamefont {{Ripperda}},\ and\
  \citenamefont {{Most}}}]{Ressler2024}%
  \BibitemOpen
  \bibfield  {author} {\bibinfo {author} {\bibfnamefont {S.~M.}\ \bibnamefont
  {{Ressler}}}, \bibinfo {author} {\bibfnamefont {L.}~\bibnamefont {{Combi}}},
  \bibinfo {author} {\bibfnamefont {B.}~\bibnamefont {{Ripperda}}},\ and\
  \bibinfo {author} {\bibfnamefont {E.~R.}\ \bibnamefont {{Most}}},\ }\href
  {https://doi.org/10.48550/arXiv.2410.10944} {\bibfield  {journal} {\bibinfo
  {journal} {arXiv e-prints}\ ,\ \bibinfo {eid} {arXiv:2410.10944}} (\bibinfo
  {year} {2024})},\ \Eprint {https://arxiv.org/abs/2410.10944}
  {arXiv:2410.10944 [astro-ph.HE]} \BibitemShut {NoStop}%
\bibitem [{\citenamefont {{Lousto}}\ and\ \citenamefont
  {{Zlochower}}(2011)}]{LoustoZlochower2011}%
  \BibitemOpen
  \bibfield  {author} {\bibinfo {author} {\bibfnamefont {C.~O.}\ \bibnamefont
  {{Lousto}}}\ and\ \bibinfo {author} {\bibfnamefont {Y.}~\bibnamefont
  {{Zlochower}}},\ }\href {https://doi.org/10.1103/PhysRevLett.107.231102}
  {\bibfield  {journal} {\bibinfo  {journal} {\prl}\ }\textbf {\bibinfo
  {volume} {107}},\ \bibinfo {eid} {231102} (\bibinfo {year} {2011})},\ \Eprint
  {https://arxiv.org/abs/1108.2009} {arXiv:1108.2009 [gr-qc]} \BibitemShut
  {NoStop}%
\bibitem [{\citenamefont {{Lousto}}\ and\ \citenamefont
  {{Healy}}(2015)}]{Lousto2015}%
  \BibitemOpen
  \bibfield  {author} {\bibinfo {author} {\bibfnamefont {C.~O.}\ \bibnamefont
  {{Lousto}}}\ and\ \bibinfo {author} {\bibfnamefont {J.}~\bibnamefont
  {{Healy}}},\ }\href {https://doi.org/10.1103/PhysRevLett.114.141101}
  {\bibfield  {journal} {\bibinfo  {journal} {\prl}\ }\textbf {\bibinfo
  {volume} {114}},\ \bibinfo {eid} {141101} (\bibinfo {year} {2015})},\ \Eprint
  {https://arxiv.org/abs/1410.3830} {arXiv:1410.3830 [gr-qc]} \BibitemShut
  {NoStop}%
\bibitem [{\citenamefont {{Lousto}}\ \emph {et~al.}(2016)\citenamefont
  {{Lousto}}, \citenamefont {{Healy}},\ and\ \citenamefont
  {{Nakano}}}]{Lousto2016}%
  \BibitemOpen
  \bibfield  {author} {\bibinfo {author} {\bibfnamefont {C.~O.}\ \bibnamefont
  {{Lousto}}}, \bibinfo {author} {\bibfnamefont {J.}~\bibnamefont {{Healy}}},\
  and\ \bibinfo {author} {\bibfnamefont {H.}~\bibnamefont {{Nakano}}},\ }\href
  {https://doi.org/10.1103/PhysRevD.93.044031} {\bibfield  {journal} {\bibinfo
  {journal} {\prd}\ }\textbf {\bibinfo {volume} {93}},\ \bibinfo {eid} {044031}
  (\bibinfo {year} {2016})},\ \Eprint {https://arxiv.org/abs/1506.04768}
  {arXiv:1506.04768 [gr-qc]} \BibitemShut {NoStop}%
\bibitem [{\citenamefont {{Kesden}}\ \emph {et~al.}(2015)\citenamefont
  {{Kesden}}, \citenamefont {{Gerosa}}, \citenamefont {{O'Shaughnessy}},
  \citenamefont {{Berti}},\ and\ \citenamefont {{Sperhake}}}]{Kesden2015}%
  \BibitemOpen
  \bibfield  {author} {\bibinfo {author} {\bibfnamefont {M.}~\bibnamefont
  {{Kesden}}}, \bibinfo {author} {\bibfnamefont {D.}~\bibnamefont {{Gerosa}}},
  \bibinfo {author} {\bibfnamefont {R.}~\bibnamefont {{O'Shaughnessy}}},
  \bibinfo {author} {\bibfnamefont {E.}~\bibnamefont {{Berti}}},\ and\ \bibinfo
  {author} {\bibfnamefont {U.}~\bibnamefont {{Sperhake}}},\ }\href
  {https://doi.org/10.1103/PhysRevLett.114.081103} {\bibfield  {journal}
  {\bibinfo  {journal} {\prl}\ }\textbf {\bibinfo {volume} {114}},\ \bibinfo
  {eid} {081103} (\bibinfo {year} {2015})},\ \Eprint
  {https://arxiv.org/abs/1411.0674} {arXiv:1411.0674 [gr-qc]} \BibitemShut
  {NoStop}%
\bibitem [{\citenamefont {{Lousto}}\ and\ \citenamefont
  {{Healy}}(2016)}]{Lousto2016FlipFlop}%
  \BibitemOpen
  \bibfield  {author} {\bibinfo {author} {\bibfnamefont {C.~O.}\ \bibnamefont
  {{Lousto}}}\ and\ \bibinfo {author} {\bibfnamefont {J.}~\bibnamefont
  {{Healy}}},\ }\href {https://doi.org/10.1103/PhysRevD.93.124074} {\bibfield
  {journal} {\bibinfo  {journal} {\prd}\ }\textbf {\bibinfo {volume} {93}},\
  \bibinfo {eid} {124074} (\bibinfo {year} {2016})},\ \Eprint
  {https://arxiv.org/abs/1601.05086} {arXiv:1601.05086 [gr-qc]} \BibitemShut
  {NoStop}%
\bibitem [{\citenamefont {{Healy}}\ and\ \citenamefont
  {{Lousto}}(2018)}]{Healy2018}%
  \BibitemOpen
  \bibfield  {author} {\bibinfo {author} {\bibfnamefont {J.}~\bibnamefont
  {{Healy}}}\ and\ \bibinfo {author} {\bibfnamefont {C.~O.}\ \bibnamefont
  {{Lousto}}},\ }\href {https://doi.org/10.1103/PhysRevD.97.084002} {\bibfield
  {journal} {\bibinfo  {journal} {\prd}\ }\textbf {\bibinfo {volume} {97}},\
  \bibinfo {eid} {084002} (\bibinfo {year} {2018})},\ \Eprint
  {https://arxiv.org/abs/1801.08162} {arXiv:1801.08162 [gr-qc]} \BibitemShut
  {NoStop}%
\bibitem [{\citenamefont {{K. Akiyama \textit{et al.} (Event Horizon Telescope
  Collaboration}}(2019)}]{EHT2019}%
  \BibitemOpen
  \bibfield  {author} {\bibinfo {author} {\bibnamefont {{K. Akiyama \textit{et
  al.} (Event Horizon Telescope Collaboration}}},\ }\href
  {https://doi.org/10.3847/2041-8213/ab0ec7} {\bibfield  {journal} {\bibinfo
  {journal} {\apjl}\ }\textbf {\bibinfo {volume} {875}},\ \bibinfo {eid} {L1}
  (\bibinfo {year} {2019})},\ \Eprint {https://arxiv.org/abs/1906.11238}
  {arXiv:1906.11238 [astro-ph.GA]} \BibitemShut {NoStop}%
\bibitem [{\citenamefont {{Narayan}}\ \emph {et~al.}(2003)\citenamefont
  {{Narayan}}, \citenamefont {{Igumenshchev}},\ and\ \citenamefont
  {{Abramowicz}}}]{Narayan2003}%
  \BibitemOpen
  \bibfield  {author} {\bibinfo {author} {\bibfnamefont {R.}~\bibnamefont
  {{Narayan}}}, \bibinfo {author} {\bibfnamefont {I.~V.}\ \bibnamefont
  {{Igumenshchev}}},\ and\ \bibinfo {author} {\bibfnamefont {M.~A.}\
  \bibnamefont {{Abramowicz}}},\ }\href {https://doi.org/10.1093/pasj/55.6.L69}
  {\bibfield  {journal} {\bibinfo  {journal} {\pasj}\ }\textbf {\bibinfo
  {volume} {55}},\ \bibinfo {pages} {L69} (\bibinfo {year} {2003})},\ \Eprint
  {https://arxiv.org/abs/astro-ph/0305029} {arXiv:astro-ph/0305029 [astro-ph]}
  \BibitemShut {NoStop}%
\bibitem [{\citenamefont {{Tchekhovskoy}}\ \emph {et~al.}(2011)\citenamefont
  {{Tchekhovskoy}}, \citenamefont {{Narayan}},\ and\ \citenamefont
  {{McKinney}}}]{Tchekhovskoy2011}%
  \BibitemOpen
  \bibfield  {author} {\bibinfo {author} {\bibfnamefont {A.}~\bibnamefont
  {{Tchekhovskoy}}}, \bibinfo {author} {\bibfnamefont {R.}~\bibnamefont
  {{Narayan}}},\ and\ \bibinfo {author} {\bibfnamefont {J.~C.}\ \bibnamefont
  {{McKinney}}},\ }\href {https://doi.org/10.1111/j.1745-3933.2011.01147.x}
  {\bibfield  {journal} {\bibinfo  {journal} {\mnras}\ }\textbf {\bibinfo
  {volume} {418}},\ \bibinfo {pages} {L79} (\bibinfo {year} {2011})},\ \Eprint
  {https://arxiv.org/abs/1108.0412} {arXiv:1108.0412 [astro-ph.HE]}
  \BibitemShut {NoStop}%
\bibitem [{\citenamefont {{Ripperda}}\ \emph {et~al.}(2022)\citenamefont
  {{Ripperda}}, \citenamefont {{Liska}}, \citenamefont {{Chatterjee}},
  \citenamefont {{Musoke}}, \citenamefont {{Philippov}}, \citenamefont
  {{Markoff}}, \citenamefont {{Tchekhovskoy}},\ and\ \citenamefont
  {{Younsi}}}]{Ripperda2022}%
  \BibitemOpen
  \bibfield  {author} {\bibinfo {author} {\bibfnamefont {B.}~\bibnamefont
  {{Ripperda}}}, \bibinfo {author} {\bibfnamefont {M.}~\bibnamefont {{Liska}}},
  \bibinfo {author} {\bibfnamefont {K.}~\bibnamefont {{Chatterjee}}}, \bibinfo
  {author} {\bibfnamefont {G.}~\bibnamefont {{Musoke}}}, \bibinfo {author}
  {\bibfnamefont {A.~A.}\ \bibnamefont {{Philippov}}}, \bibinfo {author}
  {\bibfnamefont {S.~B.}\ \bibnamefont {{Markoff}}}, \bibinfo {author}
  {\bibfnamefont {A.}~\bibnamefont {{Tchekhovskoy}}},\ and\ \bibinfo {author}
  {\bibfnamefont {Z.}~\bibnamefont {{Younsi}}},\ }\href
  {https://doi.org/10.3847/2041-8213/ac46a1} {\bibfield  {journal} {\bibinfo
  {journal} {\apjl}\ }\textbf {\bibinfo {volume} {924}},\ \bibinfo {eid} {L32}
  (\bibinfo {year} {2022})},\ \Eprint {https://arxiv.org/abs/2109.15115}
  {arXiv:2109.15115 [astro-ph.HE]} \BibitemShut {NoStop}%
\bibitem [{\citenamefont {{Most}}\ and\ \citenamefont
  {{Wang}}(2024)}]{Most2024}%
  \BibitemOpen
  \bibfield  {author} {\bibinfo {author} {\bibfnamefont {E.~R.}\ \bibnamefont
  {{Most}}}\ and\ \bibinfo {author} {\bibfnamefont {H.-Y.}\ \bibnamefont
  {{Wang}}},\ }\href {https://doi.org/10.3847/2041-8213/ad7713} {\bibfield
  {journal} {\bibinfo  {journal} {\apjl}\ }\textbf {\bibinfo {volume} {973}},\
  \bibinfo {eid} {L19} (\bibinfo {year} {2024})},\ \Eprint
  {https://arxiv.org/abs/2408.00757} {arXiv:2408.00757 [astro-ph.HE]}
  \BibitemShut {NoStop}%
\bibitem [{\citenamefont {{Stone}}\ \emph {et~al.}(2024)\citenamefont
  {{Stone}}, \citenamefont {{Mullen}}, \citenamefont {{Fielding}},
  \citenamefont {{Grete}}, \citenamefont {{Guo}}, \citenamefont {{Kempski}},
  \citenamefont {{Most}}, \citenamefont {{White}},\ and\ \citenamefont
  {{Wong}}}]{Stone2024}%
  \BibitemOpen
  \bibfield  {author} {\bibinfo {author} {\bibfnamefont {J.~M.}\ \bibnamefont
  {{Stone}}}, \bibinfo {author} {\bibfnamefont {P.~D.}\ \bibnamefont
  {{Mullen}}}, \bibinfo {author} {\bibfnamefont {D.}~\bibnamefont
  {{Fielding}}}, \bibinfo {author} {\bibfnamefont {P.}~\bibnamefont {{Grete}}},
  \bibinfo {author} {\bibfnamefont {M.}~\bibnamefont {{Guo}}}, \bibinfo
  {author} {\bibfnamefont {P.}~\bibnamefont {{Kempski}}}, \bibinfo {author}
  {\bibfnamefont {E.~R.}\ \bibnamefont {{Most}}}, \bibinfo {author}
  {\bibfnamefont {C.~J.}\ \bibnamefont {{White}}},\ and\ \bibinfo {author}
  {\bibfnamefont {G.~N.}\ \bibnamefont {{Wong}}},\ }\href
  {https://doi.org/10.48550/arXiv.2409.16053} {\bibfield  {journal} {\bibinfo
  {journal} {arXiv e-prints}\ ,\ \bibinfo {eid} {arXiv:2409.16053}} (\bibinfo
  {year} {2024})},\ \Eprint {https://arxiv.org/abs/2409.16053}
  {arXiv:2409.16053 [astro-ph.IM]} \BibitemShut {NoStop}%
\bibitem [{\citenamefont {{Tiwari}}\ \emph {et~al.}(2025)\citenamefont
  {{Tiwari}}, \citenamefont {{Chan}}, \citenamefont {{Bogdanovi{\'c}}},
  \citenamefont {{Jiang}}, \citenamefont {{Davis}},\ and\ \citenamefont
  {{Ferrel}}}]{Tiwari2025}%
  \BibitemOpen
  \bibfield  {author} {\bibinfo {author} {\bibfnamefont {V.}~\bibnamefont
  {{Tiwari}}}, \bibinfo {author} {\bibfnamefont {C.-H.}\ \bibnamefont
  {{Chan}}}, \bibinfo {author} {\bibfnamefont {T.}~\bibnamefont
  {{Bogdanovi{\'c}}}}, \bibinfo {author} {\bibfnamefont {Y.-F.}\ \bibnamefont
  {{Jiang}}}, \bibinfo {author} {\bibfnamefont {S.~W.}\ \bibnamefont
  {{Davis}}},\ and\ \bibinfo {author} {\bibfnamefont {S.}~\bibnamefont
  {{Ferrel}}},\ }\href {https://doi.org/10.48550/arXiv.2502.18584} {\bibfield
  {journal} {\bibinfo  {journal} {arXiv e-prints}\ ,\ \bibinfo {eid}
  {arXiv:2502.18584}} (\bibinfo {year} {2025})},\ \Eprint
  {https://arxiv.org/abs/2502.18584} {arXiv:2502.18584 [astro-ph.HE]}
  \BibitemShut {NoStop}%
\bibitem [{\citenamefont {Arvo}(1990)}]{Arvo1990}%
  \BibitemOpen
  \bibfield  {author} {\bibinfo {author} {\bibfnamefont {J.}~\bibnamefont
  {Arvo}},\ }\bibinfo {title} {A simple method for box-sphere intersection
  testing},\ in\ \href@noop {} {\emph {\bibinfo {booktitle} {Graphics Gems}}}\
  (\bibinfo  {publisher} {Academic Press Professional, Inc.},\ \bibinfo
  {address} {USA},\ \bibinfo {year} {1990})\ p.\ \bibinfo {pages}
  {335–339}\BibitemShut {NoStop}%
\end{thebibliography}%

\end{document}